\documentclass[11pt]{article}
\usepackage{%
  amsmath,   
  amssymb,   
  amsthm,    
  array,     
  color,     
  esvect,    
  euscript,  
  latexsym,  
  mathtools, 
  mathrsfs,  
  nccmath,   
  nicematrix,
  physics,   
  tensor,    
  upgreek,   
}

\DeclareFontFamily{OT1}{mathc}{}
\DeclareFontShape{OT1}{mathc}{m}{it}{<-> mathc10}{}
\DeclareMathAlphabet{\mathc}{OT1}{mathc}{m}{it}

\DeclareFontFamily{U}{BOONDOX-calo}{\skewchar\font=45}
\DeclareFontShape{U}{BOONDOX-calo}{m}{n}
  {<-> s*[1.0] BOONDOX-r-calo}{}
\DeclareFontShape{U}{BOONDOX-calo}{b}{n}
  {<-> s*[1.0] BOONDOX-b-calo}{}
\DeclareMathAlphabet{\mathbx}{U}{BOONDOX-calo}{m}{n}
\SetMathAlphabet{\mathbx}{bold}{U}{BOONDOX-calo}{b}{n}

\DeclareMathOperator{\Str}{Str}
\DeclareMathOperator{\Hom}{Hom}
\DeclareMathOperator{\Ber}{\mathc{Ber}}
\DeclareMathOperator{\sdim}{sdim}
\DeclareMathOperator{\sdet}{sdet}

\DeclareMathOperator{\pf}{pf}

\usepackage{%
  calc,      
  comment,   
}

\usepackage{abstract}

\numberwithin{equation}{section}

\usepackage{%
  setspace,
}
\allowdisplaybreaks[2]

\usepackage[%
a4paper,%
textwidth=165mm,
tmargin=25mm,
bmargin=40mm,
]{geometry}

\usepackage[multiple]{footmisc}

\usepackage[auth-lg,affil-it]{authblk}
\def\authlist{} 
\newcommand{\authoremail}[2]{%
  \edef\authlist{%
    \ifx\authlist\empty%
      {#1}%
    \else%
      {\authlist\ and\ #1}%
    \fi%
  }
  \author{%
    #1\thanks{%
      Email: \href{mailto:#2}{#2}
    }
  }
}

\usepackage{%
  titling,   
  titlesec,  
}

\titleformat*{\section}{\bfseries\large}
\titleformat{\subsubsection}[hang]{\itshape}{}{0pt}{}

\usepackage[inline]{enumitem}  

\setlist*[enumerate]{
  label=(\roman*),
}

\usepackage[sorting=none,%
backend=biber,%
sortcites,%
citestyle=numeric-comp,%
maxbibnames=6,%
]{biblatex}
\DeclareFieldFormat[misc]{title}{%
  {\mkbibquote{#1}}
}

\let\oldprintbibliography\printbibliography
\renewcommand{\printbibliography}{%
  \clearpage \oldprintbibliography%
}

\usepackage{%
  adjustbox, 
  graphicx,  
  subcaption,
}

\usepackage{%
  tikz,      
  tikz-cd,   
}
\usetikzlibrary{
  arrows,
  arrows.meta,
  automata,
  calc,
  decorations.pathmorphing,
  patterns.meta,
  positioning,
  shapes.misc
}
\usepackage{soul}

\newcommand{\CP}{\mathbb{CP}}
\newcommand{\WCP}{\mathrm{W}\CP}

\newcommand{\MSUSY}{\ensuremath{\mathcal{M}_{\text{SUSY}}}}

\makeatletter

\newcommand{\Rmnum}[1]{\expandafter\@slowromancap\romannumeral #1@}
\makeatother

\newcommand{\subtitle}[1]{\bigskip\noindent\textit{#1}\vspace*{0.5em}}

\DeclareDocumentCommand\Derivative{ s o m g d() }
{ 
  \begingroup
  \def\diffd{\mathrm{D}}
	\IfBooleanTF{#1}
	{\let\fractype\flatfrac}
	{\let\fractype\frac}
	\IfNoValueTF{#4}
	{
		\IfNoValueTF{#5}
		{\fractype{\diffd \IfNoValueTF{#2}{}{^{#2}}}{\diffd #3\IfNoValueTF{#2}{}{^{#2}}}}
		{\fractype{\diffd \IfNoValueTF{#2}{}{^{#2}}}{\diffd #3\IfNoValueTF{#2}{}{^{#2}}} \argopen(#5\argclose)}
	}
	{\fractype{\diffd \IfNoValueTF{#2}{}{^{#2}} #3}{\diffd #4\IfNoValueTF{#2}{}{^{#2}}}}
  \endgroup
}

\makeatletter
\newcommand{\@saveenv}[2]{
  #2
  \protected@write\@mainaux{}{%
    \string\@savedenv{#1}{%
      \unexpanded{\unexpanded{#2}}%
    }%
  }%
}
\newcommand{\@savedenv}[2]{%
  \global\@namedef{SAVEDENV@#1}{#2}%
}
\newcommand{\@repeatenv}[1]{%
  \ifcsname SAVEDENV@#1\endcsname
    \@nameuse{SAVEDENV@#1}%
  \else
    \texttt{@saveenv Error}
    \GenericError{\\@saveenv Error}{%
      Labeled env not saved: `#1'%
    }{%
      See definition of \\@saveenv in the preamble.%
    }{%
      Usage: \\@saveenv{\#1}{\#2}, where \#1 is the label and \#2 is the math.%
    }%
    \notag
  \fi
}
\newcommand{\saveboxed}[2]{%
  \@saveenv{#1}{\boxed{#2}}%
}
\newcommand{\repeatboxed}[1]{%
  \@repeatenv{#1}%
}
\newcommand{\saveenv}[2]{%
  \@saveenv{#1}{#2}%
}
\newcommand{\repeatenv}[1]{%
  \@repeatenv{#1}%
}
\makeatother

\usepackage[linktocpage,%
]{hyperref}

\makeatletter
\renewcommand{\sectionautorefname}{$\S$\@gobble}
\renewcommand{\subsectionautorefname}{$\S$\@gobble}
\renewcommand{\subsubsectionautorefname}{$\S$\@gobble}
\makeatother

\title{%
  Supergroup Gauged Linear Sigma Models \texorpdfstring{\\}{} and their Physical Mathematics
  \texorpdfstring{\vspace{2.0cm}}{}
}%

\authoremail{Arif Er}{arif.er@u.nus.edu}
\authoremail{Zhangcheng Liu}{liuzhangcheng@u.nus.edu}
\authoremail{Meng-Chwan Tan}{mctan@nus.edu.sg}

\affil{%
  Department of Physics \\%
  National University of Singapore \\%
  2 Science Drive 3, Singapore 117551%
}%
\date{}

\makeatletter
\hypersetup{
  linktocpage=true,
  breaklinks=true,
  colorlinks=true,
  citecolor=blue,
  pdftitle=\@title,
  pdfauthor=\authlist,
}
\makeatother

\begin{document}

\maketitle
\pagenumbering{gobble} 

\begin{abstract}
\normalsize \singlespacing \noindent%
We construct 2d $\mathcal{N}=(2,2)$ gauged linear sigma models with  $\mathrm{U}(1|1)^N$ supergauge group possibly with superpotential.
Despite being nonunitary, one can still study their space of supersymmetric states and explore their applications to mathematics.
In particular, we find a relation between a nonlinear sigma model on a Calabi-Yau complete intersection of hypersurfaces in a super-Grassmannian and a supergauged Landau-Ginzburg orbifold, which can reduce to a regular Calabi-Yau/Landau-Ginzburg correspondence for complete intersections.
This defines a super-Grassmannian/supergroup generalization of the correspondence proved by Clader~\cite{clader2013landauginzburgcalabiyaucorrespondencecompleteintersections} and Zhao~\cite{zhao2019landauginzburgcalabiyaucorrespondencecompleteintersection}.
Similarly, we find a relation between a nonlinear sigma model on a Calabi-Yau hypersurface in a product of super-Grassmannians and a hybrid NLSM/supergauged Landau-Ginzburg orbifold, which can reduce to a regular hybrid Calabi-Yau/Landau-Ginzburg correspondence for hypersurfaces in product space.
This defines a super-Grassmannian/supergroup generalization of the correspondence proved by Fan-Jarvis-Ruan~\cite{Fan_2017}.
We also find that Calabi-Yau supervector bundles over a super-Grassmannian can undergo a physically related mild topology change which is reducible to a regular Atiyah-type flop transition.
This defines a super-Grassmannian generalization of a birational equivalence of Calabi-Yau vector bundles in mathematics.
Similarly, we find that a Calabi-Yau complete intersection of quadrics in a super-Grassmannian can also undergo a physically related topology change which is reducible to a regular conifold transition.
This defines a super-Grassmannian generalization of  a homological projective duality for Calabi-Yau quadrics by Kuznetsov-Perry~\cite{kuznetsov-2021-homol-projec} in mathematics.



\end{abstract}


\clearpage
\pagenumbering{arabic} 

\tableofcontents

\section{Introduction, Summary, and Acknowledgements}
\vspace{0.4cm}
\setlength{\parskip}{5pt}

\subtitle{Introduction}

Supergroup gauge theories, as surveyed in~\cite{Kimura:2023iup}, are not new in physics. Despite their nonunitary nature, they can still be relevant and useful, especially when one explores their applications to mathematics.




For example, supergroups have recently been considered in 4d $\mathcal{N}=2$ gauge theory~\cite{kimura2019superinstantoncountinglocalization},  where a supergroup generalization of quantum integrable systems in mathematics was defined by a localization of the path integral onto supersymmetric configurations.

This development gave us the idea to consider supergroups in a 2d supersymmetric gauge theory, where a study of its space of supersymmetric states might have meaningful applications to mathematics. As a modest step in this direction, we were eventually led to consider a 2d $\mathcal{N} = (2,2)$ gauged linear sigma model (GLSM) with $\mathrm{U}(1|1)$ supergauge group, and its higher-rank generalization, possibly with superpotential.

Because the $\mathrm{U}(1|1)$ supergroup is non-abelian with Grassmann-even and Grassmann-odd generators, our supergroup GLSM would have the characteristics of both an ordinary non-abelian GLSM and an ordinary abelian GLSM with extra fields that are Grassmann-odd versions of the original fields.

At low energy, the former and latter ordinary GLSMs would, in one phase, respectively become a nonlinear sigma model (NLSM) on a Grassmannian~\cite{Witten:1993yc} and a supervariety~\cite{seki2005gaugedlinearsigmamodel}, while in another phase, respectively become a gauged Landau-Ginzburg (LG) orbifold~\cite{Witten:1993yc} and a bosonic-fermionic LG orbifold~\cite{seki2005gaugedlinearsigmamodel}. As such, at low energy, one can expect our supergroup GLSM to become an NLSM on a super-Grassmannian, or a supergauged LG orbifold on a supervariety, or both. Indeed it will, as we shall see.

Clearly, this will allow us to explicitly define super-Grassmannian/supergroup generalizations of various existing correspondences and topological transitions associated with ordinary GLSMs that have been established in mathematics.

Let us now give a brief plan and summary of the paper.

\subtitle{Summary and Plan of Paper}

In \autoref{sec:review:glsm}, we will review some relevant aspects of the  non-abelian GLSM  first introduced in~\cite{Witten:1993yc} and further analyzed in~\cite{Witten:1993xi} by Witten.
This would be useful because, as mentioned above, our supergroup GLSM is, aside from some extra  Grassmann-odd fields, like an ordinary non-abelian GLSM.

In \autoref{sec:review:u-1-1 superalgebra}, we will review the unitary $\mathfrak{u}(1|1)$ Lie superalgebra, focusing only on aspects most relevant to our objectives in this paper.

In \autoref{sec: U(1|1) and SGr}, we will consider the simplest case of a unitary $\mathrm{U}(1|1)$ supergroup GLSM with $m + n$ Grassmann-even and Grassmann-odd chiral matter superfields $\Phi_1,...,\Phi_m$ and $\Psi_1, \dots, \Psi_n$, respectively, that are charged identically under $\mathrm{U}(1|1)$, with Fayet-Iliopoulos (FI) parameter $t \coloneq \mathbf{i} r + \frac{\theta}{2\pi}$.
Its space $\MSUSY$ of supersymmetric states is defined in~\eqref{eq:u-1-1:m-susy} as
\begin{equation}
  \label{summary:eq:u-1-1:m-susy}
  \repeatboxed{eq:u-1-1:m-susy}
\end{equation}
which corresponds to the super-Grassmannian ${Gr}_{1|1}(\mathbb{C}^{m|n})$ of size $r$.

When the model is not scale-invariant, i.e., when $m \neq n$, the FI parameter runs as expressed in \eqref{eq:u-1-1:quantum:r-eff:form}:
\begin{equation}
  \label{summary:eq:u-1-1:quantum:r-eff:form}
  \repeatboxed{eq:u-1-1:quantum:r-eff:form}
\end{equation}
At low energy, we find that the quantum GLSM will become a quantum NLSM, with target space ${Gr}_{1|1}(\mathbb{C}^{m|n})$ of size
\begin{enumerate*}
\item $r_{\text{eff}}(\mu) > 0$ or
\item $r_{\text{eff}}(\mu) < 0$,
\end{enumerate*}
if
\begin{enumerate*}
\item $m - n > 0$ or
\item $m - n < 0$.
\end{enumerate*}

When the model is scale-invariant, i.e., when $m = n$, we can  physically derive, among other things, that super-Grassmannians of the type ${Gr}_{1|1}(\mathbb{C}^{m|n})$ are Calabi-Yau (CY) supermanifolds if and only if $m = n$. This is a \emph{novel} mathematical result that we can also mathematically verify in \autoref{app:verify cy:supergrassmannians}!

In \autoref{sec: U(1|1)^N GLSM}, we generalize our discussion in the previous section to consider a $\mathrm{U}(1|1)^N \coloneq \prod_{a=1}^N \mathrm{U}(1|1)_a$ supergroup GLSM, where the $m + n$ chiral matter superfields have corresponding non-negative gauge charges $(\tensor{Q}{_a_s}, \tensor{\mathsf{Q}}{_a_t})$ for $s \in \{1, \dots, m\}$ and $t \in \{1, \dots, n\}$.
Its space $\MSUSY$ of supersymmetric states is defined in \eqref{eq:u-1-1-N:m-susy} as
\begin{equation}
  \label{summary:eq:u-1-1-N:m-susy}
  \repeatboxed{eq:u-1-1-N:m-susy}
\end{equation}
where $\vec{r} \coloneq (r_1, \dots, r_N)$ determines its size. It corresponds to a super torified variety determined by $(\tensor{Q}{_a_s}, \tensor{\mathsf{Q}}{_a_t}, r_a)$, in the sense that its bosonized variety is a torified variety as described in~\cite{L_pez_Pe_a_2009}.

When the model is not scale-invariant, i.e., when $\sum_{s = 1}^m \tensor{Q}{_a_s} \neq \sum_{t = 1}^n \tensor{Q}{_a_t}$, the FI parameters run as expressed in \eqref{eq:u-1-1-N:quantum:r-eff:common scale parameter}:
\begin{equation}
  \label{summary:eq:u-1-1-N:quantum:r-eff:common scale parameter}
  \repeatboxed{eq:u-1-1-N:quantum:r-eff:common scale parameter}
\end{equation}
At low energy, we find that the quantum GLSM will become a quantum NLSM, with target space
$X_{\vec{r}_{\text{eff}}(\mu)}$ where $\vec{r}_{\text{eff}}(\mu) \coloneq \bigl( r_{1, \text{eff}}, \dots, r_{N, \text{eff}} \bigr)$ determines its size, such that the sign of each component in $\vec{r}_{\text{eff}}(\mu)$ follows the sign of the corresponding component in $\vec{r}$.

When the model is scale-invariant, i.e., when $\sum_{s = 1}^m \tensor{Q}{_a_s} = \sum_{t = 1}^n \tensor{Q}{_a_t}$, we can physically derive, among other things, that super torified varieties determined by $(\tensor{Q}{_a_s}, \tensor{\mathsf{Q}}{_a_t}, r_a)$ are CY supermanifolds if and only if $\sum_{s = 1}^m \tensor{Q}{_a_s} = \sum_{t = 1}^n \tensor{Q}{_a_t}$. This is yet another \emph{novel} mathematical result that we can also mathematically verify in \autoref{app:verify cy:wcp}!

In \autoref{sec: Phases of U(1|1) with W}, we consider the $\mathrm{U}(1|1)$ GLSM studied in \autoref{sec: U(1|1) and SGr}, but with a gauge-invariant superpotential defined in \eqref{eq:u-1-1 with W:W potential} as
\begin{equation}
  \label{summary:eq:u-1-1 with W:W potential}
  \repeatboxed{eq:u-1-1 with W:W potential}
\end{equation}
where $\kappa_{\alpha} = \pm$.
Here, $P^{\alpha}_+$ and $P^{\alpha}_-$, like the $\Phi$'s and $\Psi$'s, are Grassmann-even and Grassmann-odd chiral matter superfields, respectively.
Likewise,
\begin{equation}
  \label{summary:eq:u-1-1 with W:glsm:G+ alpha}
  \repeatboxed{eq:u-1-1 with W:glsm:G+ alpha}
\end{equation}
is a Grassmann-even homogeneous polynomial of degree $q^+_{\alpha}$ in
\begin{equation}
  \label{summary:eq:u-1-1 with W:glsm:Y+}
  \repeatboxed{eq:u-1-1 with W:glsm:Y+ even}
  \qand
  \repeatboxed{eq:u-1-1 with W:glsm:Y+ odd}
\end{equation}
and
\begin{equation}
  \label{summary:eq:u-1-1 with W:glsm:G- alpha}
  \repeatboxed{eq:u-1-1 with W:glsm:G- alpha}
\end{equation}
is a Grassmann-odd homogeneous polynomial of degree $q^-_{\alpha}$ in
\begin{equation}
  \label{summary:eq:u-1-1 with W:glsm:Y-}
  \repeatboxed{eq:u-1-1 with W:glsm:Y- even}
  \qand
  \repeatboxed{eq:u-1-1 with W:glsm:Y- odd}
\end{equation}
In \eqref{summary:eq:u-1-1 with W:glsm:G+ alpha}--\eqref{summary:eq:u-1-1 with W:glsm:Y-},
\begin{enumerate*}
  \item $\{s_{*}, s'_{*}\} \in \{1, \dots, m\}$ and $\{t_*, t'_{*}\} \in \{1, \dots, n\}$;

  \item $b^+_{(\alpha)}$/$b^-_{(\alpha)}$ is a Grassmann-even/odd coefficient that is symmetric/antisymmetric under the swop $(s_a t_a) \leftrightarrow (s_c t_c)$;

  and

  \item $c^+_{(\alpha)}$/$c^-_{(\alpha)}$ is a Grassmann-even/odd (resp. Grassmann-odd/even) coefficient if the integer $q^+_{\alpha}$/$q^-_{\alpha}$ is even (resp. odd), that is symmetric/antisymmetric (resp. antisymmetric/symmetric) under the swop $(t_a t'_a) \leftrightarrow (t_c t'_c)$.

\end{enumerate*}

The $\{1, \bar{1}\}$ indices in \eqref{summary:eq:u-1-1 with W:glsm:Y+} and \eqref{summary:eq:u-1-1 with W:glsm:Y-} label the $\mathrm{U}(1|1)$ color indices on the superfields.
As each $P^\alpha_+$ and $P^\alpha_-$ (resp. $G_{\alpha}^+$ and $G^{\alpha}_-$) transforms under the superdeterminantal and anti-superdeterminantal representations of $\mathrm{U}(1|1)$ with charge $- q^+_{\alpha}$ and $- q^-_{\alpha}$ (resp. $q^+_{\alpha}$ and $q^-_{\alpha}$), they \emph{effectively} transform under an even abelian subgroup $\mathrm{U}(1) \subset \mathrm{U}(1|1)$, and thus will not be labeled with $\{1, \bar{1}\}$ indices.

Its space $\MSUSY$ of supersymmetric states is defined in \eqref{eq:u-1-1 with W:glsm:M-SUSY:geometric space}--\eqref{eq:u-1-1 with W:glsm:M-SUSY:toric hypersurface -} as
\begin{equation}
  \label{summary:eq:u-1-1 with W:M-SUSY}
  \begin{gathered}
    \repeatboxed{eq:u-1-1 with W:glsm:M-SUSY:geometric space}
    \\
    \\
    \repeatboxed{eq:u-1-1 with W:glsm:M-SUSY:toric hypersurface +}
    \qand
    \repeatboxed{eq:u-1-1 with W:glsm:M-SUSY:toric hypersurface -}
  \end{gathered}
\end{equation}
Here,
\begin{enumerate*}
  \item $(G^+_1, G^+_2, \dots, G^+_{l_+})$ are the $l_+$ number of Grassmann-even polynomials given in \eqref{summary:eq:u-1-1 with W:glsm:G+ alpha} of respective degrees $(q^+_1, q^+_2, \dots, q^+_{l_+})$;

  \item $(G^-_1, G^-_2, \dots, G^-_{l_-})$ are the $l_-$ number of Grassmann-odd polynomials given in \eqref{summary:eq:u-1-1 with W:glsm:G- alpha} of respective degrees $(q^-_1,q^-_2, \dots, q^-_{l_-})$;

  \item $l_+ + l_- = l$; and

  \item the summation index $\{a, b\}$ runs over all $Y^+$ and $Y^-$ field indices.
\end{enumerate*}

The model is generically not scale-invariant, and the FI parameter runs as expressed in \eqref{eq:U(1|1) with W:quantum glsm:r-eff(mu)} as
\begin{equation}
  \label{summary:eq:U(1|1) with W:quantum glsm:r-eff(mu)}
  \repeatboxed{eq:U(1|1) with W:quantum glsm:r-eff(mu)}
\end{equation}
We find, via our analysis in \autoref{subsec: Phases U(1|1) with W} and \autoref{subsec:quantum low energy U(1|1) with W}, that there are four phases of the quantum GLSM at low energy.

(I) In the $r \gg 0$ phase, if $\phi, \psi \neq 0$ and $p^{\alpha^{\pm}}_{\pm} = 0$, the quantum GLSM would, at low energy, become a quantum NLSM on the complete intersection of $l_+$ even and $l_-$ odd hypersurfaces $\bigcap_{\alpha^+ = 1}^{l_+} \{G^+_{\alpha^+} = 0\}$ and $\bigcap_{\alpha^- = 1}^{l_-} \{G^-_{\alpha^-} = 0\}$ of degrees $q^+_{\alpha^+}$ and $q^-_{\alpha^-}$, respectively, in $Gr_{1|1}(\mathbb{C}^{m|n})$ of size $r$, defined by \eqref{eq:U(1|1) with W:phases:geometric space:r(i) > 0}--\eqref{eq:U(1|1) with W:phases:hypersurface -:r(i) > 0} as
\begin{equation}
  \label{summary:eq:U(1|1) with W:phases:r(i) > 0}
  \begin{gathered}
    \repeatboxed{eq:U(1|1) with W:phases:geometric space:r(i) > 0}
    \\
    \\
    \repeatboxed{eq:U(1|1) with W:phases:hypersurface +:r(i) > 0}
    \qand
    \repeatboxed{eq:U(1|1) with W:phases:hypersurface -:r(i) > 0}
  \end{gathered}
\end{equation}

(II) In the $r \ll 0$ phase, if $\phi, \psi = 0$ and $p^{\alpha^{\pm}}_{\pm} \neq 0$, the quantum GLSM would, at low energy, become a quantum $\mathrm{SU}(1|1) \times \mathbb{Z}_q$ supergauged LG orbifold with effective superpotential given by \eqref{eq:U(1|1) with W:phases:superpotential W:r(ii) < 0}--\eqref{eq:U(1|1) with W:phases:geometric space:r(ii) < 0:exp val} as
\begin{equation}
  \label{summary:eq:U(1|1) with W:phases:superpotential W:r(ii) < 0}
  \repeatboxed{eq:U(1|1) with W:phases:superpotential W:r(ii) < 0}
\end{equation}
where
\begin{equation}
  \label{summary:eq:U(1|1) with W:phases:geometric space:r(ii) < 0:exp val}
  \repeatboxed{eq:U(1|1) with W:phases:geometric space:r(ii) < 0:exp val}
\end{equation}
that is defined on a supervector bundle over $\WCP^{{l_+ -1}|l_-}_{\left(\vv{q^+}\middle|\vv{q^-}\right)}$ of size $\abs{r}$ described in \eqref{eq:U(1|1) with W:phases:supervector bundle:r(ii) < 0} as
\begin{equation}
  \label{summary:eq:U(1|1) with W:phases:supervector bundle:r(ii) < 0}
  \repeatboxed{eq:U(1|1) with W:phases:supervector bundle:r(ii) < 0}
\end{equation}
where $q$ is the greatest common divisor of the degrees $q^{\pm}_{\alpha^{\pm}}$ of the $G^{\pm}_{\alpha^{\pm}}$'s, and $\mathbf{\Pi}$ is a parity-reversal operator which makes the fiber Grassmann-odd.

(I$'$) In the $r \ll 0$ phase, if $\phi, \psi \neq 0$ and $p^{\alpha^{\pm}}_{\pm} = 0$, the quantum GLSM would, at low energy, become a quantum NLSM on the complete intersection of $l_+$ \emph{odd} and $l_-$ \emph{even} hypersurfaces $\bigcap_{\alpha^+ = 1}^{l_+} \{G^-_{\alpha^+ (\mathbf{\Pi})} = 0\}$ and $\bigcap_{\alpha^- = 1}^{l_-} \{G^+_{\alpha^- (\mathbf{\Pi})} = 0\}$ of degrees $q^+_{\alpha^+}$ and $q^-_{\alpha^-}$, respectively, in $Gr_{1|1}(\mathbb{C}^{n|m})$ of size $\abs{r}$, defined by \eqref{eq:U(1|1) with W:phases:geometric space:r(i) < 0}--\eqref{eq:U(1|1) with W:phases:hypersurface +:r(i) < 0} as
\begin{equation}
  \label{summary:eq:U(1|1) with W:phases:r(i) < 0}
  \begin{gathered}
    \repeatboxed{eq:U(1|1) with W:phases:geometric space:r(i) < 0}
    \\
    \\
    \repeatboxed{eq:U(1|1) with W:phases:hypersurface +:r(i) < 0}
    \qand
    \repeatboxed{eq:U(1|1) with W:phases:hypersurface -:r(i) < 0}
  \end{gathered}
\end{equation}

(II$'$) In the $r \gg 0$ phase, if $\phi, \psi = 0$ and $p^{\alpha^{\pm}}_{\pm} \neq 0$, the quantum GLSM would, at low energy, become a quantum $\mathrm{SU}(1|1) \times \mathbb{Z}_q$ supergauged LG orbifold with effective superpotential given by \eqref{eq:U(1|1) with W:phases:superpotential W:r(ii) > 0}--\eqref{eq:U(1|1) with W:phases:geometric space:r(ii) > 0:exp val} as
\begin{equation}
  \label{summary:eq:U(1|1) with W:phases:superpotential W:r(ii) > 0}
  \repeatboxed{eq:U(1|1) with W:phases:superpotential W:r(ii) > 0}
\end{equation}
where
\begin{equation}
  \label{summary:eq:U(1|1) with W:phases:geometric space:r(ii) > 0:exp val}
  \repeatboxed{eq:U(1|1) with W:phases:geometric space:r(ii) > 0:exp val}
\end{equation}
that is defined on a supervector bundle over $\WCP^{{l_- -1}|l_+}_{\left(\vv{q^-}\middle|\vv{q^+}\right)}$ of size $\abs{r}$ described in \eqref{eq:U(1|1) with W:phases:supervector bundle:r(ii) > 0} as
\begin{equation}
  \label{summary:eq:U(1|1) with W:phases:supervector bundle:r(ii) > 0}
  \repeatboxed{eq:U(1|1) with W:phases:supervector bundle:r(ii) > 0}
\end{equation}
where (i) $p^{\alpha^-}_{+ (\mathbf{\Pi})}$ and $p^{\alpha^+}_{- (\mathbf{\Pi})}$ are Grassmann-even and Grassmann-odd,
(ii) $q$ is the greatest common divisor of the degrees $q^{\pm}_{\alpha^{\pm}}$ of the $G^{\pm}_{\alpha^{\pm}}$'s, and
(iii) the ``$\mathbf{\Pi}$'' in the subscript in  \eqref{summary:eq:U(1|1) with W:phases:supervector bundle:r(ii) > 0} indicates a parity-reversal of the fiber space over the base.

Phases (I)--(II), and their parity-reversed mirrors, phases (I$'$)--(II$'$), are summarized in \autoref{fig:u-1-1-W:low energy}.

When the model is scale-invariant, i.e., when $m - n = \sum_{\alpha^+ = 1}^{l_+} q^+_{\alpha^+} - \sum_{\alpha^-}^{l_-} q^-_{\alpha^-}$, we can physically derive that a complete intersection of $l^+$ even and $l^-$ odd hypersurfaces in a super-Grassmannian of the type $Gr_{1|1}(\mathbb{C}^{m|n})$ is CY if and only if $m - n = \sum_{\alpha^+ = 1}^{l_+} q^+_{\alpha^+} - \sum_{\alpha^-}^{l_-} q^-_{\alpha^-}$.
This is a \emph{novel} mathematical result that we can also mathematically verify in \autoref{app:verify cy:hypersurfaces in Gr}!

Lastly, since the NLSM and supergauged LG orbifold in (I) and (II) are just different phases of a branch of the same underlying GLSM, they are, in the scale-invariant case whence we can freely choose $r$, actually related.
Likewise, since the NLSM and supergauged LG orbifold in (I$'$) and (II$'$) are just different phases of another branch of the same underlying GLSM, they are, in the scale-invariant case whence we can freely choose $r$, actually related.
Hence, as elaborated in \autoref{subsec: Super CY/LG for complete intersections}, we have a super-Grassmannian/supergroup generalization of the CY/LG correspondence for complete intersections established in mathematics by Clader~\cite{clader2013landauginzburgcalabiyaucorrespondencecompleteintersections} and Zhao~\cite{zhao2019landauginzburgcalabiyaucorrespondencecompleteintersection}, as well as a parity-reversed mirror version of it!

In \autoref{sec: Phases of U(1|1)^N with W}, we consider the $\mathrm{U}(1|1)^N$ GLSM studied in \autoref{sec: U(1|1)^N GLSM} where $N = 2$ with $m_1 + m_2$ identically-charged Grassmann-even chiral matter superfields $S_1, \dots, S_{m_1}, T_1, \dots, T_{m_2}$ and $n_1 + n_2$ identically-charged Grassmann-even chiral matter superfields $\mathbx{S}_1, \dots, \mathbx{S}_{m_1}, \mathbx{T}_1, \dots, \mathbx{T}_{m_2}$, but with a gauge-invariant superpotential defined in \eqref{eq:U(1|1)-N with W:W_super U(1|1)^2} as
\begin{equation}
  \label{summary:eq:U(1|1)-N with W:W_super U(1|1)^2}
  \repeatboxed{eq:U(1|1)-N with W:W_super U(1|1)^2}
\end{equation}
such that
(i) $G^+$ is a generalization of the descriptions in \eqref{summary:eq:u-1-1 with W:glsm:G+ alpha}--\eqref{summary:eq:u-1-1 with W:glsm:Y+} as a bihomogeneous polynomial of degrees $\hat{m}_1$ and $\hat{m}_2$ in $Y^+(S, \mathbx{S})$ and $Y^+(T, \mathbx{T})$,
and
(ii) $G^-$ is a generalization of the descriptions in \eqref{summary:eq:u-1-1 with W:glsm:G- alpha}--\eqref{summary:eq:u-1-1 with W:glsm:Y-} as a bihomogeneous polynomial of degrees $\hat{n}_1$ and $\hat{n}_2$ in $Y^-(S, \mathbx{S})$ and $Y^-(T, \mathbx{T})$.

Its space $\MSUSY$ of supersymmetric states is defined in \eqref{eq:U(1|1)-N with W:glsm:M-SUSY:product toric supervariety 1}--\eqref{eq:U(1|1)-N with W:glsm:M-SUSY:toric hypersurface -} as
\begin{equation}
  \label{summary:eq:U(1|1)-N with W:glsm:M-SUSY:product toric supervariety}
  \begin{gathered}
    \repeatboxed{eq:U(1|1)-N with W:glsm:M-SUSY:product toric supervariety 1}
    \\
    \text{and}
    \\
    \repeatboxed{eq:U(1|1)-N with W:glsm:M-SUSY:product toric supervariety 2}
  \end{gathered}
\end{equation}
with
\begin{equation}
  \label{summary:eq:U(1|1)-N with W:glsm:M-SUSY:toric hypersurface}
  \begin{gathered}
    \repeatboxed{eq:U(1|1)-N with W:glsm:M-SUSY:toric hypersurface +}
    \\
    \text{and}
    \\
    \repeatboxed{eq:U(1|1)-N with W:glsm:M-SUSY:toric hypersurface -}
  \end{gathered}
\end{equation}
The model is generically not scale-invariant, and the FI parameters run as expressed in \eqref{eq:U(1|1)-N with W:quantum glsm:r-eff(mu):1}-\eqref{eq:U(1|1)-N with W:quantum glsm:r-eff(mu):2} as
\begin{equation}
  \label{summary:eq:U(1|1)-N with W:quantum glsm:r-eff(mu)}
  \begin{gathered}
    \repeatboxed{eq:U(1|1)-N with W:quantum glsm:r-eff(mu):1}
    \\
    \text{and}
    \\
    \repeatboxed{eq:U(1|1)-N with W:quantum glsm:r-eff(mu):2}
  \end{gathered}
\end{equation}
We find, via our analysis in \autoref{subsec: Phases of U(1|1)^2 with W} and \autoref{subsec: Quantum low energy U(1|1)^2 with W}, that there are eight representative phases of the quantum GLSM at low energy.

(I) In the $r_1, r_2 \gg 0$ phase, if $s, \mathbx{s}, t, \mathbx{t} \neq 0$ and $p_{\pm} = 0$, the quantum GLSM would, at low energy, become a quantum NLSM on the complete intersection of an even and odd hypersurface $G^+ = 0$ and $G^- = 0$ of bidegrees $(\hat{m}_1, \hat{m}_2)$ and $(\hat{n}_1, \hat{n}_2)$ in $Gr_{1|1}(\mathbb{C}^{m_1|n_1}) \times Gr_{1|1}(\mathbb{C}^{m_2|n_2})$ of sizes $r_1$ and $r_2$, defined by \eqref{eq:U(1|1)-N with W:phases:r1(i) and r2(i) > 0:space 1}--\eqref{eq:U(1|1)-N with W:phases:r1(i) and r2(i) > 0:hypersurface -} as
\begin{equation}
  \label{summary:eq:U(1|1)-N with W:phases:r1(i) and r2(i) > 0:space 1}
  \begin{gathered}
    \repeatboxed{eq:U(1|1)-N with W:phases:r1(i) and r2(i) > 0:space 1}
    \\
    \text{and}
    \\
    \repeatboxed{eq:U(1|1)-N with W:phases:r1(i) and r2(i) > 0:space 2}
  \end{gathered}
\end{equation}
with
\begin{equation}
  \label{summary:eq:U(1|1)-N with W:phases:r1(i) and r2(i) > 0:hypersurface}
  \repeatboxed{eq:U(1|1)-N with W:phases:r1(i) and r2(i) > 0:hypersurface +}
  \qand
  \repeatboxed{eq:U(1|1)-N with W:phases:r1(i) and r2(i) > 0:hypersurface -}
\end{equation}

(II) In the $r_1 \ll 0$ and $\hat{m}_1 r_2 - \hat{m}_2 r_1 \gg 0$ phase, if $p_{\pm}, t, \mathbx{t} \neq 0$ and $s, \mathbx{s} = 0$, the quantum GLSM would, at low energy, become a hybrid between a quantum supergauged LG orbifold and a quantum NLSM.
In particular, along the $s, \mathbx{s}$-directions, it is a quantum $\mathrm{SU}(1|1) \times \mathbb{Z}_{\hat{\mathfrak{m}}_1}$ supergauged LG orbifold with effective superpotential given by \eqref{eq:U(1|1)-N with W:phases:r1(i) < 0 < r2(i):superpotential}--\eqref{eq:U(1|1)-N with W:phases:r1(i) < 0 < r2(i):wcp 2:expval} as
\begin{equation}
  \label{summary:eq:U(1|1)-N with W:phases:r1(i) < 0 < r2(i):superpotential}
  \repeatboxed{eq:U(1|1)-N with W:phases:r1(i) < 0 < r2(i):superpotential}
\end{equation}
where
\begin{equation}
  \label{summary:eq:U(1|1)-N with W:phases:r1(i) < 0 < r2(i):wcp 2:expval}
  \repeatboxed{eq:U(1|1)-N with W:phases:r1(i) < 0 < r2(i):wcp 2:expval}
\end{equation}
that is defined on a supervector bundle over $\WCP^{0|1}_{\hat{m}_1|\hat{n}_1} \times Gr_{1|1}(\mathbb{C}^{m_2|n_2})$ of sizes $\abs{r_1}$ and $r_2 - \frac{\hat{m}_2}{\hat{m}_1} r_1$ described in \eqref{eq:U(1|1)-N with W:phases:r1(i) < 0 < r2(i):space 2:supervector bundle} as
\begin{equation}
  \label{summary:eq:U(1|1)-N with W:phases:r1(i) < 0 < r2(i):space 2:supervector bundle}
  \repeatboxed{eq:U(1|1)-N with W:phases:r1(i) < 0 < r2(i):space 2:supervector bundle}
\end{equation}
where $\hat{\mathfrak{m}}_1 \coloneq \gcd(\hat{m}_1, \hat{n}_1)$, and the base superspaces $\WCP^{0|1}_{\hat{m}_1|\hat{n}_1}$ and $Gr_{1|1}(\mathbb{C}^{m_2|n_2})$ are defined by \eqref{eq:U(1|1)-N with W:phases:r1(i) < 0 < r2(i):space 2} and \eqref{eq:U(1|1)-N with W:phases:r1(i) < 0 < r2(i):space 1} as
\begin{equation}
  \label{summary:eq:U(1|1)-N with W:phases:r1(i) < 0 < r2(i):space 2}
  \repeatboxed{eq:U(1|1)-N with W:phases:r1(i) < 0 < r2(i):space 2}
\end{equation}
and
\begin{equation}
  \label{summary:eq:U(1|1)-N with W:phases:r1(i) < 0 < r2(i):space 1}
  \repeatboxed{eq:U(1|1)-N with W:phases:r1(i) < 0 < r2(i):space 1}
\end{equation}
respectively.

Along the $t, \mathbx{t}$-directions, it is a quantum NLSM on $Gr_{1|1}(\mathbb{C}^{m_2|n_2})$ of size $r_2 - \frac{\hat{m}_2}{\hat{m}_1} r_1$ defined by \eqref{summary:eq:U(1|1)-N with W:phases:r1(i) < 0 < r2(i):space 1}.

(III) In the $r_2 \ll 0$ and $\hat{m}_1 r_2 - \hat{m}_2 r_1 \ll 0$ phase, if $p_{\pm}, s, \mathbx{s} \neq 0$ and $t, \mathbx{t} = 0$, the quantum GLSM would, at low energy, become a hybrid between a quantum supergauged LG orbifold and a quantum NLSM.
In particular, along the $t, \mathbx{t}$-directions, it is a quantum $\mathrm{SU}(1|1) \times \mathbb{Z}_{\hat{\mathfrak{m}}_2}$ supergauged LG orbifold with effective superpotential given by \eqref{eq:U(1|1)-N with W:phases:r1(i) > 0 > r2(i):superpotential}--\eqref{eq:U(1|1)-N with W:phases:r1(i) > 0 > r2(i):wcp 2:expval} as
\begin{equation}
  \label{summary:eq:U(1|1)-N with W:phases:r1(i) > 0 > r2(i):superpotential}
  \repeatboxed{eq:U(1|1)-N with W:phases:r1(i) > 0 > r2(i):superpotential}
\end{equation}
where
\begin{equation}
  \label{summary:eq:U(1|1)-N with W:phases:r1(i) > 0 > r2(i):wcp 2:expval}
  \repeatboxed{eq:U(1|1)-N with W:phases:r1(i) > 0 > r2(i):wcp 2:expval}
\end{equation}
that is defined on a supervector bundle over $Gr_{1|1}(\mathbb{C}^{m_1|n_1}) \times \WCP^{0|1}_{\hat{m}_2|\hat{n}_2}$ of sizes $r_1 - \frac{\hat{m}_1}{\hat{m}_2} r_2$ and $\abs{r_2}$ described in \eqref{eq:U(1|1)-N with W:phases:r1(i) > 0 > r2(i):space 2:supervector bundle} as
\begin{equation}
  \label{summary:eq:U(1|1)-N with W:phases:r1(i) > 0 > r2(i):space 2:supervector bundle}
  \repeatboxed{eq:U(1|1)-N with W:phases:r1(i) > 0 > r2(i):space 2:supervector bundle}
\end{equation}
where $\hat{\mathfrak{m}}_2 \coloneq \gcd(\hat{m}_2, \hat{n}_2)$,
and the base superspaces $Gr_{1|1}(\mathbb{C}^{m_1|n_1})$ and $\WCP^{0|1}_{\hat{m}_2|\hat{n}_2}$ are defined by \eqref{eq:U(1|1)-N with W:phases:r1(i) > 0 > r2(i):space 1} and \eqref{eq:U(1|1)-N with W:phases:r1(i) > 0 > r2(i):space 2} as
\begin{equation}
  \label{summary:eq:U(1|1)-N with W:phases:r1(i) > 0 > r2(i):space 1}
  \repeatboxed{eq:U(1|1)-N with W:phases:r1(i) > 0 > r2(i):space 1}
\end{equation}
and
\begin{equation}
  \label{summary:eq:U(1|1)-N with W:phases:r1(i) > 0 > r2(i):space 2}
  \repeatboxed{eq:U(1|1)-N with W:phases:r1(i) > 0 > r2(i):space 2}
\end{equation}
respectively.

Along the $s, \mathbx{s}$-directions, it is a quantum NLSM on $Gr_{1|1}(\mathbb{C}^{m_1|n_1})$ of size $r_1 - \frac{\hat{m}_1}{\hat{m}_2} r_2$ defined by \eqref{summary:eq:U(1|1)-N with W:phases:r1(i) > 0 > r2(i):space 1}.

(IV) In the $r_1, r_2 \ll 0$ and $\hat{m}_1 r_2 - \hat{m}_2 r_1 = 0$ phase, if $p_{\pm} \neq 0$ and $s, \mathbx{s}, t, \mathbx{t} = 0$, the quantum GLSM would, at low energy, become a $\mathrm{U}(1|1) \times \mathrm{SU}(1|1) \times \mathbb{Z}_{\hat{\mathfrak{M}}}$ supergauged LG orbifold with effective superpotential \eqref{eq:U(1|1)-N with W:phases:r1(i) and r2(i) < 0:superpotential}--\eqref{eq:U(1|1)-N with W:phases:r1(i) and r2(i) < 0:wcp 1:expval} as
\begin{equation}
  \label{summary:eq:U(1|1)-N with W:phases:r1(i) and r2(i) < 0:superpotential}
  \repeatboxed{eq:U(1|1)-N with W:phases:r1(i) and r2(i) < 0:superpotential}
\end{equation}
where
\begin{equation}
  \label{summary:eq:U(1|1)-N with W:phases:r1(i) and r2(i) < 0:wcp 1:expval}
  \repeatboxed{eq:U(1|1)-N with W:phases:r1(i) and r2(i) < 0:wcp 1:expval}
\end{equation}
that is defined on a product of a supervector bundle over $\WCP^{0|1}_{\hat{m}_1|\hat{n}_1}$ and another over $\WCP^{0|1}_{\hat{m}_2|\hat{n}_2}$, of size $\abs{r_1}$ and $\abs{r_2}$, respectively, described in \eqref{eq:U(1|1)-N with W:phases:r1(i) and r2(i) < 0:supervector bundle over wcp} as
\begin{equation}
  \label{summary:eq:U(1|1)-N with W:phases:r1(i) and r2(i) < 0:supervector bundle over wcp}
  \repeatboxed{eq:U(1|1)-N with W:phases:r1(i) and r2(i) < 0:supervector bundle over wcp}
\end{equation}
where $\hat{\mathfrak{M}} \coloneq \gcd(\hat{m}_1, \hat{m}_2, \hat{n}_1, \hat{n}_2)$, and the base superspaces $\WCP^{0|1}_{\hat{m}_1|\hat{n}_1}$ and $\WCP^{0|1}_{\hat{m}_2|\hat{n}_2}$ are defined by \eqref{eq:U(1|1)-N with W:phases:r1(i) and r2(i) < 0:wcp 1} and \eqref{eq:U(1|1)-N with W:phases:r1(i) and r2(i) < 0:wcp 2} as
\begin{equation}
  \label{summary:eq:U(1|1)-N with W:phases:r1(i) and r2(i) < 0:wcp}
  \repeatboxed{eq:U(1|1)-N with W:phases:r1(i) and r2(i) < 0:wcp 1}
  \qand
  \repeatboxed{eq:U(1|1)-N with W:phases:r1(i) and r2(i) < 0:wcp 2}
\end{equation}
respectively

(I$'$) In the $r_1, r_2 \ll 0$ phase, if $s, \mathbx{s}, t, \mathbx{t} \neq 0$ and $p_{\pm} = 0$, the quantum GLSM would, at low energy, become a quantum NLSM which can be interpreted as a parity-reversed mirror of the quantum NLSM of phase (I).
In particular, it would be a quantum NLSM on the complete intersection of an even and odd hypersurface $G^+_{\mathbf{\Pi}|\mathbf{\Pi}} = 0$ and $G^-_{\mathbf{\Pi}|\mathbf{\Pi}} = 0$ of bidegrees $(\hat{n}_1, \hat{n}_2)$ and $(\hat{m}_1, \hat{m}_2)$ in $Gr_{1|1}(\mathbb{C}^{n_1|m_1}) \times Gr_{1|1}(\mathbb{C}^{n_2|m_2})$.

(II$'$) In the $r_1 \gg 0$ and $\hat{m}_1 r_2 - \hat{m}_2 r_1 \ll 0$ phase, if $p_{\pm}, t, \mathbx{t} \neq 0$ and $s, \mathbx{s} = 0$, the quantum GLSM would, at low energy, become a hybrid between a quantum supergauged LG orbifold and a quantum NLSM that can be interpreted as the parity-reversed mirror of the hybrid model described in (II).
In particular, along the $s, \mathbx{s}$-directions, it would be a quantum $\mathrm{SU}(1|1) \times \mathbb{Z}_{\hat{\mathfrak{m}}_1}$ supergauged LG orbifold defined on a supervector bundle over $\WCP^{0|1}_{\hat{n}_1|\hat{m}_1} \times Gr_{1|1}(\mathbb{C}^{n_2|m_2})$.
Along the $t, \mathbx{t}$-directions, it would be a quantum NLSM on $Gr_{1|1}(\mathbb{C}^{n_2|m_2})$.

(III$'$) In the $r_2 \gg 0$ and $\hat{m}_1 r_2 - \hat{m}_2 r_1 \gg 0$ phase, if $p_{\pm}, s, \mathbx{s} \neq 0$ and $t, \mathbx{t} = 0$, the quantum GLSM would, at low energy, become another hybrid between a quantum supergauged LG orbifold and a quantum NLSM that can be interpreted as the parity-reversed mirror of the hybrid model described in (III).
In particular, along the $t, \mathbx{t}$-directions, it would be a quantum $\mathrm{SU}(1|1) \times \mathbb{Z}_{\hat{m}_2}$ supergauged LG orbifold defined on a supervector bundle over $Gr_{1|1}(\mathbb{C}^{n_1|m_1}) \times \WCP^{0|1}_{\hat{n}_2|\hat{m}_2}$.
Along the $s, \mathbx{s}$-directions, it would be a quantum NLSM on $Gr_{1|1}(\mathbb{C}^{n_1|m_1})$.

(IV$'$) In the $r_1, r_2 \gg 0$ phase, if $p_{\pm} \neq 0$ and $s, \mathbx{s}, t, \mathbx{t} = 0$, the quantum GLSM would, at low energy, become a quantum supergauged LG orbifold that can be interpreted as the parity-reversed mirror of the quantum supergauged LG orbifold described in (IV).
In particular, it would be a $\mathrm{U}(1|1) \times \mathrm{SU}(1|1) \times \mathbb{Z}_{\hat{\mathfrak{M}}}$ supergauged LG orbifold defined on a product of a supervector bundle over $\WCP^{0|1}_{\hat{n}_1|\hat{m}_1}$ and another over $\WCP^{0|1}_{\hat{n}_2|\hat{m}_2}$.

Phases (I)--(IV), and their parity-reversed mirrors, phases (I$'$)--(IV$'$), are summarized in \autoref{fig:U(1|1)-N with W:low energy:first branch} and \autoref{fig:U(1|1)-N with W:low energy:second branch}, respectively.

When the model is scale-invariant, i.e., when $m_1 - n_1 = \hat{m}_1 - \hat{n}_1$ and $m_2 - n_2 = \hat{m}_2 - \hat{n}_2$, we can physically derive that a complete intersection of an even hypersurface of bidegree ($\hat{m}_1$, $\hat{m}_2$), and an odd hypersurface of bidegree ($\hat{n}_1$, $\hat{n}_2$), in a product of two super-Grassmannians of the type $Gr_{1|1}(\mathbb{C}^{m_1|n_1}) \times Gr_{1|1}(\mathbb{C}^{m_2|n_2})$, is CY if and only if $m_1 - n_1 = \hat{m}_1 - \hat{n}_1$ and $m_2 - n_2 = \hat{m}_2 - \hat{n}_2$.
This is a \emph{novel} mathematical result that we can also mathematically verify in \autoref{app:verify cy:cicy in products of SG}!

Lastly, since the NLSM, supergauged LG orbifold, and their hybrids in (I)--(IV) are just different phases of a branch of the same underlying GLSM, they are, in the scale-invariant case whence we can freely choose $r_{1,2}$, actually related.
Likewise, since the NLSM, supergauged LG orbifold, and their hybrids in (I$'$)--(IV$'$) are just different phases of another branch of the same underlying GLSM, they are, in the scale-invariant case whence we can freely choose $r_{1,2}$, actually related.
Hence, as elaborated in \autoref{subsec: Super hybrid CY/LG for hypersurfaces}, we have a super-Grassmannian/supergroup generalization of the hybrid CY/LG correspondence for hypersurfaces in a product of projective spaces established in mathematics by Fan-Jarvis-Ruan~\cite[$\S$7.3]{Fan_2017}, as well as a parity-reversed mirror version of it!

In \autoref{sec:U(1|1) phase transition}, we consider the $\mathrm{U}(1|1)$ GLSM studied in \autoref{sec: U(1|1)^N GLSM} when $N = 1$, where the $m + n$ Grassmann-even and Grassmann-odd chiral matter superfields $\Phi_1, \dots, \Phi_m$ and $\Psi_1, \dots, \Psi_n$ now have gauge charges $Q_1, \dots, Q_{l_m} > 0 > Q_{l_m + 1}, \dots, Q_n$ and $\mathsf{Q}_1, \dots, \mathsf{Q}_{l_n} > 0 > \mathsf{Q}_{l_n + 1}, \dots, \mathsf{Q}_n$, respectively.
Its space $\MSUSY$ of supersymmetric states is defined in \eqref{eq:U(1|1) phase transition:glsm:M-SUSY} as
\begin{equation}
  \label{summary:eq:U(1|1) phase transition:glsm:M-SUSY}
  \repeatboxed{eq:U(1|1) phase transition:glsm:M-SUSY}
\end{equation}
where the subscripts ``$Q$'' and ``$\mathsf{Q}$'' are defined in \eqref{eq:U(1|1) phase transition:glsm:Q} and \eqref{eq:U(1|1) phase transition:glsm:Q-serif} as
\begin{equation}
  \label{summary:eq:U(1|1) phase transition:glsm:Q}
  \repeatboxed{eq:U(1|1) phase transition:glsm:Q}
  \qand
  \repeatboxed{eq:U(1|1) phase transition:glsm:Q-serif}
\end{equation}

The model is generically not scale-invariant, and the FI parameter runs as expressed in~\eqref{eq:U(1|1) phase transition:quantum GLSM:r-eff(mu)} as
\begin{equation}
  \label{summary:eq:U(1|1) phase transition:quantum GLSM:r-eff(mu)}
  \repeatboxed{eq:U(1|1) phase transition:quantum GLSM:r-eff(mu)}
\end{equation}
We find, via our analysis in \autoref{sec:U(1|1) phase transition:low energy}, that there are two phases of the quantum GLSM at low energy.

(I) In the $r \gg 0$ phase, the target space of the quantum NLSM is a supervector bundle over a ``weighted super-Grassmannian'' defined in~\eqref{eq:phase transition topology:cy condition:positive} as
\begin{equation}
  \label{summary:eq:phase transition topology:cy condition:positive}
  \repeatboxed{eq:phase transition topology:cy condition:positive}
\end{equation}

(II) In the $r \ll 0$ phase, the target space of the quantum NLSM is another supervector bundle over a ``weighted super-Grassmannian'' defined in~\eqref{eq:phase transition topology:cy condition:negative} as
\begin{equation}
  \label{summary:eq:phase transition topology:cy condition:negative}
  \repeatboxed{eq:phase transition topology:cy condition:negative}
\end{equation}

When the model is scale-invariant, i.e., when $Q = \mathsf{Q}$, we can physically derive that such supervector bundles over these ``weighted super-Grassmannians'' are CY if and only if $Q = \mathsf{Q}$.
This is a \emph{novel} mathematical result that we can also mathematically verify in \autoref{app:verify cy:supervector over WGr}!

Since the NLSMs in (I) and (II) are just different phases of the same underlying GLSM, they are, in the scale-invariant case whence we can freely choose $r$, actually related.
Moreover, as explained in \autoref{sec:U(1|1) phase transition:topological change:worldsheet instantons}, they can, if $\theta \neq 0$ whence there are worldsheet superinstanton effects that ``shield'' the model from the  singularity at $r=0$, be smoothly-connected to each other.
Thus, as elaborated in \autoref{subsec: Super Birational Eq of CY} for a specialization where $2 l_m = m$ and $2 l_n = n$ with gauge charge assignments $Q_{s'} = 1 = \mathsf{Q}_{t'}$ and $Q_{s''} = -1 = \mathsf{Q}_{t''}$, as we go from $r > 0$ to $r < 0$ in a phase transition  which relates and smoothly connects \eqref{summary:eq:phase transition topology:cy condition:positive} and \eqref{summary:eq:phase transition topology:cy condition:negative}, we have a super-Grassmannian generalization of a birational equivalence of CYs in mathematics!

Finally, in \autoref{sec: Phase Transition Quadrics}, we consider the $\mathrm{U}(1|1)$ GLSM with superpotential studied in \autoref{sec: Phases of U(1|1) with W} where
(i) the $\Phi^{\bar{1}}$ and $\Psi^{\bar{1}}$ superfields are turned off,
and (ii) $m - n = \sum_{\alpha^+} q^+_{\alpha^+} - \sum_{\alpha^-} q^-_{\alpha^-}$ whence it is actually scale-invariant.
Specifically, we let $m = 2M$, $n = 2N$, $\alpha^+ \in \{1, \dots, M\}$, $\alpha^- \in \{1, \dots, N\}$, and $q^{\pm}_{\alpha^{\pm}} = 2$ for all $\alpha^{\pm}$.
In turn, the gauge-invariant superpotential can be expressed in \eqref{eq:quadric:superpotential:matrix} as
\begin{equation}
  \label{summary:eq:quadric:superpotential:matrix}
  \repeatboxed{eq:quadric:superpotential:matrix}
\end{equation}
where the matrix components of the
$2M \times 2M$ symmetric even matrix $B^+(P)$,
and $2N \times 2N$ antisymmetric even matrix $C^-(P)$, are given in \eqref{eq:B+(P) definition}--\eqref{eq:C-(P) definition} as
\begin{equation}
  \label{summary:quadrics:matrix definitions}
  \begin{gathered}
    \repeatboxed{eq:B+(P) definition}
    \qand
    \repeatboxed{eq:C-(P) definition}
  \end{gathered}
\end{equation}
while the components of the $([\Phi] \, \, \,[\Psi])$ row supervector are given in \eqref{eq:quadric:Y} as
\begin{equation}
  \label{summary:eq:quadric:Y}
  \repeatboxed{eq:quadric:Y}
\end{equation}
As in \autoref{sec: Phases of U(1|1) with W}, there are four phases of the quantum GLSM at low energy, as detailed in \autoref{subsec: Phase Transition N Quadrics in SGr}.

(I) In the $r > 0$ phase, if $\phi, \psi \neq 0$ and $p_{\pm} = 0$, the quantum GLSM would, at low energy, become a quantum NLSM on a CY complete intersection of $M$ even and $N$ odd quadrics (degree 2 hypersurfaces) $G^+_{\alpha^+} = 0$ ($\alpha^+ \in \{1, \dots, M\}$) and $G^-_{\alpha^-} = 0$ ($\alpha^- \in \{1, \dots, N\}$) in $Gr_{1|0}(\mathbb{C}^{2M|2N})$ of size $r$, where the equations defining the quadrics are given in \eqref{eq:quadrics:G+ and G- polynomials} as
\begin{equation}
  \label{summary:eq:quadrics:G+ and G- polynomials}
  \repeatboxed{eq:quadrics:G+ and G- polynomials}
\end{equation}

(II) In the $r < 0$ phase, if $p_{\pm} \neq 0$ and $\phi, \psi = 0$, the quantum GLSM would, at low energy, become a quantum NLSM on a double cover $\widetilde{Gr_{1|0}(\mathbb{C}^{M|N})}$, branched over the even hypersurface $\flatfrac{\det[B^+(p)]}{\det[C^-(p)]} = 0$ of degree $2M - 2N$ in $Gr_{1|0}(\mathbb{C}^{M|N})$.

(I$'$) In the $r < 0$ phase, if $\phi, \psi \neq 0$ and $p_{\pm} = 0$, the quantum GLSM would, at low energy, become a quantum NLSM that can be interpreted as the parity-reversed version of the NLSM described in (I).
In particular, it would be a quantum NLSM on a CY complete intersection of $N$ even and $M$ odd quadrics $G^+_{\alpha^- (\mathbf{\Pi})} = 0$ and $G^-_{\alpha^+ (\mathbf{\Pi})} = 0$ in $Gr_{1|0}(\mathbb{C}^{2N|2M})$ of size $\abs{r}$.

(II$'$) In the $r > 0$ phase, if $p_{\pm} \neq 0$ and $\phi, \psi = 0$, the quantum GLSM would, at low energy, become a quantum NLSM that can be interpreted as the parity-reversed version of the NLSM described in (II).
In particular, it would be a quantum NLSM on a double cover $\widetilde{Gr_{1|0}(\mathbb{C}^{N|M})}$, branched over the even hypersurface $\det[B^+(p_{\mathbf{\Pi}})] / \det[C^-(p_{\mathbf{\Pi}})] = 0$ of degree $2N - 2M$ in $Gr_{1|0}(\mathbb{C}^{N|M})$.

Since the model is scale-invariant, we are able to physically derive that double covers of super-Grassmannians of the type $Gr_{1|0}(\mathbb{C}^{M|N})$ branched over a hypersurface of degree $2M - 2N$ in $Gr_{1|0}(\mathbb{C}^{M|N})$, are CY.
This is a \emph{novel} mathematical result that we can also mathematically verify in \autoref{sec:verify cy:double cover}!

Lastly, since the NLSMs in (I) and (II) are just different phases in a branch of the same underlying GLSM, they are, in this scale-invariant case whence we can freely choose $r$, actually related.
Likewise, since the NLSMs in (I$'$) and (II$'$) are just different phases in another branch of the same underlying GLSM, they are, in this scale-invariant case whence we can freely choose $r$, actually related.
Moreover, as explained in \autoref{subsec: Worldsheet Instantons for U(1|1) GLSM with W}, they can be smoothly-connected to each other.
Thus, as elaborated in \autoref{subsec: Super HPD}, as we go from $r > 0$ to $r < 0$ (resp. $r < 0$ to $r > 0$) in a phase transition which relates and smoothly connects the target spaces in (I) and (II) (resp. (I$'$) and (II$'$)), we have a super-Grassmannian generalization of a homological projective duality for CY quadrics by Kuznetsov-Perry~\cite{kuznetsov-2021-homol-projec} in mathematics, as well as a parity-reversed mirror version of it!

\subtitle{Acknowledgements}

We would like to thank Eric Jankowski and Simone Noja for their expert knowledge on various mathematical matters, and Eric Sharpe for very helpful exchanges.
We would also like to thank Martin Rocek and Hao Zou for bringing to our attention their relevant and interesting papers.

This work is supported in part by the MOE AcRF Tier 1 grant A-8003583-00-00.

\section{A Review of the Non-Abelian GLSM \label{sec:review:glsm}}

The construction of a non-abelian GLSM is like that of an abelian one except that the vector multiplet $V$ now transforms in the adjoint representation of a \emph{non-abelian} gauge group $\mathcal{G}$, i.e., $V \coloneq V_a T^a$, where $T^a$ are generators of the Lie algebra $\mathfrak{g}$.
As for the scalars $\phi^j$ of a chiral matter multiplet, they transform in the fundamental representation of $\mathcal{G}$ as
\begin{equation}
  \label{eq:review:glsm:fundamental rep}
  \phi^i
  \rightarrow
  (\phi^{\prime})^i
  = e^{\mathbf{i} \Lambda} \phi^i
  = \tensor{(e^{\mathbf{i} \Lambda_a T^a})}{^i_j} \phi^j
  \, ,
\end{equation}
where $e^{\mathbf{i}\Lambda} \in \mathcal{G}$ and $\tensor{(T^a)}{^i_j}$ is the $(i, j)$ component of the matrix.\footnote{%
  \label{ft:review:glsm:summation}%
  In our notation, Einstein summation will be implied for indices like the color index ``$i, j$'' and the Lie algebra index ``$a$''; likewise for the spacetime index ``$\mu$'' later.
  However, we will be explicit when it comes to flavor indices such as ``$s$'' in \eqref{eq:review:glsm:potential}, and ``$t$'' later, where summation is done explicitly with a summation symbol.
}

In~\cite{Witten:1993xi}, Witten  showed that the potential energy (of the scalars) of a $\mathrm{U}(k)$ GLSM with $n$ chiral matter multiplets (in the fundamental representation of $\mathrm{U}(k)$) is
\begin{equation}
  \label{eq:review:glsm:potential}
  U_{\text{pot}}^{\mathrm{U}(k)}
  = \frac{e^2}{2} \sum_{i, j} \left[
    \sum_{s = 1}^n  \tensor{\phi}{^j_s} \tensor{\bar{\phi}}{_i^s}
    - r \tensor{\delta}{^j_i}
  \right]^2
  + \frac{1}{2e^2} \Tr \comm{\sigma}{\bar{\sigma}}^2
  + \bar{\phi} \acomm{\bar{\sigma}}{\sigma} \phi
  \, ,
\end{equation}
where
\begin{enumerate*}
  \item $\{i, j\} \in \{1, \dots, k\}$ are indices in the fundamental $\mathrm{U}(k)$ representation,

  \item $e$ is the gauge coupling, and

  \item $\sigma$ and $\bar{\sigma}$ are scalar fields from the vector superfield $V$.
\end{enumerate*}

Note that the space of physically-inequivalent supersymmetric ground  states, which corresponds to setting $U_{\text{pot}}^{\mathrm{U}(k)}=0$ modulo gauge transformations, is, for $r > 0$, the  \emph{Grassmannian} $G_k(\mathbb{C}^n)$ of size $r$.

To understand this, first, notice that to have $U_{\text{pot}}^{\mathrm{U}(k)}=0$, if $r > 0$,  we require $\phi\neq 0$ and $\sigma=0$.
Thus, we are only left with the first term of the potential energy.
In other words, $U_{pot}^{\mathrm{U}(k)} = 0$ will mean that
\begin{equation}
  \label{eq:review:glsm:orthogonality condition}
  \sum_{s = 1}^n \tensor{\phi}{^j_s} \tensor{\bar{\phi}}{_i^s}
  =  r \tensor{\delta}{^j_i}
  \, ,
  \qquad
  i, j \in \{1, \dots, k\}
  \, .
\end{equation}
Second, note that the $n$ number of $\tensor{\phi}{^j_s}$'s for fixed $j$ can be understood to correspond to the components of a vector $w^j$ in the $n$-dimensional vector space $W \cong \mathbb{C}^n$; that is, for a basis $(e^1, e^2, \dots, e^n)$ in the vector space $W$, we have the vector and covector $w^j = \sum_s \tensor{\phi}{^j_s} e^s$ and $\bar{w}_i = \sum_s \tensor{\bar{\phi}}{_i^s} e_s$.
Third, note that \eqref{eq:review:glsm:orthogonality condition} means that $w^j \cdot \bar{w}_i = r \tensor{\delta}{^j_i}$; in other words, the $k$ number of $\bar{w}_i$'s and $w^j$'s  are orthogonal to one another, and they span the $k$-dimensional subspaces in $W$ with length-squared $r$.
Finally, after modding out the  $\mathrm{U}(k)$ gauge symmetry, \eqref{eq:review:glsm:orthogonality condition} will define the $\mathrm{U}(k)$-inequivalent set of $k$-dimensional subspaces of length-squared $r$ in $W$, which is the Grassmannian $G_k(\mathbb{C}^n)$ of size $r$.

\section{A Review of the \texorpdfstring{$\mathfrak{u}(1|1)$}{u(1|1)} Lie Superalgebra \label{sec:review:u-1-1 superalgebra}}

The unitary Lie superalgebra $\mathfrak{u}(1|1)$ is the set of invertible $1|1$ Hermitian supermatrices\footnote{%
  \label{ft:review:u-1-1:hermitivity}%
  In the mathematics literature, such as in \cite[$\S$4.17]{Fioresi:2020gsf}, $\mathfrak{u}(1|1)$ is the set of invertible $1|1$ \emph{anti-Hermitian} supermatrices.
  The corresponding $\mathrm{U}(1|1)$ supergroup is obtained via a standard exponentiation, i.e., $\exp(\alpha M) \in \mathrm{U}(1|1)$ for all $\alpha \in \mathbb{R}$ and $M \in \mathfrak{u}(1|1)$ which are anti-Hermitian, and is the group of all invertible $1|1$ unitary supermatrices, i.e., $U^{\dagger} = U^{-1}$ for all $U \in \mathrm{U}(1|1)$.

  However, in the physics literature, it is convention that a Lie group $\mathcal{G}$ with Lie algebra $\mathfrak{g}$ is obtained via an exponentiation of $\mathfrak{g}$ with an imaginary number, i.e., $\exp(\mathbf{i} \alpha M) \in \mathcal{G}$ for all $\alpha \in \mathbb{R}$ and $M \in \mathfrak{g}$.
  (We refer the reader to \cite{georgi-2018-lie-algeb} for a comprehensive description of Lie algebras in the context of physics.)
  The consequence of this on unitary Lie algebras is that they are now \emph{Hermitian} matrices.
  The reason for this convention is because (unitary) Lie algebras in the physics literature correspond to physical observables, which means that their eigenvalues ought to be real.
  Thus, they have to be Hermitian.

  At any rate, we will be following the physics convention even for Lie superalgebras, which means using Hermitian supermatrices for $\mathfrak{u}(1|1)$.
  The corresponding $\mathrm{U}(1|1)$ supergroup obtained via the physics definition of exponentiation is still nevertheless the group of all invertible $1|1$ unitary supermatrices, i.e., $U^{\dagger} = U^{-1}$ for all $U \in \mathrm{U}(1|1)$.
}
\begin{equation}
  \label{eq:review:u-1-1:definition}
  \mathfrak{u}(1|1)
  = \left\{
    \mqty(A & \beta \\ - \bar{\beta} & B)
    \in \mathfrak{gl}(1|1, \mathbb{C})
    \middle\vert
    \begin{aligned}
      A
      &= \bar{A}
      \\
      B
      &= \bar{B}
    \end{aligned}
  \right\}
  \, ,
\end{equation}
where the bar implies a complex conjugation; $A$ and $B$ are Grassmann-even and real-valued; and $\beta$ is Grassmann-odd and complex-valued.

Notice from \eqref{eq:review:u-1-1:definition}, that any element $M \in \mathfrak{u}(1|1)$ is an even supermatrix, i.e., its diagonal elements are Grassmann-even while its off-diagonal elements are Grassmann-odd.
This fact will be relevant shortly.

\section{A \texorpdfstring{$\mathrm{U}(1|1)$}{U(1|1)} GLSM and Super-Grassmannians \label{sec: U(1|1) and SGr}}

In this section, we will study a $\mathrm{U}(1|1)$ GLSM with Grassmann-even and Grassmann-odd chiral matter superfields, and show that its space of supersymmetric states corresponds to a super-Grassmannian.
Going to the low energy limit of this GLSM, we will show that it will become an NLSM whose target space is the aforementioned super-Grassmannian.
Taking into account the quantum corrections to the Fayet-Iliopoulos (FI) parameter, we will have an effective scale-dependent FI parameter in the quantum GLSM.
The quantum GLSM will also become a quantum NLSM in the low energy limit.
Lastly, when the theory is scale-invariant, we will be able to (1) obtain a geometric correspondence between super-Grassmannians, and (2) physically derive a Calabi-Yau (CY) condition for super-Grassmannians that we can also mathematically verify.

\subsection{A \texorpdfstring{$\mathrm{U}(1|1)$}{U(1|1)} GLSM and its Super-Grassmannian \texorpdfstring{$\MSUSY$}{M-SUSY} \label{subsec: U(1|1) GLSM and SGr}}

\subtitle{The $\mathrm{U}(1|1)$ GLSM}

Let us now consider the $\mathrm{U}(1|1)$ unitary supergroup GLSM with $m + n$ chiral matter superfields $\Phi_1, \dots, \Phi_m$ and $\Psi_1, \dots, \Psi_n$, where the former and latter are Grassmann-even and Grassmann-odd, respectively.\footnote{%
  \label{ft:u-1-1:seki-sugiyama}%
  Such chiral matter superfields of opposite Grassmann parity were also studied by Seki-Sugiyama in \cite{seki2005gaugedlinearsigmamodel} for a $\mathrm{U}(1)$ gauge group, where they called $\Phi$ and $\Xi$ therein bosonic and fermionic chiral (matter) superfields.
  Our reason for considering such superfields will be clear shortly.
}
Its $\mathcal{N} = (2, 2)$ Lagrangian ought to be given by
\begin{equation}
  \label{eq:u-1-1:lagrangian}
  \begin{aligned}
    L^{\mathrm{U}(1|1)}
    &=
    \frac{1}{4} \int \dd[4]{\theta}
   \left( \sum_{s=1}^m \bar{\Phi}^s e^{2V} \Phi_s
    + \sum_{t=1}^n \bar{\Psi}^t e^{2V} \Psi_t \right)
    - \frac{1}{4e^2} \int \dd[4]{\theta} \Str \bar{\Sigma} \Sigma
    \\
    &\quad
    + \left(
      \frac{1}{2\sqrt{2}} \int \dd[2]{\theta} (\mathbf{i} t \Str \Sigma)
      + c.c. \
      \right)
    \, .
  \end{aligned}
\end{equation}
Here,
\begin{enumerate*}
  \item `$\Str$' is the supertrace over the ${\mathfrak{u}}(1|1)$ supermatrices,

  \item $\Sigma = \frac{1}{2\sqrt{2}} \acomm{\bar{\mathcal{D}}_+}{\mathcal{D}_-}$ is the gauge-invariant field strength of the vector superfield $V$, where $\mathcal{D}_\pm = e^{-V} \mathscr{D}_\pm e^V$ and $\bar{\mathcal{D}}_\pm = e^V \bar{\mathscr{D}}_\pm e^{-V}$,\footnote{%
    \label{ft:superspace covariant derivative}%
    Here, the superspace covariant derivatives are $\mathscr{D}_\pm \coloneq \pdv{\theta^\pm} - \mathbf{i} \bar{\theta}^\pm \left(\pdv{x^0} \pm \pdv{x^1} \right)$ and $\bar{\mathscr{D}}_\pm \coloneq - \pdv{\bar{\theta}^\pm} + \mathbf{i} \theta^\pm \left(\pdv{x^0} \pm \pdv{x^1}\right)$.
  }
  and

  \item $t \coloneq \mathbf{i} r + \frac{\vartheta}{2\pi}$ is the FI parameter.
\end{enumerate*}
Note that in writing the above, we have followed the conventions in~\cite{Witten:1993xi}.

Note that $L^{\mathrm{U}(1|1)}$ has a $\mathrm{U}(1)_A \times \mathrm{U}(1)_V$ axial and vector R-symmetry, where the charges of $\Phi_s$ and $\Psi_t$ can be arbitrary, while that of $\Sigma$ must be $(0,2)$.


\subtitle{Chiral Scalar Fields in a $1|1$ Supervector Space}

Just as in the $\mathrm{U}(1)$ case, the chiral matter multiplets ought to transform in the fundamental representation of $\mathrm{U}(1|1)$.
Consequently, they are supervectors that live in a $1|1$ supervector space.
In particular, the chiral/antichiral scalar fields (i) $\phi$/$\bar{\phi}$ and (ii) $\psi$/$\bar{\psi}$, i.e., the lowest components of (i) $\Phi$/$\bar{\Phi}$ and (ii) $\Psi$/$\bar{\Psi}$, are (i) Grassmann-even and (ii) Grassmann-odd supervectors that live in a $1|1$ supervector space.

Since a $1|1$ supervector space has a Grassmann-even and Grassmann-odd subspace with vector/covector basis $\tau_1$/$\tau^1$ and $\chi_{\bar{1}}$/$\chi^{\bar{1}}$, respectively, these supervectors can be written as
\begin{equation}
  \label{eq:u-1-1:chiral scalar fields}
  \begin{aligned}
    \phi
    &= \phi^1 \tau_1 + \phi^{\bar{1}} \chi_{\bar{1}}
    \, ,
    &\qquad
    \psi
    &= \psi^1 \tau_1 + \psi^{\bar{1}} \chi_{\bar{1}}
    \, ,
    \\
    \bar{\phi}
    &= \bar{\phi}_1 \tau^1 + \bar{\phi}_{\bar{1}} \chi^{\bar{1}}
    \, ,
    &\qquad
    \bar{\psi}
    &= \bar{\psi}_1 \tau^1 + \bar{\psi}_{\bar{1}} \chi^{\bar{1}}
    \, ,
  \end{aligned}
\end{equation}
where $(\phi^1, \bar{\phi}_1, \psi^{\bar{1}}, \bar{\psi}_{\bar{1}})$ are Grassmann-even fields, and $(\phi^{\bar{1}}, \bar{\phi}_{\bar{1}}, \psi^1, \bar{\psi}_1)$ are Grassmann-odd fields.

It would be useful to express them as column or row vectors, i.e.,
\begin{equation}
  \label{eq:u-1-1:chiral scalar fields:supervectors}
  \phi = \mqty(\phi^1 \\ \phi^{\bar{1}})
  \, ,
  \qquad
  \psi = \mqty(\psi^1 \\ \psi^{\bar{1}})
  \, ,
  \qquad
  \bar{\phi} = \mqty(\bar{\phi}_1 & \bar{\phi}_{\bar{1}})
  \, ,
  \qquad
  \bar{\psi} = \mqty(\bar{\psi}_1 & \bar{\psi}_{\bar{1}})
  \, .
\end{equation}

It would also be useful to denote the Grassmann parity of the fields by $\varsigma(i)$ and $\varrho(i)$, where $i \in \{1, \bar{1}\}$ is an index in the fundamental $\mathrm{U}(1|1)$ representation and
\begin{equation}
  \label{eq:u-1-1:grassmann parity}
  \begin{aligned}
    \varsigma(i)
    &=
    \begin{cases}
      0 \qif i = 1
      \\
      1 \qif i = \bar{1}
    \end{cases}
    \text{for Grassmann-even supervectors,}
    \\ \\
    \varrho(i)
    &=
    \begin{cases}
      1 \qif i = 1
      \\
      0 \qif i = \bar{1}
    \end{cases}
    \text{for Grassmann-odd supervectors.}
  \end{aligned}
\end{equation}

\subtitle{The Potential Energy}

Extracting the terms of the Lagrangian which are related to the potential energy (of the scalars), we have
\begin{equation}
  \label{eq:u-1-1:lagrangian:potential}
  \begin{aligned}
    L^{\mathrm{U}(1|1)}_{\text{pot}}
    &= \Str \left[
      \frac{1}{2e^2} D^2
      + D \left(
        \sum_{s = 1}^m \phi_s \bar{\phi}^s
        - \sum_{t = 1}^n \psi_t \bar{\psi}^t
        - r \mathbb{I}_{1|1}
      \right)
      - \frac{1}{2e^2} \comm{\sigma}{\bar{\sigma}}^2
    \right]
    \\
    &\quad
    - \sum_{s = 1}^m \bar{\phi}^s \acomm{\bar{\sigma}}{\sigma} \phi_s
    - \sum_{t = 1}^n \bar{\psi}^t \acomm{\bar{\sigma}}{\sigma} \psi_t
    \, ,
  \end{aligned}
\end{equation}
where
\begin{enumerate*}
  \item the auxiliary field $D$ and scalars $\sigma$ and $\bar{\sigma}$ are from the vector multiplet of $V$ and hence, are $\mathfrak{u}(1|1)$ \emph{even} supermatrices,

  \item $\phi_s \bar{\phi}^s$ and $\psi_t \bar{\psi}^t$ are \emph{even} supermatrices,\footnote{%
    \label{ft:u-1-1:psi psi-bar as grassmann even}%
    Although $\psi$ and $\bar{\psi}$ are Grassmann-odd supervectors, $\psi\bar{\psi}$ is an even supermatrix just like in \autoref{sec:review:u-1-1 superalgebra}.
  }
  and

  \item $\Str$ is the supertrace over the supermatrix indices.\footnote{
    \label{ft:u-1-1:supertrace}%
    For an even $1|1 \times 1|1$ supermatrix $M \in \text{Mat}_{1|1 \times 1|1}$ (such as a $\mathfrak{u}(1|1)$ matrix as seen in \autoref{sec:review:u-1-1 superalgebra}), its supertrace is given by $\Str M = \sum_{i} (-1)^{\varsigma(i)} \tensor{M}{^i_i} = \tensor{M}{^1_1} - \tensor{M}{^{\bar{1}}_{\bar{1}}}$, where $\tensor{M}{^1_1}$ and $\tensor{M}{^{\bar{1}}_{\bar{1}}}$ are its top and bottom Grassmann-even diagonal elements, respectively.

    In general, for an even $m|n \times m|n$ supermatrix $M \in \text{Mat}_{m|n \times m|n}$, its supertrace is given by $\Str M = \sum_{i} (-1)^{\varsigma(i)} \Tr \tensor{M}{^i_i} = \Tr M_{(m \times m)} - \Tr M_{(n \times n)}$, where
    \begin{enumerate*}
      \item $\Tr$ is the regular trace,

      \item $M_{(m \times m)}$ and $M_{(n \times n)}$ are the $(m \times m)$ upper left block and $(n \times n)$ lower right block of $M$, respectively, which are both regular Grassmann-even matrices, and

      \item $\varsigma(i) = 0$ and $1$ over all matrix indices in $M_{(m \times m)}$ and $M_{(n \times n)}$, respectively.
    \end{enumerate*}
    Also, for any two such supermatrices $M_1$ and $M_2$, $\Str M_1 M_2 = \Str M_2 M_1$~\cite[eq.~(2.18)--(2.19)]{Kimura:2023iup}.
  }
\end{enumerate*}

After eliminating the auxiliary field $D$ using its equation of motion (EOM) (see \autoref{app:u-1-1:glsm}), the potential energy will be given by
\begin{equation}
  \label{eq:u-1-1:potential}
  \begin{aligned}
    U^{\mathrm{U}(1|1)}_{\text{pot}}
    &= \frac{e^2}{2} \sum_{i, j}
    (-1)^{\varsigma(i)}
    \left[
      \sum_{s = 1}^m \phi^j{}_s \bar{\phi}_i{}^s
      - \sum_{t = 1}^n \psi^j{}_t \bar{\psi}_i{}^t
      - r \delta^j{}_i
    \right]^2
    \\
    &\quad
    + \frac{1}{2e^2} \Str \comm{\sigma}{\bar{\sigma}}^2
    + \sum_{s = 1}^m \bar{\phi}^s \acomm{\bar{\sigma}}{\sigma} \phi_s
    + \sum_{t = 1}^n \bar{\psi}^t \acomm{\bar{\sigma}}{\sigma} \psi_t
    \, .
  \end{aligned}
\end{equation}
Comparing \eqref{eq:u-1-1:potential} with \eqref{eq:review:glsm:potential}, we see that
\begin{enumerate*}
  \item there is a factor of $(-1)^{\varsigma(i)}$ which appear in the first line of \eqref{eq:u-1-1:potential} due to the supertrace (see \autoref{ft:u-1-1:supertrace}), and

  \item there are extra quartic field terms from the mixing between \emph{and} amongst the $\phi$ and $\psi$ fields.
\end{enumerate*}

\subtitle{Nonunitarity of the Model}

As mentioned in the introduction, supergauge theories are nonunitary.
In particular, due to the supertrace in the field strength term $\int \dd[4]{\theta} \Str \bar{\Sigma} \Sigma$ of the Lagrangian \eqref{eq:u-1-1:lagrangian}, there will be an opposite sign contribution from the spacetime field strength $F$ which will result in an unbounded spectrum with negative energy.
Indeed, this is also reflected in the potential energy \eqref{eq:u-1-1:potential}, where because of the overall $(-1)^{\varsigma(i)}$ factor in the first term that is otherwise absent in a regular gauge theory, it is unbounded.

\subtitle{The Space $\MSUSY$ of Supersymmetric States }

Nonetheless, one can still study concretely, its supersymmetric states that are strictly zero energy states.
Specifically, for the four supercharges and Hamiltonian of the $\mathcal{N}=(2,2)$ algebra, namely, $\mathcal{Q}_+$, $\bar{\mathcal{Q}}_+$, $\mathcal{Q}_-$, $\bar{\mathcal{Q}}_-$, and $H$, where $\mathcal{Q}_A = \bar{\mathcal{Q}}_+ + \mathcal{Q}_-$ and $\mathcal{Q}_B = \bar{\mathcal{Q}}_+ + \bar{\mathcal{Q}}_-$ are such that for $\mathcal{Q}$ either $\mathcal{Q}_A$ or $\mathcal{Q}_B$, we have $\acomm{\mathcal{Q}}{\bar{\mathcal{Q}}} = 2H$ and $\mathcal{Q}^2 = \bar{\mathcal{Q}}^2 = 0$~\cite[eq.~(13.50)--(13.51)]{Hori:2003ic},\footnote{%
  \label{ft:u-1-1:no central charge}%
  As noted in~\cite[eq.~(13.50)--(13.51)]{Hori:2003ic}, these relations that $\mathcal{Q}$ and $\bar{\mathcal{Q}}$ obey assume that the $Z$ and $\widetilde{Z}$ central charges of the algebra vanish.
  This is true only if there is a $\mathrm{U}(1)_A$ and $\mathrm{U}(1)_V$ axial and vector $R$-symmetry with conserved charges $F_A$ and $F_V$, respectively~\cite[$\S$12.3]{Hori:2003ic}.
  We shall therefore restrict ourselves to such GLSMs.
  In particular,  when we later consider the GLSM with a superpotential, we shall insist on quasi-homogeneous superpotentials, as only then can there be a $\mathrm{U}(1)_V$-symmetry~\cite[$\S$13.2]{Hori:2003ic}.
} we wish to study the supersymmetric states which are $\mathcal{Q}$ and $\bar{\mathcal{Q}}$-invariant that therefore have $H=0$ and correspond to setting $U^{\mathrm{U}(1|1)}_{\text{pot}} = 0$.

Let us denote the space of such supersymmetric states that are also physically-inequivalent as $\MSUSY$.
Due to the gauge symmetry, $\MSUSY$ should then be defined by $U^{\mathrm{U}(1|1)}_{\text{pot}} = 0$ modulo gauge transformations.

\subtitle{A Super-Grassmannian $\MSUSY$}

In the regular $\mathrm{U}(k)$ GLSM, recall that the potential $U^{\mathrm{U}(k)}_{\text{pot}}$ in \eqref{eq:review:glsm:potential} would vanish only when $r > 0$.
However, in our model, because of the minus sign that comes along with the term in the $\psi$'s in the potential $U^{\mathrm{U}(1|1)}_{\text{pot}}$ in \eqref{eq:u-1-1:potential}, this need not be the case.

From \eqref{eq:u-1-1:potential}, we see that in order to have $U^{\mathrm{U}(1|1)}_{\text{pot}} = 0$, for $\abs{r} > 0$, we require $\phi \neq 0$, $\psi \neq 0$, and $\sigma = 0$ such that
\begin{equation}
  \label{eq:u-1-1:orthogonality condition}
  \sum_{s = 1}^m \tensor{\phi}{^j_s} \tensor{\bar{\phi}}{_i^s}
  - \sum_{t = 1}^n \tensor{\psi}{^j_t} \tensor{\bar{\psi}}{_i^t}
  = r \tensor{\delta}{^j_i}
  \, ,
  \qquad
  i, j \in \{1, \bar{1}\}
  \, .
\end{equation}
The field configurations satisfying \eqref{eq:u-1-1:orthogonality condition}, modulo the $\mathrm{U}(1|1)$ gauge transformations, define a \emph{super-Grassmannian} space.

To understand this, first, note that the $m$/$n$ number of $\tensor{\phi}{^1_s}$/$\tensor{\psi}{^1_t}$ (resp. $\tensor{\bar{\phi}}{_1^s}$/$\tensor{\bar{\psi}}{_1^t}$) can be understood to correspond to the components of a Grassmann-even supervector $w^1$ (resp. supercovector $\bar{w}_1$) in the $m|n$-dimensional supervector space $\mathcal{W} \cong \mathbb{C}^{m|n}$.
That is, for an orthonormal supervector basis $(e^1, e^2, \dots, e^m)$ and $(\varepsilon^1, \varepsilon^2, \dots, \varepsilon^n)$ on the Grassmann-even and Grassmann-odd subspace, respectively, of the supervector space $\mathcal{W}$,\footnote{%
  \label{ft:u-1-1:glsm:orthonormal basis}%
  An orthonormal supervector/supercovector basis $(e^j, \varepsilon^j)$/$(e_i, \varepsilon_i)$ is one such that (i) $e^j \cdot e_i = \tensor{\delta}{^j_i} \equiv e_i \cdot e^j$, (ii) $\varepsilon^j \cdot \varepsilon_i = - \tensor{\delta}{^j_i} \equiv - \varepsilon_i \cdot \varepsilon^j$, and (iii) $e^j \cdot \varepsilon_i = 0 = \varepsilon^j \cdot e_i$.
}
we have the Grassmann-even supervector and supercovector $w^1 = \sum_s \tensor{\phi}{^1_s} e^s + \sum_t \tensor{\psi}{^1_t} \varepsilon^t$ and $\bar{w}_1 = \sum_s \tensor{\bar{\phi}}{_1^s} e_s + \sum_t \tensor{\bar{\psi}}{_1^t} \varepsilon_t$, respectively.
Likewise, from the bar-indexed fields, we also have the Grassmann-odd supervector and supercovector $w^{\bar{1}} = \sum_s \tensor{\phi}{^{\bar{1}}_s} e^s + \sum_t \tensor{\psi}{^{\bar{1}}_t} \varepsilon^t$ and $\bar{w}_{\bar{1}} = \sum_s \tensor{\bar{\phi}}{_{\bar{1}}^s} e_s + \sum_t \tensor{\bar{\psi}}{_{\bar{1}}^t} \varepsilon_t$, respectively.
Notice that $w$ and $\bar{w}$ are chiral and antichiral supervectors in the sense that their component fields are chiral and antichiral fields, respectively.
Moreover, just like how we went from \eqref{eq:u-1-1:chiral scalar fields} to \eqref{eq:u-1-1:chiral scalar fields:supervectors}, it would be useful to express them as column or row supervectors, i.e.,
\begin{equation}
  \label{eq:u-1-1:glsm:w and w-bar}
  w = \mqty(w^1 \\ w^{\bar{1}})
  \, ,
  \qquad
  \bar{w} = \mqty(\bar{w}_1 & \bar{w}_{\bar{1}})
  \, .
\end{equation}

Second, note that \eqref{eq:u-1-1:orthogonality condition} means that $w^j \cdot \bar{w}_i = r \tensor{\delta}{^j_i}$; in other words, the $1|1$ number of $\bar{w}_i$'s and $w^j$'s  are orthogonal to one another, and they span the $1|1$-dimensional subspaces in $\mathcal{W}$ with length-squared $r$.
As such, after modding out the $\mathrm{U}(1|1)$ gauge symmetry,\footnote{%
  \label{ft:u-1-1:supergroup action}%
  We can think of the chiral scalars $\tensor{\phi}{^i_s}$ and $\tensor{\psi}{^i_t}$ as components of a $(1|1 \times m|n)$ supermatrix, i.e.,
  \begin{equation*}
    Z = \mqty(
    \tensor{\phi}{^1_1} & \dots & \tensor{\phi}{^1_m} & \tensor{\psi}{^1_1} & \dots & \tensor{\psi}{^1_n}
    \\
    \tensor{\phi}{^{\bar{1}}_1} & \dots & \tensor{\phi}{^{\bar{1}}_m} & \tensor{\psi}{^{\bar{1}}_1} & \dots & \tensor{\psi}{^{\bar{1}}_n}
    )
    \, .
  \end{equation*}
  Then, the left action by an element $M \in \mathrm{U}(1|1)$ of the supergroup on the chiral scalars, which is given by
  \begin{equation*}
    \begin{aligned}
      \tensor{\phi}{^i_s}
      &\to \tensor{(\phi^{\prime})}{^i_s} = \tensor{M}{^i_j} \tensor{\phi}{^j_s}
      \, ,
      \\
      \tensor{\psi}{^i_t}
      &\rightarrow \tensor{(\psi^{\prime})}{^i_t} = \tensor{M}{^i_j} \tensor{\psi}{^j_t}
      \, ,
    \end{aligned}
  \end{equation*}
  can also be expressed as
  \begin{equation*}
    \begin{aligned}
      \phi
      &\to \phi^{\prime} = M \phi
      \, ,
      \\
      \psi
      &\to \psi^{\prime} = M \psi
      \, ,
    \end{aligned}
  \end{equation*}
  where the $1|1$ supermatrix
  \begin{equation*}
    M = \mqty(
    M^1{}_1 & M^1{}_{\bar{1}}
    \\
    M^{\bar{1}}{}_1 & M^{\bar{1}}{}_{\bar{1}}
    )
  \end{equation*}
  is even, whence $\phi^{\prime}$ and $\psi^{\prime}$ continue to be Grasmman-even and Grassmann-odd supervectors, respectively.

  Since the chiral scalars $\tensor{\phi}{^1_s}$ and $\tensor{\psi}{^1_t}$ (resp. $\tensor{\phi}{^{\bar{1}}_s}$ and $\tensor{\psi}{^{\bar{1}}_t}$) are components of the Grassmann-even supervector $w^1$ (resp. Grassmann-odd supervector $w^{\bar{1}}$), we can express the top and bottom row of Z as
  \begin{equation*}
    Z = \mqty(\omega \\ \widetilde{\omega})
    \, ,
  \end{equation*}
  where $\omega$ and $\widetilde{\omega}$ represent the components of $w^1$ and $w^{\bar{1}}$, respectively.
  Notice that $Z$ is equivalent to our $w$ from \eqref{eq:u-1-1:glsm:w and w-bar}.

  At any rate, the left action of a supergroup element $M \in \mathrm{U}(1|1)$ on the components $\omega$ and $\bar{\omega}$ of the supervectors $w^1$ and $w^{\bar{1}}$ in $\mathbb{C}^{1|1} \subset \mathcal{W}$ will be given by
  \begin{equation*}
    \mqty(\omega \\ \widetilde{\omega})
    \rightarrow \mqty(\omega^{\prime} \\ \widetilde{\omega}^{\prime})
    = M \mqty(\omega \\ \widetilde{\omega})
    \, ,
    \qquad
    \implies
    \qquad
    Z
    \rightarrow Z^{\prime}
    = M Z
    \, .
  \end{equation*}

  Similarly, we can think of the antichiral scalars $\tensor{\bar{\phi}}{_i^s}$ and $\tensor{\bar{\psi}}{_i^s}$ as components of a $(m|n \times 1|1)$ supermatrix $\overline{Z}$ (which would be equivalent to $\bar{w}$ from \eqref{eq:u-1-1:glsm:w and w-bar}), whence a supergroup element $M \in \mathrm{U}(1|1)$ would act on right as $\overline{Z} \rightarrow \overline{Z}' = \overline{Z} M^{-1}$.
}
the collection of physically-inequivalent field configurations obeying \eqref{eq:u-1-1:orthogonality condition}, which one can now express as
\begin{equation}
  \label{eq:u-1-1:m-susy}
  \saveboxed{eq:u-1-1:m-susy}{
    X_r
    = \flatfrac{
      \left\{
        \sum_{s = 1}^m \tensor{\phi}{^j_s} \tensor{\bar{\phi}}{_i^s}
        - \sum_{t = 1}^n \tensor{\psi}{^j_t} \tensor{\bar{\psi}}{_i^t}
        = r \delta^j{}_i
        \, \qcomma i, j \in \{1, \bar{1}\}
      \right\}
    }{\mathrm{U}(1|1)}
  }
\end{equation}
will define the $\mathrm{U}(1|1)$-inequivalent set of $1|1$-dimensional subspaces of length-squared $r$ in $\mathcal{W}$, which is the super-Grassmannian ${Gr}_{1|1}(\mathbb{C}^{m|n})$~\cite[ch.~4,~$\S$3]{Manin:1988ds}.

In short, $\MSUSY$, which is defined by $X_r$ in \eqref{eq:u-1-1:m-susy}, will be given by ${Gr}_{1|1}(\mathbb{C}^{m|n})$ of size $r$.

\subtitle{$\MSUSY$ when $r > 0$ and $r < 0$}

When $r > 0$, $X_{r > 0}$ in \eqref{eq:u-1-1:m-susy} defines a ${Gr}_{1|1}(\mathbb{C}^{m|n})$ of positive size $r$ which is realized by $m$ Grassmann-even and $n$ Grassmann-odd chiral scalar fields.

When $r < 0$, it appears that $X_{r < 0}$ in \eqref{eq:u-1-1:m-susy} would define a super-Grassmannian of negative size $r$. How then can one interpret  $X_{r < 0}$  sensibly?

If $r < 0$, we can multiply the LHS and RHS of \eqref{eq:u-1-1:m-susy} by $-1$ without changing the definition of $X_r$. In doing so, we would have, on the RHS, a positive number $\abs{r}$, and on
the LHS, the roles of the $\phi$ and $\psi$ fields would have been swopped -- the $m$/$n$ number of $\phi$/$\psi$ fields can now be reinterpreted as Grassmann-odd/Grassmann-even chiral scalar fields. In other words, when $r < 0$, one can sensibly reinterpret $X_r$ to be a super-Grassmannian in a parity-reversed ambient space, i.e., ${Gr}_{1|1}(\mathbb{C}^{n|m})$, of \emph{positive} size $\abs{r}$.

Let us describe this purely mathematically. Using the $w^j$ and $\bar{w}_i$ supervectors in \eqref{eq:u-1-1:glsm:w and w-bar}, \eqref{eq:u-1-1:m-susy} can, when $r > 0$, be re-expressed as
\begin{equation}
  \label{eq:u-1-1:m-susy:r > 0}
  X_{r > 0} = \flatfrac{
    \Bigl\{
      w^j \cdot \bar{w}_i = \abs{r} \tensor{\delta}{^j_i} \qcomma i, j \in \{1, \bar{1}\}
    \Bigr\}
  }{\mathrm{U}(1|1)}
  = {Gr}_{1|1}(\mathbb{C}^{m|n})
  \, .
\end{equation}
On the other hand, when $r< 0$, \eqref{eq:u-1-1:m-susy} can also be expressed as
\begin{equation}
  \label{eq:u-1-1:m-susy:r < 0}
  X_{r < 0} = \flatfrac{
    \Bigl\{
      (\mathbf{\Pi} w^j) \cdot (\mathbf{\Pi} \bar{w}_i) = \abs{r} \tensor{\delta}{^j_i} \qcomma i, j \in \{1, \bar{1}\}
    \Bigr\}
  }{\mathrm{U}(1|1)}
  = {Gr}_{1|1}(\mathbb{C}^{n|m})
  \, ,
\end{equation}
where $\mathbf{\Pi}$ denotes a Grassmann parity-reversal operation, i.e., $\mathbf{\Pi} \omega^j = \sum_s \tensor{\phi}{^j_s} (\mathbf{\Pi} e^s) + \sum_t \tensor{\psi}{^j_t} (\mathbf{\Pi} \varepsilon^t) = \sum_s \tensor{\phi}{^j_s} \varepsilon^s + \sum_t \tensor{\psi}{^j_t} e^t$, and likewise for $\mathbf{\Pi} \bar{w}_i$.

As $X_{r>0}$ and $X_{r < 0}$ are similar spaces, we shall, for brevity, focus only on the $r > 0$ case for the rest of this subsection and the next.

\subtitle{A Reduction to Seki-Sugiyama's $\mathrm{U}(1)$ GLSM in $\mathbb{CP}^{m-1|n}$}

Note that the $\mathrm{U}(1|1)$ GLSM ought to reduce to Seki-Sugiyama's $\mathrm{U}(1)$ GLSM when we turn off the $\tensor{\phi}{^{\bar{1}}_s}$ and $\tensor{\psi}{^{\bar{1}}_t}$ fields (and their corresponding bar-indexed partners in the underlying multiplets).\footnote{%
  \label{ft:u-1-1:turn off barred fields}%
  When there are only the unbarred $\tensor{\phi}{^1_s}$ and $\tensor{\psi}{^1_s}$ fields, the matrix $M$ in \autoref{ft:u-1-1:supergroup action} acts on them just with its $\tensor{M}{^1_1}$ component, whence the gauge group \emph{effectively} reduces to the subgroup $\mathrm{U}(1) \subset \mathrm{U}(1|1)$ generated by this component.%
}
Indeed, turning off these fields would mean that $w^{\bar{1}} = 0$ whence \eqref{eq:u-1-1:m-susy:r > 0} would become an orthogonality condition on $w^1$ and $\bar{w}_1$ that only span a $1|0$-dimensional subspace in $\mathcal{W}$.

In other words, $\MSUSY$ would reduce to ${Gr}_{1|0}(\mathbb{C}^{m|n}) \cong \mathbb{CP}^{m-1|n}$ \cite[$\S$4]{noja-2018-non-projec}, as it should (\textit{c.f.}~\cite[$\S$2]{seki2005gaugedlinearsigmamodel}).

\subtitle{A Reduction to the Familiar $\mathrm{U}(1)$ GLSM in $\mathbb{CP}^{m-1}$}

Note that the $\mathrm{U}(1|1)$ GLSM ought to further reduce to the familiar $\mathrm{U}(1)$ GLSM when we turn off
\begin{enumerate*}
  \item the $\tensor{\phi}{^{\bar{1}}_s}$ field (and its corresponding bar-indexed partners in the underlying multiplet), and

  \item the \emph{whole} Grassmann-odd chiral superfield $\Psi$.
\end{enumerate*}
Indeed, the remaining Grassmann-even $\tensor{\phi}{^1_s}$ fields only define an $m$-dimensional Grassmann-even subspace $W \subset \mathcal{W}$ whence the ambient vector space would reduce to $\mathbb{C}^m \cong \mathbb{C}^{m|0} \subset \mathbb{C}^{m|n}$; in other words, $\MSUSY$ would further reduce to $G_1(\mathbb{C}^m) \cong \mathbb{CP}^{m-1}$ \cite[lect.~6]{Griffiths:1994prl}, as it should (\emph{c.f.}~\cite[ch.~15]{Hori:2003ic}).

\subsection{The Low Energy NLSM on a Super-Grassmannian} \label{GLSM to NLSM}

\subtitle{The Low Energy NLSM}

The GLSM Lagrangian in~\eqref{eq:u-1-1:lagrangian} can also be expressed as
\begin{equation}
  L^{\mathrm{U}(1|1)}
  = L^{\mathrm{U}(1|1)}_{\text{gauge}}
  + L^{\mathrm{U}(1|1)}_{\text{FI}}
  + L^{\mathrm{U}(1|1)}_{\text{chiral}}
  \, ,
\end{equation}
where $L^{\mathrm{U}(1|1)}_{\text{gauge}}$, $L^{\mathrm{U}(1|1)}_{\text{FI}}$, and $L^{\mathrm{U}(1|1)}_{\text{chiral}}$ contain the vector multiplet fields, the FI parameter, and the chiral matter multiplet fields, respectively.

Let us now consider the theory in the low (positive) energy limit $e\to\infty$ whence the massive fields/modes decouple from the theory.\footnote{%
  \label{ft:u-1-1:nlsm:gauge coupling strength}%
  In 2d gauge theories such as our GLSM, the gauge coupling constant has, at the outset, a mass dimension of 1.
  Consequently, at low (positive) energies, the \emph{effective} dimensionless coupling will be large. In this sense, one can regard $e \to \infty$ to be a low (positive) energy limit.
}
In this limit, $L^{\mathrm{U}(1|1)}_{\text{gauge}} = - \frac{1}{4e^2} \int \dd[4]{\theta} \Str \bar{\Sigma} \Sigma \to 0$, and the gauge field will become an auxiliary field which can then be integrated out.
Consequently, the GLSM will become an NLSM with target space metric $\dd{\mathsf{s}}^2$, which we will now determine.

In Euclidean spacetime, the auxiliary gauge field $v$ will appear in (the purely bosonic part of) $L^{\mathrm{U}(1|1)}_{\text{chiral}}$ as\footnote{%
  \label{eq:u-1-1:nlsm:mod-square of vector}%
  In the following expression, the mod-square of a chiral supervector $\Upsilon_s$ for a fixed $s$ is defined as $\abs{\Upsilon_s}^2 \coloneq \tensor{\bar{\Upsilon}}{_i^s} \tensor{\Upsilon}{^i_s} = (-1)^{\deg(\Upsilon)} (-1)^{\varsigma(i)} \tensor{\Upsilon}{^i_s} \tensor{\bar{\Upsilon}}{_i^s} = \pm \Str (\Upsilon_s \bar{\Upsilon}^s)$ \cite[eq.~(2.26)]{Kimura:2023iup}, where $(-1)^{\deg(\Upsilon_s)} = \pm 1$ is the parity of $\Upsilon_s$, i.e., it is $+1$/$-1$ if $\Upsilon_s$ is a Grassmann-even/odd supervector.
}
\begin{equation}
  \label{eq:u-1-1:nlsm:chiral lagrangian}
  L^{U(1|1)}_{\text{chiral}}(v)
  = - \sum_{s = 1}^m \abs{D_{\mu} \phi_s}^2
  - \sum_{t = 1}^n \abs{D_{\mu} \psi_t}^2
  \, ,
\end{equation}
where
\begin{enumerate*}
  \item $D_\mu \tensor{\varphi}{^i_s} = \partial_\mu \tensor{\varphi}{^i_s} + \mathbf{i} \sum_j \tensor{v}{_\mu^i_j} \tensor{\varphi}{^j_s}$ and $D_{\mu} \tensor{\bar{\varphi}}{_i^s} = \partial_{\mu} \tensor{\bar{\varphi}}{_i^s} - \mathbf{i} \sum_j \tensor{\bar{\varphi}}{_j^s} \tensor{v}{_{\mu}^j_i}$ for $\varphi \in \{\phi, \psi\}$, with $v_\mu$ being the $\mathrm{U}(1|1)$ gauge field;

  \item $\mu$ is the 2d spacetime index; and

  \item $i, j, k \in \{1, \bar{1}\}$ are the color indices.
\end{enumerate*}
Explicitly, \eqref{eq:u-1-1:nlsm:chiral lagrangian} can be expanded as
\begin{equation}
  \label{eq:u-1-1:nlsm:chiral lagrangian:expanded}
  \begin{aligned}
    L^{\mathrm{U}(1|1)}_{\text{chiral}}(v)
    &= - \sum_{s = 1}^m \left[
      \partial^\mu \tensor{\bar{\phi}}{_i^s} \partial_\mu \tensor{\phi}{^i_s}
      - \mathbf{i} \left(
        \tensor{\bar{\phi}}{_j^s} \tensor{v}{^{\mu}^j_i} \partial_{\mu} \tensor{\phi}{^i_s}
        - \partial^{\mu} \tensor{\bar{\phi}}{_i^s} \tensor{v}{_{\mu}^i_j} \tensor{\phi}{^j_s}
      \right)
      + \tensor{\bar{\phi}}{_j^s} \tensor{v}{^{\mu}^j_i} \tensor{v}{_{\mu}^i_k} \tensor{\phi}{^k_s}
    \right]
    \\
    &\quad
    - \sum_{t = 1}^n \left[
      \partial^\mu \bar\psi_i{}^t \partial_\mu \tensor{\psi}{^i_t}
      - \mathbf{i} \left(
        \tensor{\bar{\psi}}{_j^t} \tensor{v}{^{\mu}^j_i} \partial_{\mu} \tensor{\psi}{^i_t}
        - \partial^{\mu} \tensor{\bar{\psi}}{_i^t} \tensor{v}{_{\mu}^i_j} \tensor{\psi}{^j_t}
      \right)
      + \tensor{\bar{\psi}}{_j^t} \tensor{v}{^{\mu}^j_i} \tensor{v}{_{\mu}^i_k} \tensor{\psi}{^k_t}
    \right]
    \, .
  \end{aligned}
\end{equation}

The EOM of the auxiliary gauge field $\tensor{v}{_{\mu}^i_j}$ is
\begin{equation}
  \label{eq:u-1-1:nlsm:eom of v}
  \tensor{J}{^{\mu}_i^j}
  = (-1)^{\varsigma(i)} \sum_{s = 1}^m \left(
    \tensor{\bar{\phi}}{_i^s} \tensor{v}{^{\mu}^j_k} \tensor{\phi}{^k_s}
    + \tensor{\bar{\phi}}{_k^s} \tensor{v}{^{\mu}^k_i} \tensor{\phi}{^j_s}
  \right)
  + (-1)^{\varsigma(j)} \sum_{t = 1}^n \left(
    \tensor{\bar{\psi}}{_i^t} \tensor{v}{^{\mu}^j_k} \tensor{\psi}{^k_t}
    + \tensor{\bar{\psi}}{_k^t} \tensor{v}{^{\mu}^k_i} \tensor{\psi}{^j_t}
  \right)
  \, ,
\end{equation}
where
\begin{equation}
  \label{eq:u-1-1:nlsm:current}
  \tensor{J}{^{\mu}_i^j}
  \coloneq \mathbf{i} \left[
    (-1)^{\varsigma(i)} \sum_{s = 1}^m \left(
      \tensor{\bar{\phi}}{_i^s} \partial^{\mu} \tensor{\phi}{^j_s}
      - \partial^{\mu} \tensor{\bar{\phi}}{_i^s} \tensor{\phi}{^j_s}
    \right)
    + (-1)^{\varsigma(j)} \sum_{t = 1}^n \left(
      \tensor{\bar{\psi}}{_i^t} \partial^{\mu} \tensor{\psi}{^j_t}
      - \partial^{\mu} \tensor{\bar{\psi}}{_i^t} \tensor{\psi}{^j_t}
    \right)
  \right]
  \, .
\end{equation}

From the EOM \eqref{eq:u-1-1:nlsm:eom of v} applied to \eqref{eq:u-1-1:nlsm:chiral lagrangian:expanded},\footnote{%
  \label{ft:u-1-1:nlsm:applying eom of v}%
   The $v$-quadratic terms in \eqref{eq:u-1-1:nlsm:chiral lagrangian:expanded} can be written as
  \begin{equation}
    \label{eq:u-1-1:nlsm:terms quadartic in v}
    \begin{aligned}
      - \frac{1}{2} (-1)^{\varsigma(i)\varsigma(j)} \tensor{v}{_{\mu}^i_j} \left[
        (-1)^{\varsigma(i)} \sum_{s = 1}^m \left(
          \tensor{\bar{\phi}}{_i^s} \tensor{v}{^{\mu}^j_k} \tensor{\phi}{^k_s}
          + \tensor{\bar{\phi}}{_k^s} \tensor{v}{^{\mu}^k_i} \tensor{\phi}{^j_s}
        \right)
        + (-1)^{\varsigma(j)} \sum_{t = 1}^n \left(
          \tensor{\bar{\psi}}{_i^t} \tensor{v}{^{\mu}^j_k} \tensor{\psi}{^k_t}
          + \tensor{\bar{\psi}}{_k^t} \tensor{v}{^{\mu}^k_i} \tensor{\psi}{^j_t}
        \right)
      \right]
      \, .
    \end{aligned}
  \end{equation}
  Applying the EOM of $\tensor{v}{_{\mu}^i_j}$ in \eqref{eq:u-1-1:nlsm:eom of v} to \eqref{eq:u-1-1:nlsm:terms quadartic in v}, we find that it can be written as a term \emph{linear} in $v$.
  Substituting that back into \eqref{eq:u-1-1:nlsm:chiral lagrangian:expanded}, the terms which are now linear in $v$ in \eqref{eq:u-1-1:nlsm:chiral lagrangian:expanded} can be expressed as
  \begin{equation}
    \label{eq:u-1-1:nlsm:terms linear in v}
    \frac{1}{2} (-1)^{\varsigma(i)\varsigma(j)} \tensor{v}{_{\mu}^i_j} \tensor{J}{^{\mu}_i^j}
    \, .
  \end{equation}
  Recall the definition of $\tensor{J}{^{\mu}_i^j}$ in \eqref{eq:u-1-1:nlsm:current}, which involves Grassmann-even and Grassmann-odd fields.
  Swopping the positions of these fields, \eqref{eq:u-1-1:nlsm:terms linear in v} can also be equivalently expressed as
  \begin{equation}
    \label{eq:u-1-1:nlsm:terms linear in v:swopped}
    \frac{1}{2} (-1)^{\varsigma(i)} \tensor{v}{_{\mu}^i_j} \tensor{J}{^{\mu}^j_i}
    \, ,
  \end{equation}
  where
  \begin{equation}
    \label{eq:u-1-1:nlsm:current matrix}
    \tensor{J}{^{\mu}^j_i}
    \coloneq \mathbf{i} \left[
      \sum_{s = 1}^m \left(
        \partial^{\mu} \tensor{\phi}{^j_s} \tensor{\bar{\phi}}{_i^s}
        - \tensor{\phi}{^j_s} \partial^{\mu} \tensor{\bar{\phi}}{_i^s}
      \right)
      - \sum_{t = 1}^n \left(
        \partial^{\mu} \tensor{\psi}{^j_t} \tensor{\bar{\psi}}{_i^t}
        - \tensor{\psi}{^j_t} \partial^{\mu} \tensor{\bar{\psi}}{_i^t}
      \right)
    \right]
    \equiv (-1)^{\varsigma(i) + \varsigma(i)\varsigma(j)} \tensor{J}{^{\mu}_i^j}
    \, ,
  \end{equation}
  has a more useful interpretation as an element of a supermatrix, whence \eqref{eq:u-1-1:nlsm:terms linear in v:swopped} is a supertrace.
}
we compute that

\begin{equation}
  \label{eq:u-1-1:nlsm:chiral}
  \begin{aligned}
    L^{\mathrm{U(1|1)}}_{\text{chiral}}(v)
    &= - \left[
      \sum_{s = 1}^m \partial^{\mu} \tensor{\bar{\phi}}{_i^s} \partial_{\mu} \tensor{\phi}{^i_s}
      + \sum_{t = 1}^n \partial^{\mu} \tensor{\bar{\psi}}{_i^t} \partial_{\mu} \tensor{\psi}{^i_t}
      - \frac{1}{2} (-1)^{\varsigma(i)} \tensor{v}{_{\mu}^i_j} \tensor{J}{^{\mu}^j_i}
    \right]
    \\
    &= - \Str \left[
      \partial_{\mu} w \cdot \partial^{\mu} \bar{w}
      - \frac{1}{2} v_{\mu} J^{\mu}
    \right]
    \, ,
  \end{aligned}
\end{equation}
where
\begin{enumerate*}

\item $w$ and $\bar{w}$ are the column and row supervectors previously introduced in \eqref{eq:u-1-1:glsm:w and w-bar},

\item $\tensor{J}{^{\mu}^j_i}$ is as defined in \eqref{eq:u-1-1:nlsm:current matrix}, and

\item $v_{\mu}$ appearing in the above equation is to be substituted by its on-shell constraints (see \autoref{app:u-1-1:on-shell constraints}).
\end{enumerate*}

Notice that $L^{\mathrm{U}(1|1)}_{\text{chiral}}(v)$ now describes (the bosonic kinetic term of) the Lagrangian of an NLSM, which would correspond to (the negative of) the target space metric.
In particular, the metric would be given by
\begin{equation}
  \label{eq:u-1-1:nlsm:metric}
  \dd{\mathsf{s}}^2
  = \sum_{s = 1}^m \abs{D_{\mu} \phi_s}^2
    + \sum_{t = 1}^n \abs{D_{\mu} \psi_t}^2
  = r g^{\text{SG}}
  \, ,
\end{equation}
where
\begin{equation}
  \label{eq:u-1-1:nlsm:metric:scaled}
  g^{\text{SG}}
  = \frac{1}{r} \left[
    \partial_{\mu} w^1 \partial^{\mu} \bar{w}_1
    - \partial_{\mu} w^{\bar{1}} \partial^{\mu} \bar{w}_{\bar{1}}
  \right]
  - \frac{1}{2r} \left[
    \tensor{v}{_{\mu}^1_1} \tensor{J}{^{\mu}^1_1}
    + \tensor{v}{_{\mu}^1_{\bar{1}}} \tensor{J}{^{\mu}^{\bar{1}}_1}
    - \tensor{v}{_{\mu}^{\bar{1}}_1} \tensor{J}{^{\mu}^1_{\bar{1}}}
    - \tensor{v}{_{\mu}^{\bar{1}}_{\bar{1}}} \tensor{J}{^{\mu}^{\bar{1}}_{\bar{1}}}
  \right]
  \, .
\end{equation}

Note that there are also auxiliary scalar fields $\sigma$ and $\bar{\sigma}$ that appear in $L^{\mathrm{U}(1|1)}_{\text{chiral}}$ as
\begin{equation}
  \label{eq:u-1-1:nlsm:chiral lagrangian:sigma}
  L^{\mathrm{U}(1|1)}_{\text{chiral}}(\sigma)
  = \sum_{s = 1}^m \tensor{\bar{\phi}}{_i^s} \tensor{\acomm{\bar{\sigma}}{\sigma}}{^i_j} \tensor{\phi}{^j_s}
  + \sum_{t = 1}^n \tensor{\bar{\psi}}{_i^t} \tensor{\acomm{\bar{\sigma}}{\sigma}}{^i_j} \tensor{\psi}{^j_t}
  - \sqrt{2} \sum_{s = 1}^m \tensor{(\bar{\eta}_-)}{_i^s} \tensor{\sigma}{^i_j} \tensor{(\eta_+)}{^j_s}
  - \sqrt{2} \sum_{t = 1}^n \tensor{(\bar{\lambda}_-)}{_i^t} \tensor{\sigma}{^i_j} \tensor{(\lambda_+)}{^j_t}
  \, ,
\end{equation}
where (i) $\bar{\eta}_-$/$\eta_+$ and (ii) $\bar{\lambda}_-$/$\lambda_+$ are the spin-$\frac{1}{2}$ fields in the supermultiplets associated to the antichiral/chiral superfields (i) $\bar{\Phi}$/$\Phi$ and (ii) $\bar{\Psi}$/$\Psi$.

In integrating out $\sigma$ and $\bar{\sigma}$, we will get the four-fermi term of the NLSM Lagrangian.
For brevity, we shall not elaborate on this term (and the other kinetic terms from $L^{\mathrm{U}(1|1)}_{\text{chiral}}(\eta, \lambda)$) in the resulting $\mathcal{N} = (2, 2)$ NLSM.

\subtitle{A Super-Grassmannian Target Space}

Recall that $\MSUSY$ of the underlying GLSM is the super-Grassmannian ${Gr}_{1|1}(\mathbb{C}^{m|n})$ of size $r$.
Since $\MSUSY$ is parameterized by the $(\phi, \psi, \bar{\phi}, \bar{\psi})$ fields which define the $(w, \bar{w}, v, J)$ fields appearing in \eqref{eq:u-1-1:nlsm:metric:scaled}, it would mean that $g^{\text{SG}}$ therein is the ${Gr}_{1|1}(\mathbb{C}^{m|n})$ metric of unit size.

In other words, the normalized target space metric of our NLSM is $g^{\text{SG}}$.

\subtitle{A Reduction to Garavuso-Katzarkov-Kreuzer-Noll's Low Energy NLSM of Seki-Sugiyama's $\mathrm{U}(1)$ GLSM}

As mentioned at the end of \autoref{subsec: U(1|1) GLSM and SGr}, our $\mathrm{U}(1|1)$ GLSM ought to reduce to Seki-Sugiyama's $\mathrm{U}(1)$ GLSM when we turn off the $\tensor{\phi}{^{\bar{1}}_s}$ and $\tensor{\psi}{^{\bar{1}}_t}$ fields (and their corresponding bar-indexed partners in the underlying multiplets); in the low energy limit, this ought to be Garavuso-Katzarkov-Kreuzer-Noll's NLSM.
Indeed, turning off those fields would mean that all bar-indexed fields in $g^{\text{SG}}$ \eqref{eq:u-1-1:nlsm:metric:scaled} would vanish, whence it reduces to
\begin{equation}
  \label{eq:u-1-1:nlsm:reduction to seki-sugiyama:metric}
  g^{\text{SG}}
  \quad
  \xrightarrow{\tensor{\phi}{^{\bar{1}}_s}, \tensor{\psi}{^{\bar{1}}_t}, \tensor{\bar{\phi}}{_{\bar{1}}^s}, \tensor{\bar{\psi}}{_{\bar{1}}^t} \rightarrow 0}
  \quad
  \frac{1}{r} \partial_{\mu} w^1 \cdot \partial^{\mu} \bar{w}_1
  - \frac{1}{2r} \tensor{v}{_{\mu}^1_1} \tensor{J}{^{\mu}^1_1}
  \, .
\end{equation}
Moreover, as the gauge group effectively reduces to the subgroup $\mathrm{U}(1) \subset \mathrm{U}(1|1)$ (see \autoref{ft:u-1-1:turn off barred fields}), the only non-vanishing EOM of the auxiliary gauge field $\tensor{v}{_{\mu}^i_j}$ in \eqref{eq:u-1-1:nlsm:eom of v} is that of $\tensor{v}{_{\mu}^1_1}$, from which we obtain its on-shell constraint to be\footnote{%
  \label{ft:u-1-1:nlsm:reason for not using previously derived constraint}%
  As we have turned off the bar-indexed fields, $\tensor{M}{^{\bar{1}}_{\bar{1}}}$ as defined in \eqref{eq:u-1-1:nlsm:shorthands} is 0.
  This means that we cannot simply apply the on-shell constraint of $\tensor{v}{_{\mu}^1_1}$ as computed in \eqref{eq:u-1-1:nlsm:solutions of v}, since it contradicts the necessary assumption that $\tensor{M}{^{\bar{1}}_{\bar{1}}} \neq 0$ required for its derivation (see \autoref{app:u-1-1:on-shell constraints}).
  Therefore, the on-shell constraint of $\tensor{v}{_{\mu}^1_1}$ has to be obtained by solving the now reduced EOM in \eqref{eq:u-1-1:nlsm:eom of v}.
}
\begin{equation}
  \label{eq:u-1-1:nlsm:reduction to seki-sugiyama:eom of v}
  \tensor{v}{^{\mu}^1_1}
  = \frac{\tensor{J}{^{\mu}^1_1}}{
    w^1 \cdot \bar{w}_1
  }
  \, .
\end{equation}
Substituting \eqref{eq:u-1-1:nlsm:reduction to seki-sugiyama:eom of v} in \eqref{eq:u-1-1:nlsm:reduction to seki-sugiyama:metric} while recalling the definition of $\tensor{J}{^{\mu}^1_1}$ in \eqref{eq:u-1-1:nlsm:current matrix}, we get
\begin{equation}
  \label{eq:u-1-1:nlsm:metric of seki-sugiyama}
  g^{\text{SG}}
  \quad
  \xrightarrow{\tensor{\phi}{^{\bar{1}}_s}, \tensor{\psi}{^{\bar{1}}_t}, \tensor{\bar{\phi}}{_{\bar{1}}^s}, \tensor{\bar{\psi}}{_{\bar{1}}^t} \rightarrow 0}
  \quad
  \frac{\partial_{\mu} w^1 \cdot \partial^{\mu} \bar{w}_1}{r}
  - \frac{\left( \partial^{\mu} w^1 \cdot \bar{w}_1 \right)\left( w^1 \cdot \partial_{\mu} \bar{w}_1 \right)}{r w^1 \cdot \bar{w}_1}
  \, ,
\end{equation}
which, when using the reduced orthogonality condition, i.e., $w^1 \cdot \bar{w}_1 = r$, will be the Fubini-Study metric of $\mathbb{CP}^{m-1|n}$ \cite[eq.~(2.6)]{grassi-2007-integ-super}.\footnote{%
  \label{ft:u-1-1:nlsm:disclaimer of notation in grassi}%
  Care needs to be taken when comparing with the metric which appears \textit{loc.~cit.}, as it is written in inhomogeneous coordinates on $\mathbb{CP}^{m|n}$.
}

In other words, we would, at low energy, get an NLSM whose target space is $\mathbb{CP}^{m-1|n}$, as we should (\textit{c.f.}~\cite[$\S$2]{garavuso-2011-super-landau}).

\subtitle{A Reduction to the Low Energy NLSM of the Familiar $\mathrm{U}(1)$ GLSM}

As mentioned at the end of \autoref{subsec: U(1|1) GLSM and SGr}, our $\mathrm{U}(1|1)$ GLSM ought to reduce to the familiar $\mathrm{U}(1)$ GLSM when we
\begin{enumerate*}
  \item turn off the $\phi^{\bar{1}}{}_s$ field (and its corresponding bar-indexed partners in the underlying multiplet), and

  \item turn off the whole Grassmann-odd chiral superfield $\Psi$.
\end{enumerate*}
One can see that this will further reduce $g^{\text{SG}}$ in \eqref{eq:u-1-1:nlsm:metric of seki-sugiyama} to the Fubini-Study metric of $\mathbb{CP}^{m-1}$ \cite[$\S$2]{grassi-2007-integ-super}.

In other words, we would, at low energy, get an NLSM whose target space is $\mathbb{CP}^{m-1}$, as we should (\emph{c.f.}~\cite[ch.~15]{Hori:2003ic}).

\subsection{Quantum Aspects: The GLSM \label{subsec: Quantum Prop U(1|1) GLSM}}

Of the various quantum aspects of the GLSM, we shall now discuss the one relevant to our eventual aim of exploring its applications to mathematics, which is the quantum correction to the FI parameter $r$.

\subtitle{The Effective Theory}

To derive the quantum correction to the FI parameter $r$, we need to study the quantum effective theory at some finite energy scale $\mu > 0$.
The effective theory is obtained by integrating out the modes of the physical fields with frequencies $k$ in the range $\mu \leq \abs{k} \leq \Lambda_{\text{UV}}$, where $\Lambda_{\text{UV}}$ is the UV cutoff scale.\footnote{%
  \label{ft:u-1-1:quantum:energy scale}%
  As mentioned earlier, the spectrum of our $\mathrm{U}(1|1)$ GLSM is unbounded.
  Hence, the energy scale $\mu$ can also take negative values.
  That said, the analysis for negative $\mu$ can actually be understood as an analysis for positive $\mu$ as follows.
  In the usual case of a positive $\mu$ and $\Lambda_{\text{UV}}$, we have the length scale inequality $\flatfrac{1}{\mu} > \flatfrac{1}{\Lambda_{\text{UV}}}$, but this can also be expressed as the length scale inequality $\flatfrac{1}{\Lambda^{\prime}} > \flatfrac{1}{\mu^{\prime}}$, where $\Lambda^{\prime} = -\Lambda_{\text{UV}}$ and $\mu^{\prime} = -\mu$; in other words, a length scale inequality for negative energy/cutoff scale $(\mu^{\prime}, \Lambda^{\prime})$ can also be sensibly interpreted as a length scale inequality for positive energy/cutoff scale $(\mu, \Lambda_{\text{UV}})$, albeit in the opposite sense, i.e., the theory at negative energies would have an IR instead of a UV cutoff.
  Thus, for our purpose, we shall only consider the case where $(\mu, \Lambda_{\text{UV}})$ takes positive values.
  Moreover, as our ultimate goal is to explore the applications to mathematics, and this requires us to single out the scale-invariant case, the energy scale and its sign are ultimately irrelevant.
}

\subtitle{The Quantum $D$-term}

Let us now focus on the $D$-term in the Lagrangian, as that is where the FI parameter $r$ is involved. In particular, we would like to obtain the quantum version of the $D$-term.

The $D$-term in the Lagrangian involving the auxiliary $D$ field can be read off from \eqref{eq:u-1-1:lagrangian:potential} as
\begin{equation}
  \label{eq:u-1-1:quantum:D terms}
  \Str \left[
    \frac{1}{2e^2} D^2
    + D \left(
      \sum_{s = 1}^m \phi_s \bar{\phi}^s
      - \sum_{t = 1}^n \psi_t \bar{\psi}^t
      - r \mathbb{I}_{1|1}
    \right)
  \right]
  \, .
\end{equation}
Its quantum version would involve replacing the $\phi^s \bar{\phi}_s$ and $\psi^t \bar{\psi}_t$ terms by their expectation values, i.e.,
\begin{equation}
  \label{eq:u-1-1:quantum:D terms:expectation values}
  (-1)^{\varsigma(i)} \left[
    \frac{1}{2e^2} \tensor{D}{^i_j} \tensor{D}{^j_i}
    + \tensor{D}{^i_j} \left(
      \sum_{s = 1}^m \expval{\tensor{\phi}{^j_s} \tensor{\bar{\phi}}{_i^s}}
      - \sum_{t = 1}^n \expval{\tensor{\psi}{^j_t} \tensor{\bar{\psi}}{_i^t}}
      - r \tensor{\delta}{^j_i}
    \right)
  \right]
  \, ,
\end{equation}
where we have now written the supertrace explicitly in the $\mathrm{U}(1|1)$ indices.
Notice that the expectation values are not $\expval*{\tensor{\phi}{^j_s}} \expval*{\tensor{\bar{\phi}}{_i^s}}$ and $\expval*{\tensor{\psi}{^j_t}} \expval*{\tensor{\bar{\psi}}{_i^t}}$.
This is because even though the $\expval*{\phi}$'s and $\expval*{\psi}$'s are formally nonvanishing, they are numerically vanishing.\footnote{%
  \label{ft:u-1-1:quantum:global spontaneous breaking}%
  If the $\expval*{\phi}$'s and $\expval*{\psi}$'s were numerically nonvanishing, the continuous global $\mathrm{U}(p|q)$ symmetry that acts on the $\phi$'s and $\psi$'s (via their `$s$' and `$t$' index, respectively) would be spontaneously broken.
  However, a spontaneous breaking of a continuous symmetry is impossible in two dimensions.
  This was also pointed out by Witten in the case of the closely-related non-abelian GLSM~\cite[$\S$4.4]{Witten:1993xi}.
}

\subtitle{Computing the Expectation Values}

Let us now compute the expectation values $\expval*{\tensor{\phi}{^j_s} \tensor{\bar{\phi}}{_i^s}}$ and $\expval*{\tensor{\psi}{^j_t} \tensor{\bar{\psi}}{_i^t}}$.
We shall determine their effective value to a 1-loop approximation.
That is, we treat the $\phi$'s and $\psi$'s as free fields with a mass term that can be read off from the (classical) Lagrangian.
We will do the calculation on Euclidean $\mathbb{R}^2$, by performing a standard Wick rotation of the original Lorentzian signature.

To this end, we first obtain, from the first two terms on the RHS of the Lagrangian $L^{\mathrm{U}(1|1)}$ in \eqref{eq:u-1-1:lagrangian}, the terms in the action which are quadratic in $\phi$ and $\psi$ as\footnote{%
  \label{ft:u-1-1:quantum:action normalization}%
  We have chosen to normalize the action as $S = \frac{1}{2\pi} \int \dd[2]{x} L^{\mathrm{U}(1|1)}$, for convenience.
}
\begin{equation}
  \label{eq:u-1-1:quantum:phi and psi terms}
  \begin{aligned}
    \frac{1}{2\pi} \int \dd[2]{x} \left(
      \sum_{s = 1}^m \left(
        \tensor{\bar{\phi}}{_i^s} D^{\mu} D_{\mu} \tensor{\phi}{^i_s}
        + \tensor{\bar{\phi}}{_i^s} \tensor{\acomm{\bar{\sigma}}{\sigma}}{^i_j} \tensor{\phi}{^j_s}
      \right)
      + \sum_{t = 1}^n \left(
        \tensor{\bar{\psi}}{_i^t} D^{\mu} D_{\mu} \tensor{\psi}{^i_t}
        + \tensor{\bar{\psi}}{_i^t} \tensor{\acomm{\bar{\sigma}}{\sigma}}{^i_j} \tensor{\psi}{^j_t}
      \right)
    \right)
    \, ,
  \end{aligned}
\end{equation}
where the terms involving $\sigma$ and $\bar{\sigma}$ can be understood as mass terms of the free scalar fields.
Notice that the quadratic terms  for the $\phi$ and $\psi$ fields have the same form.
This means that their aforementioned expectation values would also be the same. Thus, let us denote \emph{each} of the $m + n$ expectation values as
\begin{equation}
  \label{eq:u-1-1:quantum:expectation value:notation}
  \expval{\tensor{\phi}{^j_s} \tensor{\bar{\phi}}{_i^s}}
  \equiv \expval{\tensor{\mathcal{O}}{^j_i}}
  \equiv \expval{\tensor{\psi}{^j_t} \tensor{\bar{\psi}}{_i^t}}
  \, .
\end{equation}

Computing the expectation values (for fixed $s$ and $t$) up to 1-loop, we get
\begin{equation}
  \label{eq:u-1-1:quantum:O expectation value:divergent}
  \expval{\tensor{\mathcal{O}}{^j_i}}_{\text{1-loop}}
  = \int \frac{\dd[2]{k}}{(2\pi)^2}
  \frac{2\pi}{k^2 \tensor{\delta}{^j_i} + \tensor{\acomm{\bar{\sigma}}{\sigma}}{^j_i}}
  \, .
\end{equation}
Note however, that this integral is actually divergent. For it to be finite, we would need to introduce an upper limit on the integration by a cutoff scale $\Lambda_{\text{UV}}$. Then, to get the effective theory at energy $\mu$, we would just need to integrate out the range of frequencies $\mu \leq \abs{k} \leq \Lambda_{\text{UV}}$. That is,
\begin{equation}
  \label{eq:u-1-1:quantum:O expectation value}
  \expval{\tensor{\mathcal{O}}{^j_i}}_{\text{1-loop}}
  = \int_{\mu \leq \abs{k} \leq \Lambda_{\text{UV}}}
  \frac{\dd[2]{k}}{(2\pi)^2} \frac{2\pi}{k^2 \tensor{\delta}{^j_i} + \tensor{\acomm{\bar{\sigma}}{\sigma}}{^j_i}}
  \, .
\end{equation}

In a 1-loop approximation, the mass of a free scalar field receives no quantum corrections, i.e., it is effectively unrenormalized.
This means that the masses $\tensor{\acomm{\bar{\sigma}}{\sigma}}{^j_i}$ can also be regarded as free parameters of the theory, whence we can always set $\tensor{\acomm{\bar{\sigma}}{\sigma}}{^j_i} \ll \mu$.
Thus, we would have
\begin{equation}
  \label{eq:u-1-1:quantum:O expectation value:low mass}
  \begin{aligned}
    \expval{\tensor{\mathcal{O}}{^j_i}}_{\text{1-loop}}
    & = \int_{\mu \leq \abs{k} \leq \Lambda_{\text{UV}}}
    \frac{\dd[2]{k}}{(2\pi)^2} \frac{2\pi}{k^2} \tensor{\delta}{^j_i}
    = \log \left( \frac{\Lambda_{\text{UV}}}{\mu} \right) \tensor{\delta}{^j_i}
    \, .
  \end{aligned}
\end{equation}

\subtitle{The Effective FI Parameter $r_{\text{eff}}(\mu)$}

Substituting \eqref{eq:u-1-1:quantum:O expectation value:low mass} in \eqref{eq:u-1-1:quantum:D terms:expectation values} (via \eqref{eq:u-1-1:quantum:expectation value:notation}), we get
\begin{equation}
  \label{eq:u-1-1:quantum:D terms:expectation values:substituted}
  (-1)^{\varsigma(i)} \left[
    \frac{1}{2e^2} \tensor{D}{^i_j} \tensor{D}{^j_i}
    + \tensor{D}{^i_j} \left[
      (m - n) \log \left( \frac{\Lambda_{\text{UV}}}{\mu} \right)
      - r
    \right] \tensor{\delta}{^j_i}
  \right]
  \, .
\end{equation}

Notice that in the continuum limit $\Lambda_{\text{UV}} \to \infty$, the logarithm in \eqref{eq:u-1-1:quantum:D terms:expectation values:substituted} diverges; in turn, so does the effective action.
As such, in order to make the effective action finite, $r$ in \eqref{eq:u-1-1:quantum:D terms:expectation values:substituted} ought to be replaced with
\begin{equation}
  \label{eq:u-1-1:quantum:r constraint}
  r
  = r_{\text{eff}}
  + (m - n) \log \left( \frac{\Lambda_{\text{UV}}}{\mu} \right)
  \, ,
\end{equation}
where $r_{\text{eff}}$ is the effective FI parameter.
Notice that we can also express this as
\begin{equation}
  \label{eq:u-1-1:quantum:r-eff}
  r_{\text{eff}}
  = r
  + (m - n) \log \left( \frac{\mu}{\Lambda_{\text{UV}}} \right)
  \, ,
\end{equation}
so for a fixed $\Lambda_{\text{UV}}$ and $r$, we have
\begin{equation}
  \label{eq:u-1-1:quantum:r-eff:form}
  \saveboxed{eq:u-1-1:quantum:r-eff:form}{
    r_{\text{eff}}(\mu)
    = (m - n) \log \left( \frac{\mu}{\Lambda} \right)
  }
\end{equation}
where the $r$ is being absorbed in the redefinition of $\Lambda_{\text{UV}}$ as $\Lambda$, such that $\Lambda$ can be regarded as a dynamically-generated scale parameter of mass dimension.

\subtitle{Scale-dependence}

If $m \neq n$, from \eqref{eq:u-1-1:quantum:r-eff:form},  we see that the effective FI parameter $r_{\text{eff}} (\mu)$ runs (i.e., RG flows) as we change the energy scale $\mu$.
In particular, the dimensionless parameter $r$ of the classical theory has been replaced by the scale parameter $\Lambda$ of mass dimension in the quantum theory, whence the quantum theory would be scale-dependent.

\subtitle{Restoring Poincar\'{e} Invariance and $\abs{r} \gg 0$}

Note that the introduction of the cutoff scale $\Lambda_{\text{UV}}$ means that the quantum theory is only valid at distances larger than $\flatfrac{1}{\Lambda_{\text{UV}}}$.
This breaks the Poincar\'{e} invariance of the underlying classical theory.

To restore Poincar\'{e} invariance, one therefore needs to take the continuum limit $\Lambda_{\text{UV}} \to \infty$.
One can also regard this as going to the unrenormalized, classical limit.
When $\Lambda_{\text{UV}} \to \infty$, one can see from \eqref{eq:u-1-1:quantum:r-eff} that the magnitude of the classical (or bare) parameter $r$, must be large, i.e., $r \gg 0$ or $r \ll 0$, so that $r_{\text{eff}}$ would have (i) \emph{finite} magnitude and (ii) the \emph{same sign} as its defining classical counterpart $r$, whence it would be physically sensible.

For example, if $(m-n) > 0$, in the limit $\Lambda_{\text{UV}} \to \infty$, we see from \eqref{eq:u-1-1:quantum:r-eff} that we ought to have $r \gg 0$ to have a finite and positive $r_{\text{eff}}$. However, if $(m-n) < 0$, in the limit $\Lambda_{\text{UV}} \to \infty$, we see from \eqref{eq:u-1-1:quantum:r-eff} that we ought to have $r \ll 0$ to have a finite and negative $r_{\text{eff}}$.

In short, from the perspective of the quantum theory, although we considered $\abs{r} > 0$ in the classical theory hitherto, we ought to consider $\abs{r} \gg 0$.

\subtitle{$\MSUSY$ of the Quantum GLSM}

Last but not least, recall that classically, $\MSUSY$ was given by $X_{r>0}$ and $X_{r < 0}$ in \eqref{eq:u-1-1:m-susy:r > 0} and \eqref{eq:u-1-1:m-susy:r < 0} when $r > 0$  and $r< 0$, respectively.

In the quantum theory, when (i) $m > n$ and (ii) $m < n$, $\MSUSY$ would necessarily be given by (i) $X_{r_{\text{eff}} > 0}$ and (ii) $X_{r_{\text{eff}} < 0}$, where (i) $r \gg 0$ and (ii) $r \ll 0$.
In other words, when $m \neq n$ and the quantum GLSM is \emph{not} scale-invariant, $\MSUSY$ can \emph{only} be $X_{r_{\text{eff}} > 0}$ or $X_{r_{\text{eff}} < 0}$ (depending on the values of the positive integers $m$ and $n$).

\subsection{Quantum Aspects: The Low Energy Theory \label{subsec: Quantum Low Energy U(1|1)}}

\subtitle{$\MSUSY$ of the Quantum GLSM at Low Energy}

Recall that in our derivation of $\MSUSY$ at the classical level in \autoref{subsec: U(1|1) GLSM and SGr},  $\abs{r} > 0$. In the quantum theory, when $m\neq n$, in place of $r$, we would have $r_{\text{eff}}(\mu)$, and since we have to consider $\abs{r} \gg 0$ (as explained nearer the end of \autoref{subsec: Quantum Prop U(1|1) GLSM}), one can see from \eqref{eq:u-1-1:quantum:r-eff} that even at low energy $\mu \ll \Lambda_{\text{UV}}$, we will still have $\abs{r_{\text{eff}} (\mu)} > 0$ whence $\MSUSY$ would continue to be well-defined in the quantum GLSM at low energy.

\subtitle{The Quantum Low Energy Limit when $m \neq n$}

Recall that in the classical theory, the low energy limit was defined by $e \to \infty$ whence the gauge fields $v_\mu$ had no kinetic terms which then allowed us to integrate them out.
However, our derivation of the effective FI parameter $r_{\text{eff}} (\mu)$ in \eqref{eq:u-1-1:quantum:r-eff:form} is based on a free-field approximation which actually requires $e \ll 1$.

This conundrum is resolved by noting that the vector multiplet, in particular the gauge fields $v_\mu$, actually acquire a mass proportional to $e \sqrt{\abs{r}}$ in a Higgs mechanism of the classical theory.\footnote{%
  \label{ft:u-1-1:quantum low energy:superhiggs mechanism}%
  The explanations in \cite[$\S$15.4.1]{Hori:2003ic} for the Higgs mechanism in the $\mathrm{U}(1)$ GLSM apply equally well to our $\mathrm{U}(1|1)$ GLSM via \cite{Fayet:1974jb}.
  Also, the appearance of $\abs{r}$ and not $r$ in the mass expression is due to the fact that $\abs{r}$ is ultimately the relevant parameter in our context as can be seen from \eqref{eq:u-1-1:m-susy:r > 0}--\eqref{eq:u-1-1:m-susy:r < 0}.
}
This means that even if we have to restrict ourselves to $e \ll 1$, because $\abs{r} \gg 0$, the gauge fields $v_\mu$ would still be massive, and we can continue to integrate them out at a low energy scale $\mu \ll e \sqrt{\abs{r}}$.

\subtitle{The Quantum GLSM at Low Energy is a Quantum NLSM}

The analysis in \autoref{GLSM to NLSM} had been about the classical theory.
The question therefore, is whether at the quantum level, the GLSM will also reduce, at low energy, to an NLSM on ${Gr}_{1|1}(\mathbb{C}^{m|n})$.
To answer this question, we have to address it separately for the two possible situations: first, when $m = n$, then, when $m \neq n$.

When $m = n$, we can see from \eqref{eq:u-1-1:quantum:r-eff} that the quantum GLSM is scale-invariant with effective FI parameter still being $r$.
Therefore, without any constraints on $r$ (stemming from quantum corrections), the quantum GLSM would, at low energy, freely become a quantum NLSM on the $\MSUSY$ of the GLSM, i.e., ${Gr}_{1|1}(\mathbb{C}^{n|n})$ of size $r$.
In other words, when $m = n$, the answer is ``yes''.

What about when $m \neq n$?
First, note that as explained above, since $\abs{r} \gg 0$, we are free to consider the quantum GLSM at some low energy $\mu$, as $\MSUSY$ will remain well-defined.

Second, note that if at the quantum level, the GLSM does indeed become the NLSM on $\MSUSY$, the RG flow of the GLSM parameter $r$, which corresponds to the size of $\MSUSY$, must therefore match precisely the RG flow of the scale $\mathcal{R}$ of the target space metric $g$ of the NLSM.

Third, note that the effective target space metric $g_{\text{eff}}$ of the quantum NLSM is given by~\cite[eq.~(14.41)]{Hori:2003ic}
\begin{equation}
  \label{eq:u-1-1:quantum low:effective target metric:bare}
  g_{\text{eff}} (\mu)
  = g
  + R_{ic} \log \left( \frac{\mu}{\Lambda_{\text{UV}}} \right)
  \, ,
\end{equation}
where $R_{ic}$ is the Ricci two-tensor of $\MSUSY$, and $\Lambda_{\text{UV}}$ is again a cutoff scale which ensures that we have a finite answer.
For a fixed $\Lambda_{\text{UV}}$ and $g$, we can also express this as
\begin{equation}
  \label{g^SG_eff (u)}
  g_{\text{eff}} (\mu)
  = R_{ic} \log \left( \frac{\mu}{\Lambda} \right)
  \, ,
\end{equation}
where the $g$ is being absorbed in the redefinition of $\Lambda_{\text{UV}}$ as $\Lambda$, such that $\Lambda$ can be regarded as a dynamically-generated scale parameter of mass dimension.

Fourth, via \eqref{g^SG_eff (u)}, and the fact that $R_{ic} = \sdim(m|n) g^{\text{SG}}$,\footnote{%
  \label{ft:u-1-1:quantum low:ricci tensor and fubini-study metric}%
  Since a super-Grassmannian ${Gr}_{p|q}(\mathbb{C}^{m|n}) \cong \mathrm{U}(m|n) / [\mathrm{U}(p|q) \times \mathrm{U}(m-p|n-q)] $ is a homogeneous and symmetric superspace \cite{mohammadi-2018-super-as, zirnbauer-1996-rieman-symmet}, its Ricci tensor (which is independent of the size of ${Gr}_{p|q}(\mathbb{C}^{m|n})$) is proportional to its normalized metric \cite[$\S$2]{babichenko-2007-confor-invar}, where the proportionality factor is $\sdim(m|n)$, i.e., $R_{ic} = \sdim(m|n) g^{\text{SG}}$.
  Examples are
  (i) projective superspaces $\mathbb{CP}^{m-1|n} \cong {Gr}_{1|0}(\mathbb{C}^{m|n})$ \cite[eq.~(2.10)]{grassi-2007-integ-super},
  and (ii) CY super-Grassmannians ${Gr}_{p|q}(\mathbb{C}^{m|m}$) (see \autoref{app:verify cy:supergrassmannians}, and note that $\sdim{m|m} = 0$ is consistent with the fact that vanishing $R_{ic}$ implies a CY supermanifold).
}
where $\sdim(m|n)$ is the superdimension of the ambient $\mathbb{C}^{m|n}$, the metric $g_{\text{eff}} (\mu')$ at a lower energy scale $\mu'$ can be obtained from the metric $g_{\text{eff}} (\mu)$ at the energy scale $\mu$ as
\begin{equation}
  \label{eq:u-1-1:quantum low energy:metric relation}
  g_{\text{eff}} (\mu')
  =  g_{\text{eff}} (\mu)
  +  \sdim(m|n) g^{\text{SG}} \log \left( \frac{\mu'}{\mu} \right)
  \, .
\end{equation}
This can also be written as
\begin{equation}
  \label{r(eff, mu')g^FS}
  {\mathcal{R}}_{\text{eff}} (\mu') g^{\text{SG}}
  =  \left(
    {\mathcal{R}}_{\text{eff}} (\mu)
    + \sdim(m|n) \log \left( \frac{\mu'}{\mu} \right)
  \right) g^{\text{SG}},
\end{equation}
where ${\mathcal{R}}_{\text{eff}}(\nu)$ is the effective scale of the target space metric at an energy scale $\nu$.

Fifth, via \eqref{eq:u-1-1:quantum:r-eff:form}, and the fact that $m - n = \sdim(m|n)$, the GLSM parameter $r_{\text{eff}} (\mu')$ at a lower energy scale $\mu'$ can also be obtained from the GLSM parameter  $r_{\text{eff}} (\mu)$ at the energy scale $\mu$ as
\begin{equation}
  \label{r(eff, mu')}
  r_{\text{eff}} (\mu')
  = r_{\text{eff}} (\mu)
  + \sdim(m|n) \log \left( \frac{\mu'}{\mu} \right)
  \, .
\end{equation}

Finally, by comparing \eqref{r(eff, mu')g^FS} with \eqref{r(eff, mu')}, we find that the RG flow of the GLSM parameter $r$ indeed matches precisely the RG flow of the scale $\mathcal{R}$ of the target space metric of the NLSM!
In other words, when $m \neq n$, the answer is also ``yes''.

In sum, one can conclude that the quantum GLSM will, at low energy, become a quantum NLSM on ${Gr}_{1|1}(\mathbb{C}^{m|n})$ of size $r$ or $r_{\text{eff}}(\mu)$ if $m = n$ or $m \neq n$, respectively.

\subsection{Scale-invariance and Applications to Mathematics \label{subsec: applications to math U(1|1)}}

\subtitle{Scale-invariance of the Quantum GLSM when $m=n$}

Let us consider the case where $m = n$. From \eqref{eq:u-1-1:quantum:r-eff}, we see that $r_{\text{eff}} = r$, i.e., the effective FI parameter is still the classical parameter $r$ which does not change with the energy scale.
In other words, $r$ would persist to be a dimensionless parameter of the quantum theory, whence our quantum GLSM would be scale-invariant.

\subtitle{Scale-invariance, $\MSUSY$ of the Quantum GLSM, and a Geometric Correspondence Between Super-Grassmannians}

In our scale-invariant quantum GLSM where $r$ remains as a parameter of the theory, its $\MSUSY$ will continue to be given by $X_{r>0}$ or $X_{r < 0}$, depending on our \emph{free} choice of $r$.

Clearly, $X_{r>0}$ and $X_{r < 0}$ just describe $\MSUSY$, the target space of its low energy quantum NLSM, in different regions of its parameter space.
Thus, from the perspective of our quantum GLSM, there ought to be a physical correspondence between the quantum NLSM at $r > 0$ and $r<0$, and therefore a geometric correspondence between $X_{r>0}$ and $X_{r < 0}$.

Indeed, when $m = n$, from \eqref{eq:u-1-1:m-susy:r > 0} and \eqref{eq:u-1-1:m-susy:r < 0}, we find that $X_{r>0} \equiv X_{r < 0} = {Gr}_{1|1}(\mathbb{C}^{n|n})$!

\subtitle{Scale-invariance and a Physical Derivation of a Novel CY Condition for Super-Grassmannians}

As our supergroup quantum GLSM is scale-invariant, it would mean that its corresponding low energy quantum NLSM ought to also be scale and therefore conformally-invariant~\cite{papadopoulos2024scaleconformalinvariance2d}. As such, it would have worldsheet $\mathcal{N}=(2,2)$ superconformal symmetry. In particular, the quantum NLSM will have an unbroken $\mathrm{U}(1)_A$ symmetry, which, in turn, means that its target space, i.e., $\MSUSY$, would have vanishing first Chern class~\cite[$\S$13.2.2]{Hori:2003ic} whence it must be CY.

Thus, we have a physical derivation of a \emph{novel} mathematical result that super-Grassmannians of the type ${Gr}_{1|1}(\mathbb{C}^{m|n})$ are CY if and only if $m=n$!

Indeed, if  $\MSUSY = {Gr}_{1|1}(\mathbb{C}^{n|n})$, one can actually mathematically verify (see~\autoref{app:verify cy:supergrassmannians}) that $\MSUSY$ is a CY supermanifold!

This vindicates our application to mathematics of our otherwise nonunitary supergroup GLSM via a study of its space of supersymmetric states.

\section{A \texorpdfstring{$\mathrm{U}(1|1)^N$}{U(1|1)\textasciicircum N} GLSM and Super Torified Varieties}
\label{sec: U(1|1)^N GLSM}

In this section, we will study a $\mathrm{U}(1|1)^N$ GLSM, where there are again Grassmann-even and Grassmann-odd chiral matter superfields that are now non-identically gauge charged.
Repeating the analysis in \autoref{sec: U(1|1) and SGr}, we will show that the space of supersymmetric states of the GLSM corresponds to a super torified variety, and that in the low energy limit, the GLSM will become an NLSM whose target space is the aforementioned super torified variety.
Taking into account the quantum corrections to the FI parameters, we will have an effective scale-dependent FI parameters in the quantum GLSM.
The quantum GLSM will also become a quantum NLSM in the low energy limit.
Lastly, when the theory is scale-invariant, we will be able to (1) obtain a geometric correspondence between super torified varieties, and (2) physically derive a CY condition for super torified varieties that we can also mathematically verify.

\subsection{A \texorpdfstring{$\mathrm{U}(1|1)^N$}{U(1|1)\textasciicircum N} GLSM and its Super Torified Variety \texorpdfstring{$\MSUSY$}{M-SUSY}}
\label{subsec: U(1|1)^N GLSM and Toric SV}

\subtitle{The $\mathrm{U}(1|1)^N$ GLSM}

Let us now generalize our discussion in \autoref{sec: U(1|1) and SGr} to the gauge group $\mathrm{U}(1|1)^N \coloneq \prod_{a=1}^N \mathrm{U}(1|1)_a$, with $m + n$ Grassmann-even and Grassmann-odd chiral matter superfields $\Phi_1, \dots, \Phi_m$ and $\Psi_1, \dots, \Psi_n$, that have \emph{non-negative} integer gauge charges $\tensor{Q}{_a_1}, \dots, \tensor{Q}{_a_m}$ and $\tensor{\mathsf{Q}}{_a_1}, \dots, \tensor{\mathsf{Q}}{_a_n}$, respectively.
We shall assume that the greatest common divisor (gcd) of the gauge charges is 1 for simplicity.

The generalized $\mathcal{N} = (2, 2)$ Lagrangian ought to be given by
\begin{equation}
  \label{eq:u-1-1-N:lagrangian}
  \begin{aligned}
    L^{\mathrm{U}(1|1)^N}
    &= \frac{1}{4} \int \dd[4]{\theta} \left(
      \sum_{s = 1}^m \bar{\Phi}^s e^{2 \sum_{a = 1}^N \tensor{Q}{_a_s} V^a} \Phi_s
      + \sum_{t = 1}^n \bar{\Psi}^t e^{2 \sum_{a = 1}^N \tensor{\mathsf{Q}}{_a_t} V^a} \Psi_t
    \right)
    \\
    &\quad
    - \sum_{a = 1}^N \frac{1}{4e_a^2} \int \dd[4]{\theta} \Str \bar{\Sigma}_a \Sigma^a
    + \left(
      \frac{1}{2\sqrt{2}} \int \dd[2]{\theta} \sum_{a = 1}^N (\mathbf{i} t_a \Str \Sigma^a)
      + c.c. \
    \right)
    \, ,
  \end{aligned}
\end{equation}
where $\frac{1}{e_a^2} \bar{\Sigma}_a \Sigma^a \coloneq \sum_{b = 1}^N \frac{\delta^{ab}}{e_a e_b}\bar{\Sigma}^a \Sigma^b$ with $e_a$ being the gauge coupling associated to $\mathrm{U}(1|1)_a$, and $t_a \coloneq \mathbf{i} r_a + \frac{\vartheta_a}{2\pi}$ is the FI parameter associated to $\mathrm{U}(1|1)_a$.

Note that $L^{\mathrm{U}(1|1)^N}$ has a $\mathrm{U}(1)_A \times \mathrm{U}(1)_V$ axial and vector R-symmetry, where the charges of the $\Phi$'s and $\Psi$'s can be arbitrary, while that of the $\Sigma_a$'s must be $(0,2)$.

\subtitle{The Potential Energy}

The terms in the Lagrangian related to the potential energy (of the scalars) are
\begin{equation}
 \label{eq:u-1-1-N:lagrangian:potential energy}
 \begin{aligned}
    L^{\mathrm{U}(1|1)^N}_{\text{pot}}
    &= \sum_{a = 1}^N \Str \left[
      \frac{1}{2e_a^2} D^a D_a
      + D^a \left(
        \sum_{s = 1}^m \tensor{Q}{_a_s} \phi_s \bar{\phi}^s
        - \sum_{t = 1}^n \tensor{\mathsf{Q}}{_a_t} \psi_t \bar{\psi}^t
        - r_a \mathbb{I}_{1|1}
      \right)
      - \frac{1}{2e_a^2} \comm{\sigma^a}{\bar{\sigma}_a}^2
    \right]
    \\
    &\quad
    - \sum_{a, b = 1}^N \left[
      \sum_{s = 1}^m \tensor{Q}{_a_s} \tensor{Q}{_b_s} \bar{\phi}^s \acomm{\bar{\sigma}^a}{\sigma^b} \phi_s
      + \sum_{t = 1}^n \tensor{\mathsf{Q}}{_a_t} \tensor{\mathsf{Q}}{_b_t} \bar{\psi}^t \acomm{\bar{\sigma}^a}{\sigma^b} \psi_t
    \right]
    \, ,
 \end{aligned}
\end{equation}
where the auxiliary field $D^a$ and scalars $\sigma^a$ and $\bar{\sigma}^a$ are from the vector multiplet of $V^a$.

Via a similar calculation as that to arrive at \eqref{eq:u-1-1:potential} from \eqref{eq:u-1-1:lagrangian:potential}, we obtain, using the EOM of the $D^a$ field, the potential energy
\begin{equation}
  \label{eq:u-1-1-N:potential energy}
  \begin{aligned}
    U^{\mathrm{U}(1|1)^N}_{\text{pot}}
    &= \sum_{a = 1}^N \frac{e_a^2}{2} \sum_{i, j}
    (-1)^{\varsigma(i)}
    \left[
      \sum_{s = 1}^m \tensor{Q}{_a_s} \phi^j{}_s \bar{\phi}_i{}^s
      - \sum_{t = 1}^n \tensor{\mathsf{Q}}{_a_t} \psi^j{}_t \bar{\psi}_i{}^t
      - r_a \tensor{\delta}{^j_i}
    \right]^2
    + \sum_{a = 1}^N \frac{1}{2e_a^2} \Str \comm{\sigma^a}{\bar{\sigma}_a}^2
    \\
    &\quad
    + \sum_{a, b = 1}^N \left[
      \sum_{s = 1}^m \tensor{Q}{_a_s} \tensor{Q}{_b_s} \bar{\phi}^s \acomm{\bar{\sigma}^a}{\sigma^b} \phi_s
      + \sum_{t = 1}^n \tensor{\mathsf{Q}}{_a_t} \tensor{\mathsf{Q}}{_b_t} \bar{\psi}^t \acomm{\bar{\sigma}^a}{\sigma^b} \psi_t
    \right]
    \, .
  \end{aligned}
\end{equation}

\subtitle{A Super Torified Variety \MSUSY}

$\MSUSY$ of the theory is defined by the supersymmetric configuration $U^{\mathrm{U}(1|1)^N}_{\text{pot}} = 0$ modulo gauge transformations.
From \eqref{eq:u-1-1-N:potential energy}, we see that for $\abs{r_a} > 0$, we require $\phi \neq 0$, $\psi \neq 0$, and $\sigma = 0$.
In other words, $\MSUSY$ will be defined by
\begin{equation}
  \label{eq:u-1-1-N:m-susy}
  \saveboxed{eq:u-1-1-N:m-susy}{
    X_{\vec{r}}
    = \flatfrac{
      \left\{
        \sum_{s = 1}^m \tensor{Q}{_a_s} \tensor{\phi}{^j_s} \tensor{\bar{\phi}}{_i^s}
        - \sum_{t = 1}^n \tensor{\mathsf{Q}}{_a_t} \tensor{\psi}{^j_t} \tensor{\bar{\psi}}{_i^t}
        = r_a \tensor{\delta}{^j_i}
        \, \qcomma
        \begin{gathered}
          i, j \in \{1, \bar{1}\}
          \\
          a \in \{1, \dots, N\}
        \end{gathered}
      \right\}
    }{\mathrm{U}(1|1)^N}
  }
\end{equation}
where $\vec{r} \coloneq (r_1, \dots, r_N)$.

One can interpret this as a super torified variety determined by $(\tensor{Q}{_a_s}, \tensor{\mathsf{Q}}{_a_t}, r_a)$, in the sense that its bosonized variety in the familiar $\mathrm{U}(1)^N$ GLSM is a torified variety~\cite{L_pez_Pe_a_2009}, as we shall see shortly.

\subtitle{An Explicit Example: $\MSUSY$ as a Product of Super-Grassmannians}

Meanwhile, as an explicit example of $\MSUSY$, let us consider the case where $N = 2$, i.e., $a \in \{1, 2\}$ and $\mathrm{U}(1|1)^N = \mathrm{U}(1|1)_1 \times \mathrm{U}(1|1)_2$.
For non-negative $(m_1, m_2, n_1, n_2)$ such that $m_1 + m_2 = m$ and $n_1 + n_2 = n$, let us assume that $m_1 + n_1$ chiral matter fields are charged only under $\mathrm{U}(1|1)_1$ with unit charge, while the remaining $m_2 + n_2$ chiral matter fields are charged only under $\mathrm{U}(1|1)_2$ with unit charge.
In other words, the $\mathrm{U}(1|1)_1 \times \mathrm{U}(1|1)_2$ gauge charge assignments are
\begin{equation}
  \label{eq:u-1-1-N:charge assignment}
  \begin{aligned}
    Q_{1, s_1}
    &= 1
    = \mathsf{Q}_{1, t_1}
    \, ,
    &\quad
    Q_{2, s_1}
    &= 0
    = \mathsf{Q}_{2, t_1}
    \, ,
    &\quad
    Q_{1, s_2}
    &= 0
    = \mathsf{Q}_{1, t_2}
    \, ,
    &\quad
    Q_{2, s_2}
    &= 1
    = \mathsf{Q}_{2, t_2}
    \, ,
  \end{aligned}
\end{equation}
where $s_1 \in \{1, \dots, m_1\}$, $s_2 \in \{m_1 + 1, \dots, m\}$, $t_1 \in \{1, \dots, n_1\}$, and $t_2 \in \{n_1 + 1, \dots, n\}$.

Then, $X_{\vec{r}}$, as given in \eqref{eq:u-1-1-N:m-susy}, will become
\begin{equation}
  \label{eq:u-1-1-N:m-susy:eg 1}
  \flatfrac{
    \left\{
      \sum_{s_1 = 1}^{m_1} \tensor{\phi}{^j_{s_1}} \tensor{\bar{\phi}}{_i^{s_1}}
      - \sum_{t_1 = 1}^{n_1} \tensor{\psi}{^j_{t_1}} \tensor{\bar{\psi}}{_i^{t_1}}
      = r_1 \tensor{\delta}{^j_i}
      \, \qcomma i, j \in \{1, \bar{1}\}
    \right\}
  }{\mathrm{U}(1|1)_1}
\end{equation}
and
\begin{equation}
  \label{eq:u-1-1-N:m-susy:eg 2}
  \flatfrac{
    \left\{
      \sum_{s_2 = m_1 + 1}^m \tensor{\phi}{^j_{s_2}} \tensor{\bar{\phi}}{_i^{s_2}}
      - \sum_{t_2 = n_1 + 1}^n \tensor{\psi}{^j_{t_2}} \tensor{\bar{\psi}}{_i^{t_2}}
      = r_2 \tensor{\delta}{^j_i}
      \, \qcomma i, j \in \{1, \bar{1}\}
    \right\}
  }{\mathrm{U}(1|1)_2}.
\end{equation}

Comparing both \eqref{eq:u-1-1-N:m-susy:eg 1} and \eqref{eq:u-1-1-N:m-susy:eg 2} with \eqref{eq:u-1-1:m-susy}, it is clear that if $r_1, r_2 > 0$, $\MSUSY$ in this case will be given by ${Gr}_{1|1}(\mathbb{C}^{m_1|n_1}) \times {Gr}_{1|1}(\mathbb{C}^{m_2|n_2})$, which is a product of super-Grassmannians of sizes $r_1$ and $r_2$.

If however, (i) $r_1, r_2 < 0$; (ii) $r_1 > 0$, $r_2 < 0$; or (iii) $r_1 < 0$, $r_2 > 0$, according to our explanations leading up to \eqref{eq:u-1-1:m-susy:r > 0}--\eqref{eq:u-1-1:m-susy:r < 0}, $\MSUSY$ will be given by (i) ${Gr}_{1|1}(\mathbb{C}^{n_1|m_1}) \times {Gr}_{1|1}(\mathbb{C}^{n_2|m_2})$; (ii) ${Gr}_{1|1}(\mathbb{C}^{m_1|n_1}) \times {Gr}_{1|1}(\mathbb{C}^{n_2|m_2})$; or (iii) ${Gr}_{1|1}(\mathbb{C}^{n_1|m_1}) \times {Gr}_{1|1}(\mathbb{C}^{m_2|n_2})$.

\subtitle{A Reduction to Gu-Zou's $\mathrm{U}(1)^N$ GLSM}

Note that the $\mathrm{U}(1|1)^N$ GLSM ought to reduce to Gu-Zou's $\mathrm{U}(1)^N$ GLSM when we turn off the $\tensor{\phi}{^{\bar{1}}_s}$ and $\tensor{\psi}{^{\bar{1}}_t}$ fields (and their corresponding bar-indexed partners in the underlying multiplets).
Indeed, $\MSUSY$ in \eqref{eq:u-1-1-N:m-susy} will reduce to a set of $N$ equations with $a \in \{1, \dots, N\}$ and $i, j = 1$ therein, while the denominator would effectively reduce to the subgroup $\mathrm{U}(1)^N \subset \mathrm{U}(1|1)^N$ (see \autoref{ft:u-1-1:turn off barred fields}).

Thus, if $r_a > 0$ for all $a$, $\MSUSY$ would reduce to a super toric variety \cite[eq.~(3.6)]{gu-2019-super-local}, as it should (\emph{c.f.}~\cite[$\S$3.1]{gu-2019-super-local}).

\subtitle{A Reduction to Seki-Sugiyama's $\mathrm{U}(1)$ GLSM}

Note that the $\mathrm{U}(1|1)^N$ GLSM ought to reduce to Seki-Sugiyama's $\mathrm{U}(1)$ GLSM when we
\begin{enumerate*}

\item set $N = 1$, and

\item turn off the $\tensor{\phi}{^{\bar{1}}_s}$ and $\tensor{\psi}{^{\bar{1}}_t}$ fields (and their corresponding bar-indexed partners in the underlying multiplets).

\end{enumerate*}
Points (i) and (ii) mean that in this case, $\MSUSY$ in \eqref{eq:u-1-1-N:m-susy} would be defined by a single equation with $a = 1$ and $i, j = 1$, and that the denominator would effectively reduce to the subgroup $\mathrm{U}(1) \subset \mathrm{U}(1|1)$ (see \autoref{ft:u-1-1:turn off barred fields}).

If $r_1 > 0$, such a space
\begin{equation}
  \label{eq:u-1-1-N:m-susy eg 3}
 X_{r_1 > 0}
    = \flatfrac{
      \left\{
        \sum_{s = 1}^m \tensor{Q}{_1_s} \tensor{\phi}{^1_s} \tensor{\bar{\phi}}{_1^s}
        - \sum_{t = 1}^n \tensor{\mathsf{Q}}{_1_t} \tensor{\psi}{^1_t} \tensor{\bar{\psi}}{_1^t}
        = r_1
        \right\}
    }{\mathrm{U}(1)},
  \end{equation}
would be the weighted projective superspace $\WCP^{m-1|n}_{(\vec{Q}_1|\vec{\mathsf{Q}}_1)}$ with non-negative weights $(\vec{Q}_1|\vec{\mathsf{Q}}_1) \coloneq (Q_{1,1}, \dots, Q_{1,m}|\mathsf{Q}_{1,1}, \dots, \mathsf{Q}_{1,n})$.
In other words, if $r_1 > 0$, $\MSUSY$ would reduce to $\WCP^{m-1|n}_{(\vec{Q}_1|\vec{\mathsf{Q}}_1)}$  of size $r_1$, as it should (\textit{c.f.}~\cite[$\S$3]{seki2005gaugedlinearsigmamodel}).

If however, $r_1 < 0$, one can see from \eqref{eq:u-1-1-N:m-susy eg 3} that by multiplying both the LHS and RHS of the numerator by $-1$,
such a space would be a parity-reversed weighted projective superspace $\WCP^{n-1|m}_{(\vec{\mathsf{Q}}_1|\vec{Q}_1)}$ with non-negative weights $(\vec{\mathsf{Q}}_1|\vec{Q}_1)$.
In other words, if $r_1 < 0$, $\MSUSY$ would reduce to $\WCP^{n-1|m}_{(\vec{\mathsf{Q}}_1|\vec{Q}_1)}$ of size $\abs{r_1}$.

\subtitle{A Reduction to the Familiar $\mathrm{U}(1)^N$ GLSM}

Note that the $\mathrm{U}(1|1)^N$ GLSM ought to reduce to the familiar $\mathrm{U}(1)^N$ GLSM when we
\begin{enumerate*}

\item turn off the $\tensor{\phi}{^{\bar{1}}_s}$ field (and its corresponding bar-indexed partners in the underlying multiplet), and

\item turn off the \emph{whole} Grassmann-odd chiral superfield $\Psi$.

\end{enumerate*}
In particular, from \eqref{eq:u-1-1-N:m-susy}, we find that $\MSUSY$ would reduce to \cite[eq.~(15.81)]{Hori:2003ic}, which is a toric variety determined by $(\tensor{Q}{_a_s}, r_a)$, that is thus also a torified variety \cite[e.g.~1.3.3]{L_pez_Pe_a_2009}.

As an explicit example, consider again the case where $N = 2$ with gauge charge assignments in \eqref{eq:u-1-1-N:charge assignment}.
Then, from \eqref{eq:u-1-1-N:m-susy:eg 1} and \eqref{eq:u-1-1-N:m-susy:eg 2}, we find that $\MSUSY$ would reduce to $G_{1}(\mathbb{C}^{m_1})\times G_{1}(\mathbb{C}^{m_2}) \cong \mathbb{CP}^{m_1-1} \times \mathbb{CP}^{m_2-1}$ of sizes $r_1$ and $r_2$, respectively.
Since
\begin{enumerate*}
\item each $\mathbb{CP}^l$ is a toric variety and thus a torified variety \cite[e.g.~1.3.3]{L_pez_Pe_a_2009}, and

\item a cartesian product of torified varieties is also a torified variety \cite[lem.~1.7]{L_pez_Pe_a_2009},

\end{enumerate*}
it would mean that the toric variety $\mathbb{CP}^{m_1-1} \times \mathbb{CP}^{m_2-1}$ is also a torified variety.

In short, the bosonized variety of $\MSUSY$ in the familiar $\mathrm{U}(1)^N$ GLSM would be a toric and therefore torified variety, as it should (\emph{c.f.}~\cite[ch.~15]{Hori:2003ic}), and as claimed earlier.

\subsection{The Low Energy NLSM on a Super Torified Variety \label{subsec: GLSM to NLSM on Toric SV}}

\subtitle{The Low Energy NLSM}

Let us now do as we did in \autoref{GLSM to NLSM}, and consider the $\mathrm{U}(1|1)^N$ GLSM in the low energy limit $e_a \to \infty$ whence the massive fields/modes decouple from the theory and we get an NLSM.

By a similar calculation that led us to \eqref{eq:u-1-1:nlsm:chiral lagrangian:expanded}, we have, in place of \eqref{eq:u-1-1:nlsm:chiral lagrangian:expanded},
\begin{equation}
  \label{eq:u-1-1-N:nlsm:chiral lagrangian:expanded}
  \begin{aligned}
    L^{\mathrm{U}(1|1)^N}_{\text{chiral}}(v)
    &= - \sum_{s = 1}^m \biggl[
      \partial^\mu \bar\phi_i{}^s \partial_\mu\phi^i{}_s
      - \mathbf{i} \sum_{a = 1}^N \tensor{Q}{_a_s} \left(
        \tensor{\bar{\phi}}{_j^s} \tensor{(v^a)}{^{\mu}^j_i} \partial_{\mu} \tensor{\phi}{^i_s}
        - \partial^{\mu} \tensor{\bar{\phi}}{_i^s} \tensor{(v^a)}{_{\mu}^i_j} \tensor{\phi}{^i_s}
      \right)
      \\
      & \qquad \qquad
      + \sum_{a, b = 1}^N \tensor{Q}{_a_s} \tensor{Q}{_b_s} \tensor{\bar{\phi}}{_j^s} \tensor{(v^a)}{^{\mu}^j_i} \tensor{(v^b)}{_{\mu}^i_k} \tensor{\phi}{^k_s}
    \biggr]
    \\
    &\quad
    - \sum_{t = 1}^n \biggl[
      \partial^\mu \bar\psi_i{}^t \partial_\mu\psi^i{}_t
      - \mathbf{i} \sum_{a = 1}^N \tensor{\mathsf{Q}}{_a_t} \left(
        \tensor{\bar{\psi}}{_j^t} \tensor{(v^a)}{^{\mu}^j_i} \partial_{\mu} \tensor{\psi}{^i_t}
        - \partial^{\mu} \tensor{\bar{\psi}}{_i^t} \tensor{(v^a)}{_{\mu}^i_j} \tensor{\psi}{^i_t}
      \right)
      \\
      & \qquad \qquad
      + \sum_{a, b = 1}^N \tensor{\mathsf{Q}}{_a_t} \tensor{\mathsf{Q}}{_b_t} \tensor{\bar{\psi}}{_j^t} \tensor{(v^a)}{^{\mu}^j_i} \tensor{(v^b)}{_{\mu}^i_k} \tensor{\psi}{^k_t}
    \biggr]
    \, .
  \end{aligned}
\end{equation}

The EOM of the auxiliary gauge field $\tensor{(v^a)}{_{\mu}^i_j}$ is
\begin{equation}
  \label{eq:u-1-1-N:nlsm:eom of v}
  \begin{aligned}
    (-1)^{\varsigma(i) + \varsigma(i)\varsigma(j)} \tensor{(J_a)}{^{\mu}^j_i}
    &= (-1)^{\varsigma(i)} \sum_{s = 1}^m \tensor{Q}{_a_s} \tensor{Q}{_b_s} \left(
      \tensor{\bar{\phi}}{_i^s} \tensor{(v^b)}{^{\mu}^j_k} \tensor{\phi}{^k_s}
      + \tensor{\bar{\phi}}{_k^s} \tensor{(v^b)}{^{\mu}^k_i} \tensor{\phi}{^j_s}
    \right)
    \\
    &\quad
    + (-1)^{\varsigma(j)} \sum_{t = 1}^n \tensor{\mathsf{Q}}{_a_t} \tensor{\mathsf{Q}}{_b_t} \left(
      \tensor{\bar{\psi}}{_i^t} \tensor{(v^b)}{^{\mu}^j_k} \tensor{\psi}{^k_t}
      + \tensor{\bar{\psi}}{_k^t} \tensor{(v^b)}{^{\mu}^k_i} \tensor{\psi}{^j_t}
    \right)
    \, ,
  \end{aligned}
\end{equation}
where
\begin{equation}
  \label{eq:u-1-1-N:nlsm:current matrix}
  \tensor{(J_a)}{^{\mu}^j_i}
  \coloneq \mathbf{i} \left[
    \sum_{s = 1}^m \tensor{Q}{_a_s} \left(
      \partial^{\mu} \tensor{\phi}{^j_s} \tensor{\bar{\phi}}{_i^s}
      - \tensor{\phi}{^j_s} \partial^{\mu} \tensor{\bar{\phi}}{_i^s}
    \right)
    - \sum_{t = 1}^n \tensor{\mathsf{Q}}{_a_t} \left(
      \partial^{\mu} \tensor{\psi}{^j_t} \tensor{\bar{\psi}}{_i^t}
      - \tensor{\psi}{^j_t} \partial^{\mu} \tensor{\bar{\psi}}{_i^t}
    \right)
  \right]
  \, .
\end{equation}

By a similar computation that led us to \eqref{eq:u-1-1:nlsm:chiral} via the application of \eqref{eq:u-1-1:nlsm:eom of v} to \eqref{eq:u-1-1:nlsm:chiral lagrangian:expanded}, we compute, via the application of \eqref{eq:u-1-1-N:nlsm:eom of v} to \eqref{eq:u-1-1-N:nlsm:chiral lagrangian:expanded}, that
\begin{equation}
  \label{eq:u-1-1-N:nlsm:chiral}
  L^{\mathrm{U}(1|1)^N}_{\text{chiral}}(v)
  = - \Str \left[
    \partial_{\mu} w \cdot \partial^{\mu} \bar{w}
    - \frac{1}{2} \sum_{a = 1}^N \tensor{(v^a)}{_{\mu}} \tensor{(J_a)}{^{\mu}}
  \right]
  \, ,
\end{equation}
where
\begin{enumerate*}

\item $w$ and $\bar{w}$ are the column and row supervectors previously introduced in \eqref{eq:u-1-1:glsm:w and w-bar}, and

\item $\tensor{(v^a)}{_{\mu}}$ appearing in the above equation is to be substituted by its on-shell constraints via a similar analysis as was done in \autoref{app:u-1-1:on-shell constraints}.

\end{enumerate*}

Notice that $L^{\mathrm{U}(1|1)^N}_{\text{chiral}}(v)$ now describes (the bosonic kinetic term of) the Lagrangian of an NLSM, which corresponds to (the negative of) the target space metric.
In particular, the metric would be given by
\begin{equation}
  \label{eq:u-1-1-N:nlsm:metric}
  \dd{\mathsf{s}}^2
  = \sum_{s = 1}^m \abs{D_{\mu} \phi_s}^2 + \sum_{t = 1}^n \abs{D_{\mu} \psi_t}^2
  = \abs{\vec{r}} g^{\text{STV}}
  \, ,
\end{equation}
where
\begin{equation}
  \label{eq:u-1-1-N:nlsm:metric:scaled}
  \begin{aligned}
    g^{\text{STV}}
    &= \frac{1}{\abs{\vec{r}}} \left[
      \partial_{\mu} w^1 \partial^{\mu} \bar{w}_1
      - \partial_{\mu} w^{\bar{1}} \partial^{\mu} \bar{w}_{\bar{1}}
    \right]
    \\
    &\quad
    - \frac{1}{2 \abs{\vec{r}}} \sum_{a = 1}^N \left[
      \tensor{(v^a)}{_{\mu}^1_1} \tensor{(J_a)}{^{\mu}^1_1}
      + \tensor{(v^a)}{_{\mu}^1_{\bar{1}}} \tensor{(J_a)}{^{\mu}^{\bar{1}}_1}
      - \tensor{(v^a)}{_{\mu}^{\bar{1}}_1} \tensor{(J_a)}{^{\mu}^1_{\bar{1}}}
      - \tensor{(v^a)}{_{\mu}^{\bar{1}}_{\bar{1}}} \tensor{(J_a)}{^{\mu}^{\bar{1}}_{\bar{1}}}
    \right]
    \, ,
  \end{aligned}
\end{equation}
and $\abs{\vec{r}} \in \mathbb{R}^+$ is a function of $\vec{r}$ that corresponds to the size of the target space.

Just as in \autoref{GLSM to NLSM}, we shall not elaborate on the auxiliary scalar fields $(\sigma^a, \bar{\sigma}^a)$, as well as kinetic terms from the spin-$\frac{1}{2}$ fields $(\eta^a, \bar{\eta}^a, \lambda^a, \bar{\lambda}^a)$, which would appear in the resulting $\mathcal{N} = (2, 2)$ NLSM.

\subtitle{A Super Torified Variety Target Space}

Recall that $\MSUSY$ of the underlying GLSM is given by $X_{\vec{r}}$ of size $\abs{\vec{r}}$ in \eqref{eq:u-1-1-N:m-susy}.
Since $\MSUSY$ is parameterized by the $(\phi, \psi, \bar{\phi}, \bar{\psi})$ fields which define the $(w, \bar{w}, v_a, J^a)$ fields appearing in \eqref{eq:u-1-1-N:nlsm:metric} (via \eqref{eq:u-1-1-N:nlsm:metric:scaled}), it would mean that the target space of the corresponding low energy NLSM ought to be the super torified variety $X_{\vec{r}}$, and that $g^{\text{STV}}$ in \eqref{eq:u-1-1-N:nlsm:metric:scaled} is its metric of unit size.

In other words, the normalized target space metric of our NLSM is $g^{\text{STV}}$.

\subtitle{A Reduction to Garavuso-Katzarkov-Kreuzer-Noll's Low Energy NLSM of Seki-Sugiyama's $\mathrm{U}(1)$ GLSM}

As mentioned at the end of \autoref{subsec: U(1|1)^N GLSM and Toric SV}, our $\mathrm{U}(1|1)^N$ GLSM ought to reduce to Seki-Sugiyama's $\mathrm{U}(1)$ GLSM when we
\begin{enumerate*}

\item set $N = 1$, and

\item turn off the $\tensor{\phi}{^{\bar{1}}_s}$ and $\tensor{\psi}{^{\bar{1}}_t}$ fields (and their corresponding bar-indexed partners in the underlying multiplets);
\end{enumerate*}
in the low energy limit, it ought to become Garavuso-Katzarkov-Kreuzer-Noll's NLSM.

Indeed, from points (i) and (ii) applied to the definition of $X_{\vec{r}}$ in \eqref{eq:u-1-1-N:m-susy}, we see that $X_{\vec{r}}$ would become $\WCP^{m-1|n}_{(\vec{Q}_1|\vec{\mathsf{Q}}_1)}$; in other words, we would, at low energy, get an NLSM whose target space is $\WCP^{m-1|n}_{(\vec{Q}_1|\vec{\mathsf{Q}}_1)}$, as we should (\textit{c.f.}~\cite[$\S$2]{garavuso-2011-super-landau}).

\subtitle{A Reduction to the Low Energy NLSM of the Familiar $\mathrm{U}(1)^N$ GLSM}

As mentioned at the end of \autoref{subsec: U(1|1)^N GLSM and Toric SV}, the $\mathrm{U}(1|1)^N$ GLSM ought to reduce to the familiar $\mathrm{U}(1)^N$ GLSM when we
\begin{enumerate*}

\item turn off the $\tensor{\phi}{^{\bar{1}}_s}$ field (and its corresponding bar-indexed partners in the underlying multiplet), and

\item turn off the \emph{whole} Grassmann-odd chiral superfield $\Psi$.

\end{enumerate*}

From the definition of $X_{\vec{r}}$ in \eqref{eq:u-1-1-N:m-susy}, we see that $X_{\vec{r}}$ would become a toric variety determined by $(\tensor{Q}{_a_s}, r_a)$. In other words, we would, at low energy, get an NLSM whose target space is a toric variety determined by $(\tensor{Q}{_a_s}, r_a)$, as we should (\emph{c.f.}~\cite[ch.~15]{Hori:2003ic}).

As an explicit example, consider the case where $N = 2$ with gauge charge assignments in \eqref{eq:u-1-1-N:charge assignment}.
We find, from \eqref{eq:u-1-1-N:m-susy:eg 1} and \eqref{eq:u-1-1-N:m-susy:eg 2}, that $X_{\vec{r}}$ would become $\mathbb{CP}^{m_1 - 1} \times \mathbb{CP}^{m_2 - 1}$, of sizes $r_1$ and $r_2$, respectively.
In other words, we would, at low energy, get an NLSM whose target space is $\mathbb{CP}^{m_1 - 1} \times \mathbb{CP}^{m_2 - 1}$, as we should (\emph{c.f.}~\cite[ch.~15]{Hori:2003ic}).

\subsection{Quantum Aspects: The GLSM \label{subsec:Quantum Prop U(1|1)^N GLSM}}

Of the various quantum aspects of the GLSM, we shall now discuss the one relevant to our eventual aim of exploring its applications to mathematics, which is the quantum correction to the FI parameters $r_a$.

The analysis is similar to that in \autoref{subsec: Quantum Prop U(1|1) GLSM}, so we shall be brief.

\subtitle{The Quantum $D$-term}

The $D$-term in the Lagrangian involving the auxiliary $D_a$ fields can be read off \eqref{eq:u-1-1-N:lagrangian:potential energy} as
\begin{equation}
  \label{eq:u-1-1-N:quantum:D terms}
  \sum_{a = 1}^N \Str \left[
    \frac{1}{2e_a^2} D^a D_a
    + D^a \left(
      \sum_{s = 1}^m \tensor{Q}{_a_s} \phi_s \bar{\phi}^s
      - \sum_{t = 1}^n \tensor{\mathsf{Q}}{_a_t} \psi_t \bar{\psi}^t
      - r_a \mathbb{I}_{1|1}
    \right)
  \right]
  \, .
\end{equation}
Its quantum version would then be
\begin{equation}
  \label{eq:u-1-1-N:quantum:D terms:expectation values}
  \sum_{a = 1}^N (-1)^{\varsigma(i)} \left[
    \frac{1}{2e_a^2} \tensor{(D^a)}{^i_j} \tensor{(D_a)}{^j_i}
    + \tensor{(D^a)}{^i_j} \left(
      \sum_{s = 1}^m \tensor{Q}{_a_s} \expval{\tensor{\phi}{^j_s} \tensor{\bar{\phi}}{_i^s}}
      - \sum_{t = 1}^n \tensor{\mathsf{Q}}{_a_t} \expval{\tensor{\psi}{^j_t} \tensor{\bar{\psi}}{_i^t}}
      - r_a \tensor{\delta}{^j_i}
    \right)
  \right]
  \, .
\end{equation}

\subtitle{Computing the Expectation Values}

Let us now compute the expectation values $\expval{\tensor{\phi}{^j_s} \tensor{\bar{\phi}}{_i^s}}$ and $\expval{\tensor{\psi}{^j_t} \tensor{\bar{\psi}}{_i^t}}$ to a 1-loop (i.e., free-field) approximation.
The calculation will be done on Euclidean $\mathbb{R}^2$, by performing a standard Wick rotation of the original Lorentz signature.

To this end, first, note that from the first two terms on the RHS of the Lagrangian $L^{\mathrm{U}(1|1)}$ in \eqref{eq:u-1-1-N:lagrangian}, the terms which are quadratic in $\phi$ and $\psi$ are
\begin{equation}
  \label{eq:u-1-1-N:quantum:phi and psi terms}
  \begin{aligned}
    \frac{1}{2\pi} \int \dd[2]{x} \Biggl[
      & \sum_{s = 1}^m \Biggl(
        \tensor{\bar{\phi}}{_i^s} D^{\mu} D_{\mu} \tensor{\phi}{^i_s}
        + \sum_{a, b = 1}^N \tensor{Q}{_a_s} \tensor{Q}{_b_s} \tensor{\bar{\phi}}{_i^s} \tensor{(\acomm*{\bar{\sigma}^a}{\sigma^b})}{^i_j} \tensor{\phi}{^j_s}
      \Biggr)
      \\
      & + \sum_{t = 1}^n \Biggl(
        \tensor{\bar{\psi}}{_i^t} D^{\mu} D_{\mu} \tensor{\psi}{^i_t}
        + \sum_{a, b = 1}^N \tensor{\mathsf{Q}}{_a_t} \tensor{\mathsf{Q}}{_b_t} \tensor{\bar{\psi}}{_i^t} \tensor{(\acomm*{\bar{\sigma}^a}{\sigma^b})}{^i_j} \tensor{\psi}{^j_t}
      \Biggr)
    \Biggr]
    \, ,
  \end{aligned}
\end{equation}
where the terms involving $\sigma$ and $\bar{\sigma}$ can be understood as mass terms of the free scalar fields.

Second, by comparing \eqref{eq:u-1-1-N:quantum:phi and psi terms} with \eqref{eq:u-1-1:quantum:phi and psi terms}, we find that in place of \eqref{eq:u-1-1:quantum:O expectation value}, we would have (for fixed $s$)
\begin{equation}
  \label{eq:u-1-1-N:quantum:phi expectation value}
  \expval{\tensor{\phi}{^j_s} \tensor{\bar{\phi}}{_i^s}}_{\text{1-loop}}
  = \int_{\mu \leq \abs{k} \leq \Lambda_{\text{UV}}} \frac{\dd[2]{k}}{(2\pi)^2}
  \frac{2\pi}{k^2 \tensor{\delta}{^j_i} + \tensor{Q}{_a_s} \tensor{Q}{_b_s} \tensor{\acomm{\bar{\sigma}}{\sigma}}{^j_i}}
  \, .
\end{equation}

Third, since in a 1-loop approximation, the masses of the free scalar fields can again be regarded as free parameters of the theory, we can set them such that $\tensor{Q}{_a_s} \tensor{Q}{_b_s} \tensor{\acomm{\bar{\sigma}}{\sigma}}{^j_i} \ll \mu$, whence
\begin{equation}
  \label{eq:u-1-1-N:quantum:phi expectation value:low mass}
  \expval{\tensor{\phi}{^j_s} \tensor{\bar{\phi}}{_i^s}}_{\text{1-loop}}
  = \int_{\mu \leq \abs{k} \leq \Lambda_{\text{UV}}} \frac{\dd[2]{k}}{(2\pi)^2}
  \frac{2\pi}{k^2} \tensor{\delta}{^j_i}
  = \log \left( \frac{\Lambda_{\text{UV}}}{\mu} \right) \tensor{\delta}{^j_i}
  \, .
\end{equation}

Fourth, we can compute $\expval{\tensor{\psi}{^j_t} \tensor{\bar{\psi}}{_i^t}}$ (for fixed $t$) in a similar manner, whence it would give the following similar result,
\begin{equation}
  \label{eq:u-1-1-N:quantum:psi expectation value:low mass}
  \expval{\tensor{\psi}{^j_t} \tensor{\bar{\psi}}{_i^t}}_{\text{1-loop}}
  = \log \left( \frac{\Lambda_{\text{UV}}}{\mu} \right) \tensor{\delta}{^j_i}
  \, .
\end{equation}

\subtitle{The Effective FI Parameters $r_{a, \text{eff}}(\mu)$}

Substituting \eqref{eq:u-1-1-N:quantum:phi expectation value:low mass} and \eqref{eq:u-1-1-N:quantum:psi expectation value:low mass} in \eqref{eq:u-1-1-N:quantum:D terms:expectation values}, we get
\begin{equation}
  \label{eq:u-1-1-N:quantum:D terms:expectation values:substituted}
  \sum_{a = 1}^N (-1)^{\varsigma(i)} \left[
    \frac{1}{2e_a^2} \tensor{(D^a)}{^i_j} \tensor{(D_a)}{^j_i}
    + \tensor{(D^a)}{^i_j} \left[
      \left( \sum_{s = 1}^m \tensor{Q}{_a_s} - \sum_{t = 1}^n \tensor{\mathsf{Q}}{_a_t} \right) \log \left( \frac{\Lambda_{\text{UV}}}{\mu} \right)
      - r_a
    \right] \tensor{\delta}{^j_i}
  \right]
  \, .
\end{equation}

From \eqref{eq:u-1-1-N:quantum:D terms:expectation values:substituted}, one can see that the effective FI parameters which will ensure the finiteness of the effective action in the continuum limit $\Lambda_{\text{UV}} \rightarrow \infty$ can then be written as
\begin{equation}
  \label{eq:u-1-1-N:quantum:r-eff}
  r_{a, \text{eff}} = r_a +  \left( \sum_{s = 1}^m \tensor{Q}{_a_s} - \sum_{t = 1}^n \tensor{\mathsf{Q}}{_a_t} \right) \log \left( \frac{\mu}{\Lambda_{\text{UV}}} \right)
  \, ,
\end{equation}
such that for a fixed $\Lambda_{\text{UV}}$ and $r_a$, we have
\begin{equation}
  \label{eq:u-1-1-N:quantum:r-eff:dynamic}
  r_{a, \text{eff}}(\mu_a) = \left( \sum_{s = 1}^m \tensor{Q}{_a_s} - \sum_{t = 1}^n \tensor{\mathsf{Q}}{_a_t} \right) \log \left( \frac{\mu}{\Lambda_a} \right)
  \, ,
\end{equation}
where the $r_a$ is being absorbed in the redefinition of $\Lambda_{\text{UV}}$ as a dynamically-generated scale parameter $\Lambda_a$ of mass dimension.

Notice that we can go one step further to express $r_{a, \text{eff}} (\mu_a)$ in terms of a common scale parameter $\Lambda$ as\footnote{%
  \label{ft:u-1-1-N:quantum:common scale parameter}%
  Here, we express $\flatfrac{\mu}{\Lambda_a} = \gamma_a (\flatfrac{\mu}{\Lambda})$, and move the factor $\gamma_a$ into the definition of $\tilde{r}_a$.
}
\begin{equation}
  \label{eq:u-1-1-N:quantum:r-eff:common scale parameter}
  \saveboxed{eq:u-1-1-N:quantum:r-eff:common scale parameter}{%
    r_{a, \text{eff}}(\mu) = \left( \sum_{s = 1}^m \tensor{Q}{_a_s} - \sum_{t = 1}^n \tensor{\mathsf{Q}}{_a_t} \right) \log \left( \frac{\mu}{\Lambda} \right) + \tilde{r}_a
  }
\end{equation}

\subtitle{Scale-dependence}

If $\sum_{s = 1}^m \tensor{Q}{_a_s} \neq \sum_{t = 1}^n \tensor{\mathsf{Q}}{_a_t}$, from \eqref{eq:u-1-1-N:quantum:r-eff:common scale parameter}, we see that the effective FI parameter $r_{a, \text{eff}}(\mu)$ runs (i.e., RG flows) as we change the energy scale $\mu$.
In particular, the dimensionless parameters $r_a$ of the classical theory have been replaced by the dimensionless parameters $\tilde{r}_a$ and the scale parameter $\Lambda$ of mass dimension in the quantum theory, whence the quantum theory would be scale-dependent.

\subtitle{Restoring Poincar\'{e} Invariance and $\abs{r_a} \gg 0$}

Note that the introduction of the cutoff scale $\Lambda_{\text{UV}}$ means that the quantum theory is only valid at distances larger than $\flatfrac{1}{\Lambda_{\text{UV}}}$.
This breaks the Poincar\'{e} invariance of the underlying classical theory.

To restore Poincar\'{e} invariance, one therefore needs to take the continuum limit $\Lambda_{\text{UV}} \to \infty$.
One can also regard this as going to the unrenormalized, classical limit.
When $\Lambda_{\text{UV}} \to \infty$, one can see from \eqref{eq:u-1-1-N:quantum:r-eff} that the magnitude of the classical (or bare) parameters $r_a$, must be large, i.e., $r_a \gg 0$ or $r_a \ll 0$, so that $r_{a, \text{eff}}$ would have (i) \emph{finite} magnitude and (ii) the \emph{same sign} as its defining classical counterparts $r_a$, whence it would be physically sensible.

For example, if $\sum_{s = 1}^m \tensor{Q}{_a_s} - \sum_{t = 1}^n \tensor{\mathsf{Q}}{_a_t} > 0$, in the limit $\Lambda_{\text{UV}} \rightarrow \infty$, we see from \eqref{eq:u-1-1-N:quantum:r-eff} that we ought to have $r_a \gg 0$ to have a finite and positive $r_{a, \text{eff}}$.
On the other hand, if $\sum_{s = 1}^m \tensor{Q}{_a_s} - \sum_{t = 1}^n \tensor{\mathsf{Q}}{_a_t} < 0$, in the limit $\Lambda_{\text{UV}} \rightarrow \infty$, we see from \eqref{eq:u-1-1-N:quantum:r-eff} that we ought to have $r_a \ll 0$ to have a finite and negative $r_{a, \text{eff}}$.

In short, from the perspective of the quantum theory, although we considered $\abs{r_a} > 0$ in the classical theory hitherto, we ought to consider $\abs{r_a} \gg 0$.

\subtitle{$\MSUSY$ of the Quantum GLSM}

Last but not least, recall that classically, $\MSUSY$ would be given by $X_{\vec{r}}$ in \eqref{eq:u-1-1-N:m-susy}, where $\vec{r} \coloneq (r_1, \dots, r_N)$.

In the quantum theory, when
\begin{enumerate*}
\item $\sum_{s = 1}^m \tensor{Q}{_a_s} - \sum_{t = 1}^n \tensor{\mathsf{Q}}{_a_t} > 0$ and

\item $\sum_{s = 1}^m \tensor{Q}{_a_s} - \sum_{t = 1}^n \tensor{\mathsf{Q}}{_a_t} < 0$
\end{enumerate*}
whence we ought to have
\begin{enumerate*}
\item $r_a \gg 0$ and

\item $r_a \ll 0$,
\end{enumerate*}
$\MSUSY$ would be given by $X_{\vec{r}_{\text{eff}}(\mu)}$ where $\vec{r}_{\text{eff}}(\mu) \coloneq \bigl(r_{1, \text{eff}}(\mu), \dots, r_{N, \text{eff}}(\mu)\bigr)$ would necessarily have the same sign as $\vec{r}$ component-wise.

In other words, when $\sum_{s = 1}^m \tensor{Q}{_a_s} \neq \sum_{t = 1}^n \tensor{\mathsf{Q}}{_a_t}$ and the quantum GLSM is \emph{not} scale-invariant, $\MSUSY$ can \emph{only} be in \emph{one} specific phase (as determined by the values of $\sum_{s = 1}^m \tensor{Q}{_a_s}$ and $\sum_{t = 1}^n \tensor{\mathsf{Q}}{_a_t}$).

\subsection{Quantum Aspects: The Low Energy Theory \label{subsec: U(1|1)^N GLSM at low energy}}

\subtitle{$\MSUSY$ of the Quantum GLSM at Low Energy}

Recall that in our derivation of $\MSUSY$ at the classical level in \autoref{subsec: U(1|1)^N GLSM and Toric SV}, $\abs{r_a} > 0$.
In the quantum theory, when $\sum_{s = 1}^m \tensor{Q}{_a_s} \neq \sum_{t = 1}^n \tensor{\mathsf{Q}}{_a_t}$, in place of $r_a$, we have $r_{a, \text{eff}}(\mu)$, and since we have to consider $\abs{r_a} \gg 0$ (as explained at the end of \autoref{subsec:Quantum Prop U(1|1)^N GLSM}), one can see from \eqref{eq:u-1-1-N:quantum:r-eff} that even at low energy $\mu \ll \Lambda_{\text{UV}}$, we will still have $\abs{r_{a, \text{eff}}(\mu)} > 0$ whence $\MSUSY$ would continue to be well-defined at low energy.

\subtitle{The Quantum Low Energy Limit when $\sum_{s = 1}^m \tensor{Q}{_a_s} \neq \sum_{t = 1}^n \tensor{\mathsf{Q}}{_a_t}$}

Recall that in the classical theory, the low energy limit was defined by $e_a \to \infty$ whence the $N$ gauge fields $(v^a)_\mu$ had no kinetic terms which then allowed us to integrate them out.
However, our derivation of the effective FI parameters $r_{a, \text{eff}}(\mu)$ in \eqref{eq:u-1-1-N:quantum:r-eff:common scale parameter} is based on a free-field approximation which actually requires $e_a \ll 1$.

By adapting the relevant arguments in \autoref{subsec: Quantum Low Energy U(1|1)} to the present case, we find that there is no conundrum at all.
That is, even if we have to restrict ourselves to $e_a \ll 1$, because $\abs{r_a} \gg 0$, the $N$ gauge fields $(v^a)_\mu$, whose acquire mass is proportional to $e_a \sqrt{\abs{r_a}}$, would still be massive, whence we can continue to integrate them out at a low energy scale $\mu \ll e_a \sqrt{\abs{r_a}}$.

\subtitle{The Quantum GLSM at Low Energy is a Quantum NLSM}

The analysis in \autoref{subsec: GLSM to NLSM on Toric SV} had been about the classical theory.
The question therefore, is whether at the quantum level, the GLSM will also become, at low energy, an NLSM on a super torified variety.
To answer this question, we have to address it separately for the two possible situations: first when $\sum_{s = 1}^m \tensor{Q}{_a_s} = \sum_{t = 1}^n \tensor{\mathsf{Q}}{_a_t}$, then, when $\sum_{s = 1}^m \tensor{Q}{_a_s} \neq \sum_{t = 1}^n \tensor{\mathsf{Q}}{_a_t}$.

When $\sum_{s = 1}^m \tensor{Q}{_a_s} = \sum_{t = 1}^n \tensor{\mathsf{Q}}{_a_t}$, we can see from \eqref{eq:u-1-1-N:quantum:r-eff} that the quantum GLSM is scale-invariant with effective FI parameters still being $r_a$.
Therefore, without any constraints on $r_a$ (stemming from quantum corrections), the quantum GLSM would, at low energy, freely become a quantum NLSM on the $\MSUSY$ of the GLSM, i.e., $X_{\vec{r}}$.
In other words, when $\sum_{s = 1}^m \tensor{Q}{_a_s} = \sum_{t = 1}^n \tensor{\mathsf{Q}}{_a_t}$, the answer is ``yes''.

When $\sum_{s = 1}^m \tensor{Q}{_a_s} \neq \sum_{t = 1}^n \tensor{\mathsf{Q}}{_a_t}$, we have seen in \autoref{subsec:Quantum Prop U(1|1)^N GLSM} that the quantum GLSM is not scale-invariant with effective FI parameters $r_{a, \text{eff} }(\mu)$ that are constrained.
That said, by adapting the analysis at the end of \autoref{subsec: Quantum Low Energy U(1|1)} to the present case, we find that the quantum GLSM would nonetheless become, at low energy, a quantum NLSM on $X_{\vec{r}_{\text{eff}}(\mu)}$.
In other words, when $\sum_{s = 1}^m \tensor{Q}{_a_s} \neq \sum_{t = 1}^n \tensor{\mathsf{Q}}{_a_t}$, the answer is also ``yes''.

In sum, one can conclude that the quantum GLSM will also become, at low energy, a quantum NLSM on $X_{\vec{r}}$ or $X_{\vec{r}_{\text{eff}}(\mu)}$ if $\sum_{s = 1}^m \tensor{Q}{_a_s} = \sum_{t = 1}^n \tensor{\mathsf{Q}}{_a_t}$ or $\sum_{s = 1}^m \tensor{Q}{_a_s} \neq \sum_{t = 1}^n \tensor{\mathsf{Q}}{_a_t}$, respectively.

\subsection{Scale-invariance and Applications to Mathematics \label{subsec: applications to math U(1|1)^N}}

\subtitle{Scale-invariance of the Quantum GLSM when $\sum_{s = 1}^m \tensor{Q}{_a_s} = \sum_{t = 1}^n \tensor{\mathsf{Q}}{_a_t}$}

Let us consider the case where $\sum_{s = 1}^m \tensor{Q}{_a_s} = \sum_{t = 1}^n \tensor{\mathsf{Q}}{_a_t}$.
From \eqref{eq:u-1-1-N:quantum:r-eff}, we see that $r_{a, \text{eff}} = r_a$, i.e., the effective FI parameters are still the classical parameters $r_a$ which do not change with the energy scale.
In other words, the $r_a$'s would persist to be  dimensionless parameters of the quantum theory, whence our quantum GLSM would be scale-invariant.

\subtitle{Scale-invariance, $\MSUSY$ of the Quantum GLSM, and a Geometric Correspondence Between Super Torified Varieties}

In our scale-invariant quantum GLSM where $\vec{r} \coloneq (r_1, \dots, r_N)$ remain as parameters of the theory, its $\MSUSY$ would continue to be given by $X_{\vec{r}}$, which depends on our \emph{free} choice of $\vec{r}$.

Clearly, an $X_{\vec{r}}$ where the components of $\vec{r}$ are
\begin{enumerate*}
\item all positive,

\item all negative, or

\item mixed,

\end{enumerate*}
simply describe $\MSUSY$, the target space of its low energy quantum NLSM, in different regions of its parameter space.
Thus, from the perspective of the quantum GLSM, there ought to be a physical correspondence amongst the quantum NLSMs in these different regions of its parameter space,  and therefore a geometric correspondence amongst these different $\MSUSY$ spaces.

In other words, there ought to be an equivalence between $X_{\vec{r}}$ and $X_{\vec{r^{\prime}}}$ for all $\vec{r^{\prime}}$, where $\vec{r^{\prime}}$ has a different sign permutation component-wise compared to $\vec{r}$.

\subtitle{An Explicit Example of Our Claimed Geometric Correspondence Between Super Torified Varieties}

As an explicit example of our above-claimed geometric correspondence between super torified varieties, consider a gauge charge assignment that is similar to that in \eqref{eq:u-1-1-N:charge assignment} -- that is, for non-negative $(m_a, n_a)$ such that $\sum_{a = 1}^N m_a = m$ and $\sum_{a = 1}^N n_a = n$, assume that $m_a + n_a$ chiral matter fields are charged only under $\mathrm{U}(1|1)_a$ with unit charge.
In other words, let their $\mathrm{U}(1|1)^N = \prod_{a = 1}^N \mathrm{U}(1|1)_a$ gauge charges be
\begin{equation}
  \label{eq:u-1-1-N:quantum:charge assignment}
  \tensor{Q}{_a_{s_i}}
  =
  \begin{cases}
    1 \qif i = a \, ,
    \\
    0 \quad \text{otherwise} \, ,
  \end{cases}
  \qquad
  \tensor{\mathsf{Q}}{_a_{t_i}}
  =
  \begin{cases}
    1 \qif i = a \, ,
    \\
    0 \quad \text{otherwise} \, ,
  \end{cases}
\end{equation}
where
(i) $s_1 \in \{1, \dots, m_1\}$, $s_2 \in \{m_1 + 1, \dots, m_1 + m_2\}$, $\dots$, $s_N \in \{\sum_{a = 1}^{N -1} m_a + 1, \dots, m\}$;
and (ii) $t_1 \in \{1, \dots, n_1\}$, $t_2 \in \{n_1 + 1, \dots, n_1 + n_2\}$, $\dots$, $t_N \in \{\sum_{a = 1}^{N -1} n_a + 1, \dots, n\}$.
One can see that $X_{\vec{r}}$ will be given by a product of super-Grassmannians $\prod_{a = 1}^N {Gr}_{1|1}(\mathbb{C}^{m_a|n_a})$.

When $\sum_{s = 1}^m \tensor{Q}{_a_s} = \sum_{t = 1}^n \tensor{\mathsf{Q}}{_a_t}$, or equivalently, when $m_a = n_a$, we have $X_{\vec{r}} = \prod_{a = 1}^N X_{r_a} = \prod_{a = 1}^N {Gr}_{1|1}(\mathbb{C}^{n_a|n_a})$.
Since from \eqref{eq:u-1-1:m-susy:r > 0} and \eqref{eq:u-1-1:m-susy:r < 0}, we see that $X_{r_a > 0} \equiv X_{r_a < 0}$, one can also write $X_{\vec{r}} \equiv \prod_{a = 1}^N X_{r_a \gtrless 0} = X_{\vec{r^{\prime}}}$.
Hence, we indeed have $X_{\vec{r}} \equiv X_{\vec{r^{\prime}}}$ for all $\vec{r^{\prime}}$!

\subtitle{Scale-invariance  and a Physical Derivation of a Novel CY Condition for Super Torified Varieties}

As our supergroup quantum GLSM is scale-invariant, it would mean that its corresponding low-energy quantum NLSM ought to also be scale and therefore conformally-invariant \cite{papadopoulos2024scaleconformalinvariance2d}; in turn, according to our explanation in \autoref{subsec: applications to math U(1|1)}, this would mean that the target space of the quantum NLSM, i.e., $\MSUSY$, must be CY.

Recall also that for our $\mathrm{U}(1|1)^N$ GLSM, $\MSUSY$ is a super torified variety determined by $(\tensor{Q}{_a_s}, \tensor{\mathsf{Q}}{_a_t}, r_a)$.

Thus, we have a physical derivation of a \emph{novel} mathematical result that super torified varieties determined by $(\tensor{Q}{_a_s}, \tensor{\mathsf{Q}}{_a_t}, r_a)$ are CY if and only if $\sum_{s = 1}^m \tensor{Q}{_a_s} = \sum_{t = 1}^n \tensor{\mathsf{Q}}{_a_t}$! \footnote{%
  \label{ft:yau theorem}%
  For CY super torified varieties, one can define a vanishing Ricci two-tensor when the fermionic dimension $n > 1$ \cite{rocek-2005-calab-yau-super, zhou-2005-ricci-flat-super}, whence one can certainly interpret it as a differential-geometric supermanifold.
}

\subtitle{A Mathematical Verification of Our Novel CY Condition for Super Torified Varieties}

Let us now mathematically verify our novel CY condition for super torified varieties via two explicit examples.

For the first example, let us consider the gauge charge assignments in \eqref{eq:u-1-1-N:quantum:charge assignment}, whence $\MSUSY$, a super torified variety, will be given by a product of super-Grassmannians $\prod_{a = 1}^N {Gr}_{1|1}(\mathbb{C}^{m_a|n_a})$.

If $\sum_{s = 1}^m \tensor{Q}{_a_s} = \sum_{t = 1}^n \tensor{\mathsf{Q}}{_a_t}$, or equivalently, if $m_a = n_a$, each ${Gr}_{1|1}(\mathbb{C}^{m_a|n_a})$ multiplicand would be CY (see \autoref{app:verify cy:supergrassmannians}). Consequently, $\prod_{a = 1}^N {Gr}_{1|1}(\mathbb{C}^{m_a|n_a})$ would also be CY (see \autoref{ft:verify cy:cicy in product of SG:cy condition for product of SG}). That is, $\MSUSY$ will indeed be CY!

If, however, $\sum_{s = 1}^m \tensor{Q}{_a_s} \neq \sum_{t = 1}^n \tensor{\mathsf{Q}}{_a_t}$, or equivalently, if $m_a \neq n_a$, each ${Gr}_{1|1}(\mathbb{C}^{m_a|n_a})$ multiplicand would not be CY (see \autoref{app:verify cy:supergrassmannians}). Consequently, $\prod_{a = 1}^N {Gr}_{1|1}(\mathbb{C}^{m_a|n_a})$ would not be CY either (see \autoref{ft:verify cy:cicy in product of SG:cy condition for product of SG}).
Then, $\MSUSY$ will not be CY.

For the second example, let us
\begin{enumerate*}[label=(\roman*)]
\item set $N = 1$, i.e., $a = 1$; and

\item turn off the $\tensor{\phi}{^{\bar{1}}_s}$ and $\tensor{\psi}{^{\bar{1}}_t}$ fields (and their corresponding bar-indexed partners in the underlying multiplets).
\end{enumerate*}
Then, as we saw towards the end of \autoref{subsec: U(1|1)^N GLSM and Toric SV}, $\MSUSY$, a super torified variety, will be given by the weighted projective superspace $\WCP^{m-1|n}_{(\vec{Q}_1|\vec{\mathsf{Q}}_1)}$.

If $\sum_{s = 1}^m Q_{1,s} = \sum_{t = 1}^n \mathsf{Q}_{1,t}$, $\WCP^{m-1|n}_{(\vec{Q}_1|\vec{\mathsf{Q}}_1)}$ ought to be CY. Indeed, one can actually mathematically verify (see \autoref{app:verify cy:wcp}) that $\MSUSY$ is a CY supermanifold if and only if $\sum_{s = 1}^m Q_{1,s} = \sum_{t = 1}^n \mathsf{Q}_{1,t}$!

This again vindicates our application to mathematics of our otherwise nonunitary supergroup GLSM via a study of its space of supersymmetric states.

\section{Phases of a \texorpdfstring{$\mathrm{U}(1|1)$}{U(1|1)} GLSM with Superpotential: Complete Intersection of Hypersupersurfaces NLSMs and Supergauged LG Orbifolds}
\label{sec: Phases of U(1|1) with W}

In this section, we will study a $\mathrm{U}(1|1)$ GLSM with a superpotential, and show that its space of supersymmetric states, in its different phases, corresponds either to
\begin{enumerate*}
  \item a complete intersection of even and odd hypersurfaces in a super-Grassmannian, or

  \item a weighted projective superspace.
\end{enumerate*}
Going to the low energy limit of the GLSM, we will show that it will become either
\begin{enumerate*}
  \item an NLSM on the aforementioned complete intersection of even and odd hypersurfaces in a super-Grassmannian, or

  \item a supergauged LG orbifold on a supervector bundle over the aforementioned weighted projective superspace.
\end{enumerate*}
Taking into account the quantum corrections to the FI parameter, we will have an effective scale-dependent FI parameter in the quantum GLSM.
The quantum GLSM will also become either a quantum NLSM or quantum supergauged LG orbifold in the low energy limit.
Lastly, when the theory is scale-invariant, we will be able to (1) physically derive a CY condition for a complete intersection of even and odd hypersurfaces in a super-Grassmannian that we can also mathematically verify, (2) obtain a super-Grassmannian/supergroup generalization of the  CY/LG correspondence for complete intersections established in mathematics by Clader~\cite{clader2013landauginzburgcalabiyaucorrespondencecompleteintersections} and Zhao~\cite{zhao2019landauginzburgcalabiyaucorrespondencecompleteintersection}, and (3) obtain a parity-reversed mirror version of the aforementioned generalization.

\subsection{A \texorpdfstring{$\mathrm{U}(1|1)$}{U(1|1)} GLSM with Superpotential and its \texorpdfstring{\MSUSY}{M-SUSY}}
\label{subsection: U(1|1) with W_super}

\subtitle{The $\mathrm{U}(1|1)$ GLSM with a Superpotential}

We would now like to construct a $\mathrm{U}(1|1)$ GLSM whose $\MSUSY$ can, for some value of the FI parameter, be a complete intersection of hypersurfaces.

To this end, we ought to consider the $\mathrm{U}(1|1)$ model in \autoref{sec: U(1|1) and SGr} but with a superpotential.
Specifically, we should generalize the Lagrangian in \eqref{eq:u-1-1:lagrangian} to include $l$ extra chiral matter superfields $P_{\kappa_\alpha}^\alpha$, where $\alpha \in \{1, \dots, l\}$ and the superfields can be Grassmann-even or Grassmann-odd as indicated by $\kappa_\alpha = +$ or $-$, that transform either in the superdeterminantal (when $\kappa_{\alpha} = +$) or anti-superdeterminantal (when $\kappa_{\alpha} = -$) representation of $\mathrm{U}(1|1)$, such that the $\mathcal{N}=(2, 2)$ Lagrangian would now be given by
\begin{equation}
  \label{eq:u-1-1-W:lagrangian}
  \begin{aligned}
    L^{\mathrm{U}(1|1)}_{W_{\text{super}}}
    &= \frac{1}{4}\int \dd[4]{\theta} \left(
      \sum_{s = 1}^m \bar{\Phi}^s e^{2V} \Phi_s
      + \sum_{t = 1}^n \bar{\Psi}^t e^{2V} \Psi_t
    \right)
    - \frac{1}{4e^2} \int \dd[4]{\theta} \Str \bar{\Sigma} \Sigma
    \\
    &\quad
    + \frac{1}{2\sqrt{2}} \left(
      \int \dd[2]{\theta} (\mathbf{i}t\ \text{Str}\, \Sigma)
      + c.c.
    \right)
    \\
    &\quad
    + \frac{1}{4} \int \dd[4]{\theta} \sum_{\alpha=1}^l \bar{P}^{\kappa_\alpha}_\alpha e^{2Q^{\kappa_\alpha}_\alpha \Str V} P_{\kappa_\alpha}^\alpha
    - \frac{1}{4} \left(
      \int \dd[2]{\theta} W^{\mathrm{U}(1|1)}_{\text{super}} \left( P, \Phi, \Psi \right)
      + h.c.
    \right)
    \, .
  \end{aligned}
\end{equation}
Here, $Q^{\kappa_\alpha}_\alpha$ is the (c-number) gauge charge of $P_{\kappa_\alpha}^\alpha$, and
\begin{equation}
  \label{eq:u-1-1 with W:W potential}
  \saveboxed{eq:u-1-1 with W:W potential}{
    W^{\mathrm{U}(1|1)}_{\text{super}} \left( P, \Phi, \Psi \right)
    = \sum_{\alpha = 1}^l P_{\kappa_\alpha}^\alpha G^{\kappa_\alpha}_\alpha \left( Y(\Phi, \Psi) \right)
  }
\end{equation}
is a $\mathrm{U}(1|1)$-invariant superpotential.

Note that $ L^{\mathrm{U}(1|1)}_{W_{\text{super}}}$ has a $\mathrm{U}(1)_A \times \mathrm{U}(1)_V$ axial and vector R-symmetry, where the charges of the $\Phi$'s, $\Psi$'s, and $P$'s are $(0, *)$, while the charges of $\Sigma$ and $ W^{\mathrm{U}(1|1)}_{\text{super}}$ must be $(0,2)$.

\subtitle{More about $G^{\kappa_\alpha}_\alpha$ in $W^{\mathrm{U}(1|1)}_{\text{super}}$}

In $W^{\mathrm{U}(1|1)}_{\text{super}}$, $G^{\kappa_\alpha}_\alpha$ is, for all our purposes, a homogeneous polynomial of (integer) degree $q^{\kappa_\alpha}_\alpha > 0$ in $Y$.

As $G^{\kappa_\alpha}_\alpha$ depends on the Grassmann-even and Grassmann-odd $\phi$ and $\psi$ fields that will realize the ambient supermanifold, $G^{\kappa_\alpha}_\alpha$ can either be a Grassmann-even or Grassmann-odd polynomial whose zero-locus will realize a degree-$q^{\kappa_\alpha}_\alpha$ hypersurface which is either even or odd, as indicated by ${\kappa_\alpha} =+$ or $-$, respectively.
And since $W^{\mathrm{U}(1|1)}_{\text{super}}$, like with any term in a Lagrangian, must necessarily be Grassmann-even, it would mean that $G^{\kappa_\alpha}_\alpha$ must have the same parity as $P_{\kappa_\alpha}^\alpha$, as indicated by their identical `$\kappa_\alpha$' super(sub)scripts.
It also turns out that the `$\kappa_\alpha$' superscript of $G^{\kappa_\alpha}_\alpha$ actually determines the type of $Y$'s defining $G^{\kappa_\alpha}_\alpha$.
In particular, when $\kappa_{\alpha} = +$ (resp. $-$), the $Y$'s will transform in the superdeterminantal (resp. anti-superdeterminantal) representation of $\mathrm{U}(1|1)$.
We will see how these $Y$'s are defined very shortly.

Moreover, because $W^{\mathrm{U}(1|1)}_{\text{super}}$ ought to be $\mathrm{U}(1|1)$-invariant, it must mean that $G^{\kappa_\alpha}_\alpha$ ought to transform oppositely to $P_{\kappa_\alpha}^\alpha$.
Specifically, since by definition, $P_{\kappa_\alpha}^\alpha \to (\sdet M)^{\kappa_{\alpha} Q^{\kappa_\alpha}_\alpha} P_{\kappa_\alpha}^\alpha$ under a $\mathrm{U}(1|1)$ gauge transformation, where $M \in \mathrm{U}(1|1)$, it must mean that correspondingly, $G^{\kappa_\alpha}_\alpha \to (\sdet M)^{- \kappa_{\alpha} Q^{\kappa_\alpha}_\alpha} G^{\kappa_\alpha}_\alpha$ under a $\mathrm{U}(1|1)$ gauge transformation.

\subtitle{$G^{\kappa_{\alpha}}_{\alpha}$ Defining a Complete Intersection}

As we are interested in studying a complete intersection of hypersurfaces, it will be useful for us to consider our $G^{\kappa_{\alpha}}_{\alpha}$ polynomials to be transverse in the sense that the condition
\begin{equation}
  \label{eq:U(1|1) with W:glsm:transversality}
  \pdv{G^{\kappa_{\alpha}}_{\alpha}}{Y(\Phi, \Psi)}
  = 0
\end{equation}
holds if and only if $Y(\Phi, \Psi)$ is zero.

\subtitle{General Expression for $G^+_\alpha$}

Altogether, some thought would reveal that a general expression for the Grassmann-even $G^+_\alpha$ ought to be given by
\begin{equation}
  \label{eq:u-1-1 with W:glsm:G+ alpha}
  \saveboxed{eq:u-1-1 with W:glsm:G+ alpha}{
    \begin{aligned}
      G^+_\alpha \left( Y(\Phi, \Psi) \right)
      &\coloneq \sum_{s,t} {b^+_{(\alpha)}}^{s_1 t_1 \dots s_{q^+_\alpha} t_{q^+_\alpha}}
      \; Y^+_{s_1 t_1}(\Phi, \Psi) \cdots Y^+_{s_{q^+_\alpha} t_{q^+_\alpha}}(\Phi, \Psi)
      \\
      &\quad
      + \sum_{t, t'} {c^+_{(\alpha)}}^{t_1 t'_1 \dots t_{q^+_{\alpha}} t'_{q^+_{\alpha}}}
      \; Y^+_{t_1 t'_1}(\Psi) \cdots Y^+_{t_{q^+_{\alpha}} t'_{q^+_{\alpha}}}(\Psi)
    \end{aligned}
  }
\end{equation}
where $1\leq s_a \leq m$ and $ 1 \leq t_a < t'_a \leq n$.
The $Y^+$ fields in \eqref{eq:u-1-1 with W:glsm:G+ alpha} are defined as follows.
\begin{equation}
  \label{eq:u-1-1 with W:glsm:Y+ even}
  \saveboxed{eq:u-1-1 with W:glsm:Y+ even}{
    Y^+_{s_a t_a}(\Phi, \Psi)
    \coloneq \frac{%
      \tensor{\Phi}{^1_{s_a}} - \tensor{\Psi}{^1_{t_a}} \left[ \tensor{\Psi}{^{\bar{1}}_{t_a}} \right]^{-1} \tensor{\Phi}{^{\bar{1}}_{s_a}}
    }{\tensor{\Psi}{^{\bar{1}}_{t_a}}}
  }
\end{equation}
is Grassmann-even, whence $b^+_{(\alpha)}$ is a Grassmann-even coefficient which is symmetric under the swop $(s_a t_a) \leftrightarrow (s_c t_c)$, and
\begin{equation}
  \label{eq:u-1-1 with W:glsm:Y+ odd}
  \saveboxed{eq:u-1-1 with W:glsm:Y+ odd}{
    Y^+_{t_a t'_a}(\Psi)
    \coloneq \frac{%
      \tensor{\Psi}{^1_{t_a}} - \tensor{\Psi}{^1_{t'_a}} \left[ \tensor{\Psi}{^{\bar{1}}_{t'_a}} \right]^{-1} \tensor{\Psi}{^{\bar{1}}_{s_a}}
    }{\tensor{\Psi}{^{\bar{1}}_{t'_a}}}
  }
\end{equation}
is Grassmann-odd, whence $c^+_{(\alpha)}$ is a Grassmann-even (resp. Grassmann-odd) coefficient if the integer $q^+_{\alpha}$ is even (resp. odd), that is antisymmetric under the swop $(t_a t'_a) \leftrightarrow (t_c t'_c)$.

While the lowest component of the chiral matter superfields $(\tensor{\Phi}{^{1}_{s_a}}, \tensor{\Phi}{^{\bar{1}}_{s_a}}, \tensor{\Psi}{^1_{t_a}}, \tensor{\Psi}{^{\bar{1}}_{t_a}})$ are the holomorphic spacetime scalar fields $(\tensor{\phi}{^{1}_{s_a}}, \tensor{\phi}{^{\bar{1}}_{s_a}}, \tensor{\psi}{^1_{t_a}}, \tensor{\psi}{^{\bar{1}}_{t_a}})$ which represent coordinates on a $1|1$-plane in $\mathbb{C}^{m|n}$, i.e., the homogeneous coordinates of $Gr_{1|1}(\mathbb{C}^{m|n})$, it turns out that the lowest component of $Y^+_{s_a t_a}(\Phi, \Psi)$ and $Y^+_{t_a t'_a}(\Psi)$ actually correspond to the $+1$-weighted Grassmann-even and Grassmann-odd super-Pl\"{u}cker coordinates of $Gr_{1|1}(\mathbb{C}^{m|n})$, respectively~\cite{shemyakova-2022-super-pluec} (see \autoref{app:plucker}).

Being $+1$-weighted, it would mean that under a $\mathrm{U}(1|1)$ gauge transformation, $Y^+ \rightarrow (\sdet M) \cdot Y^+$ (as derived in \eqref{eq:plucker:Gr 1-1 2-3:tfm of Y} and \eqref{eq:plucker:Gr p-q m-n:tfm of Y}).
In turn, \eqref{eq:u-1-1 with W:glsm:G+ alpha} would mean that under a $\mathrm{U}(1|1)$ gauge transformation, we indeed have $G^+_\alpha\to (\sdet M)^{- Q^+_\alpha} \cdot G^+_\alpha$, where
\begin{equation}
  \label{eq: Q^+_alpha}
  Q^+_\alpha = - q^+_\alpha
\end{equation}
is the gauge charge of the Grassmann-even $P^+_\alpha$.

\subtitle{General Expression for $G^-_\alpha$}

Again, some thought would reveal that a general expression for the Grassmann-odd $G^-_\alpha$ ought to be given by
\begin{equation}
  \label{eq:u-1-1 with W:glsm:G- alpha}
  \saveboxed{eq:u-1-1 with W:glsm:G- alpha}{
    \begin{aligned}
      G^-_\alpha \left( Y(\Phi, \Psi) \right)
      &\coloneq \sum_{s,t} {b^-_{(\alpha)}}^{s_1 t_1 \dots s_{q^-_\alpha} t_{q^-_\alpha}}
      \; Y^-_{s_1 t_1}(\Phi, \Psi) \cdots Y^-_{s_{q^-_\alpha} t_{q^-_\alpha}}(\Phi, \Psi)
      \\
      &\quad
      + \sum_{s, s'} {c^-_{(\alpha)}}^{s_1 s'_1 \dots s_{q^-_{\alpha}} s'_{q^-_{\alpha}}}
      \; Y^-_{s_1 s'_1}(\Phi) \cdots Y^-_{s_{q^-_{\alpha}} t'_{q^-_{\alpha}}}(\Phi)
    \end{aligned}
  }
\end{equation}
where $1 \leq s_a < s'_a \leq m$;  $1 \leq t_a < t'_a \leq n$.
The $Y^-$ fields in \eqref{eq:u-1-1 with W:glsm:G- alpha} are defined as follows.
\begin{equation}
  \label{eq:u-1-1 with W:glsm:Y- even}
  \saveboxed{eq:u-1-1 with W:glsm:Y- even}{
    Y_{s_a t_a}^-(\Phi, \Psi)
    \coloneq \frac{\tensor{\Psi}{^{\bar{1}}_{t_a}} -  \tensor{\Phi}{^{\bar{1}}_{s_a}} \left[ \tensor{\Phi}{^1_{s_a}} \right]^{-1} \tensor{\Psi}{^1_{t_a}}}{\tensor{\Phi}{^1_{s_a}}}
  }
\end{equation}
is Grassmann-even, whence $b^-_{(\alpha)}$ is a Grassmann-odd coefficient which is symmetric under the swop $(s_a t_a) \leftrightarrow (s_c t_c)$;
and
\begin{equation}
  \label{eq:u-1-1 with W:glsm:Y- odd}
  \saveboxed{eq:u-1-1 with W:glsm:Y- odd}{
    Y^-_{s_a s'_a}(\Phi)
    \coloneq \frac{\tensor{\Phi}{^{\bar{1}}_{s'_a}} - \tensor{\Phi}{^{\bar{1}}_{s_a}} \left[ \tensor{\Phi}{^1_{s_a}} \right]^{-1} \tensor{\Phi}{^1_{s'_a}}}{\tensor{\Phi}{^1_{s_a}}}
  }
\end{equation}
is Grassmann-odd, whence $c^-_{(\alpha)}$ is a Grassmann-odd (resp. Grassmann-even) coefficient if the integer $q^-_{\alpha}$ is even (resp. odd), that is antisymmetric under the swop $(s_a s'_a) \leftrightarrow (s_c s'_c)$.

It turns out that the lowest component of $Y^-_{s_a t_a}(\Phi, \Psi)$ and $Y^-_{s_a s'_a}(\Phi)$ correspond to the $-1$-weighted Grassmann-even and Grassmann-odd super-Pl\"{u}cker coordinates of $Gr_{1|1}(\mathbb{C}^{m|n})$, respectively~\cite{shemyakova-2022-super-pluec} (see \autoref{app:plucker}).

Being $-1$-weighted, it would mean that under a $\mathrm{U}(1|1)$ gauge transformation, $Y^- \rightarrow (\sdet M)^{-1} \cdot Y^- $ (as derived in \eqref{eq:plucker:Gr 1-1 2-3:tfm of Y} and \eqref{eq:plucker:Gr p-q m-n:tfm of Y}).
In turn, \eqref{eq:u-1-1 with W:glsm:G- alpha} would mean that under a $\mathrm{U}(1|1)$ gauge transformation, we indeed have $G^-_\alpha\to (\sdet M)^{Q^-_\alpha} \cdot G^-_\alpha$, where
\begin{equation}
  \label{eq: Q^-_alpha}
  Q^-_\alpha = - q^-_\alpha
\end{equation}
is the gauge charge of the Grassmann-odd $P^-_\alpha$.

\subtitle{The Effective Representation of $P_{\kappa_\alpha}^\alpha$}

Since $P_{\kappa_\alpha}^\alpha$ transforms in the (anti-)superdeterminantal $\sdet^{\kappa_{\alpha} Q^{\kappa_\alpha}_\alpha}$ representation under the $\mathrm{U}(1|1)$ gauge symmetry, it will be charged under the even abelian subgroup $\mathrm{U}(1) \subset \mathrm{U}(1|1)$ with charge $Q^{\kappa_\alpha}_\alpha$.\footnote{%
  \label{P_alpha U(1) transform}%
  A straightforward calculation reveals that the superdeterminant of any unitary supergroup will always be a phase factor that is thus determined by a purely diagonal subgroup.
  In other words, if
  \begin{equation*}
    g =
    \mqty( e^{\mathbf{i} \varphi} & & 0 \\ 0 & & 1 )
  \end{equation*}
  is an element of the purely diagonal subgroup $\mathrm{U}(1) \times \mathrm{U}(1) \subset \mathrm{U}(1|1)$, the $\mathrm{U}(1|1)$ gauge transformation of $P_{\kappa_\alpha}^\alpha$ can also be computed as $ (\sdet M)^{\kappa_{\alpha} Q^{\kappa_\alpha}_\alpha} P_{\kappa_\alpha}^\alpha = (\sdet g)^{\kappa_{\alpha} Q^{\kappa_\alpha}_\alpha} P_{\kappa_\alpha}^\alpha = e^{\mathbf{i} \kappa_{\alpha} Q^{\kappa_\alpha}_\alpha \varphi} P_{\kappa_\alpha}^\alpha$.
  Hence, $P_{\kappa_\alpha}^\alpha$ is effectively charged under the even abelian subgroup $\mathrm{U}(1) \subset \mathrm{U}(1|1)$ with charge $Q^{\kappa_\alpha}_\alpha$.
}
Thus, unlike the $\Phi_s$ and $\Psi_t$ superfields, the $P_{\kappa_\alpha}^\alpha$ superfields will not have (implicit) $1, \bar 1$ indices.

\subtitle{$W^{\mathrm{U}(1|1)}_{\text{super}}$ is Quasi-Homogeneous with $\mathrm{U}_A \times \mathrm{U}(1)_V$ R-charge $(0, 2)$}

Notice also that $W^{\mathrm{U}(1|1)}_{\text{super}}$ is quasi-homogeneous is the sense that there exists some $k_{\ast}$ and $k_{\alpha}^{\pm}$ whereby
\begin{equation}
  \label{eq:U(1|1) with W:W R-charges}
  W^{\mathrm{U}(1|1)}_{\text{super}}
  \longrightarrow e^{2 \mathbf{i}\gamma} \,  W^{\mathrm{U}(1|1)}_{\text{super}}
\end{equation}
when
\begin{equation}
  \label{eq:U(1|1) with W:Y and P R-charges}
  Y_{*}^{\pm}
  \longrightarrow e^{\mathbf{i} \gamma k_{\ast}} \, Y_{*}^{\pm}
  \qand
  P_\pm^\alpha
  \longrightarrow e^{\mathbf{i} \gamma k_{\alpha}^{\pm}} \, P_\pm^{\alpha}
  \, .
\end{equation}
This is as claimed in \autoref{ft:u-1-1:no central charge}.

In fact, $W^{\mathrm{U}(1|1)}_{\text{super}}$ has $\mathrm{U}(1)_A \times \mathrm{U}(1)_V$ R-charge $(0, 2)$, as follows.

First, since the $\mathrm{U}(1)_A$ charge of the $\Phi$'s, $\Psi$'s, and $P_{\pm}$'s is 0, we find that $W^{\mathrm{U}(1|1)}_{\text{super}}$, as given by \eqref{eq:u-1-1 with W:W potential}, will also have $\mathrm{U}(1)_A$ charge equal to 0. Therefore, $W^{\mathrm{U}(1|1)}_{\text{super}}$ will not transform under the $\mathrm{U}(1)_A$-symmetry, whence \eqref{eq:U(1|1) with W:W R-charges}--\eqref{eq:U(1|1) with W:Y and P R-charges} would denote a $\mathrm{U}(1)_V$ transformation.

Second, we find that the $G^+$ and $G^-$ polynomials, as given by \eqref{eq:u-1-1 with W:glsm:G+ alpha} and \eqref{eq:u-1-1 with W:glsm:G- alpha}, have $\mathrm{U}(1)_V$ charge $q^+_{\alpha} k_*$ and $q^-_{\alpha} k_*$, respectively.

Third, since the $\mathrm{U}(1)_V$ charge of the $\Phi$'s, $\Psi$'s, and $P_{\pm}$'s is arbitrary, we can choose the charges $k_*$ and $k_{\alpha}^{\pm}$ such that $k_{\alpha}^+ + q^+_{\alpha} k_* = k_{\alpha}^- + q^-_{\alpha} k_* = 2$.

Altogether, we find that $W^{\mathrm{U}(1|1)}_{\text{super}}$, as given by \eqref{eq:u-1-1 with W:W potential}, will transform under the $\mathrm{U}(1)_V$-symmetry with charge $2$, as required.

\subtitle{The Extra Spacetime Scalars from $P_{\kappa_\alpha}^\alpha$}

Just as we denoted by $\phi_s$ and $\psi_t$ the spacetime scalars from the chiral matter superfields $\Phi_s$ and $\Psi_t$, we shall denote by $p_{\kappa_\alpha}^\alpha$ the spacetime scalars from the chiral matter superfields $P_{\kappa_\alpha}^\alpha$.

\subtitle{The Potential Energy}

After eliminating the auxiliary field, and noting \eqref{eq:u-1-1 with W:glsm:G+ alpha}, \eqref{eq: Q^+_alpha}, \eqref{eq:u-1-1 with W:glsm:G- alpha}, and \eqref{eq: Q^-_alpha}, the potential energy (of the scalars) becomes
\begin{equation}
  \label{ppot}
  \begin{aligned}
    &U^{\mathrm{U}(1|1)}_{\text{pot}, {W_{\text{super}}}}
    \\
    &= \frac{e^2}{2} \sum_{i, j}
    (-1)^{\varsigma(i)}
    \left[
      \sum_{s = 1}^m \tensor{\phi}{^j_s} \tensor{\bar{\phi}}{_i^s}
      - \sum_{t = 1}^n \tensor{\psi}{^j_t} \tensor{\bar{\psi}}{_i^t}
      - \left(
        \sum^{l_+}_{\alpha^+ = 1} q^+_{\alpha^+}  p_+^{\alpha^+} \bar{p}^{\, +}_{\alpha^+}
        - \sum^{l_-}_{\alpha^- = 1} q^-_{\alpha^-} p_-^{\alpha^-} \bar{p}^{\, -}_{\alpha^-}
        + r
      \right) \tensor{\delta}{^j_i}
    \right]^2
    \\
    &\quad
    +\left(
      \sum^{l_+}_{\alpha^+ = 1} \abs{G^+_{\alpha^+}}^2
      + \sum_{a,b} \abs{ \sum^{l_+}_{\alpha^+ = 1} p_+^{\alpha^+} \pdv{G^+_{\alpha^+}}{Y^+_{a b}} }^2
    \right)
    +
    \left(
      \sum^{l_-}_{\alpha^- = 1} \abs{G^-_{\alpha^-}}^2
      + \sum_{a, b} \abs{ \sum^{l_-}_{\alpha^- = 1} p_-^{\alpha^-} \pdv{G^-_{\alpha^-}}{Y^-_{a b}} }^2
    \right)
    \\
    &\quad
    + \frac{1}{2e^2} \Str \comm{\sigma}{\bar{\sigma}}^2
    + \sum_{s = 1}^m \bar{\phi}^s \acomm{\bar{\sigma}}{\sigma} \phi_s
    + \sum_{t = 1}^n \bar{\psi}^t \acomm{\bar{\sigma}}{\sigma} \psi_t
    \\
    &\quad
    + 2 \abs{\Str \sigma}^2 \left(
      \sum^{l_+}_{\alpha^+ = 1} (q^+_{\alpha^+})^2 \, \bar{p}^{\, +}_\alpha p_+^\alpha
      + \sum^{l_-}_{\alpha^- = 1} (q^-_{\alpha^-})^2 \, \bar{p}^{\, -}_\alpha p_-^\alpha
    \right)
    \, ,
  \end{aligned}
\end{equation}
where
\begin{enumerate*}
  \item $(G^+_1, G^+_2, \dots, G^+_{l_+})$ are the $l_+$ number of Grassmann-even polynomials given in \eqref{eq:u-1-1 with W:glsm:G+ alpha} of respective degrees $(q^+_1, q^+_2, \dots, q^+_{l_+})$;

  \item $(G^-_1, G^-_2, \dots, G^-_{l_-})$ are the $l_-$ number of Grassmann-odd polynomials given in \eqref{eq:u-1-1 with W:glsm:G- alpha} of respective degrees $(q^-_1,q^-_2, \dots, q^-_{l_-})$;

  \item $l_+ + l_- = l$; and

  \item the summation index $\{a, b\}$ runs over all $Y^+$ and $Y^-$ field indices.
\end{enumerate*}

\subtitle{$\MSUSY$}

$\MSUSY$ of the theory is defined by the supersymmetric configuration $U^{\mathrm{U}(1|1)}_{\text{pot}, W_{\text{super}}} = 0$ modulo gauge transformations.
From \eqref{ppot}, we see that generally for $\abs{r} \neq 0$, this requires $\sigma = 0$, and the vanishing of the first, second and third term on the RHS of \eqref{ppot}.
In other words, $\MSUSY$ will be defined by
\begin{equation}
  \label{eq:u-1-1 with W:glsm:M-SUSY:geometric space}
  \saveboxed{eq:u-1-1 with W:glsm:M-SUSY:geometric space}{
    \flatfrac{
      \left\{
        \sum_{s = 1}^m \tensor{\phi}{^j_s} \tensor{\bar{\phi}}{_i^s}
        - \sum_{t = 1}^n \tensor{\psi}{^j_t} \tensor{\bar{\psi}}{_i^t}
        = \left(
          \sum^{l_+}_{\alpha^+ = 1} q^+_{\alpha^+} \abs{p_+^{\alpha^+}}^2
          - \sum^{l_-}_{\alpha^- = 1} q^-_{\alpha^-} \abs{p_-^{\alpha^-}}^2
          + r
        \right) \tensor{\delta}{^j_i}
        \, , \quad
        i, j \in \{1, \bar 1\}
      \right\}
    }{\mathrm{U}(1|1)}
  }
\end{equation}
and
\begin{equation}
  \label{eq:u-1-1 with W:glsm:M-SUSY:toric hypersurface +}
  \saveboxed{eq:u-1-1 with W:glsm:M-SUSY:toric hypersurface +}{
    \sum^{l_+}_{\alpha^+ = 1} \abs{G^+_{\alpha^+}}^2
    + \sum_{a,b} \abs{ \sum^{l_+}_{\alpha^+ = 1} p_+^{\alpha^+} \pdv{G^+_{\alpha^+}}{Y^+_{a b}} }^2
    = 0
  }
\end{equation}
\begin{equation}
  \label{eq:u-1-1 with W:glsm:M-SUSY:toric hypersurface -}
  \saveboxed{eq:u-1-1 with W:glsm:M-SUSY:toric hypersurface -}{
    \sum^{l_-}_{\alpha^- = 1} \abs{G^-_{\alpha^-}}^2
    + \sum_{a, b} \abs{ \sum^{l_-}_{\alpha^- = 1} p_-^{\alpha^-} \pdv{G^-_{\alpha^-}}{Y^-_{a b}} }^2
    = 0
  }
\end{equation}

\subtitle{A Reduction to Seki-Sugiyama's $\mathrm{U}(1)$ GLSM with Superpotential}

Note that the $\mathrm{U}(1|1)$ GLSM with superpotential ought to reduce to Seki-Sugiyama's $\mathrm{U}(1)$ GLSM with superpotential when we turn off the $\tensor{\Phi}{^{\bar{1}}_s}$ and $\tensor{\Psi}{^{\bar{1}}_t}$ superfields.
Indeed, this would mean that the lowest component of the $Y^{\pm}(\Phi^1, \Psi^1)$ superfields, which corresponded to super-Pl\"{u}cker coordinates of $Gr_{1|1}(\mathbb{C}^{m|n})$, will now correspond to super-Pl\"{u}cker coordinates of $Gr_{1|0}(\mathbb{C}^{m|n}) \cong \mathbb{CP}^{m-1|n}$.
Consequently, \eqref{eq:u-1-1 with W:glsm:M-SUSY:geometric space} would reduce to an equation defining a supervector bundle over $\mathbb{CP}^{m-1|n}$.

That said, as the super-Pl\"{u}cker coordinates of $\mathbb{CP}^{m-1|n}$ are simply the coordinates on $\mathbb{CP}^{m-1|n}$ (see \autoref{sec:plucker:Gr p-0 m-n}), i.e., they are the lowest component of either the $\tensor{\Phi}{^1}$ or $\tensor{\Psi}{^1}$ superfields, we have a natural decomposition of $Y(\Phi^1, \Psi^1)$ into $Y(\Phi^1) = \Phi^1$ and $Y(\Psi^1) = \Psi^1$.
Hence, the polynomials $G^{\pm}_{\alpha^{\pm}}(Y(\Phi^1, \Psi^1))$ would become homogeneous polynomials in the $\tensor{\Phi}{^1}$'s and/or $\tensor{\Psi}{^1}$'s.

In other words, $\MSUSY$ would reduce to a supervector bundle over $\mathbb{CP}^{m-1|n}$ and \eqref{eq:u-1-1 with W:glsm:M-SUSY:toric hypersurface +}--\eqref{eq:u-1-1 with W:glsm:M-SUSY:toric hypersurface -}, where $G^{\pm}_{\alpha^{\pm}}$ is as described above (\emph{c.f.}~\cite[$\S$3]{seki2005gaugedlinearsigmamodel}).

\subtitle{A Reduction to the Familiar $\mathrm{U}(1)$ GLSM with Superpotential}

Note that the $\mathrm{U}(1|1)$ GLSM with superpotential ought to further reduce to the familiar $\mathrm{U}(1)$ GLSM with superpotential when we turn off
\begin{enumerate*}
  \item the $\tensor{\phi}{^{\bar{1}}_s}$ field (and its corresponding bar-indexed partners in the underlying multiplet),

  \item the \emph{whole} Grassmann-odd chiral superfield $\Psi$,

  and
  \item the \emph{whole} Grassmann-odd chiral superfield $P^{\alpha^-}_-$.
\end{enumerate*}
Indeed, points (i) and (ii) would mean that the lowest component of the $Y^{\pm}(\Phi, 0)$ superfields will now correspond to the (regular) Pl\"{u}cker coordinates of $Gr_{1|0}(\mathbb{C}^{m|0}) \cong Gr_1(\mathbb{C}^m) \cong \mathbb{CP}^{m-1}$; and point (iii) would mean that $G^-_{\alpha^-}$ would vanish.
Consequently, \eqref{eq:u-1-1 with W:glsm:M-SUSY:geometric space} would reduce to an equation defining a vector bundle over $\mathbb{CP}^{m-1}$.

That said, as the (regular) Pl\"{u}cker coordinates of $\mathbb{CP}^{m-1}$ are simply the coordinates on $\mathbb{CP}^{m-1}$, i.e., they are the lowest component of the $\Phi^1$ superfields, we have $Y(\Phi^1, 0) = \Phi^1$.
Hence, the homogeneous polynomial $G^+_{\alpha^+}(Y(\Phi^1, 0)) = \sum_s {b^+_{(\alpha^+)}}^{s_1 \dots s_{\alpha^+}} \Phi^1_{s_1} \cdots \Phi^1_{s_{q_{\alpha}^+}}$ is a polynomial of degree $q^+_{\alpha}$ in $\Phi^1$.

In other words, $\MSUSY$ would reduce to a vector bundle over $\mathbb{CP}^{m-1}$ and \eqref{eq:u-1-1 with W:glsm:M-SUSY:toric hypersurface +}, where $G^+_{\alpha^+}$ is as described above, as it should (\emph{c.f.}~\cite[$\S$15.4.3]{Hori:2003ic}).

\subsection{Phases of the Model: NLSMs on a Complete Intersection of Hypersurfaces in a Super-Grassmannian and Supergauged LG Orbifolds on a Supervector Bundle over Weighted Projective Superspace}
\label{subsec: Phases U(1|1) with W}

To satisfy \eqref{eq:u-1-1 with W:glsm:M-SUSY:geometric space} and \eqref{eq:u-1-1 with W:glsm:M-SUSY:toric hypersurface +}--\eqref{eq:u-1-1 with W:glsm:M-SUSY:toric hypersurface -}, we need to consider the following two possibilities that:
\begin{enumerate*}
  \item $\phi, \psi \neq 0$ but $p^{\alpha^{\pm}}_{\pm} = 0$ and $G^\pm_{\alpha^\pm} = 0$,\footnote{
    \label{dG/dphi not zero}%
    If $\phi, \psi \neq 0$, from the definition of $G^+_{\alpha^+}$ and $G^-_{\alpha^-}$ in \eqref{eq:u-1-1 with W:glsm:G+ alpha} and \eqref{eq:u-1-1 with W:glsm:G- alpha}, we have $\pdv{G^+_{\alpha^+}}{Y^+_{ab}} \neq 0$ and $\pdv{G^+_{\alpha^+}}{Y^-_{ab}} \neq 0$, whence we must have the conditions $p_\pm^{\alpha^\pm} = 0$ and $G^\pm_{\alpha^\pm} = 0$ in order to satisfy \eqref{eq:u-1-1 with W:glsm:M-SUSY:toric hypersurface +}--\eqref{eq:u-1-1 with W:glsm:M-SUSY:toric hypersurface -}.
  }

  or

  \item $p^{\alpha^{\pm}}_{\pm} \neq 0$ but $\phi, \psi = 0$.\footnote{
    \label{dG/dphi = zero}%
    If $p_{\pm}^{\alpha^{\pm}} \neq 0$, we find that \eqref{eq:u-1-1 with W:glsm:M-SUSY:toric hypersurface +}--\eqref{eq:u-1-1 with W:glsm:M-SUSY:toric hypersurface -} can never be satisfied unless $\pdv{G^+_{\alpha^+}}{Y^+_{ab}} = 0$ and $\pdv{G^+_{\alpha^+}}{Y^-_{ab}} = 0$.
    Then, from \autoref{dG/dphi not zero}, it must be that $\phi, \psi = 0$, whence \eqref{eq:u-1-1 with W:glsm:M-SUSY:toric hypersurface +}--\eqref{eq:u-1-1 with W:glsm:M-SUSY:toric hypersurface -} will be trivially satisfied, i.e., for $G^\pm_{\alpha^\pm} = \text{constant} = 0$.
  }
\end{enumerate*}
Indeed, as we shall see shortly, for different values of $r$, these field configurations will define  $\MSUSY$'s which are generally different.

\subtitle{$\MSUSY^{r(\text{i}) > 0}$ and $\MSUSY^{r(\text{ii}) > 0}$ are Disconnected}

In the \emph{same} $r > 0$ region of parameter space, the two \emph{different} field configurations (i) and (ii) will split  $\MSUSY$ into two \emph{disconnected} components, $\MSUSY^{r(\text{i}) > 0}$ and $\MSUSY^{r(\text{ii}) > 0}$.

\subtitle{$\MSUSY^{r(\text{i}) < 0}$ and $\MSUSY^{r(\text{ii}) < 0}$ are also Disconnected}

Likewise, in the \emph{same} $r < 0$ region of parameter space, the two \emph{different} field configurations (i) and (ii) will split $\MSUSY$ into two \emph{disconnected} components, $\MSUSY^{r(\text{i}) < 0}$ and $\MSUSY^{r(\text{ii}) < 0}$.

\subtitle{The $r(\text{i}) > 0$ Phase of the Model: A Complete Intersection of Hypersurfaces in ${Gr}_{1|1}(\mathbb{C}^{m|n})$ $\MSUSY^{r(\text{i}) > 0}$}

From \eqref{eq:u-1-1 with W:glsm:M-SUSY:geometric space}--\eqref{eq:u-1-1 with W:glsm:M-SUSY:toric hypersurface -}, one can see that $\MSUSY^{r(\text{i}) > 0}$ will be defined by
\begin{equation}
  \label{eq:U(1|1) with W:phases:geometric space:r(i) > 0}
  \saveboxed{eq:U(1|1) with W:phases:geometric space:r(i) > 0}{
    \flatfrac{
      \left\{
        \sum_{s = 1}^m \tensor{\phi}{^j_s} \tensor{\bar{\phi}}{_i^s}
        - \sum_{t = 1}^n \tensor{\psi}{^j_t} \tensor{\bar{\psi}}{_i^t}
        = r \tensor{\delta}{^j_i}
        \, \qcomma i, j \in \{1, \bar{1}\}
      \right\}
    }{\mathrm{U}(1|1)}
  }
\end{equation}
and
\begin{equation}
  \label{eq:U(1|1) with W:phases:hypersurface +:r(i) > 0}
  \saveboxed{eq:U(1|1) with W:phases:hypersurface +:r(i) > 0}{
    G^+_{\alpha^+} \left(Y(\phi, \psi)\right)
    = 0
    \, \qcomma
    \alpha^+ \in \{1, \dots, l_+\}
  }
\end{equation}
\begin{equation}
  \label{eq:U(1|1) with W:phases:hypersurface -:r(i) > 0}
  \saveboxed{eq:U(1|1) with W:phases:hypersurface -:r(i) > 0}{
    G^-_{\alpha^-} \left(Y(\phi, \psi)\right)
    = 0
    \, \qcomma
    \alpha^- \in \{1, \dots, l_-\}
  }
\end{equation}
where $G^+_{\alpha^+} \left(Y(\phi, \psi)\right)$ and $G^-_{\alpha^-} \left(Y(\phi, \psi)\right)$ are as given in \eqref{eq:u-1-1 with W:glsm:G+ alpha}--\eqref{eq:u-1-1 with W:glsm:Y+ even} and \eqref{eq:u-1-1 with W:glsm:G- alpha}--\eqref{eq:u-1-1 with W:glsm:Y- odd}, respectively.

One can, via our earlier description of \eqref{eq:u-1-1:m-susy}, interpret $\MSUSY^{r(\text{i}) > 0}$ as a complete intersection of $l_+$ even and $l_-$ odd hypersurfaces $G^+_{\alpha^+} = 0$ ($\alpha^+ \in \{1, \dots, l_+\}$) and $G^-_{\alpha^-} = 0$ ($\alpha^- \in \{1, \dots, l_-\}$), of degrees $q^+_{\alpha^+}$ and $q^-_{\alpha^-}$, in the super-Grassmannian $Gr_{1|1}(\mathbb{C}^{m|n})$ of size $r$.

\subtitle{The $r(\text{i}) > 0$ Phase of the Model at Low Energy: An NLSM on a Complete Intersection of Hypersurfaces in ${Gr}_{1|1}(\mathbb{C}^{m|n})$}

Staying in the $r(\text{i}) > 0$ phase, let us consider the low energy limit $e, b^{\pm}_{(\alpha^{\pm})}, c^{\pm}_{(\alpha^{\pm})} \rightarrow \infty$ where the massive fields/modes (whose masses are either of order $e \sqrt{r}$ or determined by $b^\pm_{(\alpha)}$ and $c^{\pm}_{(\alpha)}$) decouple from the theory.
Then, just as in \autoref{GLSM to NLSM}, the GLSM will become an NLSM on the above-described $\MSUSY^{r(\text{i}) > 0}$, i.e., on a complete intersection of $l_+$ even and $l_-$ odd hypersurfaces $G^+_{\alpha^+} = 0$ ($\alpha^+ \in \{1, \dots, l_+\}$) and $G^-_{\alpha^-} = 0$ ($\alpha^- \in \{1, \dots, l_-\}$), of degrees $q^+_{\alpha^+}$ and $q^-_{\alpha^-}$, in the super-Grassmannian ${Gr}_{1|1}(\mathbb{C}^{m|n})$ of size $r$.

\subtitle{The $r(\text{ii}) < 0$ Phase of the Model: A Weighted Projective Superspace $\WCP^{{l_+ -1}|l_-}_{\left(\vv{q^+}\middle|\vv{q^-}\right)}$ $\MSUSY^{r(\text{i}) < 0}$}

$\MSUSY^{r(\text{ii}) < 0}$ will be defined by \eqref{eq:u-1-1 with W:glsm:M-SUSY:geometric space} but with a negative $r$, i.e.,
\begin{equation}
  \label{eq:U(1|1) with W:phases:geometric space:r(ii) < 0}
  \saveboxed{eq:U(1|1) with W:phases:geometric space:r(ii) < 0}{
    \flatfrac{
      \left\{
        \sum^{l_+}_{\alpha^+ = 1} q^+_{\alpha^+} \abs{p_+^{\alpha^+}}^2
        - \sum^{l_-}_{\alpha^- = 1} q^-_{\alpha^-} \abs{p_-^{\alpha^-}}^2
        = \abs{r}
      \right\}
    }{\mathrm{U}(1)}
  }
\end{equation}
where we mod out by $\mathrm{U}(1)$ instead of $\mathrm{U}(1|1)$ because \autoref{P_alpha U(1) transform} tells us that $p_\pm^{\alpha^\pm}$ effectively transform under the even abelian subgroup $\mathrm{U}(1) \subset \mathrm{U}(1|1)$.

One can interpret this as a weighted projective superspace $\WCP^{l_+ -1|l_-}_{\left(\vv{q^+}\middle|\vv{q^-}\right)}$ of size $|r|$, with positive weights $\left(\vv{q^+}\middle|\vv{q^-}\right) \coloneq (q^+_1, \dots, q^+_{l_+}|q^-_1, \dots, q^-_{l_-})$.

\subtitle{The $r(\text{ii}) < 0$ Phase of the Model at Low Energy: An $\mathrm{SU}(1|1) \times \mathbb{Z}_q$ Supergauged LG Orbifold on a Supervector Bundle over $\WCP^{{l_+ -1}|l_-}_{\left(\vv{q^+}\middle|\vv{q^-}\right)}$ }

Staying in the $r(\text{ii}) < 0$ phase, note that the absence of $\phi, \psi$ in \eqref{eq:U(1|1) with W:phases:geometric space:r(ii) < 0} means that in this phase, $\phi, \psi$ would have vanishing expectation values whence the supergauge symmetry of the GLSM would not be broken at low energy, unlike the $r(\text{ii}) > 0$ phase.

That said, the presence of $p_\pm^{\alpha^\pm}$ in \eqref{eq:U(1|1) with W:phases:geometric space:r(ii) < 0}  means that there are formally nonvanishing expectation values $\expval{p_\pm^{\alpha^\pm}}$.
As a result, the $\mathrm{U}(1|1)$ gauge group will break down to a subgroup $H \subset \mathrm{U}(1|1)$ whose matrices have superdeterminant equal to the $q^{\text{th}}$ root of $1$, where the integer $q$ is the greatest common divisor of all the $q^\pm_{\alpha^\pm}$'s, such that $H \cong \mathrm{SU}(1|1) \times \mathbb{Z}_q$.\footnote{%
  \label{ft:u-1-1-W:c-number}%
  To understand this, first, note that the expectation value $\expval*{p_\pm^{\alpha^\pm}}$, being a c-number, ought to be gauge-invariant.
  Second, according to \autoref{P_alpha U(1) transform}, under a gauge transformation, we have $p_\pm^{\alpha^\pm} \to e^{\mp \mathbf{i} q^\pm_{\alpha^\pm}\varphi} p_\pm^{\alpha^\pm}$.
  Together, it will mean that to have $\expval*{p_\pm^{\alpha^\pm}} \to \expval*{p_\pm^{\alpha^\pm}}$ under a gauge transformation for all $\alpha^\pm$, it must necessarily be the case that $\varphi = \flatfrac{2 \pi m}{q}$, where the integer $q$ is the greatest common divisor of all the $q^\pm_{\alpha^\pm}$'s, and $m \in\{0, 1, \dots, q-1\}$.
  In turn, according to \autoref{P_alpha U(1) transform} again, it must mean that in the context of gauge transformations, the even abelian subgroup $\mathrm{U}(1) \subset \mathrm{U}(1|1)$ would be broken down to $\mathbb{Z}_q \subset \mathrm{U}(1|1)$.
  According to \autoref{P_alpha U(1) transform} yet again, it will mean that the superdeterminant of the matrices of $H$ must be the $q^{\text{th}}$ root of $1$, and since $\mathrm{U}(1|1) \cong \mathrm{SU}(1|1) \times \mathrm{U}(1)$ locally, we ought to have $H \cong \mathrm{SU}(1|1) \times \mathbb{Z}_q$ locally, whence the full claim follows.
}

In the low energy limit $e\to \infty$\footnote{%
  \label{ft:u-1-1-W:low energy limit}%
  Notice that here, we do not take the $b^{\pm}_{(\alpha^{\pm})}, c^{\pm}_{(\alpha^{\pm})} \rightarrow \infty$ limit.
  This is because when $\phi, \psi = 0$, there is no longer a target space for a hypersurface $G^\pm_{\alpha^\pm} = 0$ to be defined, but taking the limit $b^{\pm}_{(\alpha)}, c^{\pm}_{(\alpha)} \to \infty$ would eliminate $G^\pm_{\alpha^\pm}$ (a mass term) from the Lagrangian when we go to low energy, i.e., set it to $G^\pm_{\alpha^\pm} = 0$, which is contradictory.
}
where the relevant fields from the vector and $P_{\pm}^{\alpha^\pm}$-multiplets (of mass proportional to $e \sqrt{\abs{r}}$ acquired from the Higgs mechanism) become  massive and decouple from the theory, the GLSM will become an $\mathrm{SU}(1|1) \times \mathbb{Z}_q$ supergauged LG orbifold with effective superpotential\footnote{%
  \label{ft:u-1-1-W:explaining superpotential of LG orbifold}%
  See \eqref{eq:u-1-1 with W:W potential}, and note that the $\phi, \psi$ fields only need to be zero in the potential energy (to define \eqref{eq:U(1|1) with W:phases:geometric space:r(ii) < 0}); the $\phi, \psi$ fields are otherwise massless (as long as $q^\pm_{\alpha^\pm} > 2$) whence they survive the low energy limit.
  Also, when the massive $p_\pm^{\alpha^\pm}$ fields decouple from the low energy theory, they would be replaced by their numerically nonvanishing expectation values $\sqrt{\expval*{\abs*{p_\pm^{\alpha^\pm}}^2}}$.
}
\begin{equation}
  \label{eq:U(1|1) with W:phases:superpotential W:r(ii) < 0}
  \saveboxed{eq:U(1|1) with W:phases:superpotential W:r(ii) < 0}{
    W^{\mathrm{SU}(1|1) \times \mathbb{Z}_q}_{\text{LG}, |r(\text{ii})| > 0}
    = \sum_{\alpha^+ = 1}^{l_+} \sqrt{\expval{\abs{p_+^{\alpha^+}}^2}}
    \; G^+_{\alpha^+} \left(Y(\phi, \psi)\right)
    + \sum_{\alpha^- = 1}^{l_-} \sqrt{\expval{\abs{p_-^{\alpha^-}}^2}}
    \; G^-_{\alpha^-} \left(Y(\phi, \psi)\right)
  }
\end{equation}
where (via \eqref{eq:U(1|1) with W:phases:geometric space:r(ii) < 0})
\begin{equation}
  \label{eq:U(1|1) with W:phases:geometric space:r(ii) < 0:exp val}
  \saveboxed{eq:U(1|1) with W:phases:geometric space:r(ii) < 0:exp val}{
    \sum^{l_+}_{\alpha^+ = 1} q^+_{\alpha^+} \expval{\abs{p_+^{\alpha^+}}^2}
    - \sum^{l_-}_{\alpha^- = 1} q^-_{\alpha^-} \expval{\abs{p_-^{\alpha^-}}^2}
    = \abs{r}
  }
\end{equation}
As indicated by $W^{\mathrm{SU}(1|1) \times \mathbb{Z}_q}_{\text{LG}, |r(\text{ii})| > 0}$ which does depend on the $\phi$'s and $\psi$'s through the $G^\pm_{\alpha^\pm}$'s, we have a supergauged LG orbifold in the $(\phi, \psi)$ fields.

A relevant point to note at this juncture is that for $r < 0$, $\MSUSY$ as defined by \eqref{eq:u-1-1 with W:glsm:M-SUSY:geometric space} can be mathematically identified as (\emph{c.f.}~\cite[eq.~(15.95)]{Hori:2003ic})
\begin{equation}
  \label{eq:U(1|1) with W:phases:supervector bundle:r(ii) < 0}
  \saveboxed{eq:U(1|1) with W:phases:supervector bundle:r(ii) < 0}{
    \mathcal{S}^{\mathrm{U}(1|1)^{\abs{r} > 0}}_{q_\alpha, p^\alpha}
    = \bigoplus^m_{s=1} \mathscr{L}^{-1}_s \oplus \bigoplus^n_{t=1} \mathbf{\Pi} \mathscr{L}^{-1}_t
    \longrightarrow \WCP^{{l_+ -1}|l_-}_{\left(\vv{q^+}\middle|\vv{q^-}\right)}
  }
\end{equation}
which is a supervector bundle over a weighted projective superspace of size $\abs{r}$, where the even $\mathscr{L}_s$ and odd $\mathbf{\Pi} \mathscr{L}_t$ fibers are spanned by the Grassmann-even $\phi_s$ and Grassmann-odd $\psi_t$ supervectors, respectively.

In other words, the (zero-modes of the) $(\phi, \psi)$ fields in \eqref{eq:U(1|1) with W:phases:superpotential W:r(ii) < 0} ought to take values in the fiber space of $\mathcal{S}^{\mathrm{U}(1|1)^{\abs{r} > 0}}_{q_\alpha, p^\alpha}$ if supersymmetry were not to be completely broken.

Yet another relevant point to note is that the values of $\expval*{\abs*{p_\pm^{\alpha^\pm}}^2}$, which satisfy \eqref{eq:U(1|1) with W:phases:geometric space:r(ii) < 0:exp val}, are physically unique only up to the $\mathrm{U}(1)$ action in \eqref{eq:U(1|1) with W:phases:geometric space:r(ii) < 0}.
In other words, we can regard the $\sqrt{\expval*{\abs*{p_\pm^{\alpha^\pm}}^2}}$'s as holomorphic coordinates on the weighted projective base superspace in \eqref{eq:U(1|1) with W:phases:supervector bundle:r(ii) < 0}, where at $p_{\pm}^{\alpha^\pm} = 0$, the fiber subsuperspace defines $X_{r < 0}$ of \eqref{eq:u-1-1:m-susy:r < 0}, i.e., $Gr_{1|1}(\mathbb{C}^{n|m})$.

Altogether, this means that we have an $\mathrm{SU}(1|1) \times \mathbb{Z}_q$ supergauged LG orbifold fibered over $\WCP^{{l_+ -1}|l_-}_{\left(\vv{q^+}\middle|\vv{q^-}\right)}$ of size $|r|$, with an effective superpotential $W^{\mathrm{SU}(1|1) \times \mathbb{Z}_q}_{\text{LG}, |r|(\text{i}) > 0}$ that is a function on $\mathcal{S}^{\mathrm{U}(1|1)^{\abs{r} > 0}}_{q_\alpha, p^\alpha}$.

Thus, we have an $\mathrm{SU}(1|1) \times \mathbb{Z}_q$ supergauged LG orbifold defined on $\mathcal{S}^{\mathrm{U}(1|1)^{\abs{r} > 0}}_{q_\alpha, p^\alpha}$.

\subtitle{The $r(\text{i}) < 0$ Phase of the Model: A Complete Intersection of Hypersurfaces in ${Gr}_{1|1}(\mathbb{C}^{n|m})$ $\MSUSY^{r(\text{i}) < 0}$}

$\MSUSY^{r(\text{i}) < 0}$ is similar to $\MSUSY^{r(\text{i}) > 0}$, so we shall be brief.

$\MSUSY^{r(\text{i}) < 0}$ will be defined by \eqref{eq:U(1|1) with W:phases:geometric space:r(i) > 0}--\eqref{eq:U(1|1) with W:phases:hypersurface -:r(i) > 0} but with a negative $r$, i.e.,
\begin{equation}
  \label{eq:U(1|1) with W:phases:geometric space:r(i) < 0}
  \saveboxed{eq:U(1|1) with W:phases:geometric space:r(i) < 0}{
    \flatfrac{
      \left\{
        \sum_{t = 1}^n \tensor{\psi}{^j_t} \tensor{\bar{\psi}}{_i^t}
        - \sum_{s = 1}^m \tensor{\phi}{^j_s} \tensor{\bar{\phi}}{_i^s}
        = \abs{r} \tensor{\delta}{^j_i}
        \, \qcomma i, j \in \{1, \bar{1}\}
      \right\}
    }{\mathrm{U}(1|1)}
  }
\end{equation}
and
\begin{equation}
  \label{eq:U(1|1) with W:phases:hypersurface -:r(i) < 0}
  \saveboxed{eq:U(1|1) with W:phases:hypersurface -:r(i) < 0}{
    G^+_{\alpha^-(\mathbf{\Pi})} \left(Y(\phi, \psi)\right)
    = 0
    \, \qcomma
    \alpha^- \in \{1, \dots, l_-\}
  }
\end{equation}
\begin{equation}
  \label{eq:U(1|1) with W:phases:hypersurface +:r(i) < 0}
  \saveboxed{eq:U(1|1) with W:phases:hypersurface +:r(i) < 0}{
    G^-_{\alpha^+(\mathbf{\Pi})} \left(Y(\phi, \psi)\right)
    = 0
    \, \qcomma
    \alpha^+ \in \{1, \dots, l_+\}
  }
\end{equation}
where the ``($\mathbf{\Pi}$)'' subscript will be clear shortly.

One can, via our earlier reinterpretation of \eqref{eq:u-1-1:m-susy} for negative $r$ as a Grassmann parity-reversed version \eqref{eq:u-1-1:m-susy:r < 0} of \eqref{eq:u-1-1:m-susy:r > 0} (for positive $r$), also reinterpret \eqref{eq:U(1|1) with W:phases:geometric space:r(i) > 0}--\eqref{eq:U(1|1) with W:phases:hypersurface -:r(i) > 0} for negative $r$ as
\eqref{eq:U(1|1) with W:phases:geometric space:r(i) < 0}--\eqref{eq:U(1|1) with W:phases:hypersurface +:r(i) < 0}, which describes a complete intersection of $l_-$ \emph{even} and $l_+$ \emph{odd} hypersurfaces $G^+_{\alpha^-(\mathbf{\Pi})} = 0$ ($\alpha^- = 1, \dots, l_-$) and $G^-_{\alpha^+(\mathbf{\Pi})} = 0$ ($\alpha^+ = 1, \dots, l_+$) of \emph{even} and \emph{odd} degrees $q^-_{\alpha^-}$ and $q^+_{\alpha^+}$ in the super-Grassmannian $Gr_{1|1}(\mathbb{C}^{n|m})$ of size $|r|$.\footnote{%
  \label{ft:U(1|1) with W:phases:parity-reversed super-plucker}%
  Recall that the $Y^+(\phi, \psi)$ and $Y^-(\phi, \psi)$ fields, as super-Pl\"{u}cker coordinates of $Gr_{1|1}(\mathbb{C}^{m|n})$, transform in the superdeterminantal and anti-superdeterminantal representations of $\mathrm{U}(1|1)$, respectively (see \eqref{eq:plucker:Gr 1-1 2-3:tfm of Y} and \eqref{eq:plucker:Gr p-q m-n:tfm of Y}).
  Hence, the Grassmann-even $G^+_{\alpha^+}(Y(\phi, \psi))$ and Grassmann-odd $G^-_{\alpha^-}(Y(\phi, \psi))$ polynomials, as defined in \eqref{eq:u-1-1 with W:glsm:G+ alpha} and \eqref{eq:u-1-1 with W:glsm:G- alpha}, also transform in the superdeterminantal and anti-superdeterminantal representations of $\mathrm{U}(1|1)$, respectively.

  However, in interpreting \eqref{eq:U(1|1) with W:phases:geometric space:r(i) < 0} as the parity-reversed super-Grassmannian ${Gr}_{1|1}(\mathbb{C}^{n|m})$ of size $|r|$, the $Y(\phi, \psi)$ fields in \eqref{eq:U(1|1) with W:phases:hypersurface -:r(i) < 0} and \eqref{eq:U(1|1) with W:phases:hypersurface +:r(i) < 0} now ought to be interpreted as super-Pl\"{u}cker coordinates of the parity-reversed ${Gr}_{1|1}(\mathbb{C}^{n|m})$.
  It turns out that the super-Pl\"{u}cker coordinates of the parity-reversed ${Gr}_{1|1}(\mathbb{C}^{n|m})$ are the \emph{same} as the super-Pl\"{u}cker coordinates of $Gr_{1|1}(\mathbb{C}^{m|n})$, except that it transforms in the \emph{opposite} representation (see \autoref{sec:plucker:Gr 1-1 2-3} and \autoref{sec:plucker:Gr}).
  Hence, $G^+_{\alpha^+}(Y(\phi, \psi))$ and $G^-_{\alpha^-}(Y(\phi, \psi))$, as defined in \eqref{eq:u-1-1 with W:glsm:G+ alpha} and \eqref{eq:u-1-1 with W:glsm:G- alpha}, now transform in the anti-superdeterminantal and superdeterminantal representations of $\mathrm{U}(1|1)$, respectively, as functions in the parity-reversed ambient space.
  That is to say, $G^+_{\alpha^+}(Y(\phi, \psi))$ and $G^-_{\alpha^-}(Y(\phi, \psi))$ are to be interpreted as \emph{odd} and \emph{even} functions in the parity-reversed ambient space.

  For later convenience, we shall write $G^+_{\alpha^+}(Y(\phi, \psi))$ as $G^-_{\alpha^+(\mathbf{\Pi})}(Y(\phi, \psi))$, and $G^-_{\alpha^-}(Y(\phi, \psi))$ as $G^+_{\alpha^-(\mathbf{\Pi})}(Y(\phi, \psi))$. Then, the hypersurfaces $G^-_{\alpha^+(\mathbf{\Pi})} = 0$ and $G^+_{\alpha^-(\mathbf{\Pi})} = 0$ can thus be interpreted as \emph{odd} and \emph{even} hypersurfaces, respectively, in the parity-reversed ${Gr}_{1|1}(\mathbb{C}^{n|m})$.
}

\subtitle{The $r(i) < 0$ Phase of the Model at Low Energy: An NLSM on a Complete Intersection of Hypersurfaces in ${Gr}_{1|1}(\mathbb{C}^{n|m})$}

In the low energy  limit $e \rightarrow \infty$ and $b^\pm_{(\alpha^{\mp})}, c^{\pm}_{(\alpha^{\mp})} \rightarrow \infty$, the GLSM will become an NLSM on the above-described $\MSUSY^{r(\text{i}) < 0}$, i.e., on a complete intersection of $l_-$ \emph{even} and $l_+$ \emph{odd} hypersurfaces $G^+_{\alpha^-(\mathbf{\Pi})} = 0$ ($\alpha^- = 1, \dots, l_-$) and $G^-_{\alpha^+(\mathbf{\Pi})} = 0$ ($\alpha^+ = 1, \dots, l_+$) of \emph{even} and \emph{odd} degrees $q^-_{\alpha^-}$ and $q^+_{\alpha^+}$ in the super-Grassmannian ${Gr}_{1|1}(\mathbb{C}^{n|m})$ of size $|r|$.

\subtitle{The $r(ii) > 0$ Phase of the Model: A Weighted Projective Superspace $\WCP^{{l_- -1}|l_+}_{(\vv{q^-}|\vv{q^+})}$ $\MSUSY^{r(\text{ii}) > 0}$}

$\MSUSY^{r(\text{ii}) > 0}$ is similar to $\MSUSY^{r(\text{i}) < 0}$, so we shall again be brief.

From \eqref{eq:u-1-1 with W:glsm:M-SUSY:geometric space}, one can see that $\MSUSY^{r(\text{ii}) > 0}$ will be defined by
\begin{equation}
  \label{eq:U(1|1) with W:phases:geometric space:r(ii) > 0}
  \saveboxed{eq:U(1|1) with W:phases:geometric space:r(ii) > 0}{
    \flatfrac{
      \left\{
        \sum_{\alpha^- = 1}^{l_-} q^-_{\alpha^-} \abs{p_{+(\mathbf{\Pi})}^{\alpha^-}}^2
        - \sum_{\alpha^+ = 1}^{l_+} q^+_{\alpha^+} \abs{p_{-(\mathbf{\Pi})}^{\alpha^+}}^2
        = r
      \right\}
    }{\mathrm{U}(1)}
  }
\end{equation}
where $p_{+(\mathbf{\Pi})}^{\alpha^-}$ and $p_{-(\mathbf{\Pi})}^{\alpha^+}$ are Grassmann-even and Grassmann-odd, respectively.

Note that the Grassmann parity of $p_{\pm(\mathbf{\Pi})}^{\alpha^\mp}$ is consistent with the Grassmann parity of $G^\pm_{\alpha^\mp(\mathbf{\Pi})}$ in \eqref{eq:U(1|1) with W:phases:hypersurface -:r(i) < 0} and \eqref{eq:U(1|1) with W:phases:hypersurface +:r(i) < 0}, in the sense that if we now have $G^-_{\alpha^+(\mathbf{\Pi})}$/$G^+_{\alpha^-(\mathbf{\Pi})}$ and $p_{- (\mathbf{\Pi})}^{\alpha^+}$/$p_{+ (\mathbf{\Pi})}^{\alpha^-}$ in place of $G^+_{\alpha^+}$/$G^-_{\alpha^-}$ and $p_+^{\alpha^+}$/$p_-^{\alpha^-}$ in the superpotential $ W^{\mathrm{U}(1|1)}_{\text{super}}$ given by \eqref{eq:u-1-1 with W:W potential}, $W^{\mathrm{U}(1|1)}_{\text{super}}$ will continue to be a Grassmann-even and therefore admissible term in the Lagrangian.

As such, \eqref{eq:U(1|1) with W:phases:geometric space:r(ii) > 0} will define a weighted projective superspace $\WCP^{{l_- -1}|l_+}_{(\vv{q^-}|\vv{q^+})}$ of size $r$, with positive weights $(\vv{q^-}|\vv{q^+}) \coloneq (q^-_1, \dots, q^-_{l_-}|q^+_1, \dots, q^+_{l_+})$.

\subtitle{The $r(ii) > 0$ Phase of the Model at Low Energy: An $\mathrm{SU}(1|1) \times \mathbb{Z}_q$ Supergauged LG Orbifold on a Supervector Bundle over $\WCP^{{l_- -1}|l_+}_{(\vv{q^-}|\vv{q^+})}$}

In the low energy limit $e \to \infty$, the GLSM  will become an $\mathrm{SU}(1|1) \times \mathbb{Z}_q$ supergauged LG orbifold with effective superpotential
\begin{equation}
  \label{eq:U(1|1) with W:phases:superpotential W:r(ii) > 0}
  \saveboxed{eq:U(1|1) with W:phases:superpotential W:r(ii) > 0}{
    W^{\mathrm{SU}(1|1) \times \mathbb{Z}_q}_{\text{LG}, r(\text{ii}) > 0}
    = \sum_{\alpha^- = 1}^{l_-} \sqrt{\expval{\abs{p_{+ (\mathbf{\Pi})}^{\alpha^-}}^2}}
    \; G^+_{\alpha^- (\mathbf{\Pi})} \left(Y(\phi, \psi)\right)
    + \sum_{\alpha^+ = 1}^{l_+} \sqrt{\expval{\abs{p_{- (\mathbf{\Pi})}^{\alpha^+}}^2}}
    \; G^-_{\alpha^+ (\mathbf{\Pi})} \left(Y(\phi, \psi)\right)
  }
\end{equation}
where (via \eqref{eq:U(1|1) with W:phases:geometric space:r(ii) > 0})
\begin{equation}
  \label{eq:U(1|1) with W:phases:geometric space:r(ii) > 0:exp val}
  \saveboxed{eq:U(1|1) with W:phases:geometric space:r(ii) > 0:exp val}{
    \sum^{l_-}_{\alpha^- = 1} q^-_{\alpha^-} \expval{\abs{p_{+ (\mathbf{\Pi})}^{\alpha^-}}^2}
    - \sum^{l_+}_{\alpha^+ = 1} q^+_{\alpha^+} \expval{\abs{p_{- (\mathbf{\Pi})}^{\alpha^+}}^2}
    = r
  }
\end{equation}

A relevant point to note at this juncture is that  for $r > 0$, $\MSUSY$ as defined by \eqref{eq:u-1-1 with W:glsm:M-SUSY:geometric space}, can, in terms of $p_{+ (\mathbf{\Pi})}^{\alpha^-}$ and $p_{- (\mathbf{\Pi})}^{\alpha^+}$ now, be mathematically identified as (\emph{c.f.}~\cite[eq.~(15.93)]{Hori:2003ic})
\begin{equation}
  \label{eq:U(1|1) with W:phases:supervector bundle:r(ii) > 0}
  \saveboxed{eq:U(1|1) with W:phases:supervector bundle:r(ii) > 0}{
    \mathcal{S}^{\mathrm{U}(1|1)^{r > 0}}_{q_\alpha, p^\alpha} =
    \left[
      \bigoplus^n_{t=1} \mathbf{\Pi} \mathscr{L}^{-1}_t \oplus \bigoplus^m_{s=1} \mathscr{L}^{-1}_s
    \right]_{\mathbf{\Pi}}
    \longrightarrow \WCP^{{l_- -1}|l_+}_{\left(\vv{q^-}\middle|\vv{q^+}\right)}
  }
\end{equation}
which is a supervector bundle over a weighted projective superspace of size $r$, where the subscript ``$\mathbf{\Pi}$'' indicates a Grassmann parity-reversal in the sense discussed below \eqref{eq:u-1-1:m-susy:r < 0}.
At $p_{+ (\mathbf{\Pi})}^{\alpha^-} = 0 = p_{- (\mathbf{\Pi})}^{\alpha^+}$, the fiber subsuperspace defines $X_{r > 0}$ of \eqref{eq:u-1-1:m-susy:r > 0}, or $Gr_{1|1}(\mathbb{C}^{m|n})$ (which is indeed a Grassmann parity-reversed version of the $Gr_{1|1}(\mathbb{C}^{n|m})$ associated with \eqref{eq:U(1|1) with W:phases:supervector bundle:r(ii) < 0}).

Altogether, this means that we have an $\mathrm{SU}(1|1) \times \mathbb{Z}_q$ supergauged LG orbifold fibered over $\WCP^{{l_- -1}|l_+}_{(\vv{q^-}|\vv{q^+})}$ of size $r$, with an effective superpotential $W^{\mathrm{SU}(1|1) \times \mathbb{Z}_q}_{\text{LG}, r(\text{ii}) > 0}$ that is a function on $\mathcal{S}^{\mathrm{U}(1|1)^{r > 0}}_{q_\alpha, p^\alpha}$.

Thus, we have an $\mathrm{SU}(1|1) \times \mathbb{Z}_q$ supergauged LG orbifold defined on $\mathcal{S}^{\mathrm{U}(1|1)^{r > 0}}_{q_\alpha, p^\alpha}$.

\subtitle{$\MSUSY^{r(\text{i}) > 0}$ and $\MSUSY^{r(\text{i}) < 0}$ are also Disconnected}

The field configurations for $\MSUSY^{r(\text{i}) > 0}$ and $\MSUSY^{r(\text{i}) < 0}$ are the \emph{same}.
However, $\MSUSY^{r(\text{i}) > 0}$ and $\MSUSY^{r(\text{i}) < 0}$ are \emph{distinct} spaces.
Thus, $\MSUSY^{r(\text{i}) > 0}$ and $\MSUSY^{r(\text{i}) < 0}$ must also be \emph{disconnected}.

Incidentally, $\MSUSY^{r(\text{i}) > 0}$ and $\MSUSY^{r(\text{i}) < 0}$ are actually parity-reversed mirrors of each other.

\subtitle{$\MSUSY^{r(\text{ii}) < 0}$ and $\MSUSY^{r(\text{ii}) > 0}$ are also Disconnected}

The field configurations for $\MSUSY^{r(\text{ii}) > 0}$ and $\MSUSY^{r(\text{ii}) < 0}$ are the \emph{same}.
However, $\MSUSY^{r(\text{ii}) > 0}$ and $\MSUSY^{r(\text{ii}) < 0}$ are \emph{distinct} spaces.
Thus, $\MSUSY^{r(\text{ii}) > 0}$ and $\MSUSY^{r(\text{ii}) < 0}$ must also be \emph{disconnected}.

Incidentally, $\MSUSY^{r(\text{ii}) > 0}$ and $\MSUSY^{r(\text{ii}) < 0}$ are also parity-reversed mirrors of each other.

\subtitle{Branch (1) and Branch (2) of the $\mathrm{U}(1|1)$ GLSM with Superpotential}

Altogether, $\MSUSY^{r(\text{i}) > 0}$ is disconnected from $\MSUSY^{r(\text{i}) < 0}$ and $\MSUSY^{r(\text{ii}) > 0}$.

Also, $\MSUSY^{r(\text{ii}) < 0}$ is disconnected from $\MSUSY^{r(\text{i}) < 0}$ and $\MSUSY^{r(\text{ii}) > 0}$.

Therefore, there are two \emph{independent} branches in $\MSUSY$, with two phases each.

The phases along Branch (1) are $\MSUSY^{r(\text{i}) > 0}$ when $r > 0$ and $\MSUSY^{r(\text{ii}) < 0}$ when $r < 0$.
So in Branch (1), as we go from $r > 0$ to $r < 0$, we will go from $\MSUSY^{r(\text{i}) > 0}$ to $\MSUSY^{r(\text{ii}) < 0}$, and vice versa.

The phases along Branch (2) are $\MSUSY^{r(\text{i}) < 0}$ when $r < 0$ and $\MSUSY^{r(\text{ii}) > 0}$ when $r > 0$.
So in Branch (2), as we go from $r < 0$ to $r > 0$, we will go from $\MSUSY^{r(\text{i}) < 0}$ to $\MSUSY^{r(\text{ii}) > 0}$, and vice versa.

Moreover, since $\MSUSY^{r(\text{i}) < 0}$ is mirror to $\MSUSY^{r(\text{i}) > 0}$, and $\MSUSY^{r(\text{ii}) > 0}$ is mirror to $\MSUSY^{r(\text{ii}) < 0}$, Branch (2) is a parity-reversed mirror of Branch (1).

\subtitle{A Summary of the Phases and Branches of the $\mathrm{U}(1|1)$ GLSM with Superpotential}

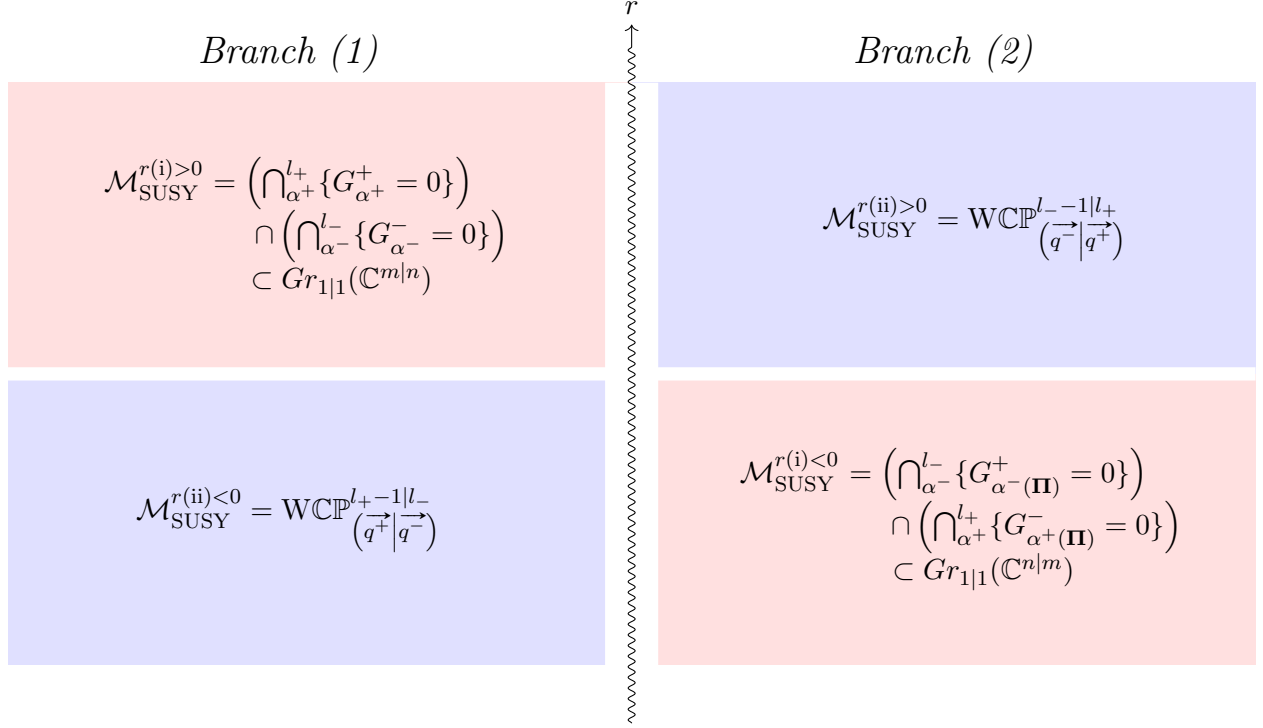
\begin{figure}
  \centering
  \begin{tikzpicture}[scale=1.1]

    \def\xmin{-7.5}
    \def\xmax{7.5}
    \def\ymin{-3.5}
    \def\ymax{3.5}

    \def\slope{0.3}

    \colorlet{phase1}{red}
    \colorlet{phase2}{blue}


    \fill[phase1!12]
    (\xmin, \ymax)
    --(0, \ymax)
    --(0, 0)
    --(\xmin, 0)
    --cycle;

    \fill[phase2!12]
    (\xmin, \ymin)
    --(0, \ymin)
    --(0, 0)
    --(\xmin, 0)
    --cycle;

    \fill[phase1!12]
    (\xmax, \ymin)
    --(0, \ymin)
    --(0, 0)
    --(\xmax, 0)
    --cycle;

    \fill[phase2!12]
    (\xmax, \ymax)
    --(0, \ymax)
    --(0, 0)
    --(\xmax, 0)
    --cycle;


    \draw[white,line width=5pt]
    (\xmin,0)--(\xmax,0);

    \draw[white,line width=20pt]
    (0,\ymin)--(0,\ymax);

    \fill[white] (0,0) circle[radius=4pt];


    \node at (\xmin*0.55,\ymax*1.1) {\Large \textit{Branch (1)}};

    \node at (\xmin*0.55,\ymax*0.5) {
      \begin{tabular}{c}
        $\MSUSY^{r(\text{i}) > 0}$
        $= \left( \bigcap_{\alpha^+}^{l_+} \{G^+_{\alpha^+} = 0\} \right) $
        \\
        \hspace*{70pt}$\cap \left( \bigcap_{\alpha^-}^{l_-} \{G^-_{\alpha^-} = 0\} \right)$
        \\
        \hspace*{40pt}$ \subset Gr_{1|1}(\mathbb{C}^{m|n})$
      \end{tabular}
    };

    \node at (\xmin*0.55,\ymin*0.5) {
      $\MSUSY^{r(\text{ii}) < 0} = \WCP^{l_+ - 1|l_-}_{\left( \vv{q^+} \middle\vert \vv{q^-} \right)}$
    };

    \node at (\xmax*0.5,\ymax*1.1) {\Large \textit{Branch (2)}};

    \node at (\xmax*0.5,\ymin*0.5) {
      \begin{tabular}{c}
        $\MSUSY^{r(\text{i}) < 0}$
        $= \left( \bigcap_{\alpha^-}^{l_-} \{G^+_{\alpha^-(\mathbf{\Pi})} = 0\} \right) $
        \\
        \hspace*{70pt}$\cap \left( \bigcap_{\alpha^+}^{l_+} \{G^-_{\alpha^+(\mathbf{\Pi})} = 0\} \right)$
        \\
        \hspace*{30pt}$ \subset Gr_{1|1}(\mathbb{C}^{n|m})$
      \end{tabular}
    };

    \node at (\xmax*0.55,\ymax*0.5) {
      $\MSUSY^{r(\text{ii}) > 0} = \WCP^{l_- - 1|l_+}_{\left( \vv{q^-} \middle\vert \vv{q^+} \right)}$
    };


    \draw[->, decorate, decoration={snake, amplitude=1.3pt, segment length=4pt, post length = 8pt}] (0,\ymin*1.2)--(0,\ymax*1.2)
    node[above] {$r$};

  \end{tikzpicture}
  \caption{$\MSUSY$ of the different phases along Branch (1) and (2) of the $\mathrm{U}(1|1)$ GLSM with superpotential.
  }
  \label{fig:u-1-1-W:m-susy:glsm}
\end{figure}

\begin{figure}
  \centering
  \begin{tikzpicture}[scale=1.1]

    \def\xmin{-7.5}
    \def\xmax{7.5}
    \def\ymin{-3.5}
    \def\ymax{3.5}

    \def\slope{0.3}

    \colorlet{phase1}{red}
    \colorlet{phase2}{blue}


    \fill[phase1!12]
    (\xmin, \ymax)
    --(0, \ymax)
    --(0, 0)
    --(\xmin, 0)
    --cycle;

    \fill[phase2!12]
    (\xmin, \ymin)
    --(0, \ymin)
    --(0, 0)
    --(\xmin, 0)
    --cycle;

    \fill[phase1!12]
    (\xmax, \ymin)
    --(0, \ymin)
    --(0, 0)
    --(\xmax, 0)
    --cycle;

    \fill[phase2!12]
    (\xmax, \ymax)
    --(0, \ymax)
    --(0, 0)
    --(\xmax, 0)
    --cycle;


    \draw[white,line width=5pt]
    (\xmin,0)--(\xmax,0);

    \draw[white,line width=20pt]
    (0,\ymin)--(0,\ymax);

    \fill[white] (0,0) circle[radius=4pt];


    \node at (\xmin*0.55,\ymax*1.1) {\Large \textit{Branch (1)}};

    \node at (\xmin*0.55,\ymax*0.5) {
      NLSM on $\MSUSY^{r(\text{i}) > 0}$
    };

    \node at (\xmin*0.55,\ymin*0.5) {
      $\mathrm{SU}(1|1) \times \mathbb{Z}_q$ LG Orbifold on $\mathcal{S}^{\mathrm{U}(1|1)^{\abs{r} > 0}}_{q_{\alpha}, p^{\alpha}}$
    };

    \node at (\xmax*0.5,\ymax*1.1) {\Large \textit{Branch (2)}};

    \node at (\xmax*0.5,\ymin*0.5) {
        NLSM on $\MSUSY^{r(\text{i}) < 0}$
    };

    \node at (\xmax*0.55,\ymax*0.5) {
      $\mathrm{SU}(1|1) \times \mathbb{Z}_q$ LG Orbifold on $\mathcal{S}^{\mathrm{U}(1|1)^{r > 0}}_{q_{\alpha}, p^{\alpha}}$
    };


    \draw[->, decorate, decoration={snake, amplitude=1.3pt, segment length=4pt, post length = 8pt}] (0,\ymin*1.2)--(0,\ymax*1.2)
    node[above] {$r$};

  \end{tikzpicture}
  \caption{The low energy limit of the $\mathrm{U}(1|1)$ GLSM with superpotential in different phases along Branch (1) and (2).}
  \label{fig:u-1-1-W:low energy}
\end{figure}
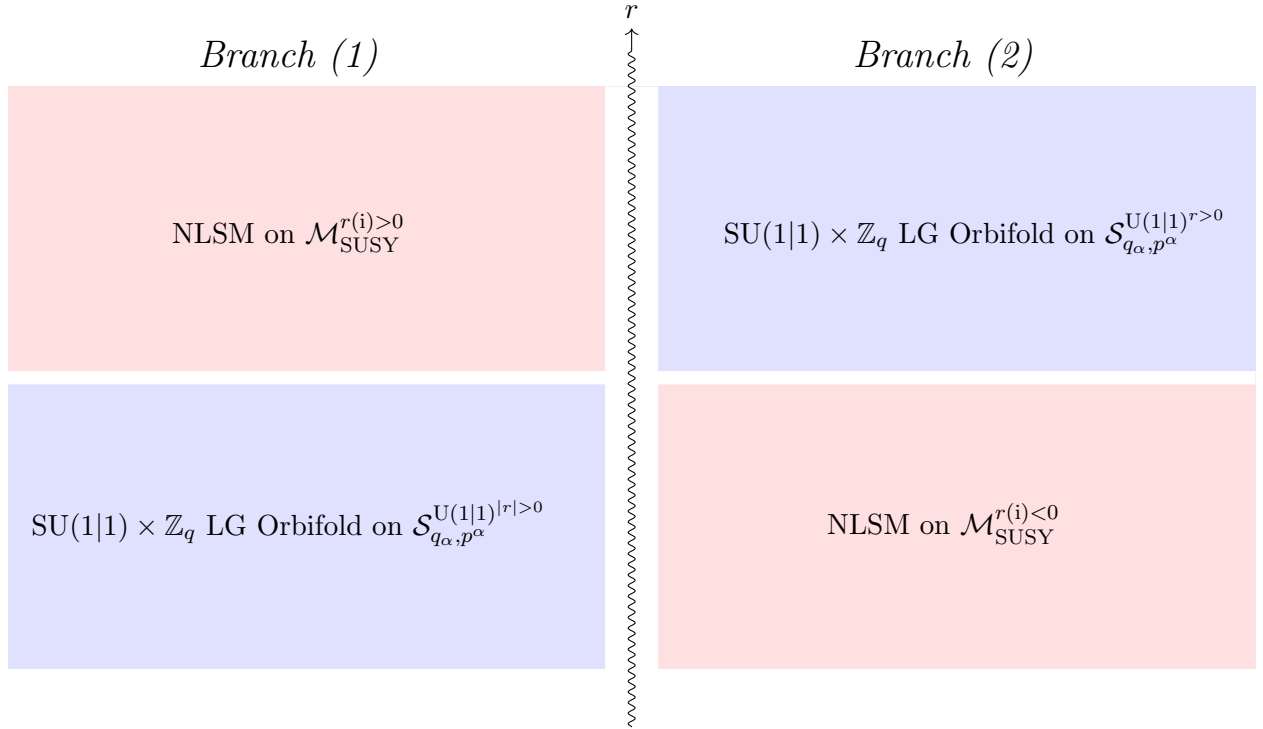

The $\MSUSY$ of the different phases along Branch (1) and (2) of the $\mathrm{U}(1|1)$ GLSM with superpotential is summarized in \autoref{fig:u-1-1-W:m-susy:glsm},
and the corresponding low energy limit of the model is summarized in \autoref{fig:u-1-1-W:low energy}.

\subsection{Phases of the Reduced \texorpdfstring{$\mathrm{U}(1)$}{U(1)} Model: An NLSM on a Complete Intersection of Hypersurfaces in Projective Space and an LG Orbifold on a Vector Bundle over Weighted Projective Space}
\label{subsec: reduced U(1) model}

Note that the $\mathrm{U}(1|1)$ GLSM with superpotential ought to reduce to the familiar $\mathrm{U}(1)$ GLSM with superpotential when we
\begin{enumerate*}
  \item turn off the Grassmann-odd $\Phi^{\bar{1}}$ chiral superfield,
  \item turn off the whole Grassmann-odd chiral superfield $\Psi$,
  and
  \item turn off the whole Grassmann-odd chiral superfield $P_-^{\alpha^-}$.
\end{enumerate*}

In doing so, as explained at the end of \autoref{subsection: U(1|1) with W_super}, there can be no Grassmann-odd $G^-_{\alpha^-}$ polynomials, while the remaining Grassmann-even $G^+_{\alpha^+}$ polynomials would now be expressed in terms of regular Pl\"{u}cker coordinates, i.e., $Y^+_{s_a}(\Phi) = \tensor{\Phi}{^1_{s_a}}$.

\subtitle{A Reduction to the $r> 0$ Phase of the Familiar $\mathrm{U}(1)$ GLSM with Superpotential and its Low Energy NLSM on a Complete Intersection of Hypersurfaces in $\mathbb{CP}^{m-1}$}

Let $r > 0$, and consider the $r(\text{ii}) > 0$ phase of the underlying $\mathrm{U}(1|1)$ GLSM with superpotential.
Then, from \eqref{eq:U(1|1) with W:phases:geometric space:r(ii) > 0} (with $q^- = 0$ and the fact that $\abs{p_{- (\mathbf{\Pi})}}^2 = \abs{p_+}^2$ is positive-definite), we find that $\MSUSY^{r(\text{ii}) > 0}$ cannot be defined.

Next, consider the $r(\text{i}) > 0$ phase of the underlying $\mathrm{U}(1|1)$ GLSM with superpotential.
Then, from \eqref{eq:U(1|1) with W:phases:geometric space:r(i) > 0} (with no $\psi$ and $i,j = 1$ only) and \eqref{eq:U(1|1) with W:phases:hypersurface +:r(i) > 0} (with $Y^+_{s_a t_a}$ replaced by $Y^+_{s_a}$), we find that $\MSUSY^{r(\text{i}) > 0}$ would reduce to a complete intersection of (even) hypersurfaces $G^+_{\alpha^+} (\tensor{\phi}{^1_s}) = 0$ $(\alpha^+ \in \{ 1, \dots, l_+\})$ in $\mathbb{CP}^{m-1}$ of size $r$, as it should (\emph{c.f.}~\cite[$\S$5.4]{Witten:1993yc}).

Consequently, at low energy, the GLSM will now become an NLSM on the above described $\MSUSY^{r(\text{i}) > 0}$, i.e., on a complete intersection of (even) hypersurfaces $G^+_{\alpha^+} (\phi^1_s) = 0$ $(\alpha^+ \in \{ 1, \dots, l_+\})$ in $\mathbb{CP}^{m-1}$ of size $r$, as it should (\emph{c.f.}~\cite[$\S$5.4]{Witten:1993yc}).

\subtitle{A Reduction to the $r < 0$ Phase of the Familiar $\mathrm{U}(1)$ GLSM with Superpotential and its Low Energy LG $\mathbb{Z}_q$-Orbifold on a Vector Bundle over  $\WCP^{l_+ - 1}_{\vv{q^+}}$}

Let $r < 0$, and consider the $r(\text{i}) < 0$ of the underlying $\mathrm{U}(1|1)$ GLSM with superpotential.
Then, from \eqref{eq:U(1|1) with W:phases:geometric space:r(i) < 0} (with no $\psi$ and the fact that $\psi^1 \bar{\psi}_1$ is positive-definite), we find that $\MSUSY^{r(\text{i}) < 0}$ cannot be defined.

Next, consider the $r(\text{ii}) < 0$ phase of the underlying $\mathrm{U}(1|1)$ GLSM with superpotential.
Then, from \eqref{eq:U(1|1) with W:phases:geometric space:r(ii) < 0} (with $q^- =0$), we find that $\MSUSY^{r(\text{i}) < 0}$ would reduce to a weighted projective space $\WCP^{l_+-1}_{\vv{q^+}}$ of size $|r|$, as it should (\emph{c.f.}~\cite[$\S$5.4]{Witten:1993yc}).

Consequently, at low energy, the GLSM will now reduce to an LG $\mathbb{Z}_q$-orbifold with effective superpotential
\begin{equation}
  \label{W^Zq_LG}
  W^{\mathbb{Z}_q}_{\text{LG}}
  = \sum_{\alpha^+=1}^{l_+} \sqrt{\expval{\abs{p_+^{\alpha^+}}^2}}
  \, G^+_{\alpha^+}(\phi^1_{s_a})
  \, .
\end{equation}
As indicated by $W^{\mathbb{Z}_q}_{\text{LG}}$ which does depend on the $\phi^1_{s_a}$'s through the $G^+_{\alpha^+}$'s, we have an LG $\mathbb{Z}_q$-orbifold in the holomorphic $\phi^1_{s_a}$ fields.

Also, the supervector bundle $\mathcal{S}^{\mathrm{U}(1|1)^{|r| > 0}}_{q_\alpha, p^\alpha, 1}$ in \eqref{eq:U(1|1) with W:phases:supervector bundle:r(ii) < 0} would now become
\begin{equation}
  \label{vector bundle over WCP - r < 0}
  \mathcal{S}^{\mathrm{U}(1)^{|r| > 0}}_{q^+_\alpha, p_+^\alpha}
  = \bigoplus^m_{s=1} \mathcal{L}^{-1}_s \longrightarrow \WCP^{l_+-1}_{\vv{q^+}}
  \, ,
\end{equation}
which is a vector bundle over a weighted projective space of size $|r|$, where the fiber $\mathcal{L}_s$ is spanned by the $\phi_s$ vector, while the holomorphic coordinates of the $\WCP^{l_+-1}_{\vv{q^+}}$ base are the  $\sqrt{\expval*{\abs*{p_+^{\alpha^+}}}^2}$'s.

Altogether, this means that we have an LG $\mathbb{Z}_q$-orbifold  fibered over $\WCP^{l_+-1}_{\vv{q^+}}$ of size $|r|$, with an effective superpotential $W^{\mathbb{Z}_q}_{\text{LG}}$ that is a holomorphic function on $\mathcal{S}^{\mathrm{U}(1)^{|r| > 0}}_{q^+_\alpha, p_+^\alpha}$.

Thus, we have an LG $\mathbb{Z}_q$-orbifold defined on $\mathcal{S}^{\mathrm{U}(1)^{|r| > 0}}_{q^+_\alpha, p_+^\alpha}$, as we should (\emph{c.f.}~\cite[$\S$5.4]{Witten:1993yc}).

\subsection{Quantum Aspects: The GLSM}
\label{subsec: Quantum Prop U(1|1) GLSM with W}

Of the various quantum aspects of the GLSM, we shall now discuss the one relevant to our eventual aim of exploring its applications to  mathematics, which is the quantum correction to the FI parameter $r$.

The analysis is just a straightforward generalization of that in \autoref{subsec: Quantum Prop U(1|1) GLSM} to include the $p_\pm^{\alpha^\pm}$ fields which contribute to the action in the same way as the $\phi$ and $\psi$ fields.
Thus, for brevity, we shall just state the results.

\subtitle{The Quantum $D$-term}

The quantum $D$-term which generalizes \eqref{eq:u-1-1:quantum:D terms:expectation values} is (omitting the $D^2$ contribution)
\begin{equation}
  \label{eq:u-1-1 with W:quantum:D terms:expectation values}
  (-1)^{\varsigma(i)} \tensor{D}{^i_j} \left(
    \sum_{s = 1}^m \expval{\tensor{\phi}{^j_s} \tensor{\bar{\phi}}{_i^s}}
    - \sum_{t = 1}^n \expval{\tensor{\psi}{^j_t} \tensor{\bar{\psi}}{_i^t}}
    - \sum^{l_+}_{\alpha^+ = 1} q^+_{\alpha^+} \expval{p_+^{\alpha^+} \bar{p}^+_{\alpha^+}} \tensor{\delta}{^j_i}
    + \sum^{l_-}_{\alpha^- = 1} q^-_{\alpha^-} \expval{p_-^{\alpha^-} \bar{p}^-_{\alpha^-}} \tensor{\delta}{^j_i}
    - r \tensor{\delta}{^j_i}
  \right)
  \, .
\end{equation}

\subtitle{The Expectation Values}

The above expectation values to a 1-loop approximation in the effective theory at energy scale $\mu$ are
\begin{equation}
  \label{eq:u-1-1 with W:quantum:phi,psi expectation values}
  \expval{\tensor{\phi}{^j_s} \tensor{\bar{\phi}}{_i^s}}_{\text{1-loop}}
  = \log \left( \frac{\Lambda_{\text{UV}}}{\mu} \right) \tensor{\delta}{^j_i}
  = \expval{\tensor{\psi}{^j_t} \tensor{\bar{\psi}}{_i^t}}_{\text{1-loop}}
\end{equation}
and
\begin{equation}
  \label{eq:u-1-1 with W:quantum:p expectation values}
  \expval{p_{\pm}^{\alpha^\pm} \bar{p}^{\pm}_{\alpha^\pm}}_{\text{1-loop}}
  = \log \left( \frac{\Lambda_{\text{UV}}}{\mu} \right)
  \, ,
\end{equation}
 where  $\Lambda_{\text{UV}}$ is the cutoff scale which ensures that we have a finite answer.

\subtitle{The Effective FI Parameter $r_{\text{eff}}(\mu)$}

By substituting \eqref{eq:u-1-1 with W:quantum:phi,psi expectation values} and \eqref{eq:u-1-1 with W:quantum:p expectation values} in \eqref{eq:u-1-1 with W:quantum:D terms:expectation values}, one can see that the effective FI parameter which will ensure the finiteness of the effective action in the continuum limit $\Lambda_{\text{UV}} \rightarrow \infty$ can then be written as
\begin{equation}
  \label{eq:U(1|1) with W:quantum glsm:r-eff}
  r_{\text{eff}}
  = r
  + \left(
    m - n
    - \sum^{l_+}_{\alpha^+ = 1} q^+_{\alpha^+}
    + \sum^{l_-}_{\alpha^- = 1} q^-_{\alpha^-}
  \right)
  \log \left( \frac{\mu}{\Lambda_{\text{UV}}} \right)
  \, ,
\end{equation}
so for a fixed $\Lambda_{\text{UV}}$ and $r$, we have
\begin{equation}
  \label{eq:U(1|1) with W:quantum glsm:r-eff(mu)}
  \saveboxed{eq:U(1|1) with W:quantum glsm:r-eff(mu)}{
    r_{\text{eff}}(\mu)
    =  \left(
      m - n
      - \sum^{l_+}_{\alpha^+ = 1} q^+_{\alpha^+}
      + \sum^{l_-}_{\alpha^- = 1} q^-_{\alpha^-}
    \right)
    \log \left( \frac{\mu}{\Lambda} \right)
  }
\end{equation}
where the $r$ is being absorbed in the redefinition of $\Lambda_{\text{UV}}$ as $\Lambda$, such that $\Lambda$ can be regarded as a dynamically-generated scale parameter of mass dimension.

\subtitle{Scale-dependence}

If $m - n \neq \sum_{\alpha^+} q^+_{\alpha^+} - \sum_{\alpha^-} q^-_{\alpha^-}$, from \eqref{eq:U(1|1) with W:quantum glsm:r-eff(mu)}, we see that the effective FI parameter $r_{\text{eff}} (\mu)$ runs (i.e., RG flows) as we change the energy scale $\mu$.
In particular, the dimensionless parameter $r$ of the classical theory has been replaced by the scale parameter $\Lambda$ of mass dimension in the quantum theory.

\subtitle{Restoring Poincar\'{e} Invariance and $\abs{r} \gg 0$}

As before, the discussion regarding the restoration of Poincar\'{e} invariance in the continuum limit $\Lambda_{\text{UV}} \to \infty$ tells us that we need to consider $|r| \gg 0$ in the quantum theory.

For example, if $m - n > \sum_{\alpha^+} q^+_{\alpha^+} - \sum_{\alpha^-} q^-_{\alpha^-}$, in the limit $\Lambda_{\text{UV}} \rightarrow \infty$, we see from \eqref{eq:U(1|1) with W:quantum glsm:r-eff} that we ought to have $r \gg 0$ so that  $r_{\text{eff}}$ would be finite and also positive.
On the other hand, if $m - n < \sum_{\alpha^+} q^+_{\alpha^+} - \sum_{\alpha^-} q^-_{\alpha^-}$, in the limit $\Lambda_{\text{UV}} \rightarrow \infty$, we see from \eqref{eq:U(1|1) with W:quantum glsm:r-eff} that we ought to have $r \ll 0$ so that $r_{\text{eff}}$ would be finite and also negative.

\subtitle{$\MSUSY = \MSUSY^{r_{\text{eff}}(\text{i})}$/$\MSUSY^{r_{\text{eff}}(\text{ii})}$ of the Quantum GLSM}

Last but not least, recall that classically, $\MSUSY$ was given by $\MSUSY^{r_{\text{eff}}(\text{i}) > 0}$/$\MSUSY^{r_{\text{eff}}(\text{ii}) > 0}$ in \eqref{eq:U(1|1) with W:phases:geometric space:r(i) > 0}--\eqref{eq:U(1|1) with W:phases:hypersurface -:r(i) > 0}/\eqref{eq:U(1|1) with W:phases:geometric space:r(ii) > 0} when $r > 0$, and $\MSUSY^{r_{\text{eff}}(\text{i}) < 0}$/$\MSUSY^{r_{\text{eff}}(\text{ii}) < 0}$ in \eqref{eq:U(1|1) with W:phases:geometric space:r(i) < 0}--\eqref{eq:U(1|1) with W:phases:hypersurface +:r(i) < 0}/\eqref{eq:U(1|1) with W:phases:geometric space:r(ii) < 0} when $r < 0$.

In the quantum theory, when
(I) $m - n > \sum_{\alpha^+} q^+_{\alpha^+} - \sum_{\alpha^-} q^-_{\alpha^-}$ and
(II) $m - n < \sum_{\alpha^+} q^+_{\alpha^+} - \sum_{\alpha^-} q^-_{\alpha^-}$,
$\MSUSY$ would necessarily be given by
(I) $\MSUSY^{r_{\text{eff}}(\text{i}) > 0}$/$\MSUSY^{r_{\text{eff}}(\text{ii}) > 0}$ and
(II) $\MSUSY^{r_{\text{eff}}(\text{i}) < 0}$/$\MSUSY^{r_{\text{eff}}(\text{ii}) < 0}$,
where
(I) $r(\text{i})/r(\text{ii}) \gg 0$ and
(II) $r(\text{i})/r(\text{ii}) \ll 0$.

In other words, when $m - n \neq \sum_{\alpha^+} q^+_{\alpha^+} - \sum_{\alpha^-} q^-_{\alpha^-}$ and the quantum GLSM is \emph{not} scale-invariant, $\MSUSY$ can \emph{only} be $\MSUSY^{r_{\text{eff}}(\text{i}) > 0}$/$\MSUSY^{r_{\text{eff}}(\text{ii}) > 0}$ or $\MSUSY^{r_{\text{eff}}(\text{i}) < 0}$/$\MSUSY^{r_{\text{eff}}(\text{ii}) < 0}$ (depending on the values of the positive integers $m, n, q^+_{\alpha^+}, q^-_{\alpha^-}$).

\subsection{Quantum Aspects: The Low Energy Theory \label{subsec:quantum low energy U(1|1) with W}}

\subtitle{$\MSUSY = \MSUSY^{r_{\text{eff}}(\text{i})}$/$\MSUSY^{r_{\text{eff}}(\text{ii})}$ of the Quantum GLSM at Low Energy}

Recall that in our derivation of $\MSUSY = \MSUSY^{r_{\text{eff}}(\text{i})}$/$\MSUSY^{r_{\text{eff}}(\text{ii})}$ at the classical level in \autoref{subsection: U(1|1) with W_super}, we needed to have $\abs{r} \neq 0$.
In the quantum theory, when $m - n \neq \sum_{\alpha^+} q^+_{\alpha^+} - \sum_{\alpha^-} q^-_{\alpha^-}$, in place of $r$, we have $r_{\text{eff}} (\mu)$, and since we have to consider either $r \gg 0$ or $r \ll 0$ when $m - n > \sum_{\alpha^+} q^+_{\alpha^+} - \sum_{\alpha^-} q^-_{\alpha^-}$  or $m - n < \sum_{\alpha^+} q^+_{\alpha^+} - \sum_{\alpha^-} q^-_{\alpha^-}$ (as explained near the end of \autoref{subsec: Quantum Prop U(1|1) GLSM with W})), one can see from \eqref{eq:U(1|1) with W:quantum glsm:r-eff} that even at low energy $\mu \ll \Lambda_{\text{UV}}$, we will still have $r_{\text{eff}} (\mu) > 0$ or $r_{\text{eff}} (\mu) < 0$ when $m - n > \sum_{\alpha^+} q^+_{\alpha^+} - \sum_{\alpha^-} q^-_{\alpha^-}$  or $m - n < \sum_{\alpha^+} q^+_{\alpha^+} - \sum_{\alpha^-} q^-_{\alpha^-}$, respectively, whence $\MSUSY$ would continue to be well-defined in the quantum GLSM at low energy.

\subtitle{The Quantum GLSM at Low Energy is a Quantum NLSM, Quantum Supergauged LG Orbifold, or Either }

The analysis in \autoref{subsec: Phases U(1|1) with W} (and therefore \autoref{subsec: reduced U(1) model}) had been about the classical theory. The question therefore, is whether at the quantum level, the GLSM will also become the NLSM and supergauged LG orbifold at low energy, as per the classical theory. The short answer is ``yes'', although to the former, latter, or either, depends on the value of  $(m - n - \sum_{\alpha^+} q^+_{\alpha^+}+ \sum_{\alpha^-} q^-_{\alpha^-} )$, and whether we are in Branch (1) or Branch (2) of $\MSUSY$.

Let us now elaborate on the three possible cases for when we are in Branch (1).

\begin{itemize}

  \item $m - n > \sum_{\alpha^+} q^+_{\alpha^+} - \sum_{\alpha^-} q^-_{\alpha^-}$ ($\MSUSY^{r_{\text{eff}}(\text{i}) > 0}$)

  In this case, as explained above, we must have $r \gg 0$ and $r_{\text{eff}}(\mu) > 0$. Also, according to our discussion in \autoref{subsec: Quantum Low Energy U(1|1)} on the Higgs mechanism, the quantum low energy limit exists at a scale $\mu \ll e \sqrt{r}$.

  Thus, since $r_{\text{eff}}(\mu) > 0$,  at low energy, the quantum GLSM with $\MSUSY^{r_{\text{eff}}(\text{i}) > 0}$ defined by \eqref{eq:U(1|1) with W:phases:geometric space:r(i) > 0}--\eqref{eq:U(1|1) with W:phases:hypersurface -:r(i) > 0}, would become a quantum NLSM whose target space is $\MSUSY^{r_{\text{eff}}(\text{i}) > 0}$, the complete intersection of $l_+$ even and $l_-$ odd hypersurfaces $G^+_{\alpha^+} = 0$ ($\alpha^+ \in \{ 1, \dots, l_+ \}$) and $G^-_{\alpha^-} = 0$ ($\alpha^- \in \{ 1, \dots, l_-\}$) of degrees $q^+_{\alpha^+}$ and $q^-_{\alpha^-}$ in the super-Grassmannian ${Gr}_{1|1}(\mathbb{C}^{m|n})$ of size $r_{\text{eff}}(\mu)$.

  At any rate, for the quantum NLSM to be well-defined, i.e., amenable to perturbation theory, at the low energy scale $\mu$, its inverse coupling, which is proportional to $r_{\text{eff}}(\mu)$, needs to be large; in other words, we need to have $r_{\text{eff}}(\mu) \gg 0$.
  In turn, from \eqref{eq:U(1|1) with W:quantum glsm:r-eff(mu)}, it would mean that $\mu \gg \Lambda$ and therefore, $\Lambda \ll \mu \ll e \sqrt{r}$.
  Thus, in place of the classical low energy limit $e, b^\pm_{\alpha} \to \infty$, we would have the quantum low energy limit $\flatfrac{e}{\Lambda}, \flatfrac{b^\pm_{(\alpha)}}{\Lambda}, \flatfrac{c^\pm_{(\alpha)}}{\Lambda} \to \infty$.

  \item $m - n < \sum_{\alpha^+} q^+_{\alpha^+} - \sum_{\alpha^-} q^-_{\alpha^-}$ ($\MSUSY^{r_{\text{eff}}(\text{ii}) < 0}$)

  In this case, as explained above, we must have $r \ll 0$ and $r_{\text{eff}}(\mu) < 0$.
  Also, according to our discussion in \autoref{subsec: Quantum Low Energy U(1|1)} on the Higgs mechanism, the quantum low energy limit exists at a scale $\mu \ll e \sqrt{|r|}$.

  Thus, since $r_{\text{eff}}(\mu) < 0$, at low energy, the quantum GLSM with $\MSUSY^{r_{\text{eff}}(\text{ii}) < 0}$ defined by \eqref{eq:U(1|1) with W:phases:geometric space:r(i) < 0}, would become a quantum $\mathrm{SU}(1|1) \times \mathbb{Z}_q$ supergauged LG orbifold with effective superpotential \eqref{eq:U(1|1) with W:phases:superpotential W:r(ii) < 0} that is defined on $\mathcal{S}^{\mathrm{U}(1|1)^{|r_{\text{eff}}| > 0}}_{q_\alpha, p^\alpha}$, where $q$ is the greatest common divisor of all the degrees $q^\pm_{\alpha^\pm}$ of the $G^\pm_{\alpha^\pm}$'s, and $\mathcal{S}^{\mathrm{U}(1|1)^{|r_{\text{eff}}| > 0}}_{q_\alpha, p^\alpha}$ is a supervector bundle over $\WCP^{{l_+ -1}|l_-}_{(\vv{q^+}|\vv{q^-})}$ of size $|r_{\text{eff}}(\mu)|$ described in \eqref{eq:U(1|1) with W:phases:supervector bundle:r(ii) < 0}.

  If we insist on being congruent with the NLSM to have $r_{\text{eff}}(\mu) \ll 0$, from \eqref{eq:U(1|1) with W:quantum glsm:r-eff(mu)}, it would mean that $\mu \gg \Lambda$ and therefore, $\Lambda \ll \mu \ll e \sqrt{|r|}$.
  Then, in place of the classical low energy limit $e \to \infty$, we would have the quantum low energy limit $\flatfrac{e}{\Lambda} \to \infty$.

  \item $m - n = \sum_{\alpha^+} q^+_{\alpha^+} - \sum_{\alpha^-} q^-_{\alpha^-}$ ($\MSUSY^{r(\text{i}) > 0}$/$\MSUSY^{r(\text{ii}) < 0}$)

  In this case, one can see from \eqref{eq:U(1|1) with W:quantum glsm:r-eff(mu)} that the FI parameter does not run, so $r$ is still a dimensionless parameter of the quantum theory, whence the quantum GLSM is scale-invariant.
  As such, we can \emph{freely} choose $r > 0$ or $r < 0$, where at low energy, the quantum GLSM with $\MSUSY^{r(\text{i}) > 0}$ or $\MSUSY^{r(\text{ii}) < 0}$ becomes the above-mentioned quantum NLSM or quantum supergauged LG orbifold, respectively.
  Since there is no dynamical scale $\Lambda$, the quantum low energy limit would instead be given by $e \sqrt{r}, b^\pm_{(\alpha)} \sqrt{r}, c^\pm_{(\alpha)} \sqrt{r} \to \infty$ and  $e \sqrt{|r|} \to \infty$, respectively.

\end{itemize}

The three possible cases for when we are in Branch (2), are just a parity-reversed mirror of the above, so we shall just state the results.

\begin{itemize}
  \item $m - n > \sum_{\alpha^+} q^+_{\alpha^+} - \sum_{\alpha^-} q^-_{\alpha^-}$ ($\MSUSY^{r_{\text{eff}}(\text{ii}) > 0}$)

  In the low energy limit $e/\Lambda \to \infty$, the quantum GLSM with $\MSUSY^{r_{\text{eff}}(\text{ii}) > 0}$ defined by \eqref{eq:U(1|1) with W:phases:geometric space:r(ii) > 0}, would become a quantum $\text{SU}(1|1) \times \mathbb{Z}_q$ supergauged LG orbifold with effective superpotential \eqref{eq:U(1|1) with W:phases:superpotential W:r(ii) > 0} that is defined on $\mathcal{S}^{\mathrm{U}(1|1)^{r_{\text{eff}} > 0}}_{q_\alpha, p^\alpha}$, where $q$ is the greatest common divisor of all the degrees $q^-_{\alpha^-}/q^+_{\alpha^+}$ of the $G^+_{\alpha^-(\mathbf{\Pi})}/G^-_{\alpha^+(\mathbf{\Pi})}$'s, and $\mathcal{S}^{\mathrm{U}(1|1)^{r_{\text{eff}} > 0}}_{q_\alpha, p^\alpha}$ is a supervector bundle over $\WCP^{{l_- -1}|l_+}_{\vec{q^-}|\vec{q^+}}$ of size $r_{\text{eff}}(\mu)$ described in \eqref{eq:U(1|1) with W:phases:supervector bundle:r(ii) > 0}.

  \item $m - n < \sum_{\alpha^+} q^+_{\alpha^+} - \sum_{\alpha^-} q^-_{\alpha^-}$ ($\MSUSY^{r_{\text{eff}}(\text{i}) < 0}$)

  In the low energy limit $\flatfrac{e}{\Lambda}, \flatfrac{b^{\pm}_{(\alpha)}}{\Lambda}, \flatfrac{c^{\pm}_{(\alpha)}}{\Lambda} \to \infty$, the quantum GLSM with $\MSUSY^{r_{\text{eff}}(\text{i}) < 0}$ defined by \eqref{eq:U(1|1) with W:phases:geometric space:r(i) < 0}--\eqref{eq:U(1|1) with W:phases:hypersurface +:r(i) < 0}, would become a quantum NLSM whose target space is $\MSUSY^{r_{\text{eff}}(\text{i}) < 0}$, the complete intersection of $l_-$ \emph{even} and $l_+$ \emph{odd} hypersurfaces $G^+_{\alpha^-(\mathbf{\Pi})} = 0$ ($\alpha^- \in \{1, \dots, l_-\}$) and $G^-_{\alpha^+(\mathbf{\Pi})} = 0$ ($\alpha^+ \in \{ 1, \dots, l_+\}$) of \emph{even} and \emph{odd} degrees $q^-_{\alpha^-}$ and $q^+_{\alpha^+}$ in the super-Grassmannian ${Gr}_{1|1}(\mathbb{C}^{n|m})$ of size $|r_{\text{eff}}(\mu)|$.

  \item $m - n = \sum_{\alpha^+} q^+_{\alpha^+} - \sum_{\alpha^-} q^-_{\alpha^-}$  ($\MSUSY^{r(\text{ii}) > 0}$/$\MSUSY^{r(\text{i}) < 0}$)

  In this case, we can \emph{freely} choose $r > 0$ or $r < 0$, where in the low energy limit $e \sqrt{r} \to \infty$ or $e \sqrt{|r|},  b^\pm_{(\alpha)} \sqrt{|r|}, c^\pm_{(\alpha)} \sqrt{|r|} \to \infty$, the quantum GLSM with $\MSUSY^{r(\text{ii}) > 0}$ or $\MSUSY^{r(\text{i}) < 0}$ becomes the above-mentioned quantum supergauged LG orbifold or quantum NLSM, respectively.

\end{itemize}

\subsection{Scale-invariance and Applications to Mathematics}
\label{subsec: applications to math U(1|1) with W}

Let us consider the case where $m - n = \sum_{\alpha^+} q^+_{\alpha^+} - \sum_{\alpha^-} q^-_{\alpha^-}$. As explained in \autoref{subsec:quantum low energy U(1|1) with W}, the quantum GLSM would be scale-invariant with free parameter $r$.

\subtitle{Scale-invariance and a Physical Derivation of a Novel CY Condition for a Complete Intersection of Hypersurfaces in a Super-Grassmannian}

As the quantum GLSM is scale-invariant, its corresponding low energy quantum NLSM ought to also be scale and therefore conformally-invariant~\cite{papadopoulos2024scaleconformalinvariance2d}; in turn, according to our explanation in \autoref{subsec: applications to math U(1|1)}, this would mean that the target space of the quantum NLSMs in \autoref{subsec:quantum low energy U(1|1) with W}, i.e., $\MSUSY^{r(\text{i}) > 0}$ and $\MSUSY^{r(\text{i}) < 0}$, must be CY.

Recall that $\MSUSY^{r(\text{i}) > 0}$ is a complete intersection of $l_+$ even and $l_-$ odd hypersurfaces $G^+_{\alpha^+} = 0$ ($\alpha^+ \in \{ 1, \dots, l_+ \}$) and $G^-_{\alpha^-} = 0$ ($\alpha^- \in \{ 1, \dots, l_-\}$) of even and odd degrees $q^+_{\alpha^+}$ and $q^-_{\alpha^-}$ in the super-Grassmannian ${Gr}_{1|1}(\mathbb{C}^{m|n})$ of size $r$.
Therefore, since $m - n = \sum_{\alpha^+} q^+_{\alpha^+} - \sum_{\alpha^-} q^-_{\alpha^-}$, the difference between the even and odd dimensions of the ambient space will match the difference between the total even and total odd degrees of the hypersurfaces.

Recall also that $\MSUSY^{r(\text{i}) < 0}$ is a complete intersection of $l_-$ even and $l_+$ odd hypersurfaces $G^+_{\alpha^- (\mathbf{\Pi})} = 0$ and $G^-_{\alpha^+ (\mathbf{\Pi})} = 0$ of even and odd degrees $q^-_{\alpha^-}$ and $q^+_{\alpha^+}$ in the super-Grassmannian $Gr_{1|1}(\mathbb{C}^{n|m})$ of size $|r|$.
Therefore, since $m - n = \sum_{\alpha^+} q^+_{\alpha^+} - \sum_{\alpha^-} q^-_{\alpha^-}$, or equivalently, since $n - m = \sum_{\alpha^-} q^-_{\alpha^-} - \sum_{\alpha^+} q^+_{\alpha^+}$, the difference between the even and odd dimensions of the ambient space will again match the difference between the total even and total odd degrees of the hypersurfaces.

Thus, we have a physical derivation of a \emph{novel} mathematical result that a complete intersection of hypersurfaces in a super-Grassmannian is CY if and only if the difference between the even and odd dimensions of the ambient space matches the difference between the total even and total odd degrees of the hypersurfaces!

Indeed, one can actually mathematically verify (see~\autoref{app:verify cy:hypersurfaces in Gr}) this statement to be true!

This yet again vindicates our application to mathematics of our otherwise nonunitary supergroup GLSM via a study of its space of supersymmetric states.

\subsection{A Super-Grassmannian/Supergroup Generalization of the CY/LG Correspondence for Complete Intersections in Mathematics}
\label{subsec: Super CY/LG for complete intersections}

Staying with the scale-invariant quantum GLSM where $m - n = \sum_{\alpha^+} q^+_{\alpha^+} - \sum_{\alpha^-} q^-_{\alpha^-}$, we shall now derive a highly-nontrivial geometric/non-geometric correspondence between a super CY and a supergauged LG orbifold, and  vice versa.

\subtitle{A Geometric/Non-Geometric Correspondence Between a Super CY and a Supergauged LG Orbifold}

Since we can freely choose $r > 0$ or $r<0$, it would mean that from the perspective of our quantum GLSM, there ought to be a physical correspondence between
(I) the quantum NLSM of \autoref{subsec:quantum low energy U(1|1) with W} with super CY target space $\MSUSY^{r(\text{i}) > 0}$ as described therein,
and
(II) the quantum $\mathrm{SU}(1|1) \times \mathbb{Z}_q$ supergauged LG orbifold of \autoref{subsec:quantum low energy U(1|1) with W} that is defined on $\mathcal{S}^{\mathrm{U}(1|1)^{|r| > 0}}_{q_\alpha, p^\alpha}$ as described therein.

Specifically, we ought to have a geometric/non-geometric correspondence between
\begin{enumerate}[label=(\Roman*)]
  \item a quantum NLSM with super CY target space being a complete intersection of $l_+$ even and $l_-$ odd hypersurfaces $G^+_{\alpha^+} = 0$ ($\alpha^+ \in \{1, \dots, l_+\}$) and $G^-_{\alpha^-}=0$ ($\alpha^- \in \{1, \dots, l_-\}$) of degrees $q^+_{\alpha^+}$ and $q^-_{\alpha^-}$ described by \eqref{eq:u-1-1 with W:glsm:G+ alpha}--\eqref{eq:u-1-1 with W:glsm:Y+ odd} and \eqref{eq:u-1-1 with W:glsm:G- alpha}--\eqref{eq:u-1-1 with W:glsm:Y- odd} in the super-Grassmannian ${Gr}_{1|1}(\mathbb{C}^{m|n})$ of size $r$, where $m - n = \sum_{\alpha^+} q^+_{\alpha^+} - \sum_{\alpha^-} q^-_{\alpha^-}$;

  and

  \item a quantum $\mathrm{SU}(1|1) \times \mathbb{Z}_q$ supergauged LG orbifold with effective superpotential \eqref{eq:U(1|1) with W:phases:superpotential W:r(ii) < 0}--\eqref{eq:U(1|1) with W:phases:geometric space:r(ii) < 0:exp val} that is defined on $\mathcal{S}^{\mathrm{U}(1|1)^{|r| > 0}}_{q_\alpha, p^\alpha}$, where $q$ is the greatest common divisor of all the degrees $q^\pm_{\alpha^\pm}$ of the $G^\pm_{\alpha^\pm}$'s, and $\mathcal{S}^{\mathrm{U}(1|1)^{|r| > 0}}_{q_\alpha, p^\alpha}$ is a supervector bundle over $\WCP^{{l_+ -1}|l_-}_{(\vv{q^+}|\vv{q^-})}$ of size $|r|$ described in \eqref{eq:U(1|1) with W:phases:supervector bundle:r(ii) < 0} with $m - n = \sum_{\alpha^+} q^+_{\alpha^+} - \sum_{\alpha^-} q^-_{\alpha^-}$!

\end{enumerate}

\subtitle{A Non-Geometric/Geometric Correspondence Between a Supergauged LG Orbifold and a Super CY}

Similarly, it would mean that from the perspective of our quantum GLSM, there ought to be a physical correspondence between
\begin{enumerate*}[label=(\Roman*$^{\prime}$)]
  \item the quantum $\mathrm{SU}(1|1) \times \mathbb{Z}_q$ supergauged LG orbifold of \autoref{subsec:quantum low energy U(1|1) with W} that is defined on $\mathcal{S}^{\mathrm{U}(1|1)^{r > 0}}_{q_\alpha, p^\alpha}$ as described therein, and

  \item the quantum NLSM of \autoref{subsec:quantum low energy U(1|1) with W} with super CY target space $\MSUSY^{r(\text{i}) < 0}$ as described therein.
\end{enumerate*}

Specifically, we ought to have a non-geometric/geometric correspondence between
\begin{enumerate}[label=(\Roman*$^{\prime}$)]
  \item a quantum $\mathrm{SU}(1|1) \times \mathbb{Z}_q$ supergauged LG orbifold with effective superpotential \eqref{eq:U(1|1) with W:phases:superpotential W:r(ii) > 0}--\eqref{eq:U(1|1) with W:phases:geometric space:r(ii) > 0:exp val} that is defined on $\mathcal{S}^{\mathrm{U}(1|1)^{r > 0}}_{q_\alpha, p^\alpha}$, where $q$ is the greatest common divisor of all the degrees $q^-_{\alpha^-}/q^+_{\alpha^+}$ of the $G^+_{\alpha^-(\mathbf{\Pi})}/G^-_{\alpha^+(\mathbf{\Pi})}$'s, and $\mathcal{S}^{\mathrm{U}(1|1)^{r > 0}}_{q_\alpha, p^\alpha}$ is a supervector bundle over $\WCP^{{l_- -1}|l_+}_{(\vv{q^-}|\vv{q^+})}$ of size $r$ described in \eqref{eq:U(1|1) with W:phases:supervector bundle:r(ii) > 0} with $n - m = \sum_{\alpha^-} q^-_{\alpha^-} - \sum_{\alpha^+} q^+_{\alpha^+}$;

  and

  \item a quantum NLSM with super CY target space being a complete intersection of $l_-$ even and $l_+$ odd hypersurfaces $G^+_{\alpha^-(\mathbf{\Pi})}=0$ ($\alpha^- \in \{ 1, \dots, l_-\}$)  and $G^-_{\alpha^+(\mathbf{\Pi})}=0$ ($\alpha^+ \in \{ 1, \dots, l_+\}$) of degrees $q^-_{\alpha^-}$ and $q^+_{\alpha^+}$ described by a parity-reversed version of \eqref{eq:u-1-1 with W:glsm:G- alpha}--\eqref{eq:u-1-1 with W:glsm:Y- odd} and \eqref{eq:u-1-1 with W:glsm:G+ alpha}--\eqref{eq:u-1-1 with W:glsm:Y+ odd}  in the super-Grassmannian ${Gr}_{1|1}(\mathbb{C}^{n|m})$ of size $|r|$, where $n - m = \sum_{\alpha^-} q^-_{\alpha^-} - \sum_{\alpha^+} q^+_{\alpha^+}$!
\end{enumerate}

Notice that this non-geometric/geometric correspondence between ($\text{I}^\prime$) and ($\text{II}^\prime$) is just a parity-reversed mirror version of the geometric/non-geometric correspondence between (I) and (II)!

\subtitle{A Super-Grassmannian/Supergroup Generalization of the CY/LG Correspondence for Complete Intersections in Mathematics}

At any rate, in carrying out the reduction to the familiar $\mathrm{U}(1)$ GLSM with superpotential at low energy (via \autoref{subsec: reduced U(1) model}) whence the scale-invariant condition is now $m = \sum_{\alpha^+} q^+_{\alpha^+}$, we find that in place of (I) and (II), we would have
\begin{enumerate*}[label=(\textbf{\roman*})]
  \item a quantum NLSM with CY target space being a complete intersection of hypersurfaces $G^+_{\alpha^+} (\tensor{\phi}{^1_s}) = 0$ $(\alpha^+ \in \{1, \dots, l_+\})$ in $\mathbb{CP}^{m-1}$ of size $r$,\footnote{%
  It is a well-known fact that an intersection of hypersurfaces in $\mathbb{CP}^{d-1}$ is CY if and only if the total degree $f$ of the hypersurfaces equals $d$, which is exactly what we have here.
} and

\item a quantum LG $\mathbb{Z}_q$-orbifold with effective superpotential $W^{\mathbb{Z}_q}_{\text{LG}}$ in \eqref{W^Zq_LG}  defined on the vector bundle $\mathcal{S}^{\mathrm{U}(1)^{|r| > 0}}_{q^+_\alpha, p_+^\alpha}$ in \eqref{vector bundle over WCP - r < 0}.
\end{enumerate*}

Note that the celebrated CY/LG correspondence for complete intersections is exactly a correspondence between the low energy models just described in (\textbf{i}) and (\textbf{ii}) above.

Thus, in deriving the above geometric/non-geometric correspondence between (I) and (II), we have a super-Grassmannian/supergroup generalization of the  CY/LG correspondence for complete intersections established in mathematics by Clader~\cite{clader2013landauginzburgcalabiyaucorrespondencecompleteintersections} and Zhao~\cite{zhao2019landauginzburgcalabiyaucorrespondencecompleteintersection}!

In fact, in deriving the above non-geometric/geometric correspondence between (II$'$) and (I$'$), we also have a parity-reversed mirror version of this generalization!

\section{Phases of a \texorpdfstring{$\mathrm{U}(1|1)^N$}{U(1|1)\textasciicircum N} GLSM with Superpotential: Complete Intersection of Hypersupersurfaces NLSMs, Supergauged LG Orbifolds, and their Hybrids}
\label{sec: Phases of U(1|1)^N with W}

In this section, we will study a $\mathrm{U}(1|1)^2$ GLSM with a superpotential, and show that its space of supersymmetric states, in its different phases, corresponds either to
\begin{enumerate*}
  \item a complete intersection of an even and odd hypersurface in a product of super-Grassmannians,

  \item a product of weighted projective superspaces,

  or

  \item a product of a weighted projective superspace and a super-Grassmannian.

\end{enumerate*}
Going to the low energy limit of the GLSM, we will show that it will become either
\begin{enumerate*}
  \item an NLSM on the aforementioned complete intersection of an even and odd hypersurface in a product of super-Grassmannians,

  \item a supergauged LG orbifold on a product of supervector bundles whose corresponding bases are the aforementioned weighted projective superspaces,

  or

  \item a hybrid NLSM/supergauged LG orbifold defined on a supervector bundle over the aforementioned product of a weighted projective superspace and a super-Grassmannian.

\end{enumerate*}
Taking into account the quantum corrections to the FI parameters, we will have effective scale-dependent FI parameters in the quantum GLSM.
The quantum GLSM will also become
\begin{enumerate*}
  \item a quantum NLSM,

  \item a quantum supergauged LG orbifold,

  or

  \item a hybrid quantum NLSM/supergauged LG orbifold,
\end{enumerate*}
in the low energy limit.
Lastly, when the theory is scale-invariant, we will be able to (1)  physically derive a CY condition for a complete intersection of an even and odd hypersurface in a product of super-Grassmannians that we can also mathematically verify, (2) obtain a super-Grassmannian/supergroup generalization of the hybrid CY/LG correspondence for hypersurfaces in a product of projective spaces established in mathematics by Fan-Jarvis-Ruan~\cite[$\S$7.3]{Fan_2017}, and (3) obtain a parity-reversed mirror version of the aforementioned generalization.

\subsection{A \texorpdfstring{$\mathrm{U}(1|1)^N$}{U(1|1)\textasciicircum N} GLSM with a Superpotential and its \texorpdfstring{$\MSUSY$}{M-SUSY}}
\label{subsec: U(1|1)^2 with W}

\subtitle{The $\mathrm{U}(1|1)_1 \times \mathrm{U}(1|1)_2$ GLSM with a Superpotential}

We would now like to construct a supergroup GLSM whose $\MSUSY$ can, for some value of the FI parameter, be a hypersurface in a product space.

To this end, we ought to consider the $\mathrm{U}(1|1)^N$ model in \autoref{sec: U(1|1)^N GLSM} but with a superpotential.
It would be simple yet illuminating enough to consider the $N = 2$ case.
Specifically, we would like to construct a $\mathrm{U}(1|1)_1\times \mathrm{U}(1|1)_2$ GLSM using two vector superfields $V_1$ and $V_2$, with gauge couplings $e_1$ and $e_2$, respectively.
We would also like to introduce the chiral matter superfields
\begin{enumerate*}
  \item $S_a$, where $a \in \{1, \dots, m_1\}$, which are Grassmann-even and identically charged in the fundamental representation of the $\mathrm{U}(1|1)_1$ gauge group;

  \item $\mathbx{S}_b$, where $b \in \{1, \dots, n_1\}$, which are Grassmann-odd and identically charged in the fundamental representation of the $\mathrm{U}(1|1)_1$ gauge group;

  \item $T_{\alpha}$, where ${\alpha} \in \{1, \dots, m_2\}$, which are Grassmann-even and identically charged in the fundamental representation of the $\mathrm{U}(1|1)_2$ gauge group;

  \item $\mathbx{T}_{\beta}$, where $\beta \in \{1, \dots, n_2\}$, which are Grassmann-odd and identically charged in the fundamental representation of the $\mathrm{U}(1|1)_2$ gauge group;

  \item $P_+$, which is Grassmann-even with gauge charges $(- \hat{m}_1, - \hat{m}_2)$ under the superdeterminantal representation of $\mathrm{U}(1|1)_1\times \mathrm{U}(1|1)_2$, where $\hat{m}_1, \hat{m}_2 \in \mathbb{Z}^+$ and $\hat{m}_2 = k \hat{m}_1$;

  and

  \item $P_-$, which is Grassmann-odd with gauge charges $(- \hat{n}_1, - \hat{n}_2)$ under the anti-superdeterminantal representation of $\mathrm{U}(1|1)_1\times \mathrm{U}(1|1)_2$, where $\hat{n}_1, \hat{n}_2 \in \mathbb{Z}^+$ and $\hat{n}_2 = k \hat{n}_1$.
\end{enumerate*}
This is summarized in \autoref{fig:U(1|1)-N with W:gauge charge assignment}.
\begin{figure}
  \centering
  \begin{tabular}{||c|c|c|c|c||}
    \hline
    Superfield & Grassmann Parity & Representation & $\mathrm{U}(1|1)_1$ charge & $\mathrm{U}(1|1)_2$ charge
    \\ \hline \hline
    $S_a$ & Even & Fundamental & 1 & 0
    \\ \hline
    $\mathbx{S}_b$ & Odd & Fundamental & 1 & 0
    \\ \hline
    $T_{\alpha}$ & Even & Fundamental & 0 & 1
    \\ \hline
    $\mathbx{T}_{\beta}$ & Odd & Fundamental & 0 & 1
    \\ \hline
    $P_+$ & Even & Superdeterminantal & $- \hat{m}_1$ & $- \hat{m}_2 = - k \hat{m}_1$
    \\ \hline
    $P_-$ & Odd & Anti-Superdeterminantal & $- \hat{n}_1$ & $- \hat{n}_2 = - k \hat{n}_1$
    \\ \hline
  \end{tabular}
  \caption{Gauge charge assignment of the chiral matter superfields.}
  \label{fig:U(1|1)-N with W:gauge charge assignment}
\end{figure}

Then, the $\mathcal{N} = (2, 2)$ Lagrangian ought to be given by
\begin{equation}
  \begin{aligned}
    L_{W_{\text{super}}}^{\mathrm{U}(1|1)_1 \times \mathrm{U}(1|1)_2}
    &= \frac{1}{4}\int \dd[4]{\theta} \Bigg(
      \sum_{a = 1}^{m_1} \bar{S}^a e^{2V_1} S_a
      + \sum_{b = 1}^{n_1} \bar{\mathbx{S}}^b e^{2V_1} \mathbx{S}_b
      + \sum_{\alpha = 1}^{m_2}\bar{T}^{\alpha} e^{2V_2} T_{\alpha}
      + \sum_{\beta = 1}^{n_2} \bar{\mathbx{T}}^{\beta} e^{2V_2} \mathbx{T}_{\beta}
      \\
      & \qquad \qquad \qquad
      + \bar{P}^+ e^{-2 \hat{m}_1 \Str V_1 - 2 \hat{m}_2 \Str V_2}P_+
      + \bar{P}^- e^{-2 \hat{n}_1 \Str V_1 - 2 \hat{n}_2 \Str V_2}P_-
    \bigg)
    \\
    &\quad
    - \frac{1}{4e_1^2}\int \dd[4]{\theta} \Str \bar{\Sigma}_1 \Sigma_1 
    - \frac{1}{4e_2^2}\int \dd[4]{\theta} \Str \bar{\Sigma}_2 \Sigma_2
    \\
    &\quad
    + \left(
      \frac{1}{2 \sqrt{2}} \int \dd[2]{\theta} (\mathbf{i} t_1 \Str \Sigma_1) + c.c
    \right)
    + \left(
      \frac{1}{2 \sqrt{2}} \int \dd[2]{\theta} (\mathbf{i} t_2 \Str \Sigma_2) + c.c
    \right)
    \\
    &\quad
    - \frac{1}{4} \left(
      \int \dd[2]{\theta} W^{\mathrm{U}(1|1)_1 \times \mathrm{U}(1|1)_2}_{\text{super}} (P, S, \mathbx{S}, T, \mathbx{T})
      + h.c.
    \right)
    \, ,
  \end{aligned}
\end{equation}
where
\begin{equation}
  \label{eq:U(1|1)-N with W:W_super U(1|1)^2}
  \saveboxed{eq:U(1|1)-N with W:W_super U(1|1)^2}{
    W^{\mathrm{U}(1|1)_1 \times \mathrm{U}(1|1)_2}_{\text{super}} (P, S, \mathbx{S}, T, \mathbx{T})
    = P_+ G^+ \left( Y^+(S, \mathbx{S}), Y^+(T, \mathbx{T}) \right)
    + P_- G^- \left( Y^-(S, \mathbx{S}), Y^-(T, \mathbx{T}) \right)
  }
\end{equation}
is a gauge-invariant $\mathrm{U}(1|1)_1 \times \mathrm{U}(1|1)_2$ superpotential, such that
\begin{enumerate*}
  \item $G^+$ is a Grassmann-even bihomogeneous polynomial of degree $\hat{m}_1$ and $\hat{m}_2$ in $Y^+(S, \mathbx{S})$ and $Y^+(T, \mathbx{T})$, respectively, described by \eqref{eq:u-1-1 with W:glsm:G+ alpha} and the equations around it;

  \item $G^-$ is a Grassmann-odd bihomogeneous polynomial of degree $\hat{n}_1$ and $\hat{n}_2$ in $Y^-(S, \mathbx{S})$ and $Y^-(T, \mathbx{T})$, respectively, described by \eqref{eq:u-1-1 with W:glsm:G- alpha} and the equations around it;

  \item $Y^+$ and $Y^-$ are as defined in \eqref{eq:u-1-1 with W:glsm:Y+ even}--\eqref{eq:u-1-1 with W:glsm:Y+ odd} and \eqref{eq:u-1-1 with W:glsm:Y- even}--\eqref{eq:u-1-1 with W:glsm:Y- odd}, respectively;

  and

  \item just as in \autoref{subsection: U(1|1) with W_super}, $G^{\pm}$ are transverse in the sense that

\end{enumerate*}
\begin{equation}
  \label{eq:U(1|1)-N with W:glsm:transversality}
  \pdv{G^+}{Y^+_{\mu\nu}(S, \mathbx{S})}
  = 0
  = \pdv{G^+}{Y^+_{\mu\nu}(T, \mathbx{T})}
  \, ,
  \qand
  \pdv{G^-}{Y^-_{\mu\nu}(S, \mathbx{S})}
  = 0
  = \pdv{G^-}{Y^-_{\mu\nu}(T, \mathbx{T})}
  \, ,
\end{equation}
where $\{\mu, \nu\}$ run over all $Y^{\pm}$ field indices, such that this condition holds if and only if \emph{either} the $Y^{\pm}_{\mu\nu}(S, \mathbx{S})$ or $Y^{\pm}_{\mu\nu}(T, \mathbx{T})$ fields are zero.

Note that $ L^{\mathrm{U}(1|1)_1 \times \mathrm{U}(1|1)_2}_{W_{\text{super}}}$ has a $\mathrm{U}(1)_A \times \mathrm{U}(1)_V$ axial and vector R-symmetry, where the charges of the $S$'s, $T$'s, and $P$'s are $(0, *)$, while the charges of $\Sigma_{\{1, 2\}}$ and $W^{\mathrm{U}(1|1)_1 \times \mathrm{U}(1|1)_2}_{\text{super}}$ must be $(0,2)$.

\subtitle{The Spacetime Scalars from $S$, $\mathbx{S}$, $T$, $\mathbx{T}$, and $P_{\pm}$}

We shall denote by $(s_a, \mathbx{s}_b, t_{\alpha}, \mathbx{t}_{\beta}, p_+, p_-)$, the spacetime scalars from the chiral matter superfields $(S_a, \mathbx{S}_b, T_{\alpha}, \mathbx{T}_{\beta}, P_+, P_-)$.

As usual, $(s_a, \mathbx{s}_b, t_{\alpha}, \mathbx{t}_{\beta})$ will be labeled by the $\mathrm{U}(1|1)$ color indices $\{1, \bar{1}\}$, but not $(p_+, p_-)$.

\subtitle{The Potential Energy}

After eliminating the auxiliary field, the potential energy (of the scalars) becomes
\begin{equation}
  \label{eq:U(1|1)-N with W:glsm:11pot}
  \begin{aligned}
    U^{\mathrm{U}(1|1)_1 \times \mathrm{U}(1|1)_2}_{\text{pot}, W_{\text{super}}}
    &
    = \frac{e_1^2}{2} \sum_{i,j} (-1)^{\sigma(i)} \left[
      \sum_{a = 1}^{m_1} \tensor{s}{^j_a} \tensor{\bar{s}}{_i^a}
      - \sum_{b = 1}^{n_1} \tensor{\mathbx{s}}{^j_b} \tensor{\bar{\mathbx{s}}}{_i^b}
      - \left(
        \hat{m}_1 p_+ \bar{p}^{\, +}
        - \hat{n}_1 p_- \bar{p}^{\, -}
        + r_1
      \right) \tensor{\delta}{^j_i}
    \right]^2
    \\
    &\quad
    + \frac{e_2^2}{2} \sum_{i,j} (-1)^{\sigma(i)} \left[
      \sum_{\alpha = 1}^{m_2} \tensor{t}{^j_{\alpha}} \tensor{\bar{t}}{_i^{\alpha}}
      - \sum_{\beta = 1}^{n_2} \tensor{\mathbx{t}}{^j_{\beta}} \tensor{\bar{\mathbx{t}}}{_i^{\beta}}
      - \left(
        \hat{m}_2 p_+ \bar{p}^{\, +}
        - \hat{n}_2 p_- \bar{p}^{\, -}
        + r_2
      \right) \tensor{\delta}{^j_i}
    \right]^2
    \\
    &\quad
    + \left(
      \abs{G^+}^2
      + \sum_{\mu, \nu} \abs{p_+ \pdv{G^+}{Y^+_{\mu\nu}(s, \mathbx{s})}}^2
      + \sum_{\mu, \nu} \abs{p_+ \pdv{G^+}{Y^+_{\mu\nu}(t, \mathbx{t})}}^2
    \right)
    \\
    &\quad
    + \left(
      \abs{G^-}^2
      + \sum_{\mu, \nu} \abs{p_- \pdv{G^-}{Y^-_{\mu\nu}(s, \mathbx{s})}}^2
      + \sum_{\mu, \nu} \abs{p_- \pdv{G^-}{Y^-_{\mu\nu}(t, \mathbx{t})}}^2
    \right)
    \\
    &\quad
    + \frac{1}{2e_1^2} \Str \comm{\sigma_1}{\bar{\sigma}_1}^2
    + \sum_{a = 1}^{m_1} \bar{s}^a \acomm{\sigma_1}{\bar{\sigma}_1} s_a
    + \sum_{b = 1}^{n_1} \bar{\mathbx{s}}^b \acomm{\sigma_1}{\bar{\sigma}_1} \mathbx{s}_b
    \\
    &\quad
    + \frac{1}{2e_2^2} \Str \comm{\sigma_2}{\bar{\sigma}_2}^2
    + \sum_{\alpha = 1}^{m_2} \bar{t}^{\alpha} \acomm{\sigma_2}{\bar{\sigma}_2} t_{\alpha}
    + \sum_{\beta = 1}^{n_2} \bar{\mathbx{t}}^{\beta} \acomm{\sigma_2}{\bar{\sigma}_2} \mathbx{t}_{\beta}
    \\
    &\quad
    + 2 \abs{\hat{m}_1 \Str \sigma_1 + \hat{m}_2 \Str \sigma_2}^2 \bar{p}^{\, +} p_+
    + 2 \abs{\hat{n}_1 \Str \sigma_1 + \hat{n}_2 \Str \sigma_2}^2 \bar{p}^{\, -} p_-
    \, ,
  \end{aligned}
\end{equation}
where the indices $i, j \in \{1, \bar{1}\}$ are the color indices of the $\mathrm{U}(1|1)_{1,2}$ gauge groups.

\subtitle{\MSUSY}

$\MSUSY$ of the theory is defined by the supersymmetric configuration $U^{\mathrm{U}(1|1)_1 \times \mathrm{U}(1|1)_2}_{\text{pot}, W_{\text{super}}} = 0$ modulo gauge transformations.
From \eqref{eq:U(1|1)-N with W:glsm:11pot},  we see that generally for $r_1, r_2 \neq 0$, this requires $\sigma_1 = 0 = \sigma_2$, and the vanishing of the first, second, third and fourth line on the RHS of \eqref{eq:U(1|1)-N with W:glsm:11pot}.
In other words, $\MSUSY$ will be defined by\footnote{%
  The alert reader would recall that $p^+$ has charge $(- \hat{m}_1, - \hat{m}_2)$ and $p^-$ has charge $(- \hat{n}_1, - \hat{n}_2)$ under $\mathrm{U}(1)_1 \times \mathrm{U}(1)_2 \subset \mathrm{U}(1|1)_1 \times \mathrm{U}(1|1)_2$.
  However, from the expression of the potential energy in \eqref{eq:U(1|1)-N with W:glsm:11pot}, notice that $p^{\pm}$ are coupled to $\mathrm{U}(1|1)_1$ and $\mathrm{U}(1|1)_2$ via the gauge couplings $e_1$ and $e_2$ in the first and second line on the RHS in \eqref{eq:U(1|1)-N with W:glsm:11pot}, respectively.
  Therefore, in defining $\MSUSY$ by setting the aforementioned first and second line on the RHS to zero, we can regard $p^{\pm}$ to transform under $\mathrm{U}(1)_1 \subset \mathrm{U}(1|1)_1$ and $\mathrm{U}(1)_2 \subset \mathrm{U}(1|1)_2$, respectively.
  As such, the following definition of $\MSUSY$ is physically consistent.
}
\begin{equation}
  \label{eq:U(1|1)-N with W:glsm:M-SUSY:product toric supervariety 1}
  \saveboxed{eq:U(1|1)-N with W:glsm:M-SUSY:product toric supervariety 1}{
    \flatfrac{
      \left\{
        \sum_{a = 1}^{m_1} \tensor{s}{^j_a} \tensor{\bar{s}}{_i^a}
        - \sum_{b = 1}^{n_1} \tensor{\mathbx{s}}{^j_b} \tensor{\bar{\mathbx{s}}}{_i^b}
        = \left(
          \hat{m}_1 \abs{p_+}^2
          - \hat{n}_1 \abs{p_-}^2
          + r_1
        \right) \tensor{\delta}{^j_i}
        \, \qcomma
        i, j \in \{1, \bar{1}\}
      \right\}
    }{\mathrm{U}(1|1)_1}
  }
\end{equation}
\begin{equation}
  \label{eq:U(1|1)-N with W:glsm:M-SUSY:product toric supervariety 2}
  \saveboxed{eq:U(1|1)-N with W:glsm:M-SUSY:product toric supervariety 2}{
    \flatfrac{
      \left\{
        \sum_{\alpha = 1}^{m_2} \tensor{t}{^j_{\alpha}} \tensor{\bar{t}}{_i^{\alpha}}
        - \sum_{\beta = 1}^{n_2} \tensor{\mathbx{t}}{^j_{\beta}} \tensor{\bar{\mathbx{t}}}{_i^{\beta}}
        = \left(
          \hat{m}_2 \abs{p_+}^2
          - \hat{n}_2 \abs{p_-}^2
          + r_2
        \right) \tensor{\delta}{^j_i}
        \, \qcomma
        i, j \in \{1, \bar{1}\}
      \right\}
    }{\mathrm{U}(1|1)_2}
  }
\end{equation}
\begin{equation}
  \label{eq:U(1|1)-N with W:glsm:M-SUSY:toric hypersurface +}
  \saveboxed{eq:U(1|1)-N with W:glsm:M-SUSY:toric hypersurface +}{
    \abs{G^+}^2
    + \sum_{\mu, \nu} \abs{p_+ \pdv{G^+}{Y^+_{\mu\nu}(s, \mathbx{s})}}^2
    + \sum_{\mu, \nu} \abs{p_+ \pdv{G^+}{Y^+_{\mu\nu}(t, \mathbx{t})}}^2
    = 0
  }
\end{equation}
and
\begin{equation}
  \label{eq:U(1|1)-N with W:glsm:M-SUSY:toric hypersurface -}
  \saveboxed{eq:U(1|1)-N with W:glsm:M-SUSY:toric hypersurface -}{
    \abs{G^-}^2
    + \sum_{\mu, \nu} \abs{p_- \pdv{G^-}{Y^-_{\mu\nu}(s, \mathbx{s})}}^2
    + \sum_{\mu, \nu} \abs{p_- \pdv{G^-}{Y^-_{\mu\nu}(t, \mathbx{t})}}^2
    = 0
  }
\end{equation}

\subtitle{A Reduction to Garavuso-Katzarkov-Kreuzer-Noll's $\mathrm{U}(1)_1 \times \mathrm{U}(1)_2$ GLSM with Superpotential}

Note that the $\mathrm{U}(1|1)_1 \times \mathrm{U}(1|1)_2$ GLSM with superpotential ought to reduce to Garavuso-Katzarkov-Kreuzer-Noll's $\mathrm{U}(1)_1 \times \mathrm{U}(1)_2$ GLSM with superpotential when we turn off
(i) the $\tensor{S}{^{\bar{1}}_a}$, $\tensor{\mathbx{S}}{^{\bar{1}}_{\alpha}}$, $\tensor{T}{^{\bar{1}}_b}$, $\tensor{\mathbx{T}}{^{\bar{1}}_{\beta}}$ superfields
and
(ii) the $P_-$ superfield.
Indeed, point (ii) would mean that the $G^-$ polynomial ought to vanish.
Moreover, when $r_1, r_2 > 0$ and $p_+ = 0$, \eqref{eq:U(1|1)-N with W:glsm:M-SUSY:product toric supervariety 1} and \eqref{eq:U(1|1)-N with W:glsm:M-SUSY:product toric supervariety 2} would define projective superspaces $\mathbb{CP}^{m_1 - 1|n_1}$ and $\mathbb{CP}^{m_2 - 1|n_2}$, respectively, while \eqref{eq:U(1|1)-N with W:glsm:M-SUSY:toric hypersurface +} tells us to look at the zero-locus of $G^+$.

In other words, when $r_1, r_2 > 0$ and $p_+ = 0$, $\MSUSY$ would reduce to the space $\{G^+ = 0\} \subset \mathbb{CP}^{m_1 - 1|n_1} \times \mathbb{CP}^{m_2 - 1|n_2}$, as it should (\emph{c.f.}~\cite[$\S$6]{garavuso-2011-super-landau}, with a gauge charge assignment similar to \autoref{fig:U(1|1)-N with W:gauge charge assignment}).

\subtitle{A Reduction to the Familiar $\mathrm{U}(1)_1 \times \mathrm{U}(1)_2$ GLSM with Superpotential}

Note that the $\mathrm{U}(1|1)_1 \times \mathrm{U}(1|1)_2$ GLSM with superpotential ought to reduce to the familiar $\mathrm{U}(1)_1 \times \mathrm{U}(1)_2$ GLSM with superpotential when we turn off
\begin{enumerate*}
  \item the $\tensor{s}{^{\bar{1}}_a}$ and $\tensor{t}{^{\bar{1}}_{\alpha}}$ fields (and their corresponding bar-indexed partners in the underlying multiplets),

  \item the \emph{whole} Grassmann-odd chiral superfields $\mathbx{S}$ and $\mathbx{T}$,

  and

  \item the \emph{whole} Grassmann-odd chiral superfield $P_-$.
\end{enumerate*}
Indeed, point (iii) would mean that the $G^-$ polynomial ought to vanish, whence, together with points (i) and (ii), the equations \eqref{eq:U(1|1)-N with W:glsm:M-SUSY:product toric supervariety 1}--\eqref{eq:U(1|1)-N with W:glsm:M-SUSY:toric hypersurface -} would consist only of the Grassmann-even $(s^1, t^1, p_+, G^+)$ fields.

In other words, $\MSUSY$ would reduce to three equations:
one involving $(s^1, p_+, r_1)$, one involving $(t^1, p_+ ,r_2)$, and one involving $(G^+, p_+)$, as it should (\emph{c.f.}~\cite[$\S$5.2]{Witten:1993yc}).

\subsection{Phases of the Model: NLSMs on a Complete Intersection of Hypersurfaces in a Product of Super-Grassmannians, Supergauged LG Orbifolds on a Supervector Bundle over an Extension of a Super-Grassmannian, and their Hybrids   }
\label{subsec: Phases of U(1|1)^2 with W}

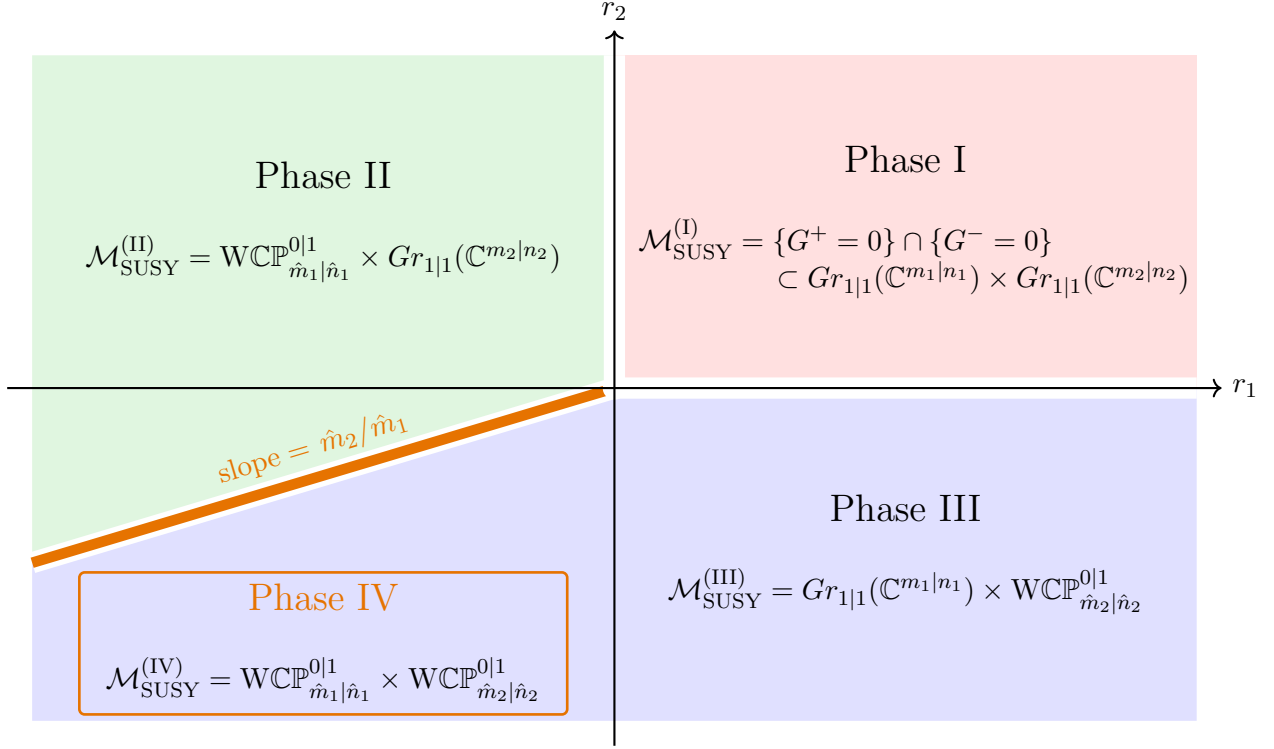
\begin{figure}
  \centering
  \begin{tikzpicture}[scale=1.1]

    \def\xmin{-7}
    \def\xmax{7}
    \def\ymin{-4}
    \def\ymax{4}

    \def\slope{0.3}

    \colorlet{phase1}{red}
    \colorlet{phase2}{blue}
    \colorlet{phase3}{green!70!black}
    \colorlet{phase4}{orange!90!black}


    \fill[phase1!12]
    (0,0)
    --(\xmax,0)
    --(\xmax,\ymax)
    --(0,\ymax)
    --cycle;

    \fill[phase2!12]
    (\xmax,0)
    --(0,0)
    --(\xmin,\slope*\xmin)
    --(\xmin,\ymin)
    --(\xmax,\ymin)
    --cycle;

    \fill[phase3!12]
    (0,\ymax)
    --(0,0)
    --(\xmin,\slope*\xmin)
    --(\xmin,\ymax)
    --cycle;


    \draw[white,line width=8pt]
    (\xmin,\slope*\xmin)--(0,0);

    \draw[white,line width=8pt]
    (0,0)--(0,\ymax);

    \draw[white,line width=8pt]
    (0,0)--(\xmax,0);


    \draw[phase4,line width=4pt]
    (\xmin,\slope*\xmin)
    --
    node[midway, above, sloped, phase4] {
      $ \text{slope} = \flatfrac{\hat{m}_2}{\hat{m}_1}$
    }
    (0,0);

    \fill[white] (0,0) circle[radius=4pt];


    \node at (\xmax*0.5,\ymax*0.5) {
      \begin{tabular}{c}
        {\Large Phase I}
        \\
        \\
        \hspace*{-45pt}$\MSUSY^{\text{(I)}} = \{G^+ = 0\} \cap \{G^- = 0\}$
        \\
        $\qquad \qquad \quad$
        $\subset Gr_{1|1}(\mathbb{C}^{m_1|n_1}) \times Gr_{1|1}(\mathbb{C}^{m_2|n_2})$
      \end{tabular}
    };

    \node at (\xmin*0.5,\ymax*0.5) {
      \begin{tabular}{c}
        {\Large Phase II}
        \\
        \\
        $\MSUSY^{\text{(II)}}
        = \WCP^{0|1}_{\hat{m}_1|\hat{n}_1}
        \times Gr_{1|1}(\mathbb{C}^{m_2|n_2})$
      \end{tabular}
    };

    \node at (\xmax*0.5,\ymin*0.5) {
      \begin{tabular}{c}
        {\Large Phase III}
        \\
        \\
        $\MSUSY^{\text{(III)}}
        = Gr_{1|1}(\mathbb{C}^{m_1|n_1})
        \times \WCP^{0|1}_{\hat{m}_2|\hat{n}_2}$
      \end{tabular}
    };

    \node[draw=phase4,
    line width=1pt,
    rounded corners=2pt,
    inner sep=4pt,
    below
    ] at (\xmin*0.5,\ymin*0.5-0.2) {
      \begin{tabular}{c}
        {\Large \textcolor{phase4}{Phase IV}}
        \\
        \\
        $\MSUSY^{\text{(IV)}}
        = \WCP^{0|1}_{\hat{m}_1|\hat{n}_1}
        \times \WCP^{0|1}_{\hat{m}_2|\hat{n}_2}$
      \end{tabular}
    };


    \draw[->,thick] (\xmin-0.3,0)--(\xmax+0.3,0)
    node[right]{$r_1$};

    \draw[->,thick] (0,\ymin-0.3)--(0,\ymax+0.3)
    node[above]{$r_2$};

  \end{tikzpicture}
  \caption{%
    $\MSUSY$ of the different phases along Branch (1) of the $\mathrm{U}(1|1)^2$ GLSM with superpotential.
    Phase I in pink, Phase II in green, Phase III in light purple, and Phase IV as the orange line.
  }
  \label{fig:U(1|1)-N with W:M-SUSY:first branch}
\end{figure}

\begin{figure}
  \centering
  \begin{tikzpicture}[scale=1.1]

    \def\xmin{-7}
    \def\xmax{7}
    \def\ymin{-4}
    \def\ymax{4}

    \def\slope{0.3}

    \colorlet{phase1}{red}
    \colorlet{phase2}{blue}
    \colorlet{phase3}{green!70!black}
    \colorlet{phase4}{orange!90!black}


    \fill[phase1!12]
    (0,0)
    --(\xmax,0)
    --(\xmax,\ymax)
    --(0,\ymax)
    --cycle;

    \fill[phase2!12]
    (\xmax,0)
    --(0,0)
    --(\xmin,\slope*\xmin)
    --(\xmin,\ymin)
    --(\xmax,\ymin)
    --cycle;

    \fill[phase3!12]
    (0,\ymax)
    --(0,0)
    --(\xmin,\slope*\xmin)
    --(\xmin,\ymax)
    --cycle;


    \draw[white,line width=8pt]
    (\xmin,\slope*\xmin)--(0,0);

    \draw[white,line width=8pt]
    (0,0)--(0,\ymax);

    \draw[white,line width=8pt]
    (0,0)--(\xmax,0);


    \draw[phase4,line width=4pt]
    (\xmin,\slope*\xmin)
    --
    node[midway, above, sloped, phase4] {
      $\text{slope} = \flatfrac{\hat{m}_2}{\hat{m}_1}$
    }
    (0,0);

    \fill[white] (0,0) circle[radius=4pt];


    \node at (\xmax*0.5,\ymax*0.5) {
      \begin{tabular}{c}
        {\Large Phase I}
        \\
        \\
        NLSM on $\MSUSY^{\text{(I)}}$
      \end{tabular}
    };

    \node at (\xmin*0.5,\ymax*0.5) {
      \begin{tabular}{c}
        {\Large Phase II}
        \\
        \\
        Hybrid
        \\
        $\mathrm{SU}(1|1)_1 \times \mathbb{Z}_{\hat{\mathfrak{m}}_1}$ LG Orbifold on $\mathcal{S}\mathcal{G}^{\mathrm{U(1|1)^2}}_{\text{III}}$$\big{/}$
        \\
        $Gr_{1|1}(\mathbb{C}^{m_2|n_2})$ NLSM
      \end{tabular}
    };

    \node at (\xmax*0.5,\ymin*0.5) {
      \begin{tabular}{c}
        {\Large Phase III}
        \\
        \\
        Hybrid
        \\
        $Gr_{1|1}(\mathbb{C}^{m_1|n_1})$ NLSM$\big{/}$
        \\
        $\mathrm{SU}(1|1)_2 \times \mathbb{Z}_{\hat{\mathfrak{m}}_2}$ LG Orbifold on $\mathcal{S}\mathcal{G}^{\mathrm{U(1|1)^2}}_{\text{II}}$
      \end{tabular}
    };

    \node[draw=phase4,
    line width=1pt,
    rounded corners=2pt,
    inner sep=4pt,
    below
    ] at (\xmin*0.5,\ymin*0.5-0.2) {
      \begin{tabular}{c}
        {\Large \textcolor{phase4}{Phase IV}}
        \\
        $\mathrm{U}(1|1) \times \mathrm{SU}(1|1) \times \mathbb{Z}_{\hat{\mathfrak{M}}}$
        \\
        LG Orbifold on $\mathcal{S}\mathcal{G}^{\mathrm{U(1|1)^2}}_{\text{IV}}$
      \end{tabular}
    };


    \draw[->,thick] (\xmin-0.3,0)--(\xmax+0.3,0)
    node[right]{$r_1$};

    \draw[->,thick] (0,\ymin-0.3)--(0,\ymax+0.3)
    node[above]{$r_2$};

  \end{tikzpicture}
  \caption{%
    The low energy limit of the $\mathrm{U}(1|1)^2$ GLSM with superpotential in different phases along Branch (1).
  }
  \label{fig:U(1|1)-N with W:low energy:first branch}
\end{figure}
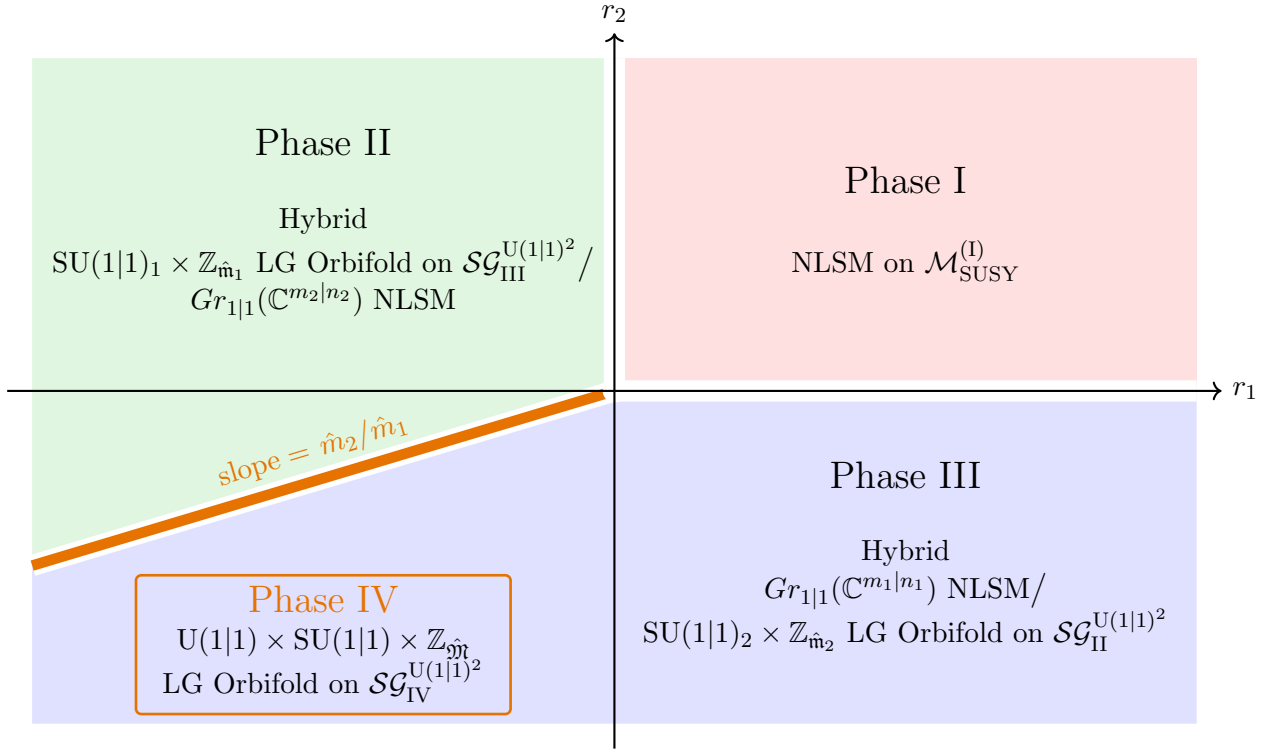

By repeating the analysis in \autoref{subsec: Phases U(1|1) with W}, we find that the $\mathrm{U}(1|1)^2$ GLSM has \emph{multiple} independent branches involving \emph{multiple} phases.
For brevity, we shall explore the two most representative branches:
\begin{enumerate*}[label=Branch (\arabic*)]
  \item with phases I, II, III, and IV (summarized in \autoref{fig:U(1|1)-N with W:M-SUSY:first branch} and \autoref{fig:U(1|1)-N with W:low energy:first branch}),

  and

  \item with phases I$'$, II$'$, III$'$, and IV$'$ (summarized in \autoref{fig:U(1|1)-N with W:M-SUSY:second branch} and \autoref{fig:U(1|1)-N with W:low energy:second branch}), where it is a parity-reversed mirror of Branch (1).

\end{enumerate*}

Let us now derive these eight phases and the two branches they lie along.

\subtitle{Four Possible Field Configurations}

To satisfy \eqref{eq:U(1|1)-N with W:glsm:M-SUSY:product toric supervariety 1}--\eqref{eq:U(1|1)-N with W:glsm:M-SUSY:toric hypersurface -}, we need to consider the following four possibilities:
\begin{enumerate*}
  \item $s, \mathbx{s}, t, \mathbx{t} \neq 0$ but $p_{\pm} = 0$ and $G^{\pm} = 0$ (see \autoref{dG/dphi not zero});

  \item $p_{\pm}, t, \mathbx{t} \neq 0$ but $s, \mathbx{s} = 0$;\footnote{%
    \label{ft:U(1|1)-N with W:phases:s is 0}%
    When $s, \mathbx{s} = 0$, the $Y^{\pm}(s, \mathbx{s})$ fields vanish.
    Then, by the transverse condition \eqref{eq:U(1|1)-N with W:glsm:transversality} of $G^{\pm}$, it must mean that $\pdv*{G^{\pm}}{Y^{\pm}_{\mu\nu}} = 0$, whence \eqref{eq:U(1|1)-N with W:glsm:M-SUSY:toric hypersurface +} and \eqref{eq:U(1|1)-N with W:glsm:M-SUSY:toric hypersurface -} will be trivially satisfied with $G^{\pm} = \text{constant} = 0$.
  }

  \item $p_{\pm}, s, \mathbx{s} \neq 0$ but $t, \mathbx{t} = 0$;\footnote{%
    \label{ft:U(1|1)-N with W:phases:t is 0}%
    When $t, \mathbx{t} = 0$, the $Y^{\pm}(t, \mathbx{t})$ fields vanish.
    Then, by the same explanation given in \autoref{ft:U(1|1)-N with W:phases:s is 0}, \eqref{eq:U(1|1)-N with W:glsm:M-SUSY:toric hypersurface +} and \eqref{eq:U(1|1)-N with W:glsm:M-SUSY:toric hypersurface -} will be trivially satisfied with $G^{\pm} = \text{constant} = 0$.
  }

  or

  \item $p_{\pm} \neq 0$ but $s, \mathbx{s}, t, \mathbx{t} = 0$ (see \autoref{dG/dphi = zero}).

\end{enumerate*}
Indeed, as we shall see, for different values of $r_1$ and $r_2$, these field configurations will define $\MSUSY$'s which are generally different.

\subtitle{$\MSUSY^{r_1(\mathrm{i}), r_2(\mathrm{i}) > 0}$, $\MSUSY^{r_1(\mathrm{ii}), r_2(\mathrm{ii}) > 0}$, $\MSUSY^{r_1(\mathrm{iii}), r_2(\mathrm{iii}) > 0}$, and $\MSUSY^{r_1(\mathrm{iv}), r_2(\mathrm{iv}) > 0}$ are Disconnected}

When $r_1, r_2 > 0$, the four different field configurations (i), (ii), (iii), and (iv) will split $\MSUSY$ into four \emph{disconnected} components, $\MSUSY^{r_1(\mathrm{i}), r_2(\mathrm{i}) > 0}$, $\MSUSY^{r_1(\mathrm{ii}), r_2(\mathrm{ii}) > 0}$, $\MSUSY^{r_1(\mathrm{iii}), r_2(\mathrm{iii}) > 0}$, and $\MSUSY^{r_1(\mathrm{iv}), r_2(\mathrm{iv}) > 0}$.

\subtitle{$\MSUSY^{r_1(\mathrm{i}) < 0 < r_2(\mathrm{i})}$, $\MSUSY^{r_1(\mathrm{ii}) < 0 < r_2(\mathrm{ii})}$, $\MSUSY^{r_1(\mathrm{iii}) < 0 < r_2(\mathrm{iii})}$, and $\MSUSY^{r_1(\mathrm{iv}) < 0 < r_2(\mathrm{iv})}$ are also Disconnected}

When $r_1 < 0 < r_2$, the four different field configurations (i), (ii), (iii), and (iv) will split $\MSUSY$ into four \emph{disconnected} components, $\MSUSY^{r_1(\mathrm{i}) < 0 < r_2(\mathrm{i})}$, $\MSUSY^{r_1(\mathrm{ii}) < 0 < r_2(\mathrm{ii})}$, $\MSUSY^{r_1(\mathrm{iii}) < 0 < r_2(\mathrm{iii})}$, and $\MSUSY^{r_1(\mathrm{iv}) < 0 < r_2(\mathrm{iv})}$.

\subtitle{$\MSUSY^{r_1(\mathrm{i}) > 0 > r_2(\mathrm{i})}$, $\MSUSY^{r_2(\mathrm{ii}) > 0 > r_2(\mathrm{ii})}$, $\MSUSY^{r_1(\mathrm{iii}) > 0 > r_2(\mathrm{iii})}$, and $\MSUSY^{r_1(\mathrm{iv}) > 0 > r_2(\mathrm{iv})}$ are also Disconnected}

When $r_1 > 0 > r_2$, the four different field configurations (i), (ii), (iii), and (iv) will split $\MSUSY$ into four \emph{disconnected} components, $\MSUSY^{r_1(\mathrm{i}) > 0 > r_2(\mathrm{i})}$, $\MSUSY^{r_1(\mathrm{ii}) > 0 > r_2(\mathrm{ii})}$, $\MSUSY^{r_1(\mathrm{iii}) > 0 > r_2(\mathrm{iii})}$, and $\MSUSY^{r_1(\mathrm{iv}) > 0 > r_1(\mathrm{iv})}$.

\subtitle{$\MSUSY^{r_1(\mathrm{i}), r_2(\mathrm{i}) < 0}$, $\MSUSY^{r_1(\mathrm{ii}), r_2(\mathrm{ii}) < 0}$, $\MSUSY^{r_1(\mathrm{iii}), r_2(\mathrm{iii}) < 0}$, and $\MSUSY^{r_1(\mathrm{iv}), r_2(\mathrm{iv}) < 0}$ are also Disconnected}

When $r_1, r_2 < 0$, the four different field configurations (i), (ii), (iii), and (iv) will split $\MSUSY$ into four \emph{disconnected} components, $\MSUSY^{r_1(\mathrm{i}), r_2(\mathrm{i}) < 0}$, $\MSUSY^{r_1(\mathrm{ii}), r_2(\mathrm{ii}) < 0}$, $\MSUSY^{r_1(\mathrm{iii}), r_2(\mathrm{iii}) < 0}$, and $\MSUSY^{r_1(\mathrm{iv}), r_2(\mathrm{iv}) < 0}$.

\subtitle{The $r_1(\text{i}), r_2(\text{i}) > 0$ Phase of the Model: A Complete Intersection of Hypersurfaces in $Gr_{1|1}(\mathbb{C}^{m_1|n_1}) \times Gr_{1|1}(\mathbb{C}^{m_2|n_2})$ $\MSUSY^{r_1(\text{i}), r_2(\text{i}) > 0}$}

From \eqref{eq:U(1|1)-N with W:glsm:M-SUSY:product toric supervariety 1}--\eqref{eq:U(1|1)-N with W:glsm:M-SUSY:toric hypersurface -}, one can see that $\MSUSY^{r_1(\text{i}), r_2(\text{i}) > 0}$ will be defined by
\begin{equation}
  \label{eq:U(1|1)-N with W:phases:r1(i) and r2(i) > 0:space 1}
  \saveboxed{eq:U(1|1)-N with W:phases:r1(i) and r2(i) > 0:space 1}{
    \flatfrac{
      \left\{
        \sum_{a = 1}^{m_1} \tensor{s}{^j_a} \tensor{\bar{s}}{_i^a}
        - \sum_{b = 1}^{n_1} \tensor{\mathbx{s}}{^j_b} \tensor{\bar{\mathbx{s}}}{_i^b}
        = r_1 \tensor{\delta}{^j_i}
        \, \qcomma
        i, j \in \{1, \bar{1}\}
      \right\}
    }{\mathrm{U}(1|1)_1}
  }
\end{equation}
\begin{equation}
  \label{eq:U(1|1)-N with W:phases:r1(i) and r2(i) > 0:space 2}
  \saveboxed{eq:U(1|1)-N with W:phases:r1(i) and r2(i) > 0:space 2}{
    \flatfrac{
      \left\{
        \sum_{\alpha = 1}^{m_2} \tensor{t}{^j_{\alpha}} \tensor{\bar{t}}{_i^{\alpha}}
        - \sum_{\beta = 1}^{n_2} \tensor{\mathbx{t}}{^j_{\beta}} \tensor{\bar{\mathbx{t}}}{_i^{\beta}}
        = r_2 \tensor{\delta}{^j_i}
        \, \qcomma
        i, j \in \{1, \bar{1}\}
      \right\}
    }{\mathrm{U}(1|1)_2}
  }
\end{equation}
\begin{equation}
  \label{eq:U(1|1)-N with W:phases:r1(i) and r2(i) > 0:hypersurface +}
  \saveboxed{eq:U(1|1)-N with W:phases:r1(i) and r2(i) > 0:hypersurface +}{
    G^+ \left( Y^+(s, \mathbx{s}), Y^+(t, \mathbx{t}) \right)
    = 0
  }
\end{equation}
and
\begin{equation}
  \label{eq:U(1|1)-N with W:phases:r1(i) and r2(i) > 0:hypersurface -}
  \saveboxed{eq:U(1|1)-N with W:phases:r1(i) and r2(i) > 0:hypersurface -}{
    G^- \left( Y^-(s, \mathbx{s}), Y^-(t, \mathbx{t}) \right)
    = 0
  }
\end{equation}

One can, like our earlier description in \eqref{eq:U(1|1) with W:phases:geometric space:r(i) > 0}--\eqref{eq:U(1|1) with W:phases:hypersurface -:r(i) > 0}, interpret $\MSUSY^{r_1(\text{i}), r_2(\text{i}) > 0}$ as a complete intersection of an even and odd hypersurface $\{G^+ = 0\}$ and $\{G^- = 0\}$ of bidegrees $(\hat{m}_1, \hat{m}_2)$ and $(\hat{n}_1, \hat{n}_2)$ in the product of super-Grassmannians $Gr_{1|1}(\mathbb{C}^{m_1|n_1}) \times Gr_{1|1}(\mathbb{C}^{m_2|n_2})$ of sizes $(r_1, r_2)$.

\subtitle{The $r_1(\text{i}), r_2(\text{i}) > 0$ Phase of the Model at Low Energy: An NLSM on a Complete Intersection of Hypersurfaces in $Gr_{1|1}(\mathbb{C}^{m_1|n_1}) \times Gr_{1|1}(\mathbb{C}^{m_2|n_2})$}

Staying in the $r_1(\text{i}), r_2(\text{i}) > 0$ phase, let us consider the low energy limit $e_{\{1, 2\}}, b^{\pm}_{\{1, 2\}}, c^{\pm}_{\{1, 2\}} \rightarrow \infty$ where the massive fields/modes (whose masses are either of order $e_1 \sqrt{r_1}$ and $e_2 \sqrt{r_2}$ or determined by $b^{\pm}_{\{1, 2\}}$ and $c^{\pm}_{\{1, 2\}}$) decouple from the theory.
Then, just as in \autoref{subsec: Phases U(1|1) with W}, the GLSM will become an NLSM on the above-described $\MSUSY^{r_1(\text{i}), r_2(\text{i}) > 0}$, which is a complete intersection of an even and odd hypersurface $\{G^+ = 0\}$ and $\{G^- = 0\}$ of bidegrees $(\hat{m}_1, \hat{m}_2)$ and $(\hat{n}_1, \hat{n}_2)$ in the product of super-Grassmannians $Gr_{1|1}(\mathbb{C}^{m_1|n_1}) \times Gr_{1|1}(\mathbb{C}^{m_2|n_2})$ of sizes $(r_1, r_2)$.

\subtitle{The $r_1(\text{ii}) < 0 < r_2(\text{ii})$ Phase of the Model: A Product Space $\WCP^{0|1}_{\hat{m}_1|\hat{n}_1} \times Gr_{1|1}(\mathbb{C}^{m_2|n_2})$ $\MSUSY^{r_1(\text{i}) < 0 < r_2(\text{i})}$}

Note that when $r_1 < 0 < r_2$ and $p_{\pm} \neq 0$, we can express \eqref{eq:U(1|1)-N with W:glsm:M-SUSY:product toric supervariety 1}--\eqref{eq:U(1|1)-N with W:glsm:M-SUSY:product toric supervariety 2} as
\begin{equation}
  \label{eq:U(1|1)-N with W:phases:r1(iii) and r2(iii) < 0:M-SUSY}
  \frac{
    \left\{
      \displaystyle
      \sum_{\alpha = 1}^{m_2} \tensor{t}{^j_{\alpha}} \tensor{\bar{t}}{_i^{\alpha}}
      - \sum_{\beta = 1}^{n_2} \tensor{\mathbx{t}}{^j_{\beta}} \tensor{\bar{\mathbx{t}}}{_i^{\beta}}
      - \frac{\hat{m}_2}{\hat{m}_1} \sum_{a = 1}^{m_1} \tensor{s}{^j_a} \tensor{\bar{s}}{_i^a}
      + \frac{\hat{m}_2}{\hat{m}_1} \sum_{b = 1}^{n_1} \tensor{\mathbx{s}}{^j_b} \tensor{\bar{\mathbx{s}}}{_i^b}
      = \left( r_2 - \frac{\hat{m}_2}{\hat{m}_1} r_1 \right) \tensor{\delta}{^j_i}
      \, \qcomma
      i, j \in \{1, \bar{1}\}
    \right\}
  }{\mathrm{U}(1|1)_1 \times \mathrm{U}(1|1)_2}
  \, .
\end{equation}
Since $\hat{m}_1, \hat{m}_2 \in \mathbb{Z}^+$ and $r_1$ is negative, $r_2 - \frac{\hat{m}_2}{\hat{m}_1} r_1$ is positive.

Then, because $\MSUSY^{r_1(\text{ii}) < 0 < r_2(\text{ii})}$ requires $p_{\pm}, t, \mathbx{t} \neq 0$ and $s, \mathbx{s} = 0$, it will be defined by following two equations.
The first equation with $p_{\pm}$ is obtained by setting $s, \mathbx{s} = 0$ in \eqref{eq:U(1|1)-N with W:glsm:M-SUSY:product toric supervariety 1}, i.e.,
\begin{equation}
  \label{eq:U(1|1)-N with W:phases:r1(i) < 0 < r2(i):space 2}
  \saveboxed{eq:U(1|1)-N with W:phases:r1(i) < 0 < r2(i):space 2}{
    \flatfrac{
      \left\{
        \hat{m}_1 \abs{p_+}^2
        - \hat{n}_1 \abs{p_-}^2
        = \abs{r_1}
      \right\}
    }{\mathrm{U}(1)_1}
  }
  \, ,
\end{equation}
where we mod out by $\mathrm{U}(1)_1$ because $p_{\pm}$ in \eqref{eq:U(1|1)-N with W:glsm:M-SUSY:product toric supervariety 1} transforms effectively under the even abelian subgroup $\mathrm{U}(1)_1 \subset \mathrm{U}(1|1)_1$.
The second equation without $p_{\pm}$ is obtained by setting $s, \mathbx{s} = 0$ in \eqref{eq:U(1|1)-N with W:phases:r1(iii) and r2(iii) < 0:M-SUSY}, i.e.,
\begin{equation}
  \label{eq:U(1|1)-N with W:phases:r1(i) < 0 < r2(i):space 1}
  \saveboxed{eq:U(1|1)-N with W:phases:r1(i) < 0 < r2(i):space 1}{
    \flatfrac{
      \left\{
        \sum_{\alpha = 1}^{m_2} \tensor{t}{^j_{\alpha}} \tensor{\bar{t}}{_i^{\alpha}}
        - \sum_{\beta = 1}^{n_2} \tensor{\mathbx{t}}{^j_{\beta}} \tensor{\bar{\mathbx{t}}}{_i^{\beta}}
        = \left( r_2 - \frac{\hat{m}_2}{\hat{m}_1} r_1 \right) \tensor{\delta}{^j_i}
        \, \qcomma
        i, j \in \{1, \bar{1}\}
      \right\}
    }{\mathrm{U}(1|1)_2}
  }
\end{equation}

Notice that \eqref{eq:U(1|1)-N with W:phases:r1(i) < 0 < r2(i):space 2} defines a weighted projective superspace $\WCP^{0|1}_{\hat{m}_1|\hat{n}_1}$ of size $\abs{r_1}$, and \eqref{eq:U(1|1)-N with W:phases:r1(i) < 0 < r2(i):space 1} defines a super-Grassmannian $Gr_{1|1}(\mathbb{C}^{m_2|n_2})$ of \emph{positive} size $r_2 - \frac{\hat{m}_2}{\hat{m}_1} r_1$.

Altogether, this means that one can interpret $\MSUSY^{r_1(\text{ii}) < 0 < r_2(\text{ii})}$ as the product $\WCP^{0|1}_{\hat{m}_1|\hat{n}_1} \times Gr_{1|1}(\mathbb{C}^{m_2|n_2})$, as defined by \eqref{eq:U(1|1)-N with W:phases:r1(i) < 0 < r2(i):space 2} and \eqref{eq:U(1|1)-N with W:phases:r1(i) < 0 < r2(i):space 1}.

Comparing \eqref{eq:U(1|1)-N with W:phases:r1(i) < 0 < r2(i):space 2} with \eqref{eq:U(1|1) with W:phases:geometric space:r(ii) < 0}, and \eqref{eq:U(1|1)-N with W:phases:r1(i) < 0 < r2(i):space 1} with \eqref{eq:U(1|1) with W:phases:geometric space:r(i) > 0}, it is clear that we have a hybrid model which has features of both the $r < 0$ and $r > 0$ phases of a standard $\mathrm{U}(1|1)$ GLSM with superpotential.
In turn, this means that at low energy, we ought to have a hybrid LG model/NLSM.
Let us investigate this now.

\subtitle{The $r_1(\text{ii}) < 0 < r_2(\text{ii})$ Phase of the Model at Low Energy: A Hybrid $\mathrm{SU}(1|1)_1 \times \mathbb{Z}_{\hat{\mathfrak{m}}_1}$ Supergauged LG Orbifold on a Supervector Bundle over $\WCP^{0|1}_{\hat{m}_1|\hat{n}_1} \times Gr_{1|1}(\mathbb{C}^{m_2|n_2})$ $\big{/}$ $Gr_{1|1}(\mathbb{C}^{m_2|n_2})$ NLSM}

Staying in the $r_1(\text{ii}) < 0 < r_2(\text{ii})$ phase, note that since $p_{\pm} \neq 0$ in \eqref{eq:U(1|1)-N with W:phases:r1(i) < 0 < r2(i):space 2}, i.e., there are formally nonvanishing expectation values of $\expval*{p_{\pm}}$, at low energy, the remaining $\mathrm{U}(1|1)_1$ supergauge symmetry (that $p_{\pm}$ also transforms under) will break down to an $H_1 \subset \mathrm{U}(1|1)_1$ symmetry, whose generator matrices have superdeterminant equal to the $\hat{\mathfrak{m}}_1^{\text{th}}$ root of 1, where the integer $\hat{\mathfrak{m}}_1 \coloneq \gcd(\hat{m}_1, \hat{n}_1)$, such that $H_1 \cong \mathrm{SU}(1|1)_1 \times \mathbb{Z}_{\hat{\mathfrak{m}}_1}$ (see \autoref{ft:u-1-1-W:c-number}).

Similarly, since $t, \mathbx{t} \neq 0$ in \eqref{eq:U(1|1)-N with W:phases:r1(i) < 0 < r2(i):space 1}, i.e., there are formally nonvanishing expectation values of $\expval{t}$ and $\expval{\mathbx{t}}$, at low energy, the $\mathrm{U}(1|1)_2$ supergauge symmetry (that $t$ and $\mathbx{t}$ transforms under) will be completely broken.\footnote{%
  \label{ft:U(1|1)-N with W:phases:broken symmetry}%
  To understand this, first, note that the expectation value $\expval{t}, \expval{\mathbx{t}}$ ought to be gauge-invariant.
  Second, under a gauge transformation, we have $t \to e^{\mathbf{i} \Lambda_a \mathsf{T}^a } t$ and $\mathbx{t} \rightarrow e^{\mathbf{i} \Lambda_a \mathsf{T}^a}$, where the $\mathsf{T}^a$'s are the generators of the $\mathrm{U}(1|1)_2$ gauge group.
  Together, it will mean that to have $\expval{t} \rightarrow \expval{t}$ and $\expval{\mathbx{t}} \rightarrow \expval{\mathbx{t}}$ under a gauge transformation, it must necessarily be the case that $\Lambda_a = 0$, whence the $\mathrm{U}(1|1)_2$ gauge transformation is trivial, i.e., the  $\mathrm{U}(1|1)_2$ gauge symmetry is completely broken.
  We actually saw this throughout the paper hitherto; in particular, in the $r_1(\text{i}), r_2(\text{i}) > 0$ phase, where $s, \mathbx{s}, t, \mathbx{t}$ were nonzero in \eqref{eq:U(1|1)-N with W:phases:r1(i) and r2(i) > 0:space 1} and \eqref{eq:U(1|1)-N with W:phases:r1(i) and r2(i) > 0:space 2} so they had formally nonvanishing expectation values, at low energy, the $\mathrm{U}(1|1)_1$ and $\mathrm{U}(1|1)_2$ gauge symmetries that they respectively transformed under were completely broken, and the GLSM reduced to an ungauged NLSM.
}

In short, at the weighted projective sector of $\MSUSY^{r_1(\text{ii}) < 0 < r_2(\text{ii})}$, which is described by \eqref{eq:U(1|1)-N with W:phases:r1(i) < 0 < r2(i):space 2}, we have, at low energy, an $\mathrm{SU}(1|1)_1 \times \mathbb{Z}_{\hat{\mathfrak{m}}_1}$ gauge theory.
Away from this sector of $\MSUSY^{r_1(\text{ii}) < 0 < r_2(\text{ii})}$, which is described by \eqref{eq:U(1|1)-N with W:phases:r1(i) < 0 < r2(i):space 1}, we have, at low energy, an NLSM.

Let us now ascertain what this gauge theory and NLSM are.

At the weighted projective sector of $\MSUSY^{r_1(\text{ii}) < 0 < r_2(\text{ii})}$, which is described by \eqref{eq:U(1|1)-N with W:phases:r1(i) < 0 < r2(i):space 2}, in the low energy limit $e_1 \rightarrow \infty$ where the relevant fields from the $V_1$ and $P_{\pm}$-multiplets become massive and decouple from the theory, based on the same reasoning that led us from \eqref{eq:U(1|1) with W:phases:geometric space:r(ii) < 0} to \eqref{eq:U(1|1) with W:phases:superpotential W:r(ii) < 0} and \eqref{eq:U(1|1) with W:phases:geometric space:r(ii) < 0:exp val}, the GLSM will become an $\mathrm{SU}(1|1)_1 \times \mathbb{Z}_{\hat{\mathfrak{m}}_1}$ supergauged LG orbifold with effective superpotential
\begin{equation}
  \label{eq:U(1|1)-N with W:phases:r1(i) < 0 < r2(i):superpotential}
  \saveboxed{eq:U(1|1)-N with W:phases:r1(i) < 0 < r2(i):superpotential}{
    W^{\mathrm{SU}(1|1)_1 \times \mathbb{Z}_{\hat{\mathfrak{m}}_1}}_{\text{LG}, \text{II}}
    = \sqrt{\expval{\abs{p_+}^2}_{\text{II}}} G^+ \left( Y(s, \mathbx{s}), Y(t, \mathbx{t}) \right)
    + \sqrt{\expval{\abs{p_-}^2}_{\text{II}}} G^- \left( Y(s, \mathbx{s}), Y(t, \mathbx{t}) \right)
  }
\end{equation}
where (via \eqref{eq:U(1|1)-N with W:phases:r1(i) < 0 < r2(i):space 2})
\begin{equation}
  \label{eq:U(1|1)-N with W:phases:r1(i) < 0 < r2(i):wcp 2:expval}
  \saveboxed{eq:U(1|1)-N with W:phases:r1(i) < 0 < r2(i):wcp 2:expval}{
    \hat{m}_1 \expval{\abs{p_+}^2}_{\text{II}}
    - \hat{n}_1 \expval{\abs{p_-}^2}_{\text{II}}
    = \abs{r_1}
  }
\end{equation}
The subscript ``$\text{II}$'' will be clear shortly.

Based on the same reasoning that led us to \eqref{eq:U(1|1) with W:phases:supervector bundle:r(ii) < 0} and our discussions thereafter, noting that in place of \eqref{eq:u-1-1 with W:glsm:M-SUSY:geometric space} discussed therein, we now have \eqref{eq:U(1|1)-N with W:phases:r1(iii) and r2(iii) < 0:M-SUSY} and \eqref{eq:U(1|1)-N with W:phases:r1(i) < 0 < r2(i):space 2}, we can conclude that for some value of $\expval*{p_{\pm}}_{\text{II}}$ in $\WCP^{0|1}_{\hat{m}_1|\hat{n}_1}$ (determined by $\hat{m}_1$, $\hat{n}_1$, and $\abs{r_1}$ via \eqref{eq:U(1|1)-N with W:phases:r1(i) < 0 < r2(i):wcp 2:expval}), we have an $\mathrm{SU}(1|1)_1 \times \mathbb{Z}_{\hat{\mathfrak{m}}_1}$ supergauged LG orbifold fibered over $Gr_{1|1}(\mathbb{C}^{m_2|n_2})$ of size $r_2 - \frac{\hat{m}_2}{\hat{m}_1} r_1$.
The effective superpotential $W^{\mathrm{SU}(1|1)_1 \times \mathbb{Z}_{\hat{\mathfrak{m}}_1}}_{\text{LG}, \text{II}}$ of the supergauged LG orbifold is thus a holomorphic function on
\begin{equation}
  \label{eq:U(1|1)-N with W:phases:r1(i) < 0 < r2(i):space 2:supervector bundle}
  \saveboxed{eq:U(1|1)-N with W:phases:r1(i) < 0 < r2(i):space 2:supervector bundle}{
    \mathcal{S}\mathcal{G}^{\mathrm{U}(1|1)^2}_{\text{II}}
    = \bigoplus_{a = 1}^{m_1} \mathscr{L}_a^{\flatfrac{\hat{m}_2}{\hat{m}_1}}
    \oplus \bigoplus_{b = 1}^{n_1} \mathbf{\Pi} \mathscr{L}_b^{\flatfrac{\hat{m}_2}{\hat{m}_1}}
    \rightarrow \WCP^{0|1}_{\hat{m}_1|\hat{n}_1} \times Gr_{1|1}(\mathbb{C}^{m_2|n_2})
  }
\end{equation}
which is a supervector bundle over $\WCP^{0|1}_{\hat{m}_1|\hat{n}_1} \times Gr_{1|1}(\mathbb{C}^{m_2|n_2})$ (if $\flatfrac{\hat{m}_2}{\hat{m}_1}$ is an integer; otherwise, it is a supervector bundle analogue of an orbifold).
Here, the even $\mathscr{L}_a$ and odd $\mathbf{\Pi} \mathscr{L}_b$ fibers are spanned by the Grassmann-even $s_a$ and Grassmann-odd $\mathbx{s}_b$ supervectors, respectively.
Also, the base can be regarded as an extension of a super-Grassmannian.

Thus, we have an $\mathrm{SU}(1|1)_1 \times \mathbb{Z}_{\hat{\mathfrak{m}}_1}$ supergauged LG orbifold defined on $\mathcal{S}\mathcal{G}^{\mathrm{U}(1|1)^2}_{\text{II}}$.

Away from the weighted projective sector of $\MSUSY^{r_1(\text{ii}) < 0 < r_2(\text{ii})}$, which is described by \eqref{eq:U(1|1)-N with W:phases:r1(i) < 0 < r2(i):space 1}, in the low energy limit $e_2 \rightarrow \infty$ where the relevant massive fields/modes (whose masses are of order $e_2 \sqrt{r_2}$) decouple from the theory, the GLSM will become an NLSM on the super-Grassmannian $Gr_{1|1}(\mathbb{C}^{m_2|n_2})$ of \emph{positive} size $r_2 - \frac{\hat{m}_2}{\hat{m}_1} r_1$ described in \eqref{eq:U(1|1)-N with W:phases:r1(i) < 0 < r2(i):space 1}.

In short, we have, at low energy, an NLSM in the $(t, \mathbx{t})$-directions, and a supergauged LG orbifold in the $(s, \mathbx{s})$-directions with an effective superpotential that depends on both the $(t, \mathbx{t}, p_{\pm})$ and $(s, \mathbx{s})$-directions along the base and fiber space of $\mathcal{S}\mathcal{G}^{\mathrm{U}(1|1)^2}_{\text{II}}$, respectively.

Simply put, we have a hybrid between an $\mathrm{SU}(1|1)_1 \times \mathbb{Z}_{\hat{\mathfrak{m}}_1}$ supergauged LG orbifold defined on $\mathcal{S}\mathcal{G}^{\mathrm{U}(1|1)^2}_{\text{II}}$ and an NLSM on $Gr_{1|1}(\mathbb{C}^{m_2|n_2})$.

\subtitle{The $r_1(\text{iii}) > 0 > r_2(\text{iii})$ Phase of the Model: A Product Space $Gr_{1|1}(\mathbb{C}^{m_1|n_1}) \times \WCP^{0|1}_{\hat{m}_2|\hat{n}_2}$ $\MSUSY^{r_1(\text{i}) > 0 > r_2(\text{i})}$}

The analysis for $\MSUSY^{r_1(\text{iii}) > 0 > r_2(\text{iii})}$ is practically identical to the analysis for $\MSUSY^{r_1(\text{ii}) < 0 < r_2(\text{ii})}$.
So, for brevity, we shall just state the results.

$\MSUSY^{r_1(\text{iii}) > 0 > r_2(\text{iii})}$ will defined by
\begin{equation}
  \label{eq:U(1|1)-N with W:phases:r1(i) > 0 > r2(i):space 1}
  \saveboxed{eq:U(1|1)-N with W:phases:r1(i) > 0 > r2(i):space 1}{
    \flatfrac{
      \left\{
        \sum_{a = 1}^{m_1} \tensor{s}{^j_a} \tensor{\bar{s}}{_i^a}
        - \sum_{b = 1}^{n_1} \tensor{\mathbx{s}}{^j_b} \tensor{\bar{\mathbx{s}}}{_i^b}
        = \left( r_1 - \frac{\hat{m}_1}{\hat{m}_2} r_2 \right) \tensor{\delta}{^j_i}
        \, \qcomma
        i, j \in \{1, \bar{1}\}
      \right\}
    }{\mathrm{U}(1|1)_1}
  }
\end{equation}
and
\begin{equation}
  \label{eq:U(1|1)-N with W:phases:r1(i) > 0 > r2(i):space 2}
  \saveboxed{eq:U(1|1)-N with W:phases:r1(i) > 0 > r2(i):space 2}{
    \flatfrac{
      \left\{
        \hat{m}_2 \abs{p_+}^2
        - \hat{n}_2 \abs{p_-}^2
        = \abs{r_2}
      \right\}
    }{\mathrm{U}(1)_2}
  }
\end{equation}
Thus, one can interpret $\MSUSY^{r_1(\text{iii}) > 0 > r_2(\text{iii})}$ as $Gr_{1|1}(\mathbb{C}^{m_1|n_1}) \times \WCP^{0|1}_{\hat{m}_2|\hat{n}_2}$, the product of $Gr_{1|1}(\mathbb{C}^{m_1|n_1})$ of \emph{positive} size $r_1 - \frac{\hat{m}_1}{\hat{m}_2} r_2$ and $\WCP^{0|1}_{\hat{m}_2|\hat{n}_2}$ of size $\abs{r_2}$.

\subtitle{The $r_1(\text{iii}) > 0 > r_2(\text{iii})$ Phase of the Model at Low Energy: A Hybrid $Gr_{1|1}(\mathbb{C}^{m_1|n_1})$ NLSM $\big{/}$ $\mathrm{SU}(1|1)_2 \times \mathbb{Z}_{\hat{\mathfrak{m}}_2}$ Supergauged LG Orbifold on a Supervector Bundle over $Gr_{1|1}(\mathbb{C}^{m_1|n_1}) \times \WCP^{0|1}_{\hat{m}_2|\hat{n}_2}$}

Staying in the $r_1(\text{iii}) > 0 > r_2(\text{iii})$ phase, at low energy, we have, in the $(t, \mathbx{t})$-directions, an $\mathrm{SU}(1|1)_2 \times \mathbb{Z}_{\hat{\mathfrak{m}}_2}$ supergauged LG orbifold with effective superpotential
\begin{equation}
  \label{eq:U(1|1)-N with W:phases:r1(i) > 0 > r2(i):superpotential}
  \saveboxed{eq:U(1|1)-N with W:phases:r1(i) > 0 > r2(i):superpotential}{
    W^{\mathrm{SU}(1|1)_2 \times \mathbb{Z}_{\hat{\mathfrak{m}}_2}}_{\text{LG}, \text{III}}
    = \sqrt{\expval{\abs{p_+}^2}_{\text{III}}} G^+ \left( Y(s, \mathbx{s}), Y(t, \mathbx{t}) \right)
    + \sqrt{\expval{\abs{p_-}^2}_{\text{III}}} G^- \left( Y(s, \mathbx{s}), Y(t, \mathbx{t}) \right)
  }
\end{equation}
where
\begin{equation}
  \label{eq:U(1|1)-N with W:phases:r1(i) > 0 > r2(i):wcp 2:expval}
  \saveboxed{eq:U(1|1)-N with W:phases:r1(i) > 0 > r2(i):wcp 2:expval}{
    \hat{m}_2 \expval{\abs{p_+}^2}_{\text{III}}
    - \hat{n}_2 \expval{\abs{p_-}^2}_{\text{III}}
    = \abs{r_2}
  }
\end{equation}
and $\hat{\mathfrak{m}}_2 \coloneq \gcd(\hat{m}_2, \hat{n}_2)$.
The subscript ``$\text{III}$'' will be clear shortly.

Note that $W^{\mathrm{SU}(1|1)_2 \times \mathbb{Z}_{\hat{\mathfrak{m}}_2}}_{\text{LG}, \text{III}}$ is a holomorphic function on
\begin{equation}
  \label{eq:U(1|1)-N with W:phases:r1(i) > 0 > r2(i):space 2:supervector bundle}
  \saveboxed{eq:U(1|1)-N with W:phases:r1(i) > 0 > r2(i):space 2:supervector bundle}{
    \mathcal{S}\mathcal{G}^{\mathrm{U}(1|1)^2}_{\text{III}}
    = \bigoplus_{\alpha = 1}^{m_2} \mathscr{L}_{\alpha}^{\flatfrac{\hat{m}_1}{\hat{m}_2}}
    \oplus \bigoplus_{\beta = 1}^{n_2} \mathbf{\Pi} \mathscr{L}_{\beta}^{\flatfrac{\hat{m}_1}{\hat{m}_2}}
    \rightarrow Gr_{1|1}(\mathbb{C}^{m_1|n_1}) \times \WCP^{0|1}_{\hat{m}_2|\hat{n}_2}
  }
\end{equation}
which is a supervector bundle over $Gr_{1|1}(\mathbb{C}^{m_1|n_1}) \times \WCP^{0|1}_{\hat{m}_2|\hat{n}_2}$ (if $\flatfrac{\hat{m}_1}{\hat{m}_2}$ is an integer; otherwise, it is a supervector bundle analogue of an orbifold).
Here, the even $\mathscr{L}_{\alpha}$ and odd $\mathbf{\Pi} \mathscr{L}_{\beta}$ fibers are spanned by the Grassmann-even $t_{\alpha}$ and Grassmann-odd $\mathbx{t}_{\beta}$ supervectors, respectively.
Again, the base can be regarded as an extension of a super-Grassmannian.

At low energy, we also have, in the $(s, \mathbx{s})$-directions, an NLSM on $Gr_{1|1}(\mathbb{C}^{m_1|n_1})$ of \emph{positive} size $r_1 - \frac{\hat{m}_1}{\hat{m}_2} r_2$ described by \eqref{eq:U(1|1)-N with W:phases:r1(i) < 0 < r2(i):space 1}.

Simply put, we have a hybrid between an NLSM on $Gr_{1|1}(\mathbb{C}^{m_1|n_1})$ and an $\mathrm{SU}(1|1)_2 \times \mathbb{Z}_{\hat{\mathfrak{m}}_2}$ supergauged LG orbifold defined on $\mathcal{S}\mathcal{G}^{\mathrm{U}(1|1)^2}_{\text{III}}$.

\subtitle{The $r_1(\text{iv}), r_2(\text{iv}) < 0$ Phase of the Model: A Product of Weighted Projective Superspaces $\WCP^{0|1}_{\hat{m}_1|\hat{n}_1} \times \WCP^{0|1}_{\hat{m}_2|\hat{n}_2}$ $\MSUSY^{r_1(\text{iv}), r_2(\text{iv}) < 0}$}

$\MSUSY^{r_1(\text{iv}), r_2(\text{iv}) < 0}$ will defined by \eqref{eq:U(1|1)-N with W:glsm:M-SUSY:product toric supervariety 1} and \eqref{eq:U(1|1)-N with W:glsm:M-SUSY:product toric supervariety 2}, but with negative $r_1$ and $r_2$, i.e.,
\begin{equation}
  \label{eq:U(1|1)-N with W:phases:r1(i) and r2(i) < 0:wcp 1}
  \saveboxed{eq:U(1|1)-N with W:phases:r1(i) and r2(i) < 0:wcp 1}{
    \flatfrac{
      \left\{
        \hat{m}_1 \abs{p_+}^2
        - \hat{n}_1 \abs{p_-}^2
        = \abs{r_1}
      \right\}
    }{\mathrm{U}(1)_1}
  }
\end{equation}
and
\begin{equation}
  \label{eq:U(1|1)-N with W:phases:r1(i) and r2(i) < 0:wcp 2}
  \saveboxed{eq:U(1|1)-N with W:phases:r1(i) and r2(i) < 0:wcp 2}{
    \flatfrac{
      \left\{
        \hat{m}_2 \abs{p_+}^2
        - \hat{n}_2 \abs{p_-}^2
        = \abs{r_2}
      \right\}
    }{\mathrm{U}(1)_2}
  }
\end{equation}
where we mod out by $\mathrm{U}(1)_1$ (resp. $\mathrm{U}(1)_2$) because $p_{\pm}$ effectively transforms under the even abelian subgroup $\mathrm{U}(1)_1 \subset \mathrm{U}(1|1)_1$ (resp. $\mathrm{U}(1)_2 \subset \mathrm{U}(1|1)_2$).

Notice that \eqref{eq:U(1|1)-N with W:phases:r1(i) and r2(i) < 0:wcp 1} defines a weighted projective superspace $\WCP^{0|1}_{\hat{m}_1|\hat{n}_1}$ of size $\abs{r_1}$ and positive weights $(\hat{m}_1|\hat{n}_1)$.
Likewise, \eqref{eq:U(1|1)-N with W:phases:r1(i) and r2(i) < 0:wcp 2} defines another weighted projective superspace $\WCP^{0|1}_{\hat{m}_2|\hat{n}_2}$ of size $\abs{r_2}$ and positive weights $(\hat{m}_2|\hat{n}_2)$.

At any rate, because $\flatfrac{\hat{m}_2}{\hat{m}_1} = k = \flatfrac{\hat{n}_2}{\hat{n}_1}$, this means from \eqref{eq:U(1|1)-N with W:phases:r1(i) and r2(i) < 0:wcp 1} and \eqref{eq:U(1|1)-N with W:phases:r1(i) and r2(i) < 0:wcp 2}, that $\abs{r_2} = k \abs{r_1}$, or, since $k$ is positive and $r_1, r_2$ are of the same sign, that $\hat{m}_1 r_2 = \hat{m}_2 r_1$.

Altogether, this means that one can interpret $\MSUSY^{r_1(\text{iv}), r_2(\text{iv}) < 0}$ as the product $\WCP^{0|1}_{\hat{m}_1|\hat{n}_1} \times \WCP^{0|1}_{\hat{m}_2|\hat{n}_2}$, as defined by \eqref{eq:U(1|1)-N with W:phases:r1(i) and r2(i) < 0:wcp 1} and \eqref{eq:U(1|1)-N with W:phases:r1(i) and r2(i) < 0:wcp 2}, whereby $\hat{m}_1 r_2 = \hat{m}_2 r_1$.

Note that the conditions $\hat{m}_1 r_2 = \hat{m}_2 r_1$ and $r_1, r_2 < 0$ do not describe an open space in the $r_1$-$r_2$ plane, but rather, a line therein.
As such, $\MSUSY^{r_1(\text{iv}), r_2(\text{iv}) < 0}$ would appear as a \emph{wall} between other phases of the model.
We will see what these other phases separated by $\MSUSY^{r_1(\text{iv}), r_2(\text{iv}) < 0}$ are, very shortly.

\subtitle{The $r_1(\text{iv}), r_2(\text{iv}) < 0$ Phase of the Model at Low Energy: A $\mathrm{U}(1|1) \times \mathrm{SU}(1|1) \times \mathbb{Z}_{\hat{\mathfrak{M}}}$ Supergauged LG Orbifold on a Supervector Bundle over $\WCP^{0|1}_{\hat{m}_1|\hat{n}_1}$}

Staying in the $r_1(\text{iv}), r_2(\text{iv}) < 0$ phase, note that the absence of $(s, \mathbx{s}, t, \mathbx{t})$ in \eqref{eq:U(1|1)-N with W:phases:r1(i) and r2(i) < 0:wcp 1} means that in this phase, $(s, \mathbx{s}, t, \mathbx{t})$ would have vanishing expectation values whence the supergauge symmetry of the GLSM would not be broken at low energy, unlike the $r_1(\text{i}), r_2(\text{i}) > 0$ phase.

Following the same analysis which led us from \eqref{eq:U(1|1) with W:phases:geometric space:r(ii) < 0} to \eqref{eq:U(1|1) with W:phases:superpotential W:r(ii) < 0} and \eqref{eq:U(1|1) with W:phases:geometric space:r(ii) < 0:exp val}, we find that in the low energy limit, the GLSM will become a $\mathrm{U}(1|1) \times \mathrm{SU}(1|1) \times \mathbb{Z}_{\hat{\mathfrak{M}}}$ supergauged LG orbifold, where $\hat{\mathfrak{M}} \coloneq \gcd(\hat{\mathfrak{m}}_1, \hat{\mathfrak{m}}_2) = \gcd(\hat{m}_1, \hat{n}_1, \hat{m}_2, \hat{n}_2)$.\footnote{%
  \label{ft:U(1|1)-N with W:phases:break one gauge group}%
  See \autoref{ft:u-1-1-W:c-number}, and note that if the gauge symmetry of one of the two $\mathrm{U}(1)$ subgroups is broken to $\mathbb{Z}_{\hat{\mathfrak{M}}}$, it would leave the other $\mathrm{U}(1)$ unbroken.
}
The effective superpotential is
\begin{equation}
  \label{eq:U(1|1)-N with W:phases:r1(i) and r2(i) < 0:superpotential}
  \saveboxed{eq:U(1|1)-N with W:phases:r1(i) and r2(i) < 0:superpotential}{
    W^{\mathrm{U}(1|1) \times \mathrm{SU}(1|1) \times \mathbb{Z}_{\hat{\mathfrak{M}}}}_{\text{LG}, \text{IV}}
    = \sqrt{\expval{\abs{p_+}^2}_{\text{IV}}} G^+ \left( Y(s, \mathbx{s}), Y(t, \mathbx{t}) \right)
    + \sqrt{\expval{\abs{p_-}^2}_{\text{IV}}} G^- \left( Y(s, \mathbx{s}), Y(t, \mathbx{t}) \right)
  }
\end{equation}
where (via \eqref{eq:U(1|1)-N with W:phases:r1(i) and r2(i) < 0:wcp 1} and \eqref{eq:U(1|1)-N with W:phases:r1(i) and r2(i) < 0:wcp 2})
\begin{equation}
  \label{eq:U(1|1)-N with W:phases:r1(i) and r2(i) < 0:wcp 1:expval}
  \saveboxed{eq:U(1|1)-N with W:phases:r1(i) and r2(i) < 0:wcp 1:expval}{
    \hat{m}_1 \expval{\abs{p_+}^2}_{\text{IV}}
    - \hat{n}_1 \expval{\abs{p_-}^2}_{\text{IV}}
    = \abs{r_1}
    = \frac{\hat{m}_1}{\hat{m}_2} \abs{r_2}
  }
\end{equation}
The subscript ``IV'' will be clear shortly.

Note that $W^{\mathrm{U}(1|1) \times \mathrm{SU}(1|1) \times \mathbb{Z}_{\hat{\mathfrak{M}}}}_{\text{LG, IV}}$ is a holomorphic function on the product of supervector bundles
\begin{equation}
  \label{eq:U(1|1)-N with W:phases:r1(i) and r2(i) < 0:supervector bundle over wcp}
  \saveboxed{eq:U(1|1)-N with W:phases:r1(i) and r2(i) < 0:supervector bundle over wcp}{
    \mathcal{SG}^{\mathrm{U}(1|1)^2}_{\text{IV}}
    = \left[
      \bigoplus_{a = 1}^{m_1} \mathscr{L}_a^{-1}
      \oplus \bigoplus_{b = 1}^{n_1} \mathbf{\Pi} \mathscr{L}_b^{-1}
      \rightarrow \WCP^{0|1}_{\hat{m}_1|\hat{n}_1}
    \right]
    \times \left[
      \bigoplus_{\alpha = 1}^{m_2} \mathscr{L}_{\alpha}^{-1}
      \oplus \bigoplus_{\beta = 1}^{n_2} \mathbf{\Pi} \mathscr{L}_{\beta}^{-1}
      \longrightarrow \WCP^{0|1}_{\hat{m}_2|\hat{n}_2}
    \right]
  }
\end{equation}
where the $(\mathscr{L}_a, \mathbf{\Pi} \mathscr{L}_b)$ fibers over $\WCP^{0|1}_{\hat{m}_1|\hat{n}_1}$ are spanned by the $(s_a, \mathbx{s}_b)$ supervectors, and the $(\mathscr{L}_{\alpha}, \mathbf{\Pi} \mathscr{L}_{\beta})$ fibers over $\WCP^{0|1}_{\hat{m}_2|\hat{n}_2}$ are spanned by the $(t_{\alpha}, \mathbx{t}_{\beta})$ supervectors.

Thus, we have a $\mathrm{U}(1|1) \times \mathrm{SU}(1|1) \times \mathbb{Z}_{\hat{\mathfrak{M}}}$ supergauged LG orbifold defined on $\mathcal{SG}^{\mathrm{U}(1|1)^2}_{\text{IV}}$.

\subtitle{The $r_1(\text{ii}), r_2(\text{ii}) < 0$ Phase of the Model: A Product Space $\WCP^{{0|1}}_{\hat{m}_1|\hat{n}_1} \times Gr_{1|1}(\mathbb{C}^{m_2|n_2})$ $\MSUSY^{r_1(\text{iii}), r_2(\text{iii}) < 0}$}

Note that when $r_1, r_2 < 0$ and $p_{\pm} \neq 0$, we can express \eqref{eq:U(1|1)-N with W:glsm:M-SUSY:product toric supervariety 1}-\eqref{eq:U(1|1)-N with W:glsm:M-SUSY:product toric supervariety 2} as
\begin{equation}
  \label{eq:U(1|1)-N with W:phases:r1(i) > 0 > r2(i):M-SUSY}
  \frac{
    \left\{
      \displaystyle
      \sum_{\alpha = 1}^{m_2} \tensor{t}{^j_{\alpha}} \tensor{\bar{t}}{_i^{\alpha}}
      - \sum_{\beta = 1}^{n_2} \tensor{\mathbx{t}}{^j_{\beta}} \tensor{\bar{\mathbx{t}}}{_i^{\beta}}
      - \frac{\hat{m}_2}{\hat{m}_1} \sum_{a = 1}^{m_1} \tensor{s}{^j_a} \tensor{\bar{s}}{_i^a}
      + \frac{\hat{m}_2}{\hat{m}_1} \sum_{b = 1}^{n_1} \tensor{\mathbx{s}}{^j_b} \tensor{\bar{\mathbx{s}}}{_i^b}
      = \left( r_2 - \frac{\hat{m}_2}{\hat{m}_1} r_1 \right) \tensor{\delta}{^j_i}
      \, \qcomma
      i, j \in \{1, \bar{1}\}
    \right\}
  }{\mathrm{U}(1|1)_1 \times \mathrm{U}(1|1)_2}
  \, .
\end{equation}
Since $\hat{m}_1, \hat{m}_2 \in \mathbb{Z}^+$ and $r_1, r_2$ are negative, $r_2 - \frac{\hat{m}_2}{\hat{m}_1} r_1$ may be positive.
Let us consider the case when it is, or equivalently, the case when $\hat{m}_1 r_2 - \hat{m}_2 r_1 > 0$.

Then, because $\MSUSY^{r_1(\text{iii}), r_2(\text{iii}) < 0}$ requires $p_{\pm}, t, \mathbx{t} \neq 0$ but $s, \mathbx{s} = 0$, it will be defined by the following two equations.
The first equation with $p_{\pm}$ is obtained by setting $s, \mathbx{s} = 0$ in \eqref{eq:U(1|1)-N with W:glsm:M-SUSY:product toric supervariety 1}, i.e.,
\begin{equation}
  \label{eq:U(1|1)-N with W:phases:r1(iii) and r2(iii) < 0:space 2}
  \flatfrac{
    \left\{
      \hat{m}_1 \abs{p_+}^2
      - \hat{n}_1 \abs{p_-}^2
      = \abs{r_1}
    \right\}
  }{\mathrm{U}(1)_1}
  \, .
\end{equation}
The second equation without $p_{\pm}$ is obtained by setting $s, \mathbx{s} = 0$ in \eqref{eq:U(1|1)-N with W:phases:r1(i) > 0 > r2(i):M-SUSY}, i.e.,
\begin{equation}
  \label{eq:U(1|1)-N with W:phases:r1(iii) and r2(iii) < 0:space 1}
  \flatfrac{
    \left\{
      \sum_{\alpha = 1}^{m_2} \tensor{t}{^j_{\alpha}} \tensor{\bar{t}}{_i^{\alpha}}
      - \sum_{\beta = 1}^{n_2} \tensor{\mathbx{t}}{^j_{\beta}} \tensor{\bar{\mathbx{t}}}{_i^{\beta}}
      = \left( r_2 - \frac{\hat{m}_2}{\hat{m}_1} r_1 \right) \tensor{\delta}{^j_i}
      \, \qcomma
      i, j \in \{1, \bar{1}\}
    \right\}
  }{\mathrm{U}(1|1)_2}
  \, .
\end{equation}
Thus, one can interpret $\MSUSY^{r_1(\text{ii}), r_2(\text{ii}) < 0}$ as $\WCP^{0|1}_{\hat{m}_1|\hat{n}_1} \times Gr_{1|1}(\mathbb{C}^{m_2|n_2})$, the product of $\WCP^{0|1}_{\hat{m}_1|\hat{n}_1}$ of size $\abs{r_1}$ and $Gr_{1|1}(\mathbb{C}^{m_2|n_2})$ of \emph{positive} size $r_2 - \frac{\hat{m}_2}{\hat{m}_1} r_1$.

Notice that the field requirements determining $\MSUSY^{r_1(\text{ii}), r_2(\text{ii}) < 0}$, i.e., $p_{\pm}, t, \mathbx{t} \neq 0$ but $s, \mathbx{s} = 0$, are exactly the same as the field requirements determining $\MSUSY^{r_1(\text{ii}) < 0 < r_2(\text{ii})}$.
Moreover, the equations \eqref{eq:U(1|1)-N with W:phases:r1(iii) and r2(iii) < 0:space 2}--\eqref{eq:U(1|1)-N with W:phases:r1(iii) and r2(iii) < 0:space 1} defining $\MSUSY^{r_1(\text{iii}), r_2(\text{iii}) < 0}$ are also the same as the equations \eqref{eq:U(1|1)-N with W:phases:r1(i) < 0 < r2(i):space 2}-\eqref{eq:U(1|1)-N with W:phases:r1(i) < 0 < r2(i):space 1} defining $\MSUSY^{r_1(\text{ii}) < 0 < r_2(\text{ii})}$.
Therefore, these two phases can be identified with each other!

In other words, when $r_1 < 0$ and $\hat{m}_1 r_2 - \hat{m}_2 r_1 > 0$, the $r_1(\text{ii}), r_2(\text{ii}) < 0$ and $r_1(\text{ii}) < 0 < r_2(\text{ii})$ phases are the same phase of the GLSM.

\subtitle{The $r_1(\text{ii}), r_2(\text{ii}) < 0$ Phase of the Model at Low Energy: A Hybrid $\mathrm{SU}(1|1)_1 \times \mathbb{Z}_{\hat{\mathfrak{m}}_1}$ Supergauged LG Orbifold on a Supervector Bundle over $\WCP^{0|1}_{\hat{m}_1|\hat{n}_1} \times Gr_{1|1}(\mathbb{C}^{m_2|n_2})$ $\big{/}$ $Gr_{1|1}(\mathbb{C}^{m_2|n_2})$ NLSM}

Since the $r_1(\text{ii}), r_2(\text{ii}) < 0$ phase is the same as the $r_1(\text{ii}) < 0 < r_2(\text{ii})$ phase, we can read off the low energy theory from the latter.

In particular, at low energy, we have a hybrid between an $\mathrm{SU}(1|1)_1 \times \mathbb{Z}_{\hat{\mathfrak{m}}_1}$ supergauged LG orbifold defined on $\mathcal{S}\mathcal{G}^{\mathrm{U}(1|1)^2}_{\text{II}}$ and an NLSM on $Gr_{1|1}(\mathbb{C}^{m_2|n_2})$, where $Gr_{1|1}(\mathbb{C}^{m_2|n_2})$ is the super-Grassmannian described by \eqref{eq:U(1|1)-N with W:phases:r1(iii) and r2(iii) < 0:space 1}, and $\mathcal{S}\mathcal{G}^{\mathrm{U}(1|1)^2}_{\text{II}}$ is the supervector bundle described by \eqref{eq:U(1|1)-N with W:phases:r1(i) < 0 < r2(i):space 2:supervector bundle}.

\subtitle{The $r_1(\text{iii}), r_2(\text{iii}) < 0$ Phase of the Model: A Product Space $Gr_{1|1}(\mathbb{C}^{m_1|n_1}) \times \WCP^{{0|1}}_{\hat{m}_2|\hat{n}_2}$ $\MSUSY^{r_1(\text{iii}), r_2(\text{iii}) < 0}$}

The analysis for the $r_1(\text{iii}), r_2(\text{iii}) < 0$ phase is practically the same as the analysis for the $r_1(\text{ii}), r_2(\text{ii}) < 0$ phase.
So, for brevity, we shall simply state the result.

When $r_2 < 0$ and $\hat{m}_2 r_1 - \hat{m}_1 r_2 > 0$, $\MSUSY^{r_1(\text{iii}), r_2(\text{iii}) < 0}$ and $\MSUSY^{r_1(\text{iii}) > 0 > r_2(\text{iii})}$ are the same.
Hence, the $r_1(\text{iii}), r_2(\text{iii}) < 0$ and $r_1(\text{iii}) > 0 > r_2(\text{iii})$ phases are the same phase of the GLSM.
In particular, it will be the product space $Gr_{1|1}(\mathbb{C}^{m_1|n_1}) \times \WCP^{0|1}_{\hat{m}_2|\hat{n}_2}$, as defined by \eqref{eq:U(1|1)-N with W:phases:r1(i) > 0 > r2(i):space 1} and \eqref{eq:U(1|1)-N with W:phases:r1(i) > 0 > r2(i):space 2}.

\subtitle{The $r_1(\text{iii}), r_2(\text{iii}) < 0$ Phase of the Model at Low Energy: A Hybrid $Gr_{1|1}(\mathbb{C}^{m_1|n_1})$ NLSM $\big{/}$ $ \mathrm{SU}(1|1)_2 \times \mathbb{Z}_{\hat{\mathfrak{m}}_2}$ Supergauged LG Orbifold on a Supervector Bundle over $Gr_{1|1}(\mathbb{C}^{m_1|n_1}) \times \WCP^{0|1}_{\hat{m}_2|\hat{n}_2}$}

Since the $r_1(\text{iii}), r_2(\text{iii}) < 0$ phase is the same as the $r_1(\text{iii}) > 0 > r_2(\text{iii})$ phase, we can read off the low energy theory from the latter.

In particular, at low energy, we have a hybrid between an NLSM on $Gr_{1|1}(\mathbb{C}^{m_1|n_1})$ and an $\mathrm{SU}(1|1)_2 \times \mathbb{Z}_{\hat{\mathfrak{m}}_2}$ supergauged LG orbifold defined on $\mathcal{S}\mathcal{G}^{\mathrm{U}(1|1)^2}_{\text{III}}$, where $Gr_{1|1}(\mathbb{C}^{m_1|n_1})$ is the super-Grassmannian described by \eqref{eq:U(1|1)-N with W:phases:r1(i) > 0 > r2(i):space 1}, and $\mathcal{S}\mathcal{G}^{\mathrm{U}(1|1)^2}_{\text{III}}$ is the supervector bundle described by \eqref{eq:U(1|1)-N with W:phases:r1(i) > 0 > r2(i):space 2:supervector bundle}.

\subtitle{Branch (1) of the $\mathrm{U}(1|1)^2$ GLSM with Superpotential}

Altogether, the phases described hitherto can be understood to be phases along Branch (1) of the $\mathrm{U}(1|1)^2$ GLSM with superpotential.
Let us refer to
the $r_1(\text{i}), r_2(\text{i}) > 0$ phase as phase I;
the $r_1(\text{ii}) < 0 < r_2(\text{ii})$ and $r_1(\text{ii}), r_2(\text{ii}) < 0$ phases as phase II;
the $r_1(\text{iii}) > 0 > r_2(\text{iii})$ and $r_1(\text{iii}), r_2(\text{iii}) < 0$ phases as phase III;
and
the $r_1(\text{iv}), r_2(\text{iv}) < 0$ phase as phase IV.

$\MSUSY$ of the different phases along Branch (1) of the $\mathrm{U}(1|1)^2$ GLSM with superpotential is summarized in \autoref{fig:U(1|1)-N with W:M-SUSY:first branch}, and the corresponding low energy limit of the model is summarized in \autoref{fig:U(1|1)-N with W:low energy:first branch}.
In both figures, one can see that phase IV is the wall between phases II and III, whose slope is determined by the value of $\hat{m}_2/\hat{m}_1$.

\subtitle{Branch (2) of the $\mathrm{U}(1|1)^2$ GLSM with Superpotential}

Repeating the analysis for the
\begin{enumerate*}[label=(\alph*)]
  \item $r_1(\text{i}), r_2(\text{i}) < 0$;

  \item $r_1(\text{ii}) > 0 > r_2(\text{ii})$;

  \item $r_1(\text{ii}), r_2(\text{ii}) > 0$;

  \item $r_1(\text{iii}) < 0 < r_2(\text{iii})$;

  \item $r_1(\text{iii}), r_2(\text{iii}) > 0$;

  and

  \item $r_1(\text{iv}), r_2(\text{iv}) > 0$
\end{enumerate*}
phases, we find that they make up the four phases along Branch (2) of the $\mathrm{U}(1|1)^2$ GLSM with superpotential.

In particular,
the $r_1(\text{i}), r_2(\text{i}) < 0$ phase would be phase I$'$;
the $r_1(\text{ii}) > 0 > r_2(\text{ii})$ and $r_1(\text{ii}), r_2(\text{ii}) > 0$ phases would be phase II$'$;
the $r_1(\text{iii}) < 0 < r_2(\text{iii})$ and $r_1(\text{iii}), r_2(\text{iii}) > 0$ phases would be phase III$'$;
and
the $r_1(\text{iv}), r_2(\text{iv}) > 0$ phase would be phase IV$'$.
Furthermore, just like how the phases along Branch (2) of the $\mathrm{U}(1|1)$ GLSM with superpotential in \autoref{subsec: Phases U(1|1) with W} are parity reversals of the respective phases along Branch (1) therein, the phases (I$'$, II$'$, III$'$, IV$'$) are parity-reversals the phases (I, II, III, IV).

In other words, Branch (2) is a parity-reversed mirror of Branch (1).

$\MSUSY$ of the different phases along Branch (2) of the $\mathrm{U}(1|1)^2$ GLSM with superpotential is summarized in \autoref{fig:U(1|1)-N with W:M-SUSY:second branch}, and the corresponding low energy limit of the model is summarized in \autoref{fig:U(1|1)-N with W:low energy:second branch}.
In both figures, one can see that phase IV$'$ is the wall between phases II$'$ and III$'$, whose slope is determined by the value of $\hat{m}_2/\hat{m}_1$.
\begin{figure}
  \centering
  \begin{tikzpicture}[scale=1.1]

    \def\xmin{-7}
    \def\xmax{7}
    \def\ymin{-4}
    \def\ymax{4}

    \def\slope{0.3}

    \colorlet{phase1}{red}
    \colorlet{phase2}{blue}
    \colorlet{phase3}{green!70!black}
    \colorlet{phase4}{orange!90!black}


    \fill[phase1!12]
    (0,0)
    --(\xmin,0)
    --(\xmin,\ymin)
    --(0,\ymin)
    --cycle;

    \fill[phase2!12]
    (\xmin,0)
    --(0,0)
    --(\xmax,\slope*\xmax)
    --(\xmax,\ymax)
    --(\xmin,\ymax)
    --cycle;

    \fill[phase3!12]
    (0,\ymin)
    --(0,0)
    --(\xmax,\slope*\xmax)
    --(\xmax,\ymin)
    --cycle;


    \draw[white,line width=8pt]
    (\xmax,\slope*\xmax)--(0,0);

    \draw[white,line width=8pt]
    (0,0)--(0,\ymin);

    \draw[white,line width=8pt]
    (0,0)--(\xmin,0);


    \draw[phase4,line width=4pt]
    (\xmax,\slope*\xmax)
    --
    node[midway, above, sloped, phase4] {
      $\text{slope} = \flatfrac{\hat{m}_2}{\hat{m}_1}$
    }
    (0,0);

    \fill[white] (0,0) circle[radius=4pt];


    \node at (\xmin*0.5,\ymin*0.5) {
      \begin{tabular}{c}
        {\Large Phase I$'$}
        \\
        \\
        \hspace*{-40pt}$\MSUSY^{(\text{I}')} = \{G^+_{\mathbf{\Pi}|\mathbf{\Pi}} = 0\} \cap \{G^-_{\mathbf{\Pi}|\mathbf{\Pi}} = 0\}$
        \\
        $\qquad \qquad$
        $\subset Gr_{1|1}(\mathbb{C}^{n_1|m_1}) \times Gr_{1|1}(\mathbb{C}^{n_2|m_2})$
      \end{tabular}
    };

    \node at (\xmax*0.5,\ymin*0.5) {
      \begin{tabular}{c}
        {\Large Phase II$'$}
        \\
        \\
        $\MSUSY^{(\text{II}')}
        = \WCP^{0|1}_{\hat{n}_1|\hat{m}_1}
        \times Gr_{1|1}(\mathbb{C}^{n_2|m_2})$
      \end{tabular}
    };

    \node at (\xmin*0.5,\ymax*0.5) {
      \begin{tabular}{c}
        {\Large Phase III$'$}
        \\
        \\
        $\MSUSY^{(\text{III}')}
        = Gr_{1|1}(\mathbb{C}^{n_1|m_1})
        \times \WCP^{0|1}_{\hat{n}_2|\hat{m}_2}$
      \end{tabular}
    };

    \node[draw=phase4,
    line width=1pt,
    rounded corners=2pt,
    inner sep=4pt,
    above
    ] at (\xmax*0.5,\ymax*0.5+0.2) {
      \begin{tabular}{c}
        {\Large \textcolor{phase4}{Phase IV$'$}}
        \\
        \\
        $\MSUSY^{(\text{IV}')}
        = \WCP^{0|1}_{\hat{n}_1|\hat{m}_1} \times \WCP^{0|1}_{\hat{n}_2|\hat{m}_2}$
      \end{tabular}
    };


    \draw[->,thick] (\xmin-0.3,0)--(\xmax+0.3,0)
    node[right]{$r_1$};

    \draw[->,thick] (0,\ymin-0.3)--(0,\ymax+0.3)
    node[above]{$r_2$};

  \end{tikzpicture}
  \caption{%
    $\MSUSY$ of the different phases  along Branch (2) of the $\mathrm{U}(1|1)^2$ GLSM with superpotential.
    Phase I$'$ in pink, Phase II$'$ in green, Phase III$'$ in light purple, and Phase IV$'$ as the orange line.
  }
  \label{fig:U(1|1)-N with W:M-SUSY:second branch}
\end{figure}
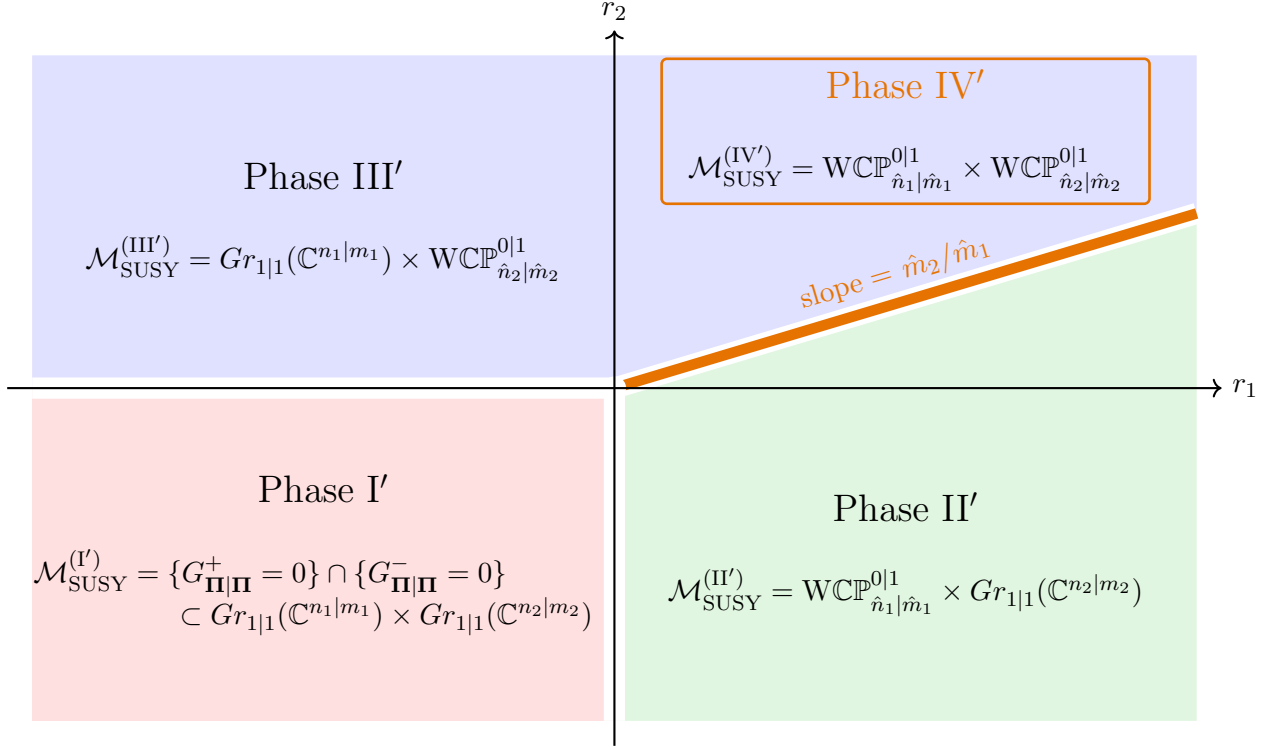
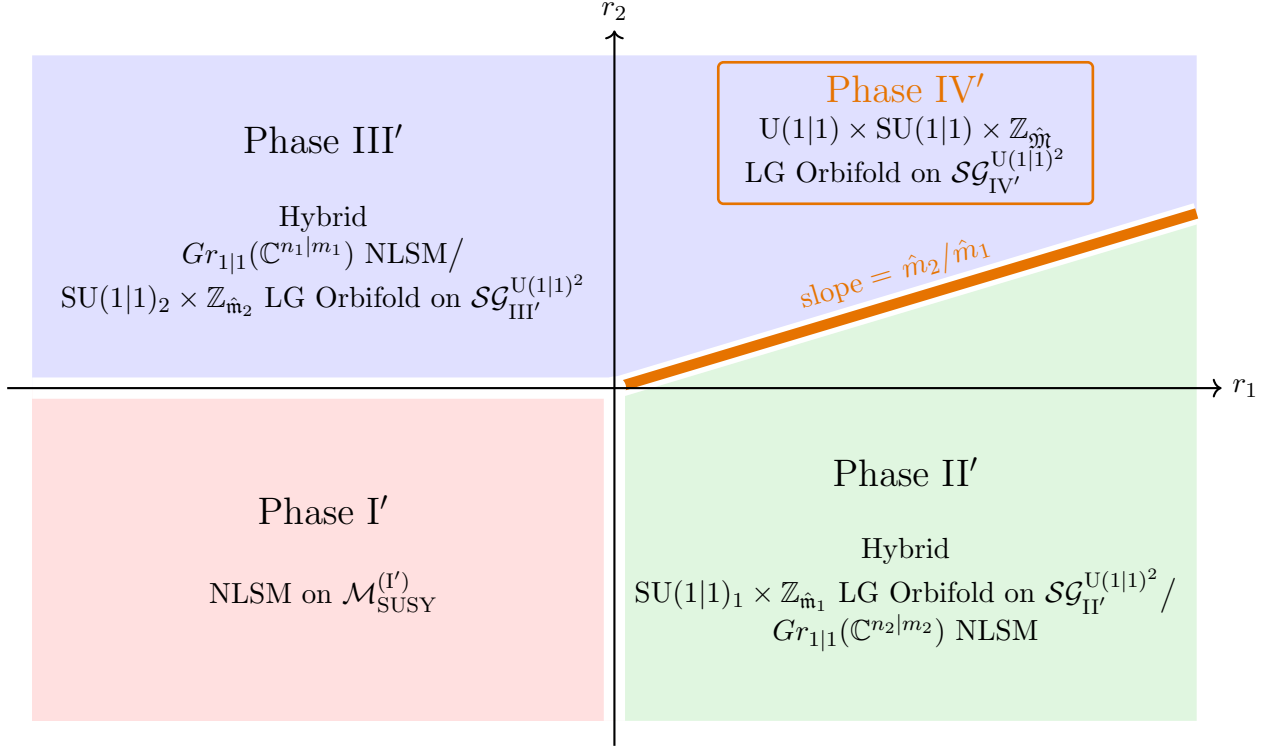
\begin{figure}
  \centering
  \begin{tikzpicture}[scale=1.1]

    \def\xmin{-7}
    \def\xmax{7}
    \def\ymin{-4}
    \def\ymax{4}

    \def\slope{0.3}

    \colorlet{phase1}{red}
    \colorlet{phase2}{blue}
    \colorlet{phase3}{green!70!black}
    \colorlet{phase4}{orange!90!black}


    \fill[phase1!12]
    (0,0)
    --(\xmin,0)
    --(\xmin,\ymin)
    --(0,\ymin)
    --cycle;

    \fill[phase2!12]
    (\xmin,0)
    --(0,0)
    --(\xmax,\slope*\xmax)
    --(\xmax,\ymax)
    --(\xmin,\ymax)
    --cycle;

    \fill[phase3!12]
    (0,\ymin)
    --(0,0)
    --(\xmax,\slope*\xmax)
    --(\xmax,\ymin)
    --cycle;


    \draw[white,line width=8pt]
    (\xmax,\slope*\xmax)--(0,0);

    \draw[white,line width=8pt]
    (0,0)--(0,\ymin);

    \draw[white,line width=8pt]
    (0,0)--(\xmin,0);


    \draw[phase4,line width=4pt]
    (\xmax,\slope*\xmax)
    --
    node[midway, above, sloped, phase4] {
      $\text{slope} = \flatfrac{\hat{m}_2}{\hat{m}_1}$
    }
    (0,0);

    \fill[white] (0,0) circle[radius=4pt];


    \node at (\xmin*0.5,\ymin*0.5) {
      \begin{tabular}{c}
        {\Large Phase I$'$}
        \\
        \\
        NLSM on $\MSUSY^{(\text{I}')}$
      \end{tabular}
    };

    \node at (\xmax*0.5,\ymin*0.5) {
      \begin{tabular}{c}
        {\Large Phase II$'$}
        \\
        \\
        Hybrid
        \\
        $\mathrm{SU}(1|1)_1 \times \mathbb{Z}_{\hat{\mathfrak{m}}_1}$ LG Orbifold on $\mathcal{S}\mathcal{G}^{\mathrm{U(1|1)^2}}_{\text{II}'}$$\big{/}$
        \\
        $Gr_{1|1}(\mathbb{C}^{n_2|m_2})$ NLSM
      \end{tabular}
    };

    \node at (\xmin*0.5,\ymax*0.5) {
      \begin{tabular}{c}
        {\Large Phase III$'$}
        \\
        \\
        Hybrid
        \\
        $Gr_{1|1}(\mathbb{C}^{n_1|m_1})$ NLSM$\big{/}$
        \\
        $\mathrm{SU}(1|1)_2 \times \mathbb{Z}_{\hat{\mathfrak{m}}_2}$ LG Orbifold on $\mathcal{S}\mathcal{G}^{\mathrm{U(1|1)^2}}_{\text{III}'}$
      \end{tabular}
    };

    \node[draw=phase4,
    line width=1pt,
    rounded corners=2pt,
    inner sep=4pt,
    above
    ] at (\xmax*0.5,\ymax*0.5+0.2) {
      \begin{tabular}{c}
        {\Large \textcolor{phase4}{Phase IV$'$}}
        \\
        $\mathrm{U}(1|1) \times \mathrm{SU}(1|1) \times \mathbb{Z}_{\hat{\mathfrak{M}}}$
        \\
        LG Orbifold on $\mathcal{S}\mathcal{G}^{\mathrm{U(1|1)^2}}_{\text{IV}'}$
      \end{tabular}
    };


    \draw[->,thick] (\xmin-0.3,0)--(\xmax+0.3,0)
    node[right]{$r_1$};

    \draw[->,thick] (0,\ymin-0.3)--(0,\ymax+0.3)
    node[above]{$r_2$};

  \end{tikzpicture}
  \caption{%
    The low energy limit of the $\mathrm{U}(1|1)^2$ GLSM with superpotential in different phases along Branch (2).
  }
  \label{fig:U(1|1)-N with W:low energy:second branch}
\end{figure}

In \autoref{fig:U(1|1)-N with W:M-SUSY:second branch}, $G^+_{\mathbf{\Pi}|\mathbf{\Pi}} = G^+(Y^+_{\mathbf{\Pi}}(s, \mathbx{s}), Y^+_{\mathbf{\Pi}}(t, \mathbx{t}))$  is a Grassmann-even bihomogeneous polynomial of degree $\hat{n}_1$ in $Y^+_{\mathbf{\Pi}}(s, \mathbx{s}) = Y^-(s, \mathbx{s})$ and $\hat{n}_2$ in $Y^+_{\mathbf{\Pi}}(t, \mathbx{t}) = Y^-(t, \mathbx{t})$, where $Y^+_{\mathbf{\Pi}}$ are super-Pl\"{u}cker coordinates of a parity-reversed super-Grassmannian (see \autoref{ft:U(1|1) with W:phases:parity-reversed super-plucker}), and the $Y^-$'s are defined in \eqref{eq:u-1-1 with W:glsm:Y- even}--\eqref{eq:u-1-1 with W:glsm:Y- odd}.

Likewise, in \autoref{fig:U(1|1)-N with W:M-SUSY:second branch}, $G^-_{\mathbf{\Pi}|\mathbf{\Pi}} = G^-(Y_{\mathbf{\Pi}}^-(s, \mathbx{s}), Y_{\mathbf{\Pi}}^-(t, \mathbx{t}))$ is a Grassmann-odd bihomogeneous polynomial of degree $\hat{m}_1$ in $Y_{\mathbf{\Pi}}^-(s, \mathbx{s}) = Y^+(s, \mathbx{s})$ and $\hat{m}_2$ in $Y_{\mathbf{\Pi}}^+(t, \mathbx{t}) = Y^-(t, \mathbx{t})$, where the $Y^+$'s are defined in \eqref{eq:u-1-1 with W:glsm:Y+ even}--\eqref{eq:u-1-1 with W:glsm:Y+ odd}.

Also, the supervector bundles $\mathcal{S}\mathcal{G}^{\mathrm{U}(1|1)^2}_{X'}$ (for $X' \in \{\text{II}', \text{III}', \text{IV}'\}$) in \autoref{fig:U(1|1)-N with W:low energy:second branch} are parity-reversals of $\mathcal{S}\mathcal{G}^{\mathrm{U}(1|1)^2}_{X}$ in Branch (1), just like how \eqref{eq:U(1|1) with W:phases:supervector bundle:r(ii) > 0} is a parity-reversal of \eqref{eq:U(1|1) with W:phases:supervector bundle:r(ii) < 0}.

\subtitle{A Summary of the Phases and Branches of the $\mathrm{U}(1|1)^2$ GLSM with Superpotential}

In summary, Branch (1) and (2) have four phases each, and are summarized in \autoref{fig:U(1|1)-N with W:M-SUSY:first branch}--\autoref{fig:U(1|1)-N with W:low energy:first branch} and \autoref{fig:U(1|1)-N with W:M-SUSY:second branch}--\autoref{fig:U(1|1)-N with W:low energy:second branch}, respectively.

\subsection{Phases of the Reduced \texorpdfstring{$\mathrm{U}(1)^N$}{U(1)\textasciicircum N} Model: An NLSM on a  Hypersurface in a Product of Projective Spaces, an LG Orbifold on a Vector Bundle over Projective Space, and their Hybrids}
\label{subsec: Phases of the reduced U(1)^2 model}

As mentioned at the end of \autoref{subsec: U(1|1)^2 with W}, the $\mathrm{U}(1|1)_1 \times \mathrm{U}(1|1)_2$ GLSM with superpotential ought to reduce to the familiar $\mathrm{U}(1)_1 \times \mathrm{U}(1)_2$ GLSM with superpotential when we turn off
\begin{enumerate*}
  \item the $\tensor{s}{^{\bar{1}}_a}$ and $\tensor{t}{^{\bar{1}}_{\alpha}}$ fields (and their corresponding bar-indexed partners in the underlying multiplets),

  \item the \emph{whole} Grassmann-odd chiral superfields $\mathbx{S}$ and $\mathbx{T}$,

  and

  \item the \emph{whole} Grassmann-odd chiral superfield $P_-$.
\end{enumerate*}

In doing so, as explained at the end of \autoref{subsec: U(1|1)^2 with W}, of the $P_{\pm}$ superfields and $G^{\pm}$ polynomials, we will only have $P_+$ and $G^+$ remaining, where the latter would now be expressed as a polynomial of bidegree $(\hat{m}_1, \hat{m}_2)$ in $Y^+(S^1) = S^1$ and $Y^+(T^1) = T^1$.
Moreover, $P_+$ and $G^+$ would transform under the determinantal representation of the reduced $\mathrm{U}(1)_1 \times \mathrm{U}(1)_2$ gauge group.

\subtitle{A Reduction to Phase I of the Familiar $\mathrm{U}(1)_1 \times \mathrm{U}(1)_2$ GLSM with Superpotential and its Low Energy NLSM on a Hypersurface in $\mathbb{CP}^{m_1 - 1} \times \mathbb{CP}^{m_2 - 1}$}

In performing the reduction, we find that $\MSUSY$ for Phase I, which was defined by \eqref{eq:U(1|1)-N with W:phases:r1(i) and r2(i) > 0:space 1}--\eqref{eq:U(1|1)-N with W:phases:r1(i) and r2(i) > 0:hypersurface -} in the $\mathrm{U}(1|1)_1 \times \mathrm{U}(1|1)_2$ model, would reduce to a hypersurface $\{G^+(s^1, t^1) = 0\}$ in $\mathbb{CP}^{m_1 - 1} \times \mathbb{CP}^{m_2 - 1}$ of sizes $r_1$ and $r_2$, as it should (\emph{c.f.}~\cite[$\S$5.2]{Witten:1993yc}).
Consequently, its low energy NLSM would have, as target space, a hypersurface $G^+(s^1, t^1) = 0$ in $\mathbb{CP}^{m_1 - 1} \times \mathbb{CP}^{m_2 - 1}$ of sizes $(r_1, r_2)$, as it should (\emph{c.f.}~\cite[$\S$5.2]{Witten:1993yc}).

\subtitle{A Reduction to Phase II of the Familiar $\mathrm{U}(1)_1 \times \mathrm{U}(1)_2$ GLSM with Superpotential and its Low Energy Hybrid LG $\mathbb{Z}_{\hat{m}_1}$-Orbifold on a Vector Bundle over $\mathbb{CP}^{m_2 - 1}$ $\big{/}$ $\mathbb{CP}^{m_2 - 1}$ NLSM}

In performing the reduction, we find that $\MSUSY$ for Phase II, which was defined by \eqref{eq:U(1|1)-N with W:phases:r1(i) < 0 < r2(i):space 2}-\eqref{eq:U(1|1)-N with W:phases:r1(i) < 0 < r2(i):space 1} in the $\mathrm{U}(1|1)_1 \times \mathrm{U}(1|1)_2$ model, would reduce to a projective space $\mathbb{CP}^{m_2 - 1}$ of size $r_2 - \frac{\hat{m}_2}{\hat{m}_1} r_1$ plus a point, as it should (\emph{c.f.}~\cite[$\S$5.2]{Witten:1993yc}).

At low energy, at the point of $\MSUSY$, we now have a $\mathbb{Z}_{\hat{m}_1}$ gauge theory.
Away from this point of $\MSUSY$, we continue to have an NLSM.

The $\mathbb{Z}_{\hat{m}_1}$ gauge theory will be an LG $\mathbb{Z}_{\hat{m}_1}$-orbifold fibered over $\mathbb{CP}^{m_2 - 1}$ of size $r_2 - \frac{\hat{m}_2}{\hat{m}_1} r_1$, with an effective superpotential
\begin{equation}
  \label{eq:U(1|1)-N with W:reduction:W^Z_n:II}
  W^{\mathbb{Z}_{\hat{m}_1}}_{\text{LG,II}}
  = \sqrt{ \flatfrac{\abs{r_1}}{\hat{m}_1}} \cdot G^+ \left( s^1, t^1 \right)
  \, ,
\end{equation}
that is a holomorphic function on the vector bundle $\mathcal{S}^{\mathrm{U}(1)^2}_{\text{II}}$, where
\begin{equation}
  \label{eq:U(1|1)-N with W:reduction:vector bundle:II}
  \mathcal{S}^{\mathrm{U}(1)^2}_{\text{II}}
  = \bigoplus^{m_1}_{a = 1} \mathcal{L}_a^{\flatfrac{\hat{m}_2}{\hat{m}_1}}
  \longrightarrow \mathbb{CP}^{m_2 - 1}
  \, ,
\end{equation}
such that the holomorphic coordinate on the fiber $\mathcal{L}_a$ is $\tensor{s}{^1_a}$, while that of the $\mathbb{CP}^{m_2 - 1}$ base is $\tensor{t}{^1_{\alpha}}$.
That is, we have an LG $\mathbb{Z}_{\hat{m}_1}$-orbifold defined on $\mathcal{S}^{\mathrm{U}(1)^2}_{\text{II}}$, as we should (\emph{c.f.}~\cite[$\S$5.2]{Witten:1993yc}).

The NLSM will now have a target space that is the projective space $\mathbb{CP}^{m_2 - 1}$ of size $r_2 - \frac{\hat{m}_2}{\hat{m}_1} r_1$.

In short, we have, at low energy, an NLSM in the $t$-direction, and an LG orbifold in the $s$-direction with an effective superpotential that depends on both the $s$ and $t$-directions along the fiber and base space of $\mathcal{S}^{\mathrm{U}(1)^2}_{\text{II}}$, respectively.

Simply put, we have a hybrid between an LG $\mathbb{Z}_{\hat{m}_1}$-orbifold defined on $\mathcal{S}^{\mathrm{U}(1)^2}_{\text{II}}$ and an NLSM on $\mathbb{CP}^{m_2 - 1}$, as we should (\emph{c.f.}~\cite[$\S$5.2]{Witten:1993yc}).

\subtitle{A Reduction to Phase III of the Familiar $\mathrm{U}(1)_1 \times \mathrm{U}(1)_2$ GLSM with Superpotential and its Low Energy Hybrid $\mathbb{CP}^{m_1 - 1}$ NLSM $\big{/}$ LG $\mathbb{Z}_{\hat{m}_2}$-Orbifold on a Vector Bundle over $\mathbb{CP}^{m_1-1}$}

The analysis of Phase III is practically identical to that of Phase II, so for brevity, we shall just state the results.

In Phase III, $\MSUSY$ is now a projective space $\mathbb{CP}^{m_1 - 1}$ of size $r_1 - \frac{\hat{m}_1}{\hat{m}_2} r_2$ plus a point, as it should (\emph{c.f.}~\cite[$\S$5.2]{Witten:1993yc}).

At low energy, we have, in the $s$-direction, an NLSM whose target space is the projective space $\mathbb{CP}^{m_1 - 1}$ of size $r_1 - \frac{\hat{m}_1}{\hat{m}_2} r_2$.

At low energy, we have, in the $t$-direction, an LG $\mathbb{Z}_{\hat{m}_2}$-orbifold with effective superpotential
\begin{equation}
  W^{\mathbb{Z}_{\hat{m}_2}}_{\text{LG,III}}
  = \sqrt{ \flatfrac{\abs{r_2}}{\hat{m}_2} } \cdot G^+ \left( s^1, t^1 \right)
  \, ,
\end{equation}
that is a holomorphic function on the vector bundle $\mathcal{S}^{\mathrm{U}(1)^2}_{\text{III}}$, where
\begin{equation}
  \mathcal{S}^{\mathrm{U}(1)^2}_{\text{III}}
  = \bigoplus^{m_2}_{\alpha = 1} \mathcal{L}^{\flatfrac{\hat{m}_1}{\hat{m}_2}}_{\alpha}
  \longrightarrow \mathbb{CP}^{m_1 - 1},
\end{equation}
such that the holomorphic coordinate on the fiber $\mathcal{L}_{\alpha}$ is $\tensor{t}{^1_{\alpha}}$, while that of the $\mathbb{CP}^{m_1 - 1}$ base is $\tensor{s}{^1_a}$.

Simply put, we have a hybrid between an NLSM on $\mathbb{CP}^{m_1 - 1}$ and an LG $\mathbb{Z}_{\hat{m}_2}$-orbifold defined on $\mathcal{S}^{\mathrm{U}(1)^2}_{\text{III}}$, as we should (\emph{c.f.}~\cite[$\S$5.2]{Witten:1993yc}).

\subtitle{A Reduction to the Boundary Between Phases II and III of the Familiar $\mathrm{U}(1)_1 \times \mathrm{U}(2)_2$ GLSM with Superpotential and its Low Energy $\mathrm{U}(1) \times \mathbb{Z}_{\hat{M}}$ Gauged LG Orbifold}

In performing the reduction, we find that $\MSUSY$ for the boundary between Phases II and III, which was defined by \eqref{eq:U(1|1)-N with W:phases:r1(i) and r2(i) < 0:wcp 1}-\eqref{eq:U(1|1)-N with W:phases:r1(i) and r2(i) < 0:wcp 2} as Phase IV of the underlying $\mathrm{U}(1|1)_1 \times \mathrm{U}(1|1)_2$ model, would reduce to a product of ``projective spaces'' $\mathbb{CP}^0 \times \mathbb{CP}^0$ of ``sizes'' $\flatfrac{\abs{r_1}}{\hat{m}_1}$ and $\flatfrac{\abs{r_2}}{\hat{m}_2}$, whereby $\hat{m}_1 r_2 = \hat{m}_2 r_1$, as it should (\emph{c.f.}~\cite[$\S$5.2]{Witten:1993yc}).

At low energy, we have a $\mathrm{U}(1) \times \mathbb{Z}_{\hat{M}}$ gauged LG orbifold, where $\hat{M} \coloneq \gcd(\hat{m}_1, \hat{m}_2)$, as we should (\emph{c.f.}~\cite[$\S$5.2]{Witten:1993yc}).

\subsection{Quantum Aspects: The GLSM}
\label{sec:U(1|1)-N with W:quantum:glsm}

Of the various quantum aspects of the GLSM, we shall now discuss the one relevant to our eventual aim of exploring its applications to mathematics, which is the quantum correction to the FI parameters $r_{1,2}$.

The analysis is just a combination of that in \autoref{subsec:Quantum Prop U(1|1)^N GLSM} and \autoref{subsec: Quantum Prop U(1|1) GLSM with W}.
Thus, for brevity, we shall just state the results.

\subtitle{The Quantum $D$-terms}

The quantum $D^{\{1,2\}}$-terms which generalizes \eqref{eq:u-1-1-N:quantum:D terms:expectation values} and \eqref{eq:u-1-1 with W:quantum:D terms:expectation values} (omitting the $D^{\{1,2\}}$-squared contributions) are
\begin{equation}
  \label{eq:U(1|1)-N with W:quantum glsm:D-term:1}
  (-1)^{\varsigma(i)} \tensor{(D^1)}{^j_i} \left(
    \sum_{a = 1}^{m_1} \expval{\tensor{s}{^j_a} \tensor{\bar{s}}{_i^a}}
    - \sum_{b = 1}^{n_1} \expval{\tensor{\mathbx{s}}{^j_b} \tensor{\bar{\mathbx{s}}}{_i^b}}
    - \hat{m}_1 \expval{p_+ \bar{p}^+} \tensor{\delta}{^j_i}
    + \hat{n}_1 \expval{p_- \bar{p}^-} \tensor{\delta}{^j_i}
    - r_1 \tensor{\delta}{^j_i}
  \right)
\end{equation}
and
\begin{equation}
  \label{eq:U(1|1)-N with W:quantum glsm:D-term:2}
  (-1)^{\varsigma(i)} \tensor{(D^2)}{^j_i} \left(
    \sum_{\alpha = 1}^{m_2} \expval{\tensor{t}{^j_{\alpha}} \tensor{\bar{t}}{_i^{\alpha}}}
    - \sum_{\beta = 1}^{n_2} \expval{\tensor{\mathbx{t}}{^j_{\beta}} \tensor{\bar{\mathbx{t}}}{_i^{\beta}}}
    - \hat{m}_2 \expval{p_+ \bar{p}^+} \tensor{\delta}{^j_i}
    + \hat{n}_2 \expval{p_- \bar{p}^-} \tensor{\delta}{^j_i}
    - r_2 \tensor{\delta}{^j_i}
  \right).
\end{equation}

\subtitle{The Expectation Values}

The above expectation values to a 1-loop approximation in the effective theory at energy scale $\mu$ are
\begin{equation}
  \label{eq:U(1|1)-N with W:quantum glsm:field exp val}
  \expval{\tensor{s}{^j_a} \tensor{\bar{s}}{_i^a}}
  = \expval{\tensor{\mathbx{s}}{^j_b} \tensor{\bar{\mathbx{s}}}{_i^b}}
  = \expval{\tensor{t}{^j_{\alpha}} \tensor{\bar{t}}{_i^{\alpha}}}
  = \expval{\tensor{\mathbx{t}}{^j_{\beta}} \tensor{\bar{\mathbx{t}}}{_i^{\beta}}}
  = \log \left( \frac{\Lambda_{\text{UV}}}{\mu} \right) \tensor{\delta}{^j_i}
\end{equation}
and
\begin{equation}
  \label{eq:U(1|1)-N with W:quantum glsm:p exp val}
  \expval{p_{\pm} \bar{p}^{\pm}}_{\text{1-loop}}
  = \log \left( \frac{\Lambda_{\text{UV}}}{\mu} \right)
  \, ,
\end{equation}
where $\Lambda_{\text{UV}}$ is the cutoff scale which ensures that we have a finite answer.

\subtitle{The Effective FI Parameters $r_{\{1, 2\}, \text{eff}}(\mu)$}

By substituting \eqref{eq:U(1|1)-N with W:quantum glsm:field exp val}--\eqref{eq:U(1|1)-N with W:quantum glsm:p exp val} in \eqref{eq:U(1|1)-N with W:quantum glsm:D-term:1}--\eqref{eq:U(1|1)-N with W:quantum glsm:D-term:2}, one can see that the effective FI parameters which will ensure the finiteness of the effective action in the continuum limit $\Lambda_{\text{UV}} \rightarrow \infty$ can then be written as
\begin{equation}
  \label{eq:U(1|1)-N with W:quantum glsm:r-eff:1}
  r_{1, \text{eff}}
  = r_1
  + \Bigl[
    (m_1 - n_1)
    - (\hat{m}_1 - \hat{n}_1)
  \Bigr]
  \log \left( \frac{\mu}{\Lambda_{\text{UV}}} \right)
  \,
\end{equation}
and
\begin{equation}
  \label{eq:U(1|1)-N with W:quantum glsm:r-eff:2}
  r_{2, \text{eff}}
  = r_2
  + \Bigl[
    (m_2 - n_2)
    - (\hat{m}_2 - \hat{n}_2)
  \Bigr]
  \log \left( \frac{\mu}{\Lambda_{\text{UV}}} \right)
  \, .
\end{equation}

So for a fixed $\Lambda_{\text{UV}}$ and $r_{\{1, 2\}}$, we have
\begin{equation}
  \label{eq:U(1|1)-N with W:quantum glsm:r-eff(mu):1}
  \saveboxed{eq:U(1|1)-N with W:quantum glsm:r-eff(mu):1}{
    r_{1, \text{eff}}(\mu)
    = \Bigl[
      (m_1 - n_1)
      - (\hat{m}_1 - \hat{n}_1)
    \Bigr]
    \log \left( \frac{\mu}{\Lambda_1} \right)
  }
\end{equation}
and
\begin{equation}
  \label{eq:U(1|1)-N with W:quantum glsm:r-eff(mu):2}
  \saveboxed{eq:U(1|1)-N with W:quantum glsm:r-eff(mu):2}{
    r_{2, \text{eff}}(\mu)
    = \Bigl[
      (m_2 - n_2)
      - (\hat{m}_2 - \hat{n}_2)
    \Bigr]
    \log \left( \frac{\mu}{\Lambda_2} \right)
  }
\end{equation}
where the $r_1$/$r_2$ is being absorbed in the redefinition of $\Lambda_{\text{UV}}$ as $\Lambda_1$/$\Lambda_2$, such that $\Lambda_1$/$\Lambda_2$ can be regarded as dynamically-generated scale parameters of mass dimension.

\subtitle{Scale-dependence}

If $m_k - n_k \neq \hat{m}_k - \hat{n}_k$ for $k \in \{1, 2\}$, from \eqref{eq:U(1|1)-N with W:quantum glsm:r-eff(mu):1} and \eqref{eq:U(1|1)-N with W:quantum glsm:r-eff(mu):2}, we see that the effective FI parameters $r_{k, \text{eff}} (\mu)$ run (i.e., RG flow) as we change the energy scale $\mu$.
In particular, the dimensionless parameters $r_k$ of the classical theory have been replaced by the scale parameters $\Lambda_k$ of mass dimension in the quantum theory.

\subtitle{Restoring Poincar\'{e} Invariance and $\abs{r_{\{1, 2\}}} \gg 0$}

As before, the discussion regarding the restoration of Poincar\'{e} invariance in the continuum limit $\Lambda_{\text{UV}} \to \infty$ tells us that we need to consider $|r_k| \gg 0$, for $k \in \{1, 2\}$, in the quantum theory.

For example, if $m_k - n_k > \hat{m}_k - \hat{n}_k$, in the limit $\Lambda_{\text{UV}} \rightarrow \infty$, we see from \eqref{eq:U(1|1)-N with W:quantum glsm:r-eff:1} and \eqref{eq:U(1|1)-N with W:quantum glsm:r-eff:2} that we ought to have $r_k \gg 0$ so that $r_{k, \text{eff}}$ would be finite and also positive.
On the other hand, if $m_k - n_k < \hat{m}_k - \hat{n}_k$, in the limit $\Lambda_{\text{UV}} \rightarrow \infty$, we see from \eqref{eq:U(1|1)-N with W:quantum glsm:r-eff:1} and \eqref{eq:U(1|1)-N with W:quantum glsm:r-eff:2} that we ought to have $r_k \ll 0$ so that $r_{k, \text{eff}}$ would be finite and also negative.

\subtitle{$\MSUSY$ of the Quantum GLSM}

Applying the same analysis as that at the end of \autoref{subsec: Quantum Prop U(1|1) GLSM with W}, we find that when
\begin{enumerate}[label=(\Alph*)]
  \item $m_1 - n_1 > \hat{m}_1 - \hat{n}_1$ and $m_2 - n_2 > \hat{m}_2 - \hat{n}_2$,

  \item $m_1 - n_1 < \hat{m}_1 - \hat{n}_1$ and $m_2 - n_2 > \hat{m}_2 - \hat{n}_2$,

  \item $m_1 - n_1 > \hat{m}_1 - \hat{n}_1$ and $m_2 - n_2 < \hat{m}_2 - \hat{n}_2$,

  \item $m_1 - n_1 < \hat{m}_1 - \hat{n}_1$ and $m_2 - n_2 < \hat{m}_2 - \hat{n}_2$,

\end{enumerate}
$\MSUSY$ would necessarily be given by
\begin{enumerate}[label=(\Alph*)]
  \item $\MSUSY^{r_{1, \text{eff}}(\text{i}), r_{2, \text{eff}}(\text{i}) > 0}$/$\MSUSY^{r_{1, \text{eff}}(\text{ii}), r_{2, \text{eff}}(\text{ii}) > 0}$/$\MSUSY^{r_{1, \text{eff}}(\text{iii}), r_{2, \text{eff}}(\text{iii}) > 0}$/$\MSUSY^{r_{1, \text{eff}}(\text{iv}), r_{2, \text{eff}}(\text{iv}) > 0}$,

  \item $\MSUSY^{r_{1, \text{eff}}(\text{i}) < 0 < r_{2, \text{eff}}(\text{i})}$/$\MSUSY^{r_{1, \text{eff}}(\text{ii}) < 0 < r_{2, \text{eff}}(\text{ii})}$/$\MSUSY^{r_{1, \text{eff}}(\text{iii}) < 0 < r_{2, \text{eff}}(\text{iii})}$/$\MSUSY^{r_{1, \text{eff}}(\text{iv}) < 0 < r_{2, \text{eff}}(\text{iv})}$,

  \item $\MSUSY^{r_{1, \text{eff}}(\text{i}) > 0 > r_{2, \text{eff}}(\text{i})}$/$\MSUSY^{r_{1, \text{eff}}(\text{ii}) > 0 > r_{2, \text{eff}}(\text{ii})}$/$\MSUSY^{r_{1, \text{eff}}(\text{iii}) > 0 > r_{2, \text{eff}}(\text{iii})}$/$\MSUSY^{r_{1, \text{eff}}(\text{iv}) > 0 > r_{2, \text{eff}}(\text{iv})}$,

  \item $\MSUSY^{r_{1, \text{eff}}(\text{i}), r_{2, \text{eff}}(\text{i}) < 0}$/$\MSUSY^{r_{1, \text{eff}}(\text{ii}), r_{2, \text{eff}}(\text{ii}) < 0}$/$\MSUSY^{r_{1, \text{eff}}(\text{iii}), r_{2, \text{eff}}(\text{iii}) < 0}$/$\MSUSY^{r_{1, \text{eff}}(\text{iv}), r_{2, \text{eff}}(\text{iv}) < 0}$,
\end{enumerate}
respectively.

In other words, when $m_k - n_k \neq \hat{m}_k - \hat{n}_k$ (for $k \in \{1, 2\}$) and the quantum GLSM is \emph{not} scale-invariant, $\MSUSY$ can \emph{only} be (A), (B), (C), or (D) (depending on the values of the positive integers $m_k, n_k ,\hat{m}_k, \hat{n}_k$).

\subsection{Quantum Aspects: The Low Energy Theory}
\label{subsec: Quantum low energy U(1|1)^2 with W}

\subtitle{$\MSUSY$ of the Quantum GLSM at Low Energy}

Recall that in our derivation of $\MSUSY$ at the classical level in \autoref{subsec: U(1|1)^2 with W}, $r_k \neq 0$ (for $k \in \{1, 2\}$).
In the quantum theory, in place of $r_k$, we have $r_{k, \text{eff}} (\mu)$.
As explained towards the end of \autoref{sec:U(1|1)-N with W:quantum:glsm}, we have to consider either $r_k \gg 0$ when $m_k - n_k > \hat{m}_k - \hat{n}_k$, or $r_k \ll 0$ when $m_k - n_k < \hat{m}_k - \hat{n}_k$.
Then, one can deduce from \eqref{eq:U(1|1)-N with W:quantum glsm:r-eff:1}-\eqref{eq:U(1|1)-N with W:quantum glsm:r-eff:2} that even at low energy $\mu \ll \Lambda_{\text{UV}}$, we will always have $r_{k, \text{eff}} (\mu) > 0$ when $m_k - n_k > \hat{m}_k - \hat{n}_k$, or $r_{k, \text{eff}} (\mu) < 0$ when $m_k - n_k < \hat{m}_k - \hat{n}_k$, whence supersymmetry will be unbroken.

Notice at this point that \eqref{eq:U(1|1)-N with W:quantum glsm:r-eff:1}-\eqref{eq:U(1|1)-N with W:quantum glsm:r-eff:2} can also be expressed as the linear combination
\begin{equation}
  \label{eq:U(1|1)-N with W:quantum low energy:r-eff}
  m_1 r_{2, \text{eff}} - m_2 r_{1, \text{eff}}
  = (m_1 r_2 - m_2 r_1) + \mho \log \left( \frac{\mu}{\Lambda_{\text{UV}}} \right)
  \, ,
\end{equation}
where the integer $\mho \coloneq m_1 [m_2 - n_2 - \hat{m}_2 + \hat{n}_2] - m_2 [m_1 - n_1 - \hat{m}_1 + \hat{n}_1]$.
Similarly, we have to consider $m_1 r_2 - m_2 r_1 \gg 0$ when $\mho > 0$, or $m_1 r_2 - m_2 r_1 \ll 0$ when $\mho < 0$.
Also, we have to consider $m_1 r_2 - m_2 r_1 = 0$ when $\mho = 0$ such that the LHS of \eqref{eq:U(1|1)-N with W:quantum low energy:r-eff} is finite and unambiguous.\footnote{%
  This third consideration is reasonable in the sense that although $r_1 \neq 0 \neq r_2$, their linear combination can be zero.
}
Altogether, one can deduce from \eqref{eq:U(1|1)-N with W:quantum low energy:r-eff} that even at low energy $\mu \ll \Lambda_{\text{UV}}$, we will always have $m_1 r_{2, \text{eff}}(\mu) - m_2 r_{1, \text{eff}}(\mu) > 0$ when $\mho > 0$, $m_1 r_{2, \text{eff}}(\mu) - m_2 r_{1, \text{eff}}(\mu) < 0$ when $\mho < 0$, or $m_1 r_{2, \text{eff}}(\mu) - m_2 r_{1, \text{eff}}(\mu) = 0$ when $\mho = 0$, whence supersymmetry will be unbroken.

\subtitle{The Quantum GLSM at Low Energy is a Quantum NLSM, Quantum Supergauged LG Orbifold, their Hybrids, or Either}

The analysis in \autoref{subsec: Phases of U(1|1)^2 with W} (and therefore \autoref{subsec: Phases of the reduced U(1)^2 model}) had been about the classical theory.
The question therefore, is whether at the quantum level, the GLSM will also become the NLSM, supergauged LG orbifold, their hybrids, or either, at low energy, as per the classical theory.
The short answer is ``yes'', although which it becomes, depends on the values of $(m_k - n_k) - (\hat{m}_k - \hat{n}_k)$ and possibly $\mho$.

Let us elaborate on the seven possible cases for when we are in Branch (1).

\begin{itemize}
  \item $m_1 - n_1 > \hat{m}_1 - \hat{n}_1$ and $m_2 - n_2 > \hat{m}_2 - \hat{n}_2$ ($\MSUSY^{r_{1, \text{eff}}(\text{i}), r_{2, \text{eff}}(\text{i}) > 0}$)

  In this case, as explained above, we have $r_{\{1,2\}} \gg 0$ and $r_{\{1, 2\}, \text{eff}}(\mu) > 0$.
  Also, a $\mathrm{U}(1|1)_1 \times \mathrm{U}(1|1)_2$ generalization of our discussion in \autoref{subsec: Quantum Low Energy U(1|1)} on the Higgs mechanism tells us that the quantum low energy limit exists at a scale $\mu \ll e_1 \sqrt{r_1}$ and $\mu \ll e_2 \sqrt{r_2}$.

  Thus, since $r_{\{1, 2\}, \text{eff}}(\mu) > 0$, at low energy, the quantum GLSM with $\MSUSY^{r_{1, \text{eff}}(\text{i}), r_{2, \text{eff}}(\text{i}) > 0}$ defined by \eqref{eq:U(1|1)-N with W:phases:r1(i) and r2(i) > 0:space 1}--\eqref{eq:U(1|1)-N with W:phases:r1(i) and r2(i) > 0:hypersurface -}, would become a quantum NLSM on $\MSUSY^{r_{1, \text{eff}}(\text{i}), r_{2, \text{eff}}(\text{i}) > 0}$, the complete intersection of an even and odd hypersurface $\{G^+ = 0\}$ and $\{G^- = 0\}$ of bidegrees $(\hat{m}_1, \hat{m}_2)$ and $(\hat{n}_1, \hat{n}_2)$ in $Gr_{1|1}(\mathbb{C}^{m_1|n_1}) \times Gr_{1|1}(\mathbb{C}^{m_2|n_2})$ of sizes $r_{1,\text{eff}}(\mu)$ and $r_{2, \text{eff}}(\mu)$.

  At any rate, for the quantum NLSM to be well-defined, i.e., amenable to perturbation theory, at the low energy scale $\mu$, its inverse coupling, which is proportional to $r_{\{1, 2\}, \text{eff}}(\mu)$, needs to be large; in other words, we need to have $r_{\{1,2\}, \text{eff}}(\mu) \gg 0$.
  In turn, from \eqref{eq:U(1|1)-N with W:quantum glsm:r-eff(mu):1}--\eqref{eq:U(1|1)-N with W:quantum glsm:r-eff(mu):2}, it would mean that $\mu \gg \Lambda_{\{1, 2\}}$ and therefore, $\Lambda_1 \ll \mu \ll e_1 \sqrt{r_1}$ and $\Lambda_2 \ll \mu \ll e_2 \sqrt{r_2}$.
  Thus, in place of the classical low energy limit $e_{\{1, 2\}}, b^{\pm}_{\{1, 2\}}, c^{\pm}_{\{1, 2\}} \to \infty$, we would have the quantum low energy limit $\flatfrac{e_1}{\Lambda_1}, \flatfrac{e_2}{\Lambda_2}, \flatfrac{b^{\pm}_1}{\Lambda_1}, \flatfrac{b^{\pm}_2}{\Lambda_2}, \flatfrac{c^{\pm}_1}{\Lambda_1}, \flatfrac{c^{\pm}_2}{\Lambda_2} \to \infty$.

  \item $m_1 - n_1 < \hat{m}_1 - \hat{n}_1$ and $m_2 - n_2 < \hat{m}_2 - \hat{n}_2$ but $\mho = 0$ ($\MSUSY^{r_{1, \text{eff}}(\text{iv}), r_{2, \text{eff}}(\text{iv}) < 0}$)

  In this case, as explained above, we must have $r_{\{1, 2\}} \ll 0$ and $m_1 r_2 - m_2 r_1 = 0$, as well as $r_{\{1, 2\}, \text{eff}}(\mu) < 0$ and $m_1 r_{2, \text{eff}}(\mu) - m_2 r_{1, \text{eff}}(\mu) = 0$.
  Also, according to our discussion in \autoref{subsec: Quantum Low Energy U(1|1)} on the Higgs mechanism, the quantum low energy limit exists at a scale $\mu \ll e_1 \sqrt{\abs{r_1}}$ and $\mu \ll e_2 \sqrt{\abs{r_2}}$.

  Thus, since $r_{\{1, 2\}, \text{eff}}(\mu) < 0$, at low energy, the quantum GLSM with $\MSUSY^{r_{1, \text{eff}}(\text{iv}), r_{2, \text{eff}}(\text{iv}) < 0}$ defined by \eqref{eq:U(1|1)-N with W:phases:r1(i) and r2(i) < 0:wcp 1}--\eqref{eq:U(1|1)-N with W:phases:r1(i) and r2(i) < 0:wcp 2}, would become a quantum $\mathrm{U}(1|1) \times \mathrm{SU}(1|1) \times \mathbb{Z}_{\hat{\mathfrak{M}}}$ supergauged LG orbifold with effective superpotential \eqref{eq:U(1|1)-N with W:phases:r1(i) and r2(i) < 0:superpotential} that is defined on $\mathcal{S}\mathcal{G}^{\mathrm{U}(1|1)^2}_{\text{IV}}$, where $\hat{\mathfrak{M}}$ is the gcd of the degrees $(\hat{m}_1, \hat{m}_2, \hat{n}_1, \hat{n}_2)$ of $G^{\pm}$, and $\mathcal{S}\mathcal{G}^{\mathrm{U}(1|1)^2}_{\text{IV}}$ is a product of a supervector bundle over $\WCP^{0|1}_{\hat{m}_1|\hat{n}_1}$ and another over $\WCP^{0|1}_{\hat{m}_2|\hat{n}_2}$, of sizes $\abs{r_{1, \text{eff}}(\mu)}$ and $\abs{r_{2, \text{eff}}(\mu)}$, respectively, described in \eqref{eq:U(1|1)-N with W:phases:r1(i) and r2(i) < 0:supervector bundle over wcp}.

  If we insist on being congruent with the NLSM to have $r_{\{1, 2\}, \text{eff}}(\mu) \ll 0$, from \eqref{eq:U(1|1)-N with W:quantum glsm:r-eff(mu):1}--\eqref{eq:U(1|1)-N with W:quantum glsm:r-eff(mu):2}, it would mean that $\mu \gg \Lambda_{\{1, 2\}}$ and therefore, $\Lambda_1 \ll \mu \ll \sqrt{\abs{r_1}}$ and $\Lambda_2 \ll \mu \ll \sqrt{\abs{r_2}}$.
  Then, in place of the classical low energy limit $e_{\{1, 2\}} \rightarrow \infty$, we would have the quantum low energy limit $\flatfrac{e_1}{\Lambda_1}, \flatfrac{e_2}{\Lambda_2} \rightarrow \infty$.

  \item $m_1 - n_1 < \hat{m}_1 - \hat{n}_1$ and $m_2 - n_2 > \hat{m}_2 - \hat{n}_2$
  ($\MSUSY^{r_{1, \text{eff}}(\text{ii}) < 0 < r_{2, \text{eff}}(\text{ii})}$)

  In this case, as explained above, we have $r_1 \ll 0$ and $r_{1, \text{eff}}(\mu) < 0$, as well as $r_2 \gg 0$ and $r_{2, \text{eff}}(\mu) > 0$.
  Also, a $\mathrm{U}(1|1)_1 \times \mathrm{U}(1|1)_2$ generalization of our discussion in \autoref{subsec: Quantum Low Energy U(1|1)} on the Higgs mechanism tells us that the quantum low energy limit exists at a scale $\mu \ll e_1 \sqrt{r_1}$ and $\mu \ll e_2 \sqrt{r_2}$.

  Thus, since $r_{1, \text{eff}}(\mu) < 0 < r_{2, \text{eff}}(\mu)$, at low energy, the quantum GLSM with $\MSUSY^{r_{1, \text{eff}}(\text{ii}) < 0 < r_{2, \text{eff}}(\text{ii})}$ defined by \eqref{eq:U(1|1)-N with W:phases:r1(i) < 0 < r2(i):space 2}--\eqref{eq:U(1|1)-N with W:phases:r1(i) < 0 < r2(i):space 1}, would become a hybrid between
  \begin{enumerate*}[label=(\Alph*)]
    \item a quantum $\mathrm{SU}(1|1)_1 \times \mathbb{Z}_{\hat{\mathfrak{m}}_1}$ supergauged LG orbifold with effective superpotential \eqref{eq:U(1|1)-N with W:phases:r1(i) < 0 < r2(i):superpotential} that is defined on a supervector bundle $\mathcal{S}\mathcal{G}^{\mathrm{U}(1|1)^2}_{\text{II}}$, where $\hat{\mathfrak{m}}_1$ is the gcd of the degrees $\hat{m}_1$ and $\hat{n}_1$ of $G^+$ and $G^-$, respectively, and $\mathcal{S}\mathcal{G}^{\mathrm{U}(1|1)^2}_{\text{II}}$ is a supervector bundle over $\WCP^{0|1}_{\hat{m}_1|\hat{n}_1} \times Gr_{1|1}(\mathbb{C}^{m_2|n_2})$ as described in \eqref{eq:U(1|1)-N with W:phases:r1(i) < 0 < r2(i):space 2:supervector bundle};

    and

    \item a quantum NLSM on $Gr_{1|1}(\mathbb{C}^{m_2|n_2})$ of size $r_2 - \frac{\hat{m}_2}{\hat{m}_1} r_1$ as described in \eqref{eq:U(1|1)-N with W:phases:r1(i) < 0 < r2(i):space 1}.

  \end{enumerate*}

  If we insist on being congruent with the NLSM and the supergauged LG orbifold in the previous cases to have $r_{1, \text{eff}}(\mu) \ll 0 \ll r_{2, \text{eff}}(\mu)$, from \eqref{eq:U(1|1)-N with W:quantum glsm:r-eff(mu):1}--\eqref{eq:U(1|1)-N with W:quantum glsm:r-eff(mu):2}, it would mean that $\mu \gg \Lambda_{\{1, 2\}}$ and therefore, $\Lambda_1 \ll \mu \ll \sqrt{\abs{r_1}}$ and $\Lambda_2 \ll \mu \ll \sqrt{r_2}$.
  Then, in place of the classical low energy limit $e_{\{1, 2\}}, b^{\pm}_2, c^{\pm}_2 \rightarrow \infty$, we would have the quantum low energy limit $\flatfrac{e_1}{\Lambda_1}, \flatfrac{e_2}{\Lambda_2}, \flatfrac{b^{\pm}_2}{\Lambda_2}, \flatfrac{c^{\pm}_2}{\Lambda_2} \rightarrow \infty$.

  \item $m_1 - n_1 < \hat{m}_1 - \hat{n}_1$ and $m_2 - n_2 < \hat{m}_2 - \hat{n}_2$ but $\mho > 0$ ($\MSUSY^{r_{1, \text{eff}}(\text{ii}), r_{2, \text{eff}}(\text{ii}) < 0}$)

  This case is identical to the previous one.
  For brevity, we shall not elaborate on it.

  \item $m_1 - n_1 > \hat{m}_1 - \hat{n}_1$ and $m_2 - n_2 < \hat{m}_2 - \hat{n}_2$ ($\MSUSY^{r_{1, \text{eff}}(\text{iii}) > 0 > r_{2, \text{eff}}(\text{iii})}$)

  This case is identical to the previous two cases up to the swops $(m_1, n_1, \hat{m}_1, \hat{n}_1) \leftrightarrow (m_2, n_2, \hat{m}_2, \hat{n}_2)$, $(r_1, \Lambda_1, e_1, b^{\pm}_1, c^{\pm}_1) \leftrightarrow (r_2, \Lambda_2, e_2, b^{\pm}_2, c^{\pm}_2)$, and $r_{1, \text{eff}}(\mu) \leftrightarrow r_{2, \text{eff}}(\mu)$.
  For brevity, we shall not elaborate on it.

  \item $m_1 - n_1 < \hat{m}_1 - \hat{n}_1$ and $m_2 - n_2 < \hat{m}_2 - \hat{n}_2$ but $\mho < 0$ ($\MSUSY^{r_{1, \text{eff}}(\text{iii}), r_{2, \text{eff}}(\text{iii}) < 0}$)

  This case is identical to the previous one.
  For brevity, we shall not elaborate on it.

  \item $m_1 - n_1 = \hat{m}_1 - \hat{n}_1$ and $m_2 - n_2 = \hat{m}_2 - \hat{n}_2$ ($\MSUSY^{r_{1, \text{eff}}(\text{i}), r_{2, \text{eff}}(\text{i}) > 0}$ / $\MSUSY^{r_{1, \text{eff}}(\text{iv}), r_{2, \text{eff}}(\text{iv}) < 0}$ / $\MSUSY^{r_{1, \text{eff}}(\text{ii}) < 0 < r_{2, \text{eff}}(\text{ii})}$ / $\MSUSY^{r_{1, \text{eff}}(\text{ii}), r_{2, \text{eff}}(\text{ii}) < 0}$ / $\MSUSY^{r_{1, \text{eff}}(\text{iii}) > 0 > r_{2, \text{eff}}(\text{iii})}$ / $\MSUSY^{r_{1, \text{eff}}(\text{iii}), r_{2, \text{eff}}(\text{iii}) < 0}$)

  In this case, one can see from \eqref{eq:U(1|1)-N with W:quantum glsm:r-eff:1}--\eqref{eq:U(1|1)-N with W:quantum glsm:r-eff:2} that the FI parameters do not run, so $r_k$ for $k \in \{1, 2\}$ are still dimensionless parameters of the quantum theory, whence the quantum GLSM is scale-invariant.
  As such, we can \emph{freely} choose $r_k > 0$ or $r_k < 0$, where at low energy, the quantum GLSM with the above-listed $\MSUSY$'s becomes the aforementioned quantum NLSM, quantum supergauged LG orbifold, or their hybrids.
  Since there are no dynamical scales $\Lambda_k$, the different quantum low energy limits would instead be given by (some or all of) $e_k \sqrt{r_k}, b^{\pm}_k \sqrt{r_k}, c^{\pm}_k \sqrt{r_k} \rightarrow \infty$.

\end{itemize}

The seven possible cases for when we are in Branch (2), are just a parity-reversed mirror of the above, so we shall just state the results.

\begin{itemize}
  \item $m_1 - n_1 < \hat{m}_1 - \hat{n}_1$ and $m_2 - n_2 < \hat{m}_2 - \hat{n}_2$ ($\MSUSY^{r_{1, \text{eff}}(\text{i}), r_{2, \text{eff}}(\text{i}) < 0}$)

  In the low energy limit $\flatfrac{e_1}{\Lambda_1}, \flatfrac{e_2}{\Lambda_2}, \flatfrac{b^{\pm}_1}{\Lambda_1}, \flatfrac{b^{\pm}_2}{\Lambda_2}, \flatfrac{c^{\pm}_1}{\Lambda_1}, \flatfrac{c^{\pm}_2}{\Lambda_2} \to \infty$, the quantum GLSM with $\MSUSY^{r_{1, \text{eff}}(\text{i}), r_{2, \text{eff}}(\text{i}) < 0}$ would become a quantum NLSM on $\MSUSY^{r_{1, \text{eff}}(\text{i}), r_{2, \text{eff}}(\text{i}) < 0}$, the complete intersection of an even and odd hypersurface $\{G^+_{\mathbf{\Pi}|\mathbf{\Pi}} = 0\}$ and $\{G^-_{\mathbf{\Pi}|\mathbf{\Pi}} = 0\}$ of bidegrees $(\hat{n}_1, \hat{n}_2)$ and $(\hat{m}_1, \hat{m}_2)$ in $Gr_{1|1}(\mathbb{C}^{n_1|m_1}) \times Gr_{1|1}(\mathbb{C}^{n_2|m_2})$ of sizes $\abs{r_{1,\text{eff}}(\mu)}$ and $\abs{r_{2, \text{eff}}(\mu)}$.

  \item $m_1 - n_1 > \hat{m}_1 - \hat{n}_1$ and $m_2 - n_2 > \hat{m}_2 - \hat{n}_2$ but $\mho = 0$ ($\MSUSY^{r_{1, \text{eff}}(\text{iv}), r_{2, \text{eff}}(\text{iv}) > 0}$)

  In the low energy limit $\flatfrac{e_1}{\Lambda_1}, \flatfrac{e_2}{\Lambda_2} \rightarrow \infty$, the quantum GLSM with $\MSUSY^{r_{1, \text{eff}}(\text{iv}), r_{2, \text{eff}}(\text{iv}) > 0}$ would become a quantum $\mathrm{U}(1|1) \times \mathrm{SU}(1|1) \times \mathbb{Z}_{\hat{\mathfrak{M}}}$ supergauged LG orbifold defined on the product $\mathcal{S}\mathcal{G}^{\mathrm{U}(1|1)^2}_{\text{IV}'}$ of a supervector bundle over $\WCP^{0|1}_{\hat{n}_1|\hat{m}_1}$ and another over $\WCP^{0|1}_{\hat{n}_2|\hat{m}_2}$, of sizes $r_{1, \text{eff}}(\mu)$ and $r_{2, \text{eff}}(\mu)$, respectively.

  \item $m_1 - n_1 > \hat{m}_1 - \hat{n}_1$ and $m_2 - n_2 < \hat{m}_2 - \hat{n}_2$
  ($\MSUSY^{r_{1, \text{eff}}(\text{ii}) > 0 > r_{2, \text{eff}}(\text{ii})}$)

  In the low energy limit $\flatfrac{e_1}{\Lambda_1}, \flatfrac{e_2}{\Lambda_2}, \flatfrac{b^{\pm}_2}{\Lambda_2}, \flatfrac{c^{\pm}_2}{\Lambda_2} \rightarrow \infty$, the quantum GLSM with $\MSUSY^{r_{1, \text{eff}}(\text{ii}) > 0 > r_{2, \text{eff}}(\text{ii})}$ would become a hybrid between
  \begin{enumerate*}[label=(\Alph*)]
    \item a quantum $\mathrm{SU}(1|1)_1 \times \mathbb{Z}_{\hat{\mathfrak{m}}_1}$ supergauged LG orbifold defined on $\mathcal{S}\mathcal{G}^{\mathrm{U}(1|1)^2}_{\text{II}'}$, where $\hat{\mathfrak{m}}_1$ is the gcd of the degrees $\hat{n}_1$ and $\hat{m}_1$ of $G^+_{\mathbf{\Pi}|\mathbf{\Pi}}$ and $G^-_{\mathbf{\Pi}|\mathbf{\Pi}}$, respectively, and $\mathcal{S}\mathcal{G}^{\mathrm{U}(1|1)^2}_{\text{II}'}$ is a supervector bundle over $\WCP^{0|1}_{\hat{n}_1|\hat{m}_1} \times Gr_{1|1}(\mathbb{C}^{n_2|m_2})$;

    and

    \item a quantum NLSM on $Gr_{1|1}(\mathbb{C}^{n_2|m_2})$ of size $r_1 - \frac{\hat{m}_1}{\hat{m}_2} r_2$.

  \end{enumerate*}

  \item $m_1 - n_1 > \hat{m}_1 - \hat{n}_1$ and $m_2 - n_2 > \hat{m}_2 - \hat{n}_2$ but $\mho < 0$ ($\MSUSY^{r_{1, \text{eff}}(\text{ii}), r_{2, \text{eff}}(\text{ii}) > 0}$)

  This case is identical to the previous one.
  For brevity, we shall not elaborate on it.

  \item $m_1 - n_1 < \hat{m}_1 - \hat{n}_1$ and $m_2 - n_2 > \hat{m}_2 - \hat{n}_2$ ($\MSUSY^{r_{1, \text{eff}}(\text{iii}) < 0 < r_{2, \text{eff}}(\text{iii})}$)

  This case is identical to the previous two cases up to the swops $(m_1, n_1, \hat{m}_1, \hat{n}_1) \leftrightarrow (m_2, n_2, \hat{m}_2, \hat{n}_2)$, $(r_1, \Lambda_1, e_1, b^{\pm}_1, c^{\pm}_1) \leftrightarrow (r_2, \Lambda_2, e_2, b^{\pm}_2, c^{\pm}_2)$, and $r_{1, \text{eff}}(\mu) \leftrightarrow r_{2, \text{eff}}(\mu)$.
  For brevity, we shall not elaborate on it.

  \item $m_1 - n_1 > \hat{m}_1 - \hat{n}_1$ and $m_2 - n_2 > \hat{m}_2 - \hat{n}_2$ but $\mho > 0$ ($\MSUSY^{r_{1, \text{eff}}(\text{iii}), r_{2, \text{eff}}(\text{iii}) > 0}$)

  This case is identical to the previous one.
  For brevity, we shall not elaborate on it.

  \item $m_1 - n_1 = \hat{m}_1 - \hat{n}_1$ and $m_2 - n_2 = \hat{m}_2 - \hat{n}_2$ ($\MSUSY^{r_{1, \text{eff}}(\text{i}), r_{2, \text{eff}}(\text{i}) < 0}$ / $\MSUSY^{r_{1, \text{eff}}(\text{iv}), r_{2, \text{eff}}(\text{iv}) > 0}$ / $\MSUSY^{r_{1, \text{eff}}(\text{ii}) > 0 > r_{2, \text{eff}}(\text{ii})}$ / $\MSUSY^{r_{1, \text{eff}}(\text{ii}), r_{2, \text{eff}}(\text{ii}) > 0}$ / $\MSUSY^{r_{1, \text{eff}}(\text{iii}) < 0 < r_{2, \text{eff}}(\text{iii})}$ / $\MSUSY^{r_{1, \text{eff}}(\text{iii}), r_{2, \text{eff}}(\text{iii}) > 0}$)

  In this case, we can \emph{freely} choose $r_k > 0$ or $r_k < 0$ (for $k \in \{1, 2\}$), where in the low energy limit $e_k \sqrt{r_k}$ and/or $b^{\pm}_k \sqrt{r_k}$ and/or $c^{\pm}_k \sqrt{r_k} \rightarrow \infty$, the quantum GLSM with the above-listed $\MSUSY$'s becomes the aforementioned quantum NLSM, quantum supergauged LG orbifold, or their hybrids.
\end{itemize}

\subsection{Scale-invariance and Applications to Mathematics}
\label{sec:U(1|1)-N with W:scale-invariance}

Let us consider the case where $m_1 - n_1 = \hat{m}_1 - \hat{n}_1$ and $m_2 - n_2 = \hat{m}_2 - \hat{n}_2$.
As explained in \autoref{subsec: Quantum low energy U(1|1)^2 with W}, the quantum GLSM would be scale-invariant with free parameters $r_1$ and $r_2$.

\subtitle{Scale-invariance and a Physical Derivation of a Novel CY Condition for a Complete Intersection in a Product of Super-Grassmannians}

As the quantum GLSM is scale-invariant, its corresponding low energy quantum NLSM ought to also be scale and therefore conformally-invariant \cite{papadopoulos2024scaleconformalinvariance2d}; in turn, according to our explanation in \autoref{subsec: applications to math U(1|1)}, this would mean that the target space of the quantum NLSMs in \autoref{subsec: Quantum low energy U(1|1)^2 with W}, i.e., $\MSUSY^{r_1(\text{i}), r_2(\text{i}) > 0}$ and $\MSUSY^{r_1(\text{i}), r_2(\text{i}) < 0}$, must be CY.

Recall that $\MSUSY^{r_1(\text{i}), r_2(\text{i}) > 0}$ is a complete intersection of an even and odd hypersurface $G^+ = 0$ and $G^- = 0$ of even and odd bidegrees $(\hat{m}_1, \hat{m}_2)$ and $(\hat{n}_1, \hat{n}_2)$ in the product of super-Grassmannians $Gr_{1|1}(\mathbb{C}^{m_1|n_1}) \times Gr_{1|1}(\mathbb{C}^{m_2|n_2})$ of sizes $r_1$ and $r_2$.
Therefore, since $m_1 - n_1 = \hat{m}_1 - \hat{n}_1$, the difference between the even and odd dimensions of the ambient space of the \emph{first} super-Grassmannian will match the difference between the even and odd degrees of the hypersurfaces in the the \emph{first} super-Grassmannian.
Likewise, since $m_2 - n_2 = \hat{m}_2 - \hat{n}_2$, the difference between the even and odd dimensions of the ambient space of the \emph{second} super-Grassmannian will match the difference between the  even and odd degrees of the hypersurfaces in the \emph{second} super-Grassmannian.

Recall also that $\MSUSY^{r_1(\text{i}), r_2(\text{i}) < 0}$ is a complete intersection of an even and odd hypersurface $G^+_{\mathbf{\Pi}|\mathbf{\Pi}} = 0$ and $G^-_{\mathbf{\Pi}|\mathbf{\Pi}} = 0$ of even and odd bidegrees $(\hat{n}_1, \hat{n}_2)$ and $(\hat{m}_1, \hat{m}_2)$ in the product of super-Grassmannians $Gr_{1|1}(\mathbb{C}^{n_1|m_1}) \times Gr_{1|1}(\mathbb{C}^{n_2|m_2})$ of sizes $\abs{r_1}$ and $\abs{r_2}$.
Therefore, since $m_1 - n_1 = \hat{m}_1 - \hat{n}_1$, or equivalently, $n_1 - m_1 = \hat{n}_1 - \hat{m}_1$, the difference between the even and odd dimensions of the ambient space of the \emph{first} super-Grassmannian will match the difference between the even and odd degrees of the hypersurfaces in the \emph{first} super-Grassmannian.
Likewise, since $m_2 - n_2 = \hat{m}_2 - \hat{n}_2$, or equivalently, $n_2 - m_2 = \hat{n}_2 - \hat{m}_2$, the difference between the even and odd dimensions of the ambient space of the \emph{second} super-Grassmannian will match the difference between the even and odd degrees of the hypersurfaces in the  \emph{second} super-Grassmannian.

Thus, we have a physical derivation of a \emph{novel} mathematical result that a complete intersection of an even and odd hypersurface in a product of two super-Grassmannians is CY, if and only if the difference between the even and odd dimensions of the ambient space of the first and second super-Grassmannians match the difference between the even and odd degrees of the hypersurfaces in the first and second super-Grassmannians, respectively!

Indeed, one can actually mathematically verify (see \autoref{app:verify cy:cicy in products of SG}) this statement to be true!

This once again vindicates our application to mathematics of our otherwise nonunitary supergroup GLSM via a study of its space of supersymmetric states.

\subsection{A Super-Grassmannian/Supergroup Generalization of the Hybrid CY/LG Correspondence for Hypersurfaces in Product Space in Mathematics}
\label{subsec: Super hybrid CY/LG for hypersurfaces}

Staying with the scale-invariant quantum GLSM where $m_k - n_k = \hat{m}_k - \hat{n}_k$ for $k \in \{1, 2\}$, we shall now derive a highly-nontrivial geometric/hybrid correspondence between a super CY and a hybrid NLSM/supergauged LG orbifold, and vice versa.

\subtitle{A Geometric/Hybrid Correspondence between a Super CY and a Hybrid NLSM$\big{/}$Supergauged LG Orbifold}

Since we can freely choose $r_{\{1, 2\}} > 0$ and $r_{\{1, 2\}} < 0$, it would mean that from the perspective of our quantum GLSM, there ought to be a physical correspondence between
\begin{enumerate*}[label=(\Roman*)]
  \item the quantum NLSM of \autoref{subsec: Quantum low energy U(1|1)^2 with W} with super CY target space $\MSUSY^{r_1(\text{i}), r_2(\text{i}) > 0}$ as described therein,

  and

  \item the hybrid quantum NLSM/$\mathrm{SU}(1|1)_1 \times \mathbb{Z}_{\hat{\mathfrak{m}}_1}$ supergauged LG orbifold of \autoref{subsec: Quantum low energy U(1|1)^2 with W} that is defined on $\mathcal{S}\mathcal{G}^{\mathrm{U}(1|1)^2}_{\text{II}}$ as described therein.

\end{enumerate*}

Specifically, we ought to have a geometric/hybrid correspondence between

\begin{enumerate}[label=(\Roman*)]
  \item a quantum NLSM with super CY target space being a complete intersection of an even and odd hypersurface $G^+ = 0$ and $G^- = 0$ of even and odd bidegrees $(\hat{m}_1, \hat{m}_2)$ and $(\hat{n}_1, \hat{n}_2)$ in the product of super-Grassmannians $Gr_{1|1}(\mathbb{C}^{m_1|n_1}) \times Gr_{1|1}(\mathbb{C}^{m_2|n_2})$ of sizes $r_1$ and $r_2$, where $m_1 - n_1 = \hat{m}_1 - \hat{n}_1$ and $m_2 - n_2 = \hat{m}_2 - \hat{n}_2$;

  and

  \item a hybrid between a quantum NLSM on a super-Grassmannian $Gr_{1|1}(\mathbb{C}^{m_2|n_2})$ of size $r_2 - \frac{\hat{m}_2}{\hat{m}_1} r_1$ and a quantum $\mathrm{SU}(1|1)_1 \times \mathbb{Z}_{\hat{\mathfrak{m}}_1}$ supergauged LG orbifold with effective superpotential \eqref{eq:U(1|1)-N with W:phases:r1(i) < 0 < r2(i):superpotential} that is defined on $\mathcal{S}\mathcal{G}^{\mathrm{U}(1|1)^2}_{\text{II}}$, where $\hat{\mathfrak{m}}_1 \coloneq \gcd(\hat{m}_1, \hat{n}_1)$, and $\mathcal{S}\mathcal{G}^{\mathrm{U}(1|1)^2}_{\text{II}}$ is a supervector bundle over $\WCP^{0|1}_{\hat{m}_1|\hat{n}_1} \times Gr_{1|1}(\mathbb{C}^{m_2|n_2})$ of sizes $\abs{r_1}$ and $r_2 - \frac{\hat{m}_2}{\hat{m}_1} r_1$ described in \eqref{eq:U(1|1)-N with W:phases:r1(i) < 0 < r2(i):space 2:supervector bundle}, with $m_1 - n_1 = \hat{m}_1 - \hat{n}_1$ and $m_2 - n_2 = \hat{m}_2 - \hat{n}_2$!
\end{enumerate}

\subtitle{A Hybrid/Geometric Correspondence Between a Hybrid NLSM$\big{/}$Supergauged LG Orbifold and a Super CY}

Similarly, it would mean that from the perspective of our quantum GLSM, there ought to be a physical correspondence between
\begin{enumerate*}[label=(\Roman*$'$)]
  \setcounter{enumi}{1}

  \item the hybrid quantum NLSM/$\mathrm{SU}(1|1)_1 \times \mathbb{Z}_{\hat{\mathfrak{m}}_1}$ supergauged LG orbifold of \autoref{subsec: Quantum low energy U(1|1)^2 with W} that is defined on $\mathcal{S}\mathcal{G}^{\mathrm{U}(1|1)^2}_{\text{II}'}$ as described therein,

  and%
  \addtocounter{enumi}{-2}%

  \item the quantum NLSM of \autoref{subsec: Quantum low energy U(1|1)^2 with W} with super CY target space $\MSUSY^{r_1(\text{i}), r_2(\text{i}) < 0}$ as described therein.

\end{enumerate*}

Specifically, we ought to have a hybrid/geometric correspondence between

\begin{enumerate}[label=(\Roman*$'$)]
  \setcounter{enumi}{1}

  \item a hybrid between a quantum NLSM on a super-Grassmannian $Gr_{1|1}(\mathbb{C}^{n_2|m_2})$ of size $r_1 - \frac{\hat{m}_1}{\hat{m}_2} r_2$ and a quantum $\mathrm{SU}(1|1)_1 \times \mathbb{Z}_{\hat{\mathfrak{m}}_1}$ supergauged LG orbifold that is defined on $\mathcal{S}\mathcal{G}^{\mathrm{U}(1|1)^2}_{\text{II}'}$, where $\hat{\mathfrak{m}}_1 \coloneq \gcd(\hat{m}_1, \hat{n}_1)$, and $\mathcal{S}\mathcal{G}^{\mathrm{U}(1|1)^2}_{\text{II}'}$ is a supervector bundle over $\WCP^{0|1}_{\hat{n}_1|\hat{m}_1} \times Gr_{1|1}(\mathbb{C}^{n_2|m_2})$ of sizes $\abs{r_2}$ and $r_1 - \frac{\hat{m}_1}{\hat{m}_2} r_2$, with $m_1 - n_1 = \hat{m}_1 - \hat{n}_1$ and $m_2 - n_2 = \hat{m}_2 - \hat{n}_2$;

  and
  \addtocounter{enumi}{-2}

  \item a quantum NLSM with super CY target space being a complete intersection of an even and odd hypersurface $G^+_{\mathbf{\Pi}|\mathbf{\Pi}} = 0$ and $G^-_{\mathbf{\Pi}|\mathbf{\Pi}} = 0$ of even and odd bidegrees $(\hat{n}_1, \hat{n}_2)$ and $(\hat{m}_1, \hat{m}_2)$ in the product of super-Grassmannians $Gr_{1|1}(\mathbb{C}^{n_1|m_1}) \times Gr_{1|1}(\mathbb{C}^{n_2|m_2})$ of sizes $\abs{r_1}$ and $\abs{r_2}$, where $m_1 - n_1 = \hat{m}_1 - \hat{n}_1$ and $m_2 - n_2 = \hat{m}_2 - \hat{n}_2$!
\end{enumerate}

Notice that this hybrid/geometric correspondence between (II$'$) and (I$'$) is just a parity-reversed mirror version of the geometric/hybrid correspondence between (I) and (II)!

\subtitle{A Super-Grassmannian/Supergroup Generalization of the Hybrid CY/LG Correspondence for Hypersurfaces in Product Space in Mathematics}

At any rate, in carrying out the reduction to the familiar $\mathrm{U}(1)$ GLSM with superpotential at low energy (via \autoref{subsec: Phases of the reduced U(1)^2 model}) whence the scale-invariant condition is now $m_1 = \hat{m}_1$ and $m_2 = \hat{m}_2$, we find that in place of (I) and (II), we would have
\begin{enumerate*}[label=(\textbf{\roman*})]
  \item a quantum NLSM with CY target space being a hypersurface $G^+(s^1, t^1)$ in $\mathbb{CP}^{m_1 - 1} \times \mathbb{CP}^{m_2 - 1}$ of sizes $r_1$ and $r_2$,\footnote{%
    \label{ft:U(1|1)-N with W:CY-LG:hypersurface is CY}%
    The hypersurface is CY because $G (s^1, t^1)$ is of bidegree $(m_1, m_2)$ in $\mathbb{CP}^{m_1 - 1} \times \mathbb{CP}^{m_2 - 1}$, and it is a well-known fact that a smooth hypersurface of degree $d$ in $\mathbb{CP}^{d-1}$ is CY.
  }

  and

  \item a hybrid that, in the $t$-direction, is a quantum NLSM whose target space is the projective superspace $\mathbb{CP}^{m_2 - 1}$ of size $r_2 - \frac{\hat{m}_2}{\hat{m}_1} r_1$, and in the $s$-direction, is a quantum LG $\mathbb{Z}_{\hat{M}}$-orbifold with effective superpotential $W^{\mathbb{Z}_{\hat{m}_1}}_{\text{LG,II}}$ \eqref{eq:U(1|1)-N with W:reduction:W^Z_n:II} defined on a vector bundle $\mathcal{S}^{\mathrm{U}(1)^2}_{\text{II}}$ in \eqref{eq:U(1|1)-N with W:reduction:vector bundle:II}.
\end{enumerate*}

Note that the celebrated hybrid CY/LG correspondence for hypersurfaces in product space is exactly the correspondence between the low energy models just described in (\textbf{i}) and (\textbf{ii}) above.

Thus, in deriving the above geometric/hybrid correspondence between (I) and (II), we have a super-Grassmannian/supergroup generalization of the hybrid CY/LG correspondence for hypersurfaces in product space established in mathematics by Fan-Jarvis-Ruan~\cite[$\S$7.3]{Fan_2017}!

In fact, in deriving the above hybrid/geometric correspondence between (II$'$) and (I$'$), we also have a parity-reversed mirror version of this generalization!

\section{Phase Transition of a Certain \texorpdfstring{$\mathrm{U}(1|1)$}{U(1|1)} GLSM: Mild Topology Change of a Supervector Bundle over a Super-Grassmannian}
\label{sec:U(1|1) phase transition}

In this section, we will generalize the $N = 1$ setting of \autoref{sec: U(1|1)^N GLSM} by studying a $\mathrm{U}(1|1)$ GLSM where the non-identical gauge charges of the Grassmann-even and Grassmann-odd chiral matter superfields may be negative.
Repeating the analysis in \autoref{sec: U(1|1) and SGr}, we will show that the space of supersymmetric states of the GLSM corresponds to a supervector bundle over a ``weighted super-Grassmannian'', and that in the low energy limit, the GLSM will become an NLSM on the aforementioned supervector bundle.
Taking into account the quantum corrections to the FI parameter, we will have an effective scale-dependent FI parameter in the quantum GLSM.
The quantum GLSM will also become a quantum NLSM in the low energy limit.
When the theory is scale-invariant, we will be able to
\begin{enumerate*}[label=(\arabic*)]
  \item physically derive a CY condition for supervector bundles over ``weighted super-Grassmannians'' that we can also mathematically verify,

  \item explain how worldsheet superinstanton effects would ``shield'' the NLSM from a singularity when the FI parameter passes through zero,

  and

  \item obtain a super-Grassmannian generalization of a birational equivalence of CYs in mathematics.

\end{enumerate*}

\subsection{A Certain \texorpdfstring{$\mathrm{U}(1|1)$}{U(1|1)} GLSM and its \texorpdfstring{\MSUSY}{M-SUSY}}
\label{sec:U(1|1) phase transition:glsm}

\subtitle{A Certain $\mathrm{U}(1|1)$ GLSM}

Let us now consider a $\mathrm{U}(1|1)$  GLSM with $m + n$ chiral matter superfields $\Phi_1,...,\Phi_m$ and $\Psi_1, \dots, \Psi_n$, where the former and latter are Grassmann-even and Grassmann-odd, whose integer gauge charges are $Q_1,...,Q_{l_m}, Q_{l_m+1} \dots Q_{m}$ and $\mathsf{Q}_1,...,\mathsf{Q}_{l_n}, \mathsf{Q}_{l_n+1} \dots \mathsf{Q}_{n}$, respectively.

Its $\mathcal{N} = (2,2)$ Lagrangian ought to be given by
\begin{equation}
  \label{eq:U(1|1) phase transition:glsm:lagrangian}
  \begin{aligned}
    L^{\mathrm{U}(1|1)}_{Q_s, \mathsf{Q}_t}
    &=
    \frac{1}{4} \int \dd[4]{\theta}
    \left( \sum_{s=1}^m \bar{\Phi}^s e^{2Q_sV} \Phi_s
      + \sum_{t=1}^n \bar{\Psi}^t e^{2\mathsf{Q}_tV} \Psi_t \right)
    - \frac{1}{4e^2} \int \dd[4]{\theta} \Str \bar{\Sigma} \Sigma
    \\
    &\quad
    + \left(
      \frac{1}{2\sqrt{2}} \int \dd[2]{\theta} (\mathbf{i} t \Str \Sigma)
      + c.c. \
    \right)
    \, .
  \end{aligned}
\end{equation}

\subtitle{Potential Energy}

After eliminating the auxiliary field, the potential energy (of the scalars) becomes
\begin{equation}
 \label{eq:U(1|1) phase transition:glsm:potential}
  \begin{aligned}
    U^{\mathrm{U}(1|1)}_{\text{pot}, Q_s, \mathsf{Q}_t}
    &= \frac{e^2}{2} \sum_{i, j}
    (-1)^{\varsigma(i)}
    \Bigg[
      \sum_{s' = 1}^{l_m} Q_{s'} \tensor{\phi}{^j_{s'}} \tensor{\bar{\phi}}{_i^{s'}}
      + \sum_{s'' = l_m + 1}^m Q_{s''} \tensor{\phi}{^j_{s''}} \tensor{\bar{\phi}}{_i^{s''}}
      \\
      &\qquad \qquad \qquad \qquad
      - \sum_{t' = 1}^{l_n} \mathsf{Q}_{t'} \tensor{\psi}{^j_{t'}} \tensor{\bar{\psi}}{_i^{t'}}
      - \sum_{t'' = l_n +1}^n \mathsf{Q}_{t''} \tensor{\psi}{^j_{t''}} \tensor{\bar{\psi}}{_i^{t''}}
      - r \tensor{\delta}{^j_i}
    \Bigg]^2
    \\
    &\quad
    + \frac{1}{2e^2} \Str \comm{\sigma}{\bar{\sigma}}^2
    + \sum_{s' = 1}^{l_m} (Q_{s'})^2\bar{\phi}^{s'} \acomm{\bar{\sigma}}{\sigma} \phi_{s'}
    + \sum_{t' = 1}^{l_n} (\mathsf{Q}_{t'})^2\bar{\psi}^{t'} \acomm{\bar{\sigma}}{\sigma} \psi_{t'}
    \\
    &\quad
    + \sum_{s'' = l_m + 1}^m (Q_{s''})^2 \bar{\phi}^{s''} \acomm{\bar{\sigma}}{\sigma} \phi_{s''}
    + \sum_{t'' = l_n + 1}^n (\mathsf{Q}_{t''})^2 \bar{\psi}^{t''} \acomm{\bar{\sigma}}{\sigma} \psi_{t''}
    \, .
  \end{aligned}
\end{equation}

\subtitle{\MSUSY}

$\MSUSY$ of the theory is defined by the supersymmetric configuration $U^{\mathrm{U}(1|1)}_{\text{pot}, Q_s, \mathsf{Q}_t} = 0$ modulo gauge transformations.
From \eqref{eq:U(1|1) phase transition:glsm:potential}, we see that for $\abs{r} > 0$, we require $\phi \neq 0$, $\psi \neq 0$, and $\sigma = 0$.
In other words, if $Q_1,...,Q_{l_m} > 0 > Q_{l_m+1} \dots Q_{m}$ and $\mathsf{Q}_1,...,\mathsf{Q}_{l_n} > 0 > \mathsf{Q}_{l_n+1} \dots \mathsf{Q}_{n}$, $\MSUSY$ will be defined by
\begin{equation}
  \label{eq:U(1|1) phase transition:glsm:M-SUSY}
  \saveboxed{eq:U(1|1) phase transition:glsm:M-SUSY}{
    \mathcal{S}^{\mathrm{U}(1|1)}_{Q,\mathsf{Q}, r}
    = \flatfrac{\left\{
        \begin{gathered}
          \sum_{s' = 1}^{l_m} Q_{s'} \tensor{\phi}{^j_{s'}} \tensor{\bar{\phi}}{_i^{s'}}
          - \sum_{t' = 1}^{l_n} \mathsf{Q}_{t'} \tensor{\psi}{^j_{t'}} \tensor{\bar{\psi}}{_i^{t'}}
          \\
          = r \tensor{\delta}{^j_i}
          + \sum_{s'' = l_m +1}^m \abs{Q_{s''}} \tensor{\phi}{^j_{s''}} \tensor{\bar{\phi}}{_i^{s''}}
          - \sum_{t'' = l_n +1}^n \abs{\mathsf{Q}_{t''}} \tensor{\psi}{^j_{t''}} \tensor{\bar{\psi}}{_i^{t''}}
        \end{gathered}
        \, \qcomma
        i,j \in \{1, \bar{1}\}
      \right\}}
    {\mathrm{U}(1|1)}
  }
\end{equation}
where the subscripts ``$Q$'' and ``$\mathsf{Q}$'' in $\mathcal{S}^{\mathrm{U}(1|1)}_{Q, \mathsf{Q}, r}$ are
\begin{equation}
  \label{eq:U(1|1) phase transition:glsm:Q}
  \saveboxed{eq:U(1|1) phase transition:glsm:Q}{
    Q
    \coloneq \sum^m_{s = 1} Q_s
    = \sum^{l_m}_{s' = 1} Q_{s'}
    + \sum^m_{s'' = l_m + 1} Q_{s''}
  }
\end{equation}
and
\begin{equation}
  \label{eq:U(1|1) phase transition:glsm:Q-serif}
  \saveboxed{eq:U(1|1) phase transition:glsm:Q-serif}{
    \mathsf{Q}
    \coloneq \sum^n_{t = 1} \mathsf{Q}_t
    = \sum^{l_n}_{t' = 1} \mathsf{Q}_{t'}
    + \sum^n_{t'' = l_n + 1} \mathsf{Q}_{t''}
  }
\end{equation}

\subtitle{$\MSUSY$ when $r > 0$ and $r < 0$}

When $r > 0$, $\mathcal{S}^{\mathrm{U}(1|1)}_{Q,\mathsf{Q}, r > 0}$ in \eqref{eq:U(1|1) phase transition:glsm:M-SUSY} can be mathematically identified as (\emph{c.f.}~\cite[eq.~(15.94)]{Hori:2003ic})
\begin{equation}
  \label{eq:phase transition topology:cy condition:positive}
  \saveboxed{eq:phase transition topology:cy condition:positive}{
    \mathcal{S}^{\mathrm{U}(1|1)}_{Q, \mathsf{Q}, r > 0}
    = \bigoplus_{s'' = l_m + 1}^m \mathscr{L}_{s''}^{Q_{s''}}
    \oplus \bigoplus_{t'' = l_n + 1}^n \mathbf{\Pi} \mathscr{L}_{t''}^{\mathsf{Q}_{t''}}
    \rightarrow \text{W}Gr_{1|1} \left( \mathbb{C}^{l_m|l_n}_{(\vec{Q'}|\vec{\mathsf{Q}'})} \right)_{s',t'}
  }
\end{equation}
a supervector bundle over a ``weighted super-Grassmannian'' $\text{W}Gr_{1|1} \left( \mathbb{C}^{l_m|l_n}_{(\vec{Q'}|\vec{\mathsf{Q}'})} \right)_{s',t'}$.\footnote{%
  \label{ft:phase transition topology:weighted super-grassmannian}%
  Such a space is the super torified variety defined by \eqref{eq:u-1-1-N:m-susy}, but with $N = 1$ and $r_1 =r$, summed over $(Q_{s'}, \mathsf{Q}_{t'})$ instead.
  It is a ``weighted super-Grassmannian'' in the sense that it reduces to a super-Grassmannian of size $r$ when the weights become 1, i.e., $(\vec{Q'}|\vec{\mathsf{Q}'}) = (\vec{1}|\vec{1})$.
}
Here,
\begin{enumerate*}

  \item the even $\mathscr{L}_{s''}$ and odd $\mathbf{\Pi} \mathscr{L}_{t''}$ fibers are spanned by the Grassmann-even $\phi_{s''}$ and Grassmann-odd $\psi_{t''}$ supervectors,

  \item the coordinates on the $\text{W}Gr_{1|1} \left( \mathbb{C}^{l_m|l_n}_{(\vec{Q'}|\vec{\mathsf{Q}'})} \right)_{s',t'}$ base are the components of the Grassmann-even $\phi_{s'}$ and Grassmann-odd $\psi_{t'}$ supervectors, and

  \item $(\vec{Q'}|\vec{\mathsf{Q}'}) \coloneq (Q_1, \dots, Q_{l_m}|\mathsf{Q}_1, \dots, \mathsf{Q}_{l_n})$ are positive weights.

\end{enumerate*}

When $r < 0$, $\mathcal{S}^{\mathrm{U}(1|1)}_{Q, \mathsf{Q}, r < 0}$ in \eqref{eq:U(1|1) phase transition:glsm:M-SUSY} can be mathematically identified as (\emph{c.f.}~\cite[eq.~(15.95)]{Hori:2003ic})
\begin{equation}
  \label{eq:phase transition topology:cy condition:negative}
  \saveboxed{eq:phase transition topology:cy condition:negative}{
    \mathcal{S}^{\mathrm{U}(1|1)}_{Q, \mathsf{Q}, r < 0}
    = \bigoplus_{s' = 1}^{l_m} \mathscr{L}_{s'}^{- Q_{s'}}
    \oplus \bigoplus_{t' = 1}^{l_n} \mathbf{\Pi} \mathscr{L}_{t'}^{- \mathsf{Q}_{t'}}
    \rightarrow \text{W} Gr_{1|1} \left( \mathbb{C}^{m-l_m|n-l_n}_{(\vec{\abs{Q''}}|\vec{\abs{\mathsf{Q}''}})} \right)_{s'',t''}
  }
\end{equation}
a supervector bundle over a ``weighted super-Grassmannian'' $\text{W} Gr_{1|1} \big( \mathbb{C}^{m-l_m|n-l_n}_{(\vec{\abs{Q''}}|\vec{\abs{\mathsf{Q}''}})} \big)_{s'',t''}$.
Here,
\begin{enumerate*}

  \item the even $\mathscr{L}_{s'}$ and odd $\mathbf{\Pi}\mathscr{L}_{t'}$ fibers are spanned by the Grassmann-even $\phi_{s'}$ and Grassmann-odd $\psi_{t'}$ supervectors,

  \item the coordinates on the $\text{W} Gr_{1|1} \big( \mathbb{C}^{m-l_m|n-l_n}_{(\vec{\abs{Q''}}|\vec{\abs{\mathsf{Q}''}})} \big)_{s'',t''}$ base are the components of the Grassmann-even $\phi_{s''}$ and Grassmann-odd $\psi_{t''}$ supervectors, and

  \item $(\vec{\abs{Q''}}|\vec{\abs{\mathsf{Q}''}}) \coloneq (\abs{Q_{l_m + 1}}, \dots, \\ \abs{Q_m}|\abs{\mathsf{Q}_{l_n + 1}},  \dots, \abs{\mathsf{Q}_n})$ are positive weights.

\end{enumerate*}

\subtitle{A Reduction to Garavuso-Katzarkov-Kreuzer-Noll's $\mathrm{U}(1)$ GLSM}

Note that the $\mathrm{U}(1|1)$ GLSM  ought to reduce to Garavuso-Katzarkov-Kreuzer-Noll's $\mathrm{U}(1)$ GLSM when we
\begin{enumerate*}
  \item turn off the $\tensor{\phi}{^{\bar{1}}_s}$ and $\tensor{\psi}{^{\bar{1}}_t}$ fields (and their corresponding bar-indexed partners in the underlying multiplets),

  and

  \item set $l_m = m - 1$ and $l_n = n$ (i.e., only a single $\Phi$ chiral matter superfield has a negative integer gauge charge, while all other chiral matter superfields have positive integer gauge charges).

\end{enumerate*}
Indeed, when $r > 0$, point (i) would mean that the base of $\mathcal{S}^{\mathrm{U}(1|1)}_{Q, \mathsf{Q}, r}$ in \eqref{eq:phase transition topology:cy condition:positive} reduces to $\WCP^{m-2|n}_{(\vec{Q'}|\vec{\mathsf{Q}'})}$, and point (ii) would mean that when $r > 0$, we would only have a \emph{single} even $\mathrm{U}(1)$ line fiber over the base.

In other words, when $r > 0$, $\MSUSY$ would reduce to a line bundle over $\WCP^{m-2|n}_{(\vec{Q'}|\vec{\mathsf{Q}'})}$, as it should (\emph{c.f.}~\cite[$\S$3]{garavuso-2011-super-landau}).

\subtitle{A Reduction to the Familiar $\mathrm{U}(1)$ GLSM}

Note that the $\mathrm{U}(1|1)$ GLSM ought to reduce to the familiar $\mathrm{U}(1)$ GLSM when we turn off
\begin{enumerate*}
  \item the $\tensor{\phi}{^{\bar{1}}_s}$ field (and its corresponding bar-index partners in the underlying multiplet),

  and

  \item the \emph{whole} Grassmann-odd chiral matter superfield $\Psi$.
\end{enumerate*}
Indeed, with only Grassmann-even $\tensor{\phi}{^1_s}$ fields, the base of $\MSUSY$ would reduce to a weighted projective space, and the fiber space over the base would be a sum of even $\mathrm{U}(1)$ line fibers.

In other words, $\MSUSY$ would reduce to a vector bundle over a weighted projective space, as it should (\emph{c.f.}~\cite[$\S$15.4.2]{Hori:2003ic}).
We will encounter this reduced model again in \autoref{sec:U(1|1) phase transition:topological change:V over WCP}.

\subtitle{The Low Energy NLSM on $\mathcal{S}^{\mathrm{U}(1|1)}_{Q, \mathsf{Q}, r}$}

In the low energy limit $e \to \infty$, following the steps in \autoref{GLSM to NLSM}, we would get an NLSM on $\mathcal{S}^{\mathrm{U}(1|1)}_{Q, \mathsf{Q}, r}$.

\subsection{Quantum Aspects: The GLSM}
\label{sec:U(1|1) phase transition:quantum glsm}

Comparing the Lagrangian \eqref{eq:U(1|1) phase transition:glsm:lagrangian} with the Lagrangian \eqref{eq:u-1-1:lagrangian}, one can see that the present GLSM is of the same type as that in \autoref{sec: U(1|1) and SGr}, where in the latter, the charges are $Q_1 = Q_2 = \dots = Q_m = 1$ and $\mathsf{Q}_1 = \mathsf{Q}_2 = \dots = \mathsf{Q}_n = 1$, whence from \eqref{eq:U(1|1) phase transition:glsm:Q}--\eqref{eq:U(1|1) phase transition:glsm:Q-serif}, we have $Q = m$ and $\mathsf{Q} = n$, respectively.
As such, the quantum aspects of the GLSM would be as discussed in \autoref{subsec: Quantum Prop U(1|1) GLSM}, except that in place of $m$ and $n$ therein, we have $Q$ and $\mathsf{Q}$ here which need not be equal to $m$ and $n$.

\subtitle{The Effective FI Parameter $r_{\text{eff}}(\mu)$}

In particular, the effective FI parameter at an energy scale $\mu$ would be given by
\begin{equation}
  \label{eq:U(1|1) phase transition:quantum GLSM:r-eff(mu)}
  \saveboxed{eq:U(1|1) phase transition:quantum GLSM:r-eff(mu)}{
    r_{\text{eff}} (\mu)
    =  \big(Q - \mathsf{Q} \big) \log \left( \frac{\mu}{\Lambda} \right)
  }
\end{equation}
where the dimensionless parameter $r$ of the classical theory has now been replaced by the scale parameter $\Lambda$ of mass dimension in the quantum theory.

Thus, depending on whether $Q - \mathsf{Q}$ is zero or nonzero, we have an effective FI parameter $r_{\text{eff}} (\mu)$ that does not run or runs (i.e., RG flows) as we change the energy scale $\mu$.

In short, the quantum GLSM may or may not be scale-invariant, where $r$ may or may no longer be a parameter of the theory.

\subsection{Quantum Aspects: The Low Energy Theory}
\label{sec:U(1|1) phase transition:low energy}

\subtitle{$\MSUSY$ of the Quantum GLSM at Low Energy}

Recall that in our derivation of $\MSUSY$ at the classical level in \autoref{sec:U(1|1) phase transition:glsm}, $|r| \neq 0$.
In the quantum theory, when $Q \neq \mathsf{Q}$, in place of $r$, we would have $r_{\text{eff}}(\mu)$.

From \eqref{eq:u-1-1:quantum:r-eff}, with $m$ and $n$ therein replaced with $Q$ and $\mathsf{Q}$, notice that  even at low energy $\mu \ll \Lambda_{\text{UV}}$, if for (i) $Q - \mathsf{Q} > 0$ or (ii) $Q - \mathsf{Q} < 0$, we consider (i) $r \gg 0$ or (ii) $r \ll 0$, we will always have (i)   $r_{\text{eff}} (\mu) > 0$ or (ii)  $r_{\text{eff}} (\mu) < 0$.
That is, $\abs{r_{\text{eff}} (\mu)} > 0$, whence $\MSUSY$ would continue to be well-defined in the quantum GLSM at low energy.

\subtitle{The Quantum GLSM at Low Energy is a Quantum NLSM on $\mathcal{S}^{\mathrm{U}(1|1)}_{Q, \mathsf{Q}, r_{\text{eff}}(\mu) > 0 }$, $\mathcal{S}^{\mathrm{U}(1|1)}_{Q, \mathsf{Q}, r_{\text{eff}}(\mu)< 0}$, or Either}

The question again, is whether at the quantum level, the GLSM will also become the NLSM on $\mathcal{S}^{\mathrm{U}(1|1)}_{Q, \mathsf{Q}, r_{\text{eff}}(\mu) > 0 }$ or $\mathcal{S}^{\mathrm{U}(1|1)}_{Q, \mathsf{Q}, r_{\text{eff}}(\mu) < 0 }$ at low energy, as per the classical theory.
The short answer is ``yes'', although to the former, latter, or either, depends on the value of $Q - \mathsf{Q}$.

Let us now elaborate on the three possible cases.

\begin{itemize}
  \item $Q > \mathsf{Q}$

  In this case, as explained above, we have $r \gg 0$ and $r_{\text{eff}}(\mu) > 0$.
  Also, according to our discussion in \autoref{subsec: Quantum Low Energy U(1|1)} on the Higgs mechanism, the quantum low energy limit exists at a scale $\mu \ll e \sqrt{r}$.

  Thus, since $r_{\text{eff}}(\mu) > 0$,  at low energy, the quantum GLSM would become a quantum NLSM on $\mathcal{S}^{\mathrm{U}(1|1)}_{Q > \mathsf{Q}, r_{\text{eff}}(\mu) > 0}$.

  As before, in place of the classical low energy limit $e \to \infty$, we would have the quantum low energy limit $\flatfrac{e}{\Lambda} \to \infty$.

  \item $Q < \mathsf{Q}$

  In this case, as explained above, we have $r \ll 0$ and $r_{\text{eff}}(\mu) < 0$.
  Also, according to our discussion in \autoref{subsec: Quantum Low Energy U(1|1)} on the Higgs mechanism, the quantum low energy limit exists at a scale $\mu \ll e \sqrt{|r|}$.

  Thus, since $r_{\text{eff}}(\mu) < 0$,  at low energy, the quantum GLSM would become a quantum NLSM on $\mathcal{S}^{\mathrm{U}(1|1)}_{Q < \mathsf{Q}, r_{\text{eff}}(\mu) < 0}$.

  As before, in place of the classical low energy limit $e \to \infty$, we would have the quantum low energy limit $\flatfrac{e}{\Lambda}  \to\infty$.

  \item $Q = \mathsf{Q}$

  In this case, one can see from \eqref{eq:U(1|1) phase transition:quantum GLSM:r-eff(mu)} that the FI parameter does not run, and the quantum GLSM is scale-invariant.
  Hence, $r$ is still a parameter.
  As such, we can choose either $r > 0$ or $r < 0$, where at low energy, the quantum GLSM becomes either the quantum NLSM on $\mathcal{S}^{\mathrm{U}(1|1)}_{Q = \mathsf{Q}, r > 0}$ or $\mathcal{S}^{\mathrm{U}(1|1)}_{Q = \mathsf{Q}, r < 0}$.
  Since there is no dynamical scale $\Lambda$, the quantum low energy limit would instead be given by $e \sqrt{r} \to \infty$ and   $e \sqrt{|r|} \to \infty$, respectively.

\end{itemize}

\subsection{Scale-invariance and Applications to Mathematics}
\label{sec:U(1|1) phase transition:scale-invariance}

\subtitle{Scale-invariance of the Quantum GLSM when $Q = \mathsf{Q}$}

Let us consider the case where $Q = \mathsf{Q}$.
As explained in \autoref{sec:U(1|1) phase transition:low energy}, the quantum GLSM would be scale-invariant with free parameter $r$.

\subtitle{Scale-invariance, $\MSUSY$ of the Quantum GLSM, and a Geometric Correspondence}

In our scale-invariant quantum GLSM where $r$ remains a parameter of the theory, its $\MSUSY$ will continue to be given by $\mathcal{S}^{\mathrm{U}(1|1)}_{Q =\mathsf{Q}, r > 0}$ or $\mathcal{S}^{\mathrm{U}(1|1)}_{Q = \mathsf{Q}, r < 0}$, depending on our \emph{free} choice of $r$.

Clearly, $\mathcal{S}^{\mathrm{U}(1|1)}_{Q = \mathsf{Q}, r > 0}$ and $\mathcal{S}^{\mathrm{U}(1|1)}_{Q = \mathsf{Q}, r < 0}$ just describe $\MSUSY$, the target space of its low-energy quantum NLSM, in different regions of the parameter space.
Thus, from the perspective of our quantum GLSM, there ought to be a physical correspondence between the quantum NLSM when $r > 0$ and $r < 0$, and therefore a geometric correspondence between $\mathcal{S}^{\mathrm{U}(1|1)}_{Q = \mathsf{Q}, r > 0}$ and $\mathcal{S}^{\mathrm{U}(1|1)}_{Q = \mathsf{Q}, r < 0}$.

Indeed, as we shall explain in \autoref{subsec: Super Birational Eq of CY}, this correspondence can, for a particular example, reduce to a birational equivalence of CYs, that, moreover, also generalizes the latter to super-Grassmannians!

\subtitle{Scale-invariance and a Physical Derivation of a Novel CY Condition for Supervector Bundles over ``Weighted Super-Grassmannians''}

As the quantum GLSM is scale-invariant, its corresponding low energy quantum NLSM ought to also be scale and therefore conformally-invariant~\cite{papadopoulos2024scaleconformalinvariance2d}; in turn, according to our explanation in \autoref{subsec: applications to math U(1|1)}, this would mean that the target spaces of the low energy quantum NLSM in \autoref{sec:U(1|1) phase transition:low energy}, i.e., $\mathcal{S}^{\mathrm{U}(1|1)}_{Q = \mathsf{Q}, r > 0}$ and $\mathcal{S}^{\mathrm{U}(1|1)}_{Q = \mathsf{Q}, r < 0}$, must be CY.

As described in \eqref{eq:phase transition topology:cy condition:positive} and \eqref{eq:phase transition topology:cy condition:negative}, $\mathcal{S}^{\mathrm{U}(1|1)}_{Q = \mathsf{Q}, r > 0}$ and $\mathcal{S}^{\mathrm{U}(1|1)}_{Q = \mathsf{Q}, r < 0}$ are supervector bundles over ``weighted super-Grassmannians'', whose weights on the fibers and bases are determined by $(Q, \mathsf{Q})$.

Thus, we have a physical derivation of a \emph{novel} mathematical result that supervector bundles over ``weighted super-Grassmannians'', whose weights on the fibers and bases are determined by $(Q, \mathsf{Q})$, are CY if and only if $Q = \mathsf{Q}$!

Indeed, one can actually mathematically verify (see \autoref{app:verify cy:supervector over WGr}) this statement to be true!

This once more vindicates our application to mathematics of our otherwise nonunitary supergroup GLSM via a study of its space of supersymmetric states.

\subsection{Phase Transition of the Quantum Model: Mild Topology Change of a Supervector Bundle over a Super-Grassmannian}
\label{sec:U(1|1) phase transition:topological change:SV over Gr}

Let us stay with the scale-invariant quantum GLSM, i.e., $Q = \mathsf{Q}$, so $r$ remains a free parameter of the theory.

For simplicity, let us consider the case where $Q_1 = Q_2 \cdots = Q_{l_m} = 1$ while $Q_{l_m + 1} = Q_{l_m + 2} \cdots = Q_m = -1$, and $\mathsf{Q}_1 = \mathsf{Q}_2 \cdots = \mathsf{Q}_{l_n} = 1$ while $\mathsf{Q}_{l_n + 1} = \mathsf{Q}_{l_n + 2} \cdots = \mathsf{Q}_n = -1$.
Hence, from \eqref{eq:U(1|1) phase transition:glsm:Q}--\eqref{eq:U(1|1) phase transition:glsm:Q-serif}, we have $Q = l_m - (m - l_m) = 2 l_m - m$ and $\mathsf{Q} = l_n - (n - l_n) = 2 l_n - n$.

\subtitle{The $r > 0$ Phase of the Quantum Model: A Supervector Bundle over $Gr_{1|1}(\mathbb{C}^{l_m|l_n})_{s',t'}$ $\MSUSY$ and its Low Energy NLSM on $\MSUSY$}

Note that for $r > 0$, $\MSUSY$ in \eqref{eq:U(1|1) phase transition:glsm:M-SUSY} can be mathematically identified as \eqref{eq:phase transition topology:cy condition:positive} with $Q_{s''} = -1 = \mathsf{Q}_{t''}$ and $(\vec{Q'}|\vec{\mathsf{Q}'}) = (\vec{1}| \vec{1})$, i.e.,
\begin{equation}
  \label{eq:U(1|1) phase transition:SV over Gr:r > 0}
  \saveboxed{eq:U(1|1) phase transition:SV over Gr:r > 0}{
    \mathcal{S}^{\mathrm{U}(1|1)}_{2 l_m - m = 2 l_n - n, r > 0}
    = \bigoplus_{s'' = l_m + 1}^m \mathscr{L}^{-1}_{s''}
    \oplus \bigoplus_{t'' = l_n + 1}^n \mathbf{\Pi} \mathscr{L}^{-1}_{t''}
    \rightarrow Gr_{1|1}(\mathbb{C}^{l_m|l_n})_{s',t'}
  }
\end{equation}
a supervector bundle over a super-Grassmannian $Gr_{1|1}(\mathbb{C}^{l_m|l_n})_{s',t'}$ of size $r$.
Here,
\begin{enumerate*}

  \item the even $\mathscr{L}_{s''}$ and odd $\mathbf{\Pi} \mathscr{L}_{t''}$ fibers are spanned by the Grassmann-even $\phi_{s''}$ and Grassmann-odd $\psi_{t''}$ supervectors, and

  \item the coordinates on the $Gr_{1|1}(\mathbb{C}^{l_m|l_n})_{s',t'}$ base are the components of the Grassmann-even $\phi_{s'}$ and Grassmann-odd $\psi_{t'}$ supervectors.

\end{enumerate*}

In the low energy limit $e\sqrt{r} \to \infty$, following the steps at the end of \autoref{subsec: Quantum Low Energy U(1|1)}, we find that the quantum GLSM would become a quantum NLSM on $\mathcal{S}^{\mathrm{U}(1|1)}_{2 l_m - m = 2 l_n - n, r > 0}$.

\subtitle{The $r < 0$ Phase of the Quantum Model: A Supervector Bundle over $Gr_{1|1}(\mathbb{C}^{m - l_m|n - l_n})_{s'', t''}$ $\MSUSY$ and its Low Energy NLSM on $\MSUSY$}

On the other hand, for $r < 0$, $\MSUSY$ in \eqref{eq:U(1|1) phase transition:glsm:M-SUSY} can be mathematically identified as \eqref{eq:phase transition topology:cy condition:negative} with $Q_{s'} = 1 = \mathsf{Q}_{t'}$ and $(\vec{\abs{Q''}}|\vec{\abs{\mathsf{Q}''}}) = (\vec{1}| \vec{1})$, i.e.,
\begin{equation}
  \label{eq:U(1|1) phase transition:SV over Gr:r < 0}
  \saveboxed{eq:U(1|1) phase transition:SV over Gr:r < 0}{
    \mathcal{S}^{\mathrm{U}(1|1)}_{2 l_m - m = 2 l_n - n, r < 0}
    = \bigoplus_{s' = 1}^{l_m} \mathscr{L}_{s'}^{-1}
    \oplus \bigoplus_{t' = 1}^{l_n} \mathbf{\Pi} \mathscr{L}^{-1}_{t'}
    \rightarrow Gr_{1|1}(\mathbb{C}^{m - l_m|n - l_n})_{s'',t''}
  }
\end{equation}
a supervector bundle over a super-Grassmannian $Gr_{1|1}(\mathbb{C}^{m - l_m|n - l_n})_{s'',t''}$ of size $r$.
Here,
\begin{enumerate*}

  \item the even $\mathscr{L}_{s'}$ and odd $\mathbf{\Pi}\mathscr{L}_{t'}$ fibers are spanned by the Grassmann-even $\phi_{s'}$ and Grassmann-odd $\psi_{t'}$ supervectors, and

  \item the coordinates on the $Gr_{1|1}(\mathbb{C}^{m - l_m|n - l_n})_{s'',t''}$ base are the components of the Grassmann-even $\phi_{s''}$ and Grassmann-odd $\psi_{t''}$ supervectors.

\end{enumerate*}

Similarly, in the low energy limit $e\sqrt{|r|} \to \infty$, following the steps at the end of \autoref{subsec: Quantum Low Energy U(1|1)}, we find that the quantum GLSM would become a quantum NLSM on $\mathcal{S}^{\mathrm{U}(1|1)}_{2 l_m - m = 2 l_n - n, r < 0}$.

\subtitle{A Mild Topology Change from a $r > 0$ to $r < 0$ Phase Transition}

Thus, as we are free to go from $r > 0$ to $r < 0$ in a phase transition of the quantum $\mathrm{U}(1|1)$ GLSM,\footnote{%
  This possibility of going smoothly from the $r > 0$ to $r < 0$ phase of the quantum $\mathrm{U}(1|1)$ GLSM will be justified in \autoref{sec:U(1|1) phase transition:topological change:worldsheet instantons}.
}
at low energy, we will have a quantum NLSM on $\mathcal{S}^{\mathrm{U}(1|1)}_{2 l_m - m = 2 l_n - n, r > 0}$ that goes to a quantum NLSM on $\mathcal{S}^{\mathrm{U}(1|1)}_{2 l_m - m = 2 l_n - n, r < 0}$.

Notice that when $2 l_m = m$ and $2 l_n = n$, $\mathcal{S}^{\mathrm{U}(1|1)}_{0 = 0, r > 0}$ and $\mathcal{S}^{\mathrm{U}(1|1)}_{0 = 0, r < 0}$ are isomorphic.
That said, the two spaces are actually \emph{mildly topologically distinct} from each other --- via \eqref{eq:U(1|1) phase transition:SV over Gr:r > 0} and \eqref{eq:U(1|1) phase transition:SV over Gr:r < 0}, one can see that the directions of the fiber and base of $\mathcal{S}^{\mathrm{U}(1|1)}_{0 = 0, r > 0}$ and $\mathcal{S}^{\mathrm{U}(1|1)}_{0 = 0, r < 0}$ are swopped around, whence their intersection form governing their classical homology
ring differs.

In other words, we have a mild topology change in the target space of the low energy quantum NLSM from $\mathcal{S}^{\mathrm{U}(1|1)}_{0 = 0, r > 0}$ to $\mathcal{S}^{\mathrm{U}(1|1)}_{0 = 0, r < 0}$ as we freely go from $r > 0$ to $r < 0$!

\subsection{Phase Transition of the Familiar Quantum \texorpdfstring{$\mathrm{U}(1)$}{U(1)} Model: Mild Topology Change of a CY Vector Bundle over Weighted Projective Space}
\label{sec:U(1|1) phase transition:topological change:V over WCP}

As mentioned at the end of \autoref{sec:U(1|1) phase transition:glsm}, the $\mathrm{U}(1|1)$ GLSM ought to reduce to the familiar $\mathrm{U}(1)$ GLSM when we
\begin{enumerate*}

\item turn off the $\tensor{\phi}{^{\bar{1}}_s}$ field (and its corresponding bar-indexed partners in the underlying multiplet), and

\item turn off the \emph{whole} Grassmann-odd chiral superfield $\Psi$.

\end{enumerate*}

\subtitle{$\MSUSY$}

In doing so, the $\MSUSY$ of the $\mathrm{U}(1|1)$ model defined by \eqref{eq:U(1|1) phase transition:glsm:M-SUSY}, would reduce to the $\MSUSY$ of the $\mathrm{U}(1)$ model defined by
\begin{equation}
  \label{eq:U(1|1) phase transition:V over WCP:M-SUSY}
  \mathcal{S}^{\mathrm{U}(1)}_{Q, r}
  = \flatfrac{\left\{
      \sum^{l_m}_{s' = 1} Q_{s'} \tensor{\phi}{^1_{s'}} \tensor{\bar{\phi}}{_1^{s'}}
      = r + \sum^m_{s'' = l_m + 1} \abs{Q_{s''}} \tensor{\phi}{^1_{s''}} \tensor{\bar{\phi}}{_1^{s''}}
    \right\}}{\mathrm{U}(1)}
  \, .
\end{equation}

\subtitle{The $r > 0$ Phase and its Low Energy NLSM on a Vector Bundle over $\WCP^{l_m - 1}_{(\vec{Q'})_{s'}}$}

When $r > 0$, the $\MSUSY$ of the $\mathrm{U}(1)$ model in \eqref{eq:U(1|1) phase transition:V over WCP:M-SUSY} can be mathematically identified as (\emph{c.f.}~\cite[eq.~(15.94)]{Hori:2003ic})
\begin{equation}
  \label{eq:U(1|1) phase transition:V over WCP:r > 0}
  \mathcal{S}^{\mathrm{U}(1)}_{Q, r > 0}
  = \bigoplus_{s'' = l_m + 1}^m \mathcal{L}^{Q_{s''}}_{s''}
  \longrightarrow \WCP^{l_m-1}_{(\vec{Q'})_{s'}}
  \, ,
\end{equation}
a vector bundle over a weighted projective space of size $r$, where
\begin{enumerate*}
  \item the fibers $\mathcal{L}_{s''}$ are spanned by the $\tensor{\phi}{^1_{s''}}$ fields,

  and

  \item the coordinates on the $\WCP^{l_m - 1}_{(\vec{Q'})_{s'}}$ base are the $\tensor{\phi}{^1_{s'}}$ fields with positive weights $(\vec{Q'})_{s'} \coloneq (Q_1, \dots, Q_{l_m})$.

\end{enumerate*}

In the low energy limit $e \to \infty$, following the steps in \autoref{GLSM to NLSM}, we would get an NLSM on $\mathcal{S}^{\mathrm{U}(1)}_{Q, r > 0}$.

\subtitle{The $r < 0$ Phase and its Low Energy NLSM on a Geometrically Distinct Vector Bundle over $\WCP^{m - l_m -1}_{(\abs{\vec{Q''}})_{s''}}$}

When $r < 0$, $\MSUSY$ of the $\mathrm{U}(1)$ model in \eqref{eq:U(1|1) phase transition:V over WCP:M-SUSY} can be mathematically identified as (\emph{c.f.}~\cite[eq.~(15.95)]{Hori:2003ic})
\begin{equation}
  \label{eq:U(1|1) phase transition:V over WCP:r < 0}
  \mathcal{S}^{\mathrm{U}(1)}_{Q, r < 0}
  =  \bigoplus_{s' = 1}^{l_m} \mathcal{L}^{- Q_{s'}}_{s'}
  \longrightarrow  \WCP^{m - l_m - 1}_{(\abs{\vec{Q''}})_{s''}},
\end{equation}
a vector bundle over a weighted projective space of size $|r|$, where
\begin{enumerate*}
  \item the fibers $\mathcal{L}_{s'}$ are spanned by the $\tensor{\phi}{^1_{s'}}$ fields,

  and

  \item the coordinates on the $\WCP^{m - l_m -1}_{(\abs{\vec{Q''}})_{s''}}$ base are the components of the $\tensor{\phi}{^1_{s''}}$ fields with negative weights whose absolute values are $(\abs*{\vec{Q''}})_{s''} \coloneq (|Q_{l_m + 1}|, \dots, |Q_m|)$.
\end{enumerate*}

In the low energy limit $e \to \infty$, following the steps in \autoref{GLSM to NLSM}, we would get an NLSM on $\mathcal{S}^{\mathrm{U}(1)}_{Q, r < 0}$.

\subtitle{Quantum Aspects}

The quantum aspects of the $\mathrm{U}(1)$ model most relevant to us at this point are that
\begin{enumerate*}

  \item the model is scale-invariant where $r$ remains a parameter of the quantum theory if and only if $Q =0$ in \eqref{eq:U(1|1) phase transition:glsm:Q};

  \item for a well-defined $\MSUSY$ at the quantum level, if $Q \neq 0$, we need to consider $r \gg 0$ or $r \ll 0$;

  and

  \item if $Q=0$, in the low energy limit $e \sqrt{r} \to \infty$ or $e \sqrt{|r|} \to \infty$, the model will become a quantum NLSM on \eqref{eq:U(1|1) phase transition:V over WCP:r > 0} or \eqref{eq:U(1|1) phase transition:V over WCP:r < 0}, respectively.

\end{enumerate*}

\subtitle{A Mild Topology Change from a $r > 0$ to $r < 0$ Phase Transition}

Let us now consider the scale-invariant quantum model, i.e., $Q = 0$ in \eqref{eq:U(1|1) phase transition:glsm:Q}.
Then, $\mathcal{S}^{\mathrm{U}(1)}_{0, r > 0}$ in \eqref{eq:U(1|1) phase transition:V over WCP:r > 0} would be a (non-compact) CY manifold; likewise, $\mathcal{S}^{\mathrm{U}(1)}_{0, r < 0}$ in \eqref{eq:U(1|1) phase transition:V over WCP:r < 0} would also be a (non-compact) CY manifold.

Thus, as we are free go from $r > 0$ to $r < 0$ in a phase transition of the $\mathrm{U}(1)$ model, at low energy, we will have a quantum NLSM on $\mathcal{S}^{\mathrm{U}(1)}_{0, r > 0}$ that goes to a quantum NLSM on $\mathcal{S}^{\mathrm{U}(1)}_{0, r < 0}$.

Let us now also consider $Q_{s'} = 1$ and $Q_{s''} = -1$, whence $2 l_m = m$.
From our discussion at the end of \autoref{sec:U(1|1) phase transition:topological change:SV over Gr}, it should also be clear from \eqref{eq:U(1|1) phase transition:V over WCP:r > 0} and \eqref{eq:U(1|1) phase transition:V over WCP:r < 0} that $\mathcal{S}^{\mathrm{U}(1)}_{0, r > 0}$ and $\mathcal{S}^{\mathrm{U}(1)}_{0, r < 0}$, though isomorphic, are \emph{mildly topologically distinct} from each other.

In other words, we have a mild topology change in the CY target space of the low energy quantum NLSM from $\mathcal{S}^{\mathrm{U}(1)}_{0, r > 0}$ to $\mathcal{S}^{\mathrm{U}(1)}_{0, r < 0}$ as we freely go from $r > 0$ to $r < 0$, as we should \cite[$\S$15.5.3]{Hori:2003ic}.

\subsection{The ``Singularity'' at \texorpdfstring{$r = 0$}{r = 0}, a Smooth Phase Transition, and Worldsheet Superinstanton Effects}
\label{sec:U(1|1) phase transition:topological change:worldsheet instantons}

\subtitle{The ``Singularity'' at $r = 0$ and a Smooth Phase Transition from $r > 0$ to $r < 0$}

Notice from the expression of the potential energy in \eqref{eq:U(1|1) phase transition:glsm:potential} that when $r=0$, as long as $\sigma$ is valued in a commuting subalgebra of the underlying $\mathfrak{u}(1|1)$ superalgebra, $\MSUSY$ would develop a new \emph{non-compact} branch where the $\phi$'s and $\psi$'s must vanish while its coordinate $\sigma$ is no longer constrained and can even be infinitely large.

This non-compactness of $\MSUSY$ leads to a singularity in the theory where the $r > 0$ and $r< 0$ regions of parameter space seem completely disconnected by the singular  $r=0$ point with different physics.
This suggests that the phase transition in \autoref{sec:U(1|1) phase transition:topological change:SV over Gr} when we go from $r > 0$ to $r < 0$ may not be smooth and therefore well-defined after all.

The question therefore, is whether $\MSUSY$ actually develops a new non-compact branch at $r = 0$,  or at some value of $t = \mathbf{i} r + \frac{\vartheta}{2\pi}$, in the \emph{quantum} theory.
If it is the latter, then the transition from $r > 0$ to $r < 0$ passing through $r = 0$ would actually be smooth.

Let us now ascertain the answer to this question.
To this end, since we are concerned with the non-compactness of $\MSUSY$, let us consider its coordinate $\sigma$ to be large.

When $\sigma$ is large, the $\phi$ and $\psi$ matter fields would be massive (see \eqref{eq:U(1|1) phase transition:glsm:potential}), and they can be integrated out (along with the other fields in their chiral matter multiplets because of supersymmetry).
We are then left with a pure theory in the gauge multiplet fields, where $\sigma$ is the lowest component scalar field.

Note at this point from \eqref{eq:U(1|1) phase transition:glsm:potential} that the classical non-compact branch of $\MSUSY$ at $r = 0$ is defined by the condition $\comm{\sigma}{\bar{\sigma}} = 0$.
Thus, as mentioned, $\sigma$ ought to be valued in a commuting subalgebra of $\mathfrak{u}(1|1)$ that corresponds to a purely diagonal element in \eqref{eq:review:u-1-1:definition}, or explicitly,
\begin{equation}
  \label{eq:U(1|1) phase transition:worldsheet instantons:diagonal sigma}
  \sigma = \mqty(\dmat[0]{\sigma_1, \sigma_{\bar{1}}}) \, .
\end{equation}
This means that $\sigma$ ought to transform under the commuting $\mathrm{U}(1)_1 \times \mathrm{U}(1)_{\bar{1}}$ bosonic subgroup of the underlying non-abelian $\mathrm{U}(1|1)$ supergauge group, and because of supersymmetry, the other fields in the gauge multiplet, in particular, the gauge field $v_\mu$, ought to transform under the $\mathrm{U}(1)_1 \times \mathrm{U}(1)_{\bar{1}}$ bosonic subgroup, too.
In other words, the non-abelian $\mathrm{U}(1|1)$ supergauge group would be broken to its maximal torus $\mathrm{U}(1)_1 \times \mathrm{U}(1)_{\bar{1}}$, whence this classical non-compact branch of $\MSUSY$ at $r = 0$ is actually a Coulomb branch.

As we effectively have, from an underlying non-abelian GLSM, a $\mathrm{U}(1)_1 \times \mathrm{U}(1)_{\bar{1}}$ maximal torus GLSM with large $\Sigma$ and $m + n$ massive $\Phi_s$ and $\Psi_t$ superfields transforming in the fundamental representation along the Coulomb branch, we can simply adapt the calculation leading up to~\cite[eq.~(2.16)]{Hori:2006dk}, with no $P$'s but with Grassmann-even $\Phi_s$'s and Grassmann-odd $\Psi_t$'s having gauge charges $Q_s$ and $\mathsf{Q}_t$, to arrive at the relation that $t$ must obey, from which one can determine the value of $t$ where a non-compact Coulomb branch appears in the quantum theory.

Specifically, in integrating out the massive fields, we would have an effective twisted superpotential for $\Sigma_1$ and $\Sigma_{\bar{1}}$ (corresponding to the $\mathrm{U}(1)_1 \times \mathrm{U}(1)_{\bar{1}}$ gauge group) given by
\begin{equation}
  \label{eq:U(1|1) phase transition:W-eff}
  \widetilde{W}
  = \sum_i (-1)^{\varsigma(i)} \Sigma_i \left(
    2 \pi \mathbf{i} t(\mu)
    - \sum^m_{s = 1} Q_s \left[
      \log \left( \frac{Q_s \Sigma_i}{\mu} \right) - 1
    \right]
    + \sum^n_{t = 1} \mathsf{Q}_t \left[
      \log \left( \frac{\mathsf{Q}_t \Sigma_i}{\mu} \right) - 1
    \right]
  \right)
  \, ,
\end{equation}
where $i \in \{1, \bar{1}\}$, from which the Coulomb branch would be defined by setting
\begin{equation}
  \label{eq:U(1|1) phase transition:coulomb branch}
  \pdv{\widetilde{W}}{\Sigma_i} = 0
  \, \qcomma
  i \in \{1, \bar{1}\}
  \, ,
\end{equation}
that, via
\begin{equation}
  \label{eq:U(1|1) phase transition:coulomb branch:exponent}
  \exp \left(
    \pdv{\widetilde{W}}{\Sigma_i}
  \right)
  = 1
  \, \qcomma
  i \in \{1, \bar{1}\}
  \, ,
\end{equation}
would give us
\begin{equation}
  \label{eq:U(1|1) phase transition:coulomb branch:exponent:scale-invariant}
  \left(
    \prod_{s = 1}^m Q_s^{Q_s}
  \right) \left(
    \prod_{t = 1}^n \mathsf{Q}_t^{-\mathsf{Q}_t}
  \right)
  = e^{2 \pi \mathbf{i}t}
  \, ,
\end{equation}
after we use the scale-invariant condition $\sum_s Q_s = \sum_t \mathsf{Q}_t$.

From this relation that $t$ must obey, it must mean that in general, the solutions are such that $t \neq 0$.

In conclusion, in the quantum theory, $\MSUSY$ actually develops a new non-compact Coulomb branch not at $r = 0$, but at some typically nonzero value of $t = \mathbf{i}r + \frac{\vartheta}{2\pi}$!
Therefore, the phase transition in \autoref{sec:U(1|1) phase transition:topological change:SV over Gr} when we go from $r > 0$ to $r < 0$ passing through $r = 0$ (along the imaginary axis in the $t$-plane), would actually be \emph{smooth} and therefore well-defined.

\subtitle{The $\vartheta$-angle and Worldsheet Superinstanton Effects at Low Energy}

Clearly, in the quantum GLSM, the singular point has been shifted from $r = 0$ to $t = \mathbf{i} r + \frac{\vartheta}{2\pi}$; one can therefore attribute this shift to the $\vartheta$-angle.
In other words, a nonzero $\vartheta$-angle allows us to avoid a singularity whence we would have a smooth phase transition.

One can also understand this from the perspective of  worldsheet superinstantons in the low energy NLSM that ``shield'' the model from the  singularity at $r = 0$ when there is a nonzero $\vartheta$-angle, whence we would have a non-catastrophic topology change of its CY target space from \eqref{eq:U(1|1) phase transition:SV over Gr:r > 0} to \eqref{eq:U(1|1) phase transition:SV over Gr:r < 0}.

Firstly, note that the last term on the RHS of the Lagrangian in \eqref{eq:U(1|1) phase transition:glsm:lagrangian} would give us, in Euclidean signature, the $\vartheta$-term
\begin{equation}
  \label{eq:U(1|1) phase transition:superinstanton:F term}
  - \frac{\mathbf{i} \vartheta}{2 \pi} \int_{\mathcal{C}}  \Str F
  \, ,
\end{equation}
where the field strength $F = Dv$ is the covariant derivative of the $\mathrm{U}(1|1)$ gauge field $v$ on the worldsheet $\mathcal{C}$.
From the properties of the supertrace (see \autoref{ft:u-1-1:supertrace}), we can also write this as
\begin{equation}
  \label{eq:U(1|1) phase transition:superinstanton:F term:explicit}
  - \frac{\mathbf{i} \vartheta}{2 \pi} \left(
    \int_{\mathcal{C}} [F]_{11} -  \int_{\mathcal{C}} [F]_{\bar{1} \bar{1}}
  \right)
  \, ,
\end{equation}
where $[F]_{11}$ and $[F]_{\bar{1} \bar{1}}$ are the upper-left and lower-right diagonal entries of the $2 \times 2$ matrix that corresponds to the $\mathfrak{u}(1|1)$-valued $F$.
From the explicit form of $[F]_{11}$ and $[F]_{\bar{1} \bar{1}}$ given in~\cite[eq.~(4.4)]{Kimura:2023iup}, this works out to be\footnote{%
  To arrive at this, we note that \emph{loc.~cit.}, the one-forms $\psi$ and $\psi^\dagger$, being \emph{fermionic}, obey $\psi \wedge \psi^\dagger = \psi^\dagger \wedge \psi$.
}
\begin{equation}
  \label{eq:U(1|1) phase transition:superinstanton:F_1 - F_2}
  - \frac{\mathbf{i} \vartheta}{2 \pi} \left(
    \int_{\mathcal{C}} F_1 -  \int_{\mathcal{C}} F_{\bar{1}}
  \right)
  \, ,
\end{equation}
where $F_1$ and $F_{\bar{1}}$ are the field strengths of the  $\mathrm{U}(1)_1$ and $\mathrm{U}(1)_{\bar{1}}$ gauge fields (that are associated with $\sigma_1$ and $\sigma_{\bar{1}}$ above).

Secondly, recall that in the low energy limit, the gauge field $v$ would become an auxiliary field that can be integrated out via its EOM in \eqref{eq:u-1-1:nlsm:solutions of v}.
In particular, by substituting $v$ as given in \eqref{eq:u-1-1:nlsm:solutions of v} back into \eqref{eq:U(1|1) phase transition:superinstanton:F_1 - F_2}, we would get an expression that no longer contains $v$.
To determine what this expression is, note that
\begin{enumerate*}
  \item the EOM of $v$ in \eqref{eq:u-1-1:nlsm:solutions of v} and its accompanying relation \eqref{eq:u-1-1:nlsm:current matrix} mean that at low energy, the gauge fields $v_1$ and $v_{\bar{1}}$ in the first and second term of \eqref{eq:U(1|1) phase transition:superinstanton:F_1 - F_2} can be interpreted as a pullback of a connection one-form on a $\mathrm{U}(1)_1$ and $\mathrm{U}(1)_{\bar{1}}$ principal line bundle over the $\MSUSY$ target space \eqref{eq:U(1|1) phase transition:SV over Gr:r > 0} or \eqref{eq:U(1|1) phase transition:SV over Gr:r < 0} of the NLSM;

  \item $\flatfrac{F_1}{2\pi}$ and $\flatfrac{F_2}{2\pi}$, being the curvature of the aforementioned line bundles (whose fibers are locally $\mathbb{C}^{1|0}$), would correspond to their first Chern class that generates the de Rham cohomology $H^2_{\text{dR}}(\MSUSY, \mathbb{Z})$;

  \item for a supervariety $\MSUSY$, we can identify $H^2_{\text{dR}}(\MSUSY, \mathbb{Z})$ as $H^2_{\text{dR}}(\MSUSY^{\text{bos}}, \mathbb{Z})$, where $\MSUSY^{\text{bos}}$ is the bosonization of $\MSUSY$~\cite[Theorem~1.1]{polishchuk2023rhamcohomologysupervarieties};

  and

  \item because $\MSUSY^{\text{bos}}$ is, by inspecting \eqref{eq:U(1|1) phase transition:SV over Gr:r > 0} or \eqref{eq:U(1|1) phase transition:SV over Gr:r < 0}, the (non-compact) CY vector bundle $\mathcal{O}(-1)^{\oplus M} \to \mathbb{CP}^{M-1}$, which is an integral symplectic manifold, its symplectic two-form $\flatfrac{\omega}{2 \pi}$ would generate $H^2_{\text{dR}}(\MSUSY^{\text{bos}}, \mathbb{Z})$.

\end{enumerate*}
In sum, the low energy expression for \eqref{eq:U(1|1) phase transition:superinstanton:F_1 - F_2} that no longer contains $v$ would read
\begin{equation}
  \label{eq:U(1|1) phase transition:superinstanton:w_1 - w_2}
  - \frac{\mathbf{i}\vartheta}{2 \pi} \left(
    \int_{\mathcal{C}} \phi^\ast(\omega_1)
    - \int_{\mathcal{C}} \phi^\ast(\omega_{\bar{1}})
  \right)
  \, ,
\end{equation}
where the (possibly negative) integer
\begin{equation}
  \label{eq:U(1|1) phase transition:superinstanton:n_1 - n_2}
  \frac{1}{2\pi} \int_{\mathcal{C}} \phi^\ast(\omega_1)
  - \frac{1}{2\pi} \int_{\mathcal{C}} \phi^\ast(\omega_{\bar{1}})
  = n_1 - n_{\bar{1}}
\end{equation}
can be interpreted as the NLSM worldsheet superinstanton number, where $n_1$ and $n_{\bar{1}}$ count the positive and negative degree worldsheet instantons in $\MSUSY^{\text{bos}}$.
This is just a 2d NLSM analog of the 4d superinstanton number $\frac{1}{8 \pi^2} \int_{\mathcal{M}_4} \text{Str} \, F \wedge F = \frac{1}{8 \pi^2} \int_{\mathcal{M}_4}  F_1 \wedge F_1  - \frac{1}{8 \pi^2} \int_{\mathcal{M}_4} F_{\bar{1}} \wedge F_{\bar{1}} = k_1 - k_{\bar{1}}$, where  $k_1$ and $k_{\bar{1}}$ count the positive and negative charge gauge instantons on $\mathcal{M}_4$ (see~\cite[eq.~(5.28)]{Kimura:2023iup}).

Finally, the GLSM $\vartheta$-term, which at low energy, corresponds to a worldsheet superinstanton term described by \eqref{eq:U(1|1) phase transition:superinstanton:w_1 - w_2}--\eqref{eq:U(1|1) phase transition:superinstanton:n_1 - n_2}, will manifest in the Euclidean path integral of the low energy NLSM as

\begin{equation}
  \label{eq:U(1|1) phase transition:superinstanton:n_1 - n_2:path integral}
  \int \text{D}X \, e^{- S_{\text{NLSM}}(X)} \sum_{n_1, n_{\bar{1}} \in \mathbb{Z}}  e^{\mathbf{i} \vartheta (n_1 - n_{\bar{1}})}
  \, ,
\end{equation}
a superposition of oscillatory phase factors that will shield the path integral from vanishing at the singularity $r = 0$ where the target space metric and hence $S_{\text{NLSM}}(X) \to \infty$.

\subsection{A Super-Grassmannian Generalization of a  Birational Equivalence of CY Vector Bundles in Mathematics}
\label{subsec: Super Birational Eq of CY}

\subtitle{An Atiyah-Type Flop Transition of CY Vector Bundles over a Projective Space}

If we satisfy the scale-invariant condition $Q = 0$ at the end of \autoref{sec:U(1|1) phase transition:topological change:V over WCP} by letting $m = 2M$ and $l_m = M$ such that $Q_1 = Q_2 \cdots = Q_{M} = 1$ and $Q_{M+1} = Q_{M+2} \cdots = Q_{2M} = -1$, as we freely go from $r > 0$ to $r < 0$, we would get a change in the target space of the low energy quantum NLSM from
\begin{equation}
  \label{eq:U(1|1) phase transition:atiyah flop:U(1) r > 0}
  \mathcal{S}^{\mathrm{U}(1)}_{0, r > 0}
  = \bigoplus^{2M}_{s'' = M + 1} \mathcal{L}^{-1}_{s''} \longrightarrow \mathbb{CP}^{M-1}_{s'}
  \, ,
\end{equation}
a CY vector bundle over a projective space of size $r$, to
\begin{equation}
  \label{eq:U(1|1) phase transition:atiyah flop:U(1) r < 0}
  \mathcal{S}^{\mathrm{U}(1)}_{0, r < 0}
  = \bigoplus^M_{s' = 1} \mathcal{L}^{-1}_{s'} \longrightarrow \mathbb{CP}^{M-1}_{s''}
  \, ,
\end{equation}
an isomorphic but \emph{mildly topologically distinct} CY vector bundle over a projective space of size $\abs{r}$.

The mild topology change as we go from \eqref{eq:U(1|1) phase transition:atiyah flop:U(1) r > 0} to \eqref{eq:U(1|1) phase transition:atiyah flop:U(1) r < 0} is well-recognized as an Atiyah-type flop transition of CY vector bundles over a projective space.

\subtitle{A Flop Transition of CY Supervector Bundles over a Super-Grassmannian}

Consider the scale-invariant charge assignment at the end of \autoref{sec:U(1|1) phase transition:topological change:SV over Gr} with $m = 2M$ and $l_m = M$, as well as $n = 2N$ and $l_n = N$, whence $Q = \sum_s Q_s = 0$ and $\mathsf{Q} = \sum_t \mathsf{Q}_t = 0$.

Then, \eqref{eq:U(1|1) phase transition:SV over Gr:r > 0} becomes
\begin{equation}
  \label{eq:U(1|1) phase transition:atiyah flop:U(1|1) r > 0}
  \mathcal{S}^{\mathrm{U}(1|1)}_{0 = 0, r > 0}
  = \bigoplus_{s'' = M + 1}^{2M} \mathscr{L}^{-1}_{s''}
  \oplus \bigoplus_{t'' = N + 1}^{2N} \mathbf{\Pi} \mathscr{L}^{-1}_{t''}
  \rightarrow Gr_{1|1}(\mathbb{C}^{M|N})_{s',t'}
  \, ,
\end{equation}
and \eqref{eq:U(1|1) phase transition:SV over Gr:r < 0} becomes
\begin{equation}
  \label{eq:U(1|1) phase transition:atiyah flop:U(1|1) r < 0}
  \mathcal{S}^{\mathrm{U}(1|1)}_{0 = 0, r < 0}
  = \bigoplus_{s' = 1}^{M} \mathscr{L}_{s'}^{-1}
  \oplus \bigoplus_{t' = 1}^{N} \mathbf{\Pi} \mathscr{L}^{-1}_{t'}
  \rightarrow Gr_{1|1}(\mathbb{C}^{M|N})_{s'',t''}
  \, .
\end{equation}
These are isomorphic but \emph{mildly topologically distinct} CY supervector bundles over a super-Grassmannian of size $\abs{r}$.

In freely going from $r> 0$ to $r < 0$, we have a mild topology change from \eqref{eq:U(1|1) phase transition:atiyah flop:U(1|1) r > 0} to \eqref{eq:U(1|1) phase transition:atiyah flop:U(1|1) r < 0} that we can also recognize as a flop transition of CY supervector bundles over a super-Grassmannian.

\subtitle{A Super-Grassmannian Generalization of a Birational Equivalence of CY Vector Bundles in Mathematics}

Notice that the above assignments of the $Q_s$'s such that $Q = \sum_s Q_s = 0$ are the same in examples \eqref{eq:U(1|1) phase transition:atiyah flop:U(1) r > 0}--\eqref{eq:U(1|1) phase transition:atiyah flop:U(1) r < 0} and \eqref{eq:U(1|1) phase transition:atiyah flop:U(1|1) r > 0}--\eqref{eq:U(1|1) phase transition:atiyah flop:U(1|1) r < 0}.
In other words, the $\mathrm{U}(1|1)$ model generalization of the reduced $\mathrm{U}(1)$ model mild topology change from $ \mathcal{S}^{\mathrm{U}(1)}_{0, r > 0}$  to $ \mathcal{S}^{\mathrm{U}(1)}_{0, r < 0}$ of CY vector bundles over a projective space, is the mild topology change from $\mathcal{S}^{\mathrm{U}(1|1)}_{0 = 0, r > 0}$  to $\mathcal{S}^{\mathrm{U}(1|1)}_{0 = 0, r < 0}$  of CY supervector bundles over a super-Grassmannian.
In particular, the latter reduces to the former when we specialize the $\mathrm{U}(1|1)$ GLSM to the $\mathrm{U}(1)$ GLSM.

Since according to \autoref{sec:U(1|1) phase transition:topological change:SV over Gr}, the low energy quantum NLSMs at $r > 0$ and $r < 0$ on \eqref{eq:U(1|1) phase transition:atiyah flop:U(1|1) r > 0} and \eqref{eq:U(1|1) phase transition:atiyah flop:U(1|1) r < 0} are just different phases of the \emph{same} underlying quantum $\mathrm{U}(1|1)$ GLSM, the low energy quantum NLSMs are actually related whence $\mathcal{S}^{\mathrm{U}(1|1)}_{0 = 0, r > 0}$  and $\mathcal{S}^{\mathrm{U}(1|1)}_{0 = 0, r < 0}$ are also \emph{related} and even \emph{smoothly-connected} as we go from $r > 0$ to $r< 0$ (because the singularity at $r =0$ is non-catastrophic due to worldsheet superinstantons, as explained in \autoref{sec:U(1|1) phase transition:topological change:worldsheet instantons}).

A relevant fact is also that $\mathcal{S}^{\mathrm{U}(1)}_{0, r > 0}$ and $\mathcal{S}^{\mathrm{U}(1)}_{0, r < 0}$ are birational to each other.

In sum, we have, via the phase transition in \autoref{sec:U(1|1) phase transition:topological change:SV over Gr} which relates and smoothly connects $\mathcal{S}^{\mathrm{U}(1|1)}_{0 = 0, r > 0}$   in \eqref{eq:U(1|1) phase transition:atiyah flop:U(1|1) r > 0} to  $\mathcal{S}^{\mathrm{U}(1|1)}_{0 = 0, r < 0}$ in \eqref{eq:U(1|1) phase transition:atiyah flop:U(1|1) r < 0}, a super-Grassmannian generalization of a birational equivalence of CY vector bundles in mathematics!

\section{Phase Transition of a \texorpdfstring{$\mathrm{U}(1|1)$}{U(1|1)} GLSM with a Certain Superpotential: Topology Change of a CY Complete Intersection of Quadrics in a Super-Grassmannian}
\label{sec: Phase Transition Quadrics}

In this section, we will specialize to the scale-invariant setting of \autoref{sec: Phases of U(1|1) with W} where the hypersurfaces are quadrics, i.e., of degree 2.
We will show that in the low energy limit, the quantum GLSM will become a quantum NLSM on either (i) a CY complete intersection of quadrics in a super-Grassmannian, or (ii) a CY double cover of a super-Grassmannian branched over a hypersurface in the super-Grassmannian.
We will also
(1) explain how the NLSM remains well-defined over a singularity when the FI parameter passes through zero,
(2) obtain a super-Grassmannian generalization of a homological projective duality for CY quadrics in mathematics, and
(3) obtain a parity-reversed mirror version of the aforementioned generalization.

\subsection{A \texorpdfstring{$\mathrm{U}(1|1)$}{U(1|1)} GLSM with a Certain Superpotential}
\label{subsec: U(1|1) with certain W}

\subtitle{A $\mathrm{U}(1|1)$ GLSM with a Certain Superpotential}

Let us now consider the $\mathrm{U}(1|1)$ GLSM with superpotential from \autoref{sec: Phases of U(1|1) with W} and turn off the $\tensor{\Phi}{^{\bar{1}}_s}$ and $\tensor{\Psi}{^{\bar{1}}_t}$ superfields.
Since the $\mathrm{U}(1|1)$ color index $i \in \{1, \bar{1}\}$ is now just $i=1$, we shall henceforth omit it from our notation in this section.
For our purposes, we would like to study the model when it is scale-invariant, i.e., when $m - n = \sum_{\alpha^+} q^+_{\alpha^+} - \sum_{\alpha^-} q^-_{\alpha^-}$, where we shall assume that $m \neq n$ for generality.

Specifically, we would like to have $m = 2M$, $n = 2N$, $\alpha^+ \in \{1, \dots, M\}$, $\alpha^- \in \{1, \dots, N\}$, and $q^{\pm}_{\alpha^{\pm}} = 2$ for all $\alpha^{\pm}$.
That would give us
(i) $2M$ Grassmann-even chiral matter superfields $\Phi_s$ (where $s \in \{1, \dots, 2M\}$),
(ii) $2N$ Grassmann-odd chiral matter superfields $\Psi_t$ (where $t \in \{1, \dots, 2N\}$),
(iii) $M$ Grassmann-even chiral matter superfields $P_+^{\alpha^+}$ and homogeneous polynomials $G^+_{\alpha^+}$ of degree 2,
and
(iv) $N$ Grassmann-odd chiral matter superfields $P_-^{\alpha^-}$ and homogeneous polynomials $G^-_{\alpha^-}$ of degree 2.

The $2M + 2N$ chiral matter superfields $\Phi_s$ and $\Psi_t$ would all have gauge charge of 1.
The $M + N$ chiral matter superfields $P_{\pm}^{\alpha_{\pm}}$ would all have gauge charge of $- 2$.

An admissible superpotential in the Lagrangian, i.e., one that is Grassmann-even and gauge-invariant, would, in this case, be
\begin{equation}
  \label{eq:quadrics:superpotential:standard}
  \mathcal{W}_{\text{super}}(P, \Phi, \Psi)
  = \sum_{\alpha^+}^M P_+^{\alpha^+} G^+_{\alpha^+}(\Phi, \Psi)
  + \sum_{\alpha^-}^N P_-^{\alpha^-} G^-_{\alpha^-}(\Phi, \Psi),
\end{equation}
where, in a convenient basis of $\Phi_s$'s and $\Psi_t$'s, the $G^{\pm}$ polynomials are
\begin{equation}
  \label{eq:quadrics:G+ and G- polynomials}
  \saveboxed{eq:quadrics:G+ and G- polynomials}{
    \begin{aligned}
      G^+_{\alpha^+}(\Phi, \Psi)
      &= \sum_{s, s'} {b^+_{(\alpha^+)}}^{s s'} \Phi_s \Phi_{s'}
      \\
      G^-_{\alpha^-}(\Phi, \Psi)
      &= \sum_{t, t'} {c^-_{(\alpha^-)}}^{t t'} \Psi_t \Psi_{t'}
    \end{aligned}
  }
\end{equation}
with
\begin{enumerate*}
  \item $s, s' \in \{1, \dots, 2 M\}$;

  \item $t, t' \in \{1, \dots, 2 N\}$;

  \item Grassmann-even $b^+$ coefficients being symmetric under the swop $s \leftrightarrow s'$;

  and

  \item Grassmann-odd $c^-$ coefficients being antisymmetric under the swop $t \leftrightarrow t'$.

\end{enumerate*}

Note that \eqref{eq:quadrics:superpotential:standard} can also be written as
\begin{equation}
  \label{eq:quadrics:superpotential}
  \mathcal{W}_{\text{super}}(P, \Phi, \Psi)
  = \sum_{s, s'} B^+(P)^{s s'} \Phi_s \Phi_{s'}
  + \sum_{t, t'} C^-(P)^{t t'} \Psi_t \Psi_{t'}
  \, ,
\end{equation}
where
\begin{equation}
  \label{eq:B+(P) definition}
  \saveboxed{eq:B+(P) definition}{
    B^+(P)^{s s'}
    \coloneq \sum_{\alpha^+ = 1}^M {b^+_{(\alpha^+)}}^{s s'} P_+^{\alpha^+}
  }
\end{equation}
can be regarded as the $(s, s')$ component of a $2M \times 2M$ symmetric Grassmann-even matrix $B^+(P)$ linear in the $P_+$'s;
and
\begin{equation}
  \label{eq:C-(P) definition}
  \saveboxed{eq:C-(P) definition}{
    C^-(P)^{t t'}
    \coloneq \sum_{\alpha^- = 1}^N {c^-_{(\alpha^-)}}^{t t'} P_-^{\alpha^-}
  }
\end{equation}
can be regarded as the $(t, t')$ component of a $2N \times 2N$ antisymmetric Grassmann-even matrix $C^-(P)$ linear in the $P_-$'s.

This means that we can also express the superpotential in \eqref{eq:quadrics:superpotential} as
\begin{equation}
  \label{eq:quadric:superpotential:matrix}
  \saveboxed{eq:quadric:superpotential:matrix}{
    \mathcal{W}_{\text{super}}
    = \mqty( [\Phi] & [\Psi] )
    \mqty(\dmat[0]{B^+(P), C^-(P)})
    \mqty( [\Phi]^{\intercal} \\ [\Psi]^{\intercal} )
  }
\end{equation}
where the components of the row supervector $( [\Phi] \, \, \, [\Psi] )$ are
\begin{equation}
  \label{eq:quadric:Y}
  \saveboxed{eq:quadric:Y}{
    \begin{aligned}
      \left[ \Phi \right]
      &\coloneq \left[ \Phi_1, \Phi_2, \dots, \Phi_{2M} \right]
      \\
      \left[ \Psi \right]
      &\coloneq \left[ \Psi_1, \Psi_2, \dots, \Psi_{2N} \right]
    \end{aligned}
  }
\end{equation}
and the matrix is a $2M|2N \times 2M|2N$ supermatrix.

\subtitle{The Mass-squared Supermatrix and Massless Superfields}

Notice that the supermatrix in \eqref{eq:quadric:superpotential:matrix} can actually be interpreted as a mass-squared supermatrix for the $\Phi$'s and $\Psi$'s.
Hence, any particular set of $\Phi$'s and $\Psi$'s in \eqref{eq:quadric:Y} which is massless would correspond to a set of zero eigenvalues of the mass-squared supermatrix.

In other words, the space of massless $\Phi$ and $\Psi$ superfields would correspond to the space of mass-squared supermatrices with zero eigenvalues, i.e., non-invertible supermatrices.
In this case, the conditions for the non-invertibility of the supermatrix in \eqref{eq:quadric:superpotential:matrix} are either
(i) the non-invertibility of $B^+(P)$ \emph{and} $C^-(P)$,
(ii) the vanishing of the superdeterminant of the supermatrix if $C^-(P)$ is invertible (see \eqref{eq:plucker:prelim:sdet}),
or
(iii) the vanishing of the \emph{inverse} superdeterminant of the supermatrix if $B^+(P)$ is invertible (see \eqref{eq:plucker:prelim:inverse sdet}).
Let us look at these conditions individually.

Condition (i) is easy to analyze.
For symmetric matrices such as $B^+(P)$, its non-invertibility corresponds to
\begin{equation}
  \label{eq:quadrics:det B+ = 0}
  \det[B^+(P)] = 0
  \, .
\end{equation}
This is the zero-locus of a Grassmann-even polynomial of degree $2M$ in $P_+$.
For antisymmetric matrices such as $C^-(P)$, its non-invertibility corresponds to
\begin{equation}
  \label{eq:quadrics:pf C- = 0}
  \pf[C^-(P)] = 0
  \, ,
\end{equation}
where $\pf$ is the Pfaffian.
This is the zero-locus of a Grassmann-even polynomial of degree $2N$ in $P_-$.
Together, condition (i) would be the intersection of \eqref{eq:quadrics:det B+ = 0} \emph{and} \eqref{eq:quadrics:pf C- = 0}, i.e., supermatrices that satisfy
\begin{equation}
  \label{eq:7}
  \left\{ \det[B^+(P)] = 0 \right\}
  \cap
  \left\{ \pf[C^-(P)] = 0 \right\}
  \, .
\end{equation}
Note that condition (i) corresponds physically to the condition that there is at least one $\Phi$ \emph{and} one $\Psi$ superfield that is massless.

Condition (ii) assumes that $\pf[C^-(P)] \neq 0$, i.e., \emph{none} of the $\Psi$ superfields are massless.
Then, the vanishing of the superdeterminant of the supermatrix would, via~\eqref{eq:plucker:prelim:sdet} while noting that $\pf[C^-(P)]^2 = \det[C^-(P)] \neq 0$, be given by
\begin{equation}
  \label{eq:3}
  \flatfrac{\det[B^+(P)]}{\det[C^-(P)]} = 0
  \, .
\end{equation}
This is the zero-locus of a Grassmann-even polynomial of degree $2M - 2N$ in the $P$'s.

Condition (iii) assumes that $\det[B^+(P)] \neq 0$, i.e., \emph{none} of the $\Phi$ superfields are massless.
Then, the vanishing of the inverse superdeterminant of the supermatrix would, via~\eqref{eq:plucker:prelim:inverse sdet}, be given by
\begin{equation}
  \label{eq:5}
  \flatfrac{\det[C^-(P)]}{\det[B^+(P)]} = 0
  \, .
\end{equation}
This is the zero-locus of a Grassmann-even polynomial of degree $2N - 2M$ in the $P$'s.

\subtitle{The Space of Massless Superfields is Determined by $\flatfrac{\det[B^+(P)]}{\det[C^-(P)]} = 0$}

The question now is, which of these three conditions do we have when our model is scale-invariant?
To answer this question, first, recall the scale-invariant condition from \autoref{sec: Phases of U(1|1) with W}, i.e., $m - n = \sum_{\alpha^+} q^+_{\alpha^+} - \sum_{\alpha^-} q^-_{\alpha^-}$.
Here, $m$ and $n$ are the even and odd dimensions of a supermanifold $\mathcal{M}$ defined by the $\Phi_s$'s and $\Psi_t$'s; each $q^+_{\alpha^+}$ corresponds to the degree of an even hypersurface in $\mathcal{M}$; and each $q^-_{\alpha^-}$ corresponds to the degree of an odd hypersurface in $\mathcal{M}$.

Second, let us assume that $q^-_{\alpha^-} = 0$ for all $\alpha^-$, whence there are only even hypersurfaces in $\mathcal{M}$, of which we shall also restrict our analysis to at most two even hypersurfaces, as this is all we will need.
The scale-invariant condition then implies that the degrees of the even hypersurfaces depend on the values of the integers $m$ and $n$ in the following manner.
If $m > n$, the sum of the degrees of the even hypersurfaces is a positive integer -- when there is only \emph{one} hypersurface, its degree is positive; when there are \emph{two} hypersurfaces, the sum of the degrees is positive.
If $m < n$, the sum of the degrees of the even hypersurfaces is a negative integer -- when there is only \emph{one} hypersurface, its degree is negative; when there are \emph{two} hypersurfaces, the sum of the degrees is negative.

In other words, the sign of the (sum of the) degree(s) of the even hypersurface(s) follows the sign of $m - n$, i.e., the superdimension of $\mathcal{M}$.

Third, notice that the space of massless $\Phi$ and $\Psi$ superfields as given by the equation defined by either of the three conditions (i) \eqref{eq:7}, (ii) \eqref{eq:3}, or (iii) \eqref{eq:5}, can be understood as (intersections of) the zero-locus of Grassmann-even polynomials in the space of $P_{\pm}$ superfields.
Such a space of $P_{\pm}$ superfields would be a supermanifold $\mathcal{P}$, that would live in an ambient superspace endowed with $M$ even and $N$ odd coordinates corresponding to the $M$ even $P_+$ and $N$ odd $P_-$ superfields, respectively.

In other words, the space of massless $\Phi$ and $\Psi$ superfields can be understood as hypersurfaces in $\mathcal{P}$ which are specifically even.

In particular, condition (i) defined by \eqref{eq:7} would be an intersection of two even hypersurfaces in $\mathcal{P}$ of degrees $2M$ and $N$, where the sum of the degrees is $2M + N$ is always positive no matter the sign of $M - N$.

Condition (ii) defined by \eqref{eq:3} would be an even hypersurface in $\mathcal{P}$ of degree $2M - 2N$, and is always the same sign as $M - N$.

Condition (iii) defined by \eqref{eq:5} would be an even hypersurface in $\mathcal{P}$ of degree $2N - 2M$, and is always opposite in sign to $M - N$.

Notice that only condition (ii) is consistent with our earlier observation that the sign of the (sum of the) degree(s) of the even hypersurface(s) in $\mathcal{P}$ must follow the sign of the superdimension of $\mathcal{P}$, i.e., $M - N$.

Thus, the answer to our question is that, in our scale-invariant case, we would only have condition (ii).

In other words, the space of massless $\Phi$ and $\Psi$ superfields would correspond to the hypersurface in $\mathcal{P}$ defined by \eqref{eq:3}.

\subsection{Quantum Aspects: The Low Energy Theory}
\label{sec:quadrics:quantum low energy}

Applying a similar analysis to that in \autoref{subsec: Phases U(1|1) with W} and \autoref{subsec:quantum low energy U(1|1) with W}, we find that our scale-invariant quantum GLSM with superpotential $\mathcal{W}_{\text{super}}$ has four phases and two independent branches.
Branch 1 is composed of the $r(\text{i}) > 0$ and $r(\text{ii}) < 0$ phases, while Branch 2 is composed of the $r(\text{ii}) > 0$ and $r(\text{i}) < 0$ phases.
Let us look at these four phases in detail.

\subtitle{The $r(\text{i}) > 0$ Phase of the Quantum Model at Low Energy: A Quantum NLSM on a CY Complete Intersection of $M + N$ Quadrics in $Gr_{1|0}(\mathbb{C}^{2M|2N})$}

Via a similar analysis to that in \autoref{subsec: Phases U(1|1) with W} and \autoref{subsec:quantum low energy U(1|1) with W}, when $r(\text{i}) > 0$, at low energy, our scale-invariant quantum GLSM with superpotential $\mathcal{W}_{\text{super}}$ in \eqref{eq:quadric:superpotential:matrix} will become a quantum NLSM on a CY complete intersection of $M$ even and $N$ odd quadrics (degree-2 hypersurfaces) $G^+_{\alpha^+}(\phi, \psi) = 0$ ($\alpha^+ \in \{1, \dots, M\}$) and $G^-_{\alpha^-}(\phi, \psi) = 0$ ($\alpha^- \in \{1, \dots, N\}$), as defined in \eqref{eq:quadrics:G+ and G- polynomials}, in $Gr_{1|0}(\mathbb{C}^{2M|2N})$ of size $r$.

\subtitle{The $r(\text{ii}) < 0$ Phase of the Quantum Model at Low Energy: A Quantum NLSM on $\widetilde{Gr_{1|0}(\mathbb{C}^{M|N})}$ Branched over the Hypersurface $\flatfrac{\det[B^+(p)]}{\det[C^-(p)]} = 0$ in $Gr_{1|0}(\mathbb{C}^{M|N})$}

Via a similar analysis to that in \autoref{subsec: Phases U(1|1) with W} and \autoref{subsec:quantum low energy U(1|1) with W}, when $r(\text{ii}) < 0$, at low energy, our scale-invariant quantum GLSM with superpotential $\mathcal{W}_{\text{super}}$ in \eqref{eq:quadric:superpotential:matrix} will become a quantum LG $\mathbb{Z}_2$-orbifold with effective superpotential defined on a supervector bundle $\mathcal{S}\mathcal{G}^{\mathrm{U}(1)}_{\vec{2}|\vec{2}}$ over a weighted projective superspace $\WCP^{M-1|N}_{(\vec{2}|\vec{2})}$ of size $|r|$, where the even and odd coordinates on the superbase $\WCP^{M-1|N}_{(\vec{2}|\vec{2})}$ are spanned by $p_+^{\alpha^+}$ and $p_-^{\alpha^-}$, and the even and odd fibers over the superbase are spanned by $\phi_s$ and $\psi_t$.

Recall at this point, that in order to arrive at this LG $\mathbb{Z}_2$-orbifold, one had to go to an energy scale that was lower than the mass scale of the $P^{\alpha^{\pm}}_\pm$'s, whence they could be integrated out of the action and be replaced by their vev's where appropriate (such as in \eqref{eq:U(1|1) with W:phases:superpotential W:r(ii) < 0}).

Note also at this point, that in this case where $\mathcal{W}_{\text{super}}$ in \eqref{eq:quadric:superpotential:matrix} can be interpreted as a mass-squared term for the $\Phi_s$'s and $\Psi_t$'s, at an energy scale that is lower than their mass scale, they can be integrated out.

Since our quantum GLSM is scale-invariant, we are free to consider an energy scale that is lower than the mass scale of the $\Phi_s$'s and $\Psi_t$'s, but still higher than the mass scale of the $P^{\alpha^{\pm}}_\pm$'s.
In doing so, we can integrate out the massive $\Phi_s$'s and $\Psi_t$'s, while retaining the $P^{\alpha^{\pm}}_\pm$'s.
Consequently, at this particular energy scale, our quantum GLSM would be one in just the $P^{\alpha^{\pm}}_\pm$'s, but without the $\Phi_s$'s, $\Psi_t$'s, or the superpotential.
Thus, at low energy, it would become a quantum $\mathbb{Z}_2$ gauged NLSM in the $p_{\pm}$ fields on  $\WCP^{M-1|N}_{(\vec{2}|\vec{2})} \cong Gr_{1|0}(\mathbb{C}^{M|N})$.

As the $\mathbb{Z}_2$ gauge symmetry means that $p(z,\bar z)$ and its distinct gauge-transformed $p^{\prime}(z, \bar z)$ are \emph{physically-equivalent} scalar fields on the NLSM worldsheet $\mathcal{C}$ with coordinates $(z, \bar z)$, it would mean that if $p(z,\bar z): \mathcal{C}_{z, \bar z} \to p \in Gr_{1|0}(\mathbb{C}^{M|N})$, we would have $p^{\prime}(z,\bar z) \equiv p(z, \bar z): \mathcal{C}_{z, \bar z} \to p \in Gr_{1|0}(\mathbb{C}^{M|N})$.
Then, some thought would reveal that one can reinterpret the quantum $\mathbb{Z}_2$ gauged NLSM as a quantum \emph{ungauged} NLSM on a double cover $\widetilde{Gr_{1|0}(\mathbb{C}^{M|N})}$ of $Gr_{1|0}(\mathbb{C}^{M|N})$.\footnote{%
  Indeed, in the ungauged NLSM, $p(z,\bar z)$ and   $p^{\prime}(z,\bar z)$ would map to two different points in $\widetilde{Gr_{1|0}(\mathbb{C}^{M|N})}$ that can ultimately be identified as the same point in $Gr_{1|0}(\mathbb{C}^{M|N})$, which is exactly the description of the $\mathbb{Z}_2$ gauged NLSM.
}

All that we have said so far assumes that $\mathcal{W}_{\text{super}}$ is such that \emph{all} the $\Phi_s$'s and $\Psi_t$'s are massive.
However, one could possibly have, in our scale-invariant set-up, specific $P^{\alpha^{\pm}}_{\pm}$'s where $\mathcal{W}_{\text{super}}$ is such that some of the $\Phi_s$'s are actually massless; these $\Phi_s$'s would then not be integrated out.
In this instance, at the aforementioned energy scale, our quantum GLSM would then be one in these specific $P^{\alpha^{\pm}}_\pm$'s and the remaining $\Phi_s$'s.
Thus, at low energy, since the $\phi$'s have no vev, it would become a quantum $\mathbb{Z}_2$ gauged NLSM in the specific $p$ fields on $\mathscr{H}$, where $\mathscr{H}$ is defined by $\flatfrac{\det[B^+(p)]}{\det[C^-(p)]} = 0$.

Similarly, one can reinterpret this quantum $\mathbb{Z}_2$ gauged NLSM as a quantum \emph{ungauged} NLSM in the specific $p$ fields on $\mathscr{H}$.\footnote{%
  \label{ft:quadrics:hypersurface no double cover}%
  The alert reader would notice that the target space of the ungauged NLSM continues to be $\mathscr{H}$ and not its double cover $\widetilde{\mathscr{H}}$, unlike earlier.
  The reason is because $\flatfrac{\det[B^+(p)]}{\det[C^-(p)]} = 0$ which defines $\mathscr{H}$, involves polynomials in $p$ of degree $2\mathbb{Z}$; hence, the $\mathbb{Z}_2$-action on $p$ would leave $\mathscr{H}$ invariant.
  In other words, in the ungauged NLSM, $p(z,\bar z)$ and   $p^{\prime}(z,\bar z)$ would now map to the same point in $\mathscr{H}$, whence there is no double cover.
}

In sum, if we consider the quantum GLSM at the aforementioned energy scale, which is equivalent to choosing a certain value of $|r|$, at low energy, it would become a quantum NLSM on $\widetilde{Gr_{1|0}(\mathbb{C}^{M|N})}$ branched over the hypersurface $\flatfrac{\det[B^+(p)]}{\det[C^-(p)]} = 0$ in $Gr_{1|0}(\mathbb{C}^{M|N})$.\footnote{%
  A double cover $\widetilde{\mathcal{X}}$ of a space $\mathcal{X}$, branched over a hypersurface $\mathcal{H} \subset \mathcal{X}$, is a space where for a point $p \in \mathcal{X}$,
  \begin{enumerate*}
    \item if $p \in \mathcal{H} \subset \mathcal{X}$, it would correspond to a \emph{single} point in $\mathcal{H} \subset \widetilde{\mathcal{X}}$,

    and

    \item if $p \notin \mathcal{H} \subset \mathcal{X}$, it would correspond to \emph{two} different points in $\widetilde{\mathcal{X}}$.
  \end{enumerate*}
}

\subtitle{The $r(\text{ii}) > 0$ Phase of the Quantum Model at Low Energy: A Quantum NLSM on $\widetilde{Gr_{1|0}(\mathbb{C}^{N|M})}$ Branched over the Hypersurface $\flatfrac{\det[B^+(p_{(\mathbf{\Pi})})]}{\det[C^-(p_{(\mathbf{\Pi})})]} = 0$ in $Gr_{1|0}(\mathbb{C}^{N|M})$}

By applying the same analysis as the above for when $r(\text{ii}) > 0$, we find that our scale-invariant quantum GLSM with superpotential $\mathcal{W}_{\text{super}}$ in \eqref{eq:quadric:superpotential:matrix} will, at low energy, become a quantum NLSM on $\widetilde{Gr_{1|0}(\mathbb{C}^{N|M})}$ branched over the hypersurface $\flatfrac{\det[B^+(p_{(\mathbf{\Pi})})]}{\det[C^-(p_{(\mathbf{\Pi})})]} = 0$ in $Gr_{1|0}(\mathbb{C}^{N|M})$.
Here,
(i) $B^+(p_{(\mathbf{\Pi})})$ is a $2N \times 2N$ symmetric Grassmann-even matrix linear in the $p_{+ (\mathbf{\Pi})} = \mathbf{\Pi} p_-$ fields,
and (ii) $C^-(p_{(\mathbf{\Pi})})$ is a $2M \times 2M$ antisymmetric Grassmann-even matrix linear in the $p_{- (\mathbf{\Pi})} = \mathbf{\Pi} p_+$ fields.

\subtitle{The $r(\text{i}) < 0$ Phase of the Quantum Model at Low Energy: A Quantum NLSM on a CY Complete Intersection of $N + M$ Quadrics in $Gr_{1|0}(\mathbb{C}^{2N|2M})$}

By applying the same analysis as that at the start of this subsection for when $r(\text{i}) < 0$, we find that our scale-invariant quantum GLSM with superpotential $\mathcal{W}_{\text{super}}$ in \eqref{eq:quadric:superpotential:matrix} will, at low energy, become a quantum NLSM on a CY complete intersection of $N$ even and $M$ odd quadrics $G^+_{\alpha^-(\mathbf{\Pi})}(\phi, \psi) = 0$ ($\alpha^- \in \{1, \dots, N\}$) and $G^-_{\alpha^+(\mathbf{\Pi})}(\phi, \psi) = 0$ ($\alpha^+ \in \{1, \dots, M\}$) in $Gr_{1|0}(\mathbb{C}^{2N|2M})$ of size $\abs{r}$.
Here, the $G^{\pm}_{\alpha^{\mp}(\mathbf{\Pi})}$ polynomials are the parity-reversed version of $G^{\pm}_{\alpha^{\pm}}$, just like the $p_{(\mathbf{\Pi})}$ fields above.

\subsection{Scale-invariance and Applications to Mathematics}
\label{sec:quadrics:cy}

As explained in \autoref{subsec: applications to math U(1|1) with W}, since our our quantum GLSM is scale-invariant, its corresponding low energy quantum NLSM ought to also be scale and therefore conformally-invariant~\cite{papadopoulos2024scaleconformalinvariance2d}.
We already know from \autoref{subsec: applications to math U(1|1) with W} that this means that the target spaces of the quantum NLSMs in the $r(\text{i}) > 0$ and $r(\text{i}) < 0$ phases are CY; in fact, we even mathematically verified this (see~\autoref{app:verify cy:hypersurfaces in Gr}).
Now that we also have quantum NLSMs in the remaining two $r(\text{ii}) < 0$ and $r(\text{ii}) > 0$ phases described in \autoref{sec:quadrics:quantum low energy}, the scale-invariance would also mean that the target spaces of these quantum NLSMs must also be CY.

Recall that the target space of the low energy quantum NLSM of the $r(\text{ii}) < 0$ phase is $\widetilde{Gr_{1|0}(\mathbb{C}^{M|N})}$ branched over the hypersurface $\flatfrac{\det[B^+(p)]}{\det[C^-(p)]} = 0$ of degree $2M - 2N$ in $Gr_{1|0}(\mathbb{C}^{M|N})$.
Notice that the degree of the even hypersurface, i.e., $2(M - N)$, is \emph{twice} the difference between the even and odd dimensions of the ambient space that is doubly covered and branched over it.
Equivalently, this means that the difference between the even and odd dimensions of the ambient space that is doubly covered matches \emph{half} of the degree of the even hypersurface it is branched over.

Recall also that the target space of the low energy quantum NLSM of the $r(\text{ii}) > 0$ phase is $\widetilde{Gr_{1|0}(\mathbb{C}^{N|M})}$ branched over the hypersurface $\flatfrac{\det[B^+(p_{\mathbf{\Pi}})]}{\det[C^-(p_{\mathbf{\Pi}})]} = 0$ of degree $2N - 2M$ in $Gr_{1|0}(\mathbb{C}^{N|M})$.
Notice again that the degree of the even hypersurface, i.e., $2(N - M)$, is \emph{twice} the difference between the even and odd dimensions of the ambient space that is doubly cover and branched over it.
Equivalently, this means that the difference between the even and odd dimensions of the ambient space that is doubly covered matches \emph{half} of the degree of the even hypersurface it is branched over.

Thus, we have a physical derivation of a \emph{novel} mathematical result that the double cover of a super-Grassmannian branched over an even hypersurface is CY, if and only if the difference between the even and odd dimensions of the super-Grassmannian is \emph{half} the degree of the even hypersurface branched over.

Indeed, one can actually mathematically verify (see \autoref{sec:verify cy:double cover}) this statement to be true!

This vindicates, for the final time, our application to mathematics of our otherwise nonunitary supergroup GLSM via a study of its space of supersymmetric states.

\subsection{Phase Transition of the Quantum Model: Topology Change of a CY Complete Intersection of Quadrics in a Super-Grassmannian}
\label{subsec: Phase Transition N Quadrics in SGr}

\subtitle{A Topology Change from an $r(\text{i}) > 0$ to $r(\text{ii}) < 0$ Phase Transition}

In Branch 1, as we are free to go from $r(\text{i}) > 0$ to $r(\text{ii}) < 0$ in a phase transition of the quantum GLSM with superpotential $\mathcal{W}_{\text{super}}$,\footnote{%
  This possibility of going smoothly from the $r(\text{i}) > 0$ to $r(\text{ii}) < 0$ phase of the quantum GLSM with superpotential  will be justified in \autoref{subsec: Worldsheet Instantons for U(1|1) GLSM with W}.
}
at low energy, we will have a quantum NLSM on
(I) a CY complete intersection of $M + N$ quadrics in $Gr_{1|0}(\mathbb{C}^{2M|2N})$,
that goes to a quantum NLSM on
(II) a CY double cover  $\widetilde{Gr_{1|0}(\mathbb{C}^{M|N})}$ branched over the hypersurface $\flatfrac{\det[B^+(p)]}{\pf[C^-(p)]} = 0$ in $Gr_{1|0}(\mathbb{C}^{M|N})$.

In other words, since the spaces (I) and (II) are \emph{topologically distinct}, we have a topology change in the target space of the low energy quantum NLSM from (I) to (II) as we freely go from $r(\text{i}) > 0$ to $r(\text{ii}) < 0$!

\subtitle{A Topology Change from an $r(\text{i}) < 0$ to $r(\text{ii}) > 0$ Phase Transition}

Similarly, in Branch 2, as we are free to go from $r(\text{i}) < 0$ to $r(\text{ii}) > 0$ in a phase transition of the quantum GLSM with superpotential $\mathcal{W}_{\text{super}}$, at low energy, we will have a quantum NLSM on
(I$'$) a CY complete intersection of $N + M$ quadrics in $Gr_{1|0}(\mathbb{C}^{2N|2M})$,
that goes to a quantum NLSM on
(II$'$) a CY double cover $\widetilde{Gr_{1|0}(\mathbb{C}^{N|M})}$ branched over the hypersurface $\flatfrac{\det[B^+(p_{(\mathbf{\Pi})})]}{\det[C^-(p_{(\mathbf{\Pi})})]} = 0$ in $Gr_{1|0}(\mathbb{C}^{N|M})$.

In other words, since the spaces (I$'$) and (II$'$) are \emph{topologically distinct}, we have a topology change in the target space of the low energy quantum NLSM from (I$'$) to (II$'$) as we freely go from $r(\text{i}) < 0$ to $r(\text{ii}) > 0$!

\subsection{Phase Transition of the Familiar Quantum \texorpdfstring{$\mathrm{U}(1)$}{U(1)} Model: Topology Change of a CY Complete Intersection of Quadrics in Projective Space}
\label{subsec: transition reduced U(1) quadrics}

Just like in \autoref{subsec: reduced U(1) model}, the quantum GLSM with superpotential $\mathcal{W}_{\text{super}}(P, \Phi, \Psi)$ ought to reduce to the familiar quantum $\mathrm{U}(1)$ GLSM with superpotential $\mathcal{W}^{\mathrm{U}(1)}_{\text{super}}(P_+, \Phi^1)$ when we (i) turn off the $\tensor{\Psi}{^{1}_t}$ superfields, i.e., et $N = 0$,
and (ii) turn off the $P_-^{\alpha^-}$ superfields.

Since $\mathcal{W}^{\mathrm{U}(1)}_{\text{super}}$ will now consist only of the $P_+$ and $\Phi$ superfields, one can see that matrix in the superpotential \eqref{eq:quadric:superpotential:matrix} will just be the $B^+(P)$ matrix.

Since as explained in \autoref{subsec: reduced U(1) model}, the phases in Branch 2 cannot be defined in the reduced model, we shall only discuss the phases in Branch 1, where we will simply write $r(\text{i}) > 0$ as $r > 0$ and $r(\text{ii}) < 0$ as $r < 0$.

\subtitle{The $r > 0$ Phase at Low Energy}

In the $r > 0$ phase of the $\mathrm{U}(1)$ GLSM with superpotential $\mathcal{W}^{\mathrm{U}(1)}_{\text{super}}(P^+, \Phi)$, at low energy, we will have a quantum NLSM on a CY complete intersection of $M$ quadrics in $\mathbb{CP}^{2M-1}$.

\subtitle{The $r < 0$ Phase at Low Energy}

In the $r < 0$ phase of the $\mathrm{U}(1)$ GLSM with superpotential $\mathcal{W}^{\mathrm{U}(1)}_{\text{super}}(P^+, \Phi)$, when all the $\Phi_s$'s are massive, at low energy, we will have a $\mathbb{Z}_2$ gauged quantum NLSM on $\mathbb{CP}^{M-1}$, which can be reinterpreted as a quantum \emph{ungauged} NLSM on a double cover $\widetilde{\CP^{M-1}}$ of $\CP^{M-1}$.

When there are massless $\Phi_s$'s, at low energy, we will have a $\mathbb{Z}_2$ gauged quantum NLSM on the hypersurface $\det[B^+(p)] = 0$ of degree $2M$ in $\mathbb{CP}^{M-1}$, which can be reinterpreted as a quantum \emph{ungauged} NLSM on the hypersurface $\det[B^+(p)] = 0$ of degree $2M$ in $\CP^{M-1}$ (see \autoref{ft:quadrics:hypersurface no double cover}).

In sum, at low energy, we have a quantum NLSM on $\widetilde{\CP^{M-1}}$ branched over the hypersurface $\det[B^+(p)] = 0$ of degree $2M$ in $\CP^{M-1}$.

\subtitle{A Topology Change from an $r > 0$ to $r < 0$ Phase Transition}

Again, as we are free go from $r > 0$ to $r < 0$ in a phase transition of the quantum $\mathrm{U}(1)$ GLSM with superpotential $\mathcal{W}^{\mathrm{U}(1)}_{\text{super}}(P^+, \Phi)$, at low energy, we will have a quantum NLSM on
(I) a CY complete intersection of $M$ quadrics in $\mathbb{CP}^{2M-1}$,
that goes to a quantum NLSM on
(II) a double cover $\widetilde{\CP^{M-1}}$ branched over a hypersurface $\det[B^+(p)] = 0$ of degree $2M$ in $\CP^{M-1}$.

In other words, since the spaces (I) and (II) are \emph{topologically distinct}, we have a topology change in the target space of the low energy quantum NLSM from (I) to (II) as we freely go from $r > 0$ to $r < 0$, as we should~\cite[$\S$2.8]{Caldararu:2010ljp}.

\subsection{The ``Singularity'' at \texorpdfstring{$r=0$}{r = 0} and a Smooth Phase Transition}
\label{subsec: Worldsheet Instantons for U(1|1) GLSM with W}

\subtitle{The ``Singularity'' at $r=0$ and a Smooth Phase Transition from $r > 0$ to $r < 0$}

Notice from the expression of the potential energy in \eqref{ppot} with the $\tensor{\phi}{^{\bar{1}}_s}$ and $\tensor{\psi}{^{\bar{1}}_t}$ fields turned off as per this section, that at $r = 0$, $\MSUSY$ would develop a new \emph{non-compact} branch where the remaining $\phi_s$'s, $\psi_t$'s, and $p^{\pm}_{\alpha^{\pm}}$'s must vanish while its coordinate $\sigma$ is no longer constrained and can even be infinitely large.

This non-compactness of $\MSUSY$ leads to a singularity in the theory where the $r > 0$ and $r < 0$ regions of parameter space seem completely disconnected by the singular $r = 0$ point with different physics.
This suggests that the phase transition in \autoref{subsec: Phase Transition N Quadrics in SGr} when we go from $r > 0$ to $r < 0$ may not be smooth and therefore well-defined after all.

The question therefore, is whether $\MSUSY$ actually develops a new non-compact branch at $r = 0$, or at some value of $t = \mathbf{i} r + \frac{\theta}{2\pi}$, in the \emph{quantum} theory.
If it is the latter, then, like the model without superpotential as discussed in \autoref{sec:U(1|1) phase transition:topological change:worldsheet instantons}, the transition from $r > 0$ to $r < 0$ passing through $r = 0$ would actually be smooth.

Let us now ascertain the answer to this question.
Since we are again concerned with the non-compactness of $\MSUSY$, let us consider its coordinate $\sigma$ to be large.

When $\sigma$ is large, the $\phi_s$, $\psi_t$, and $p_{\pm}^{\alpha^{\pm}}$ matter fields would be massive (see \eqref{ppot}), and they can be integrated out (along with the other fields in their chiral matter multiplets because of supersymmetry).
We are then left with a pure theory in the gauge multiplet fields, where $\sigma$ is the lowest component scalar field.

As we effectively have a $\mathrm{U}(1)$ GLSM with
\begin{enumerate*}
  \item a superpotential,

  \item large $\Sigma$,

  \item $2M + 2N$ massive Grassmann-even $\Phi_s$ and Grassmann-odd $\Phi_t$ superfields of gauge charge 1,

  and

  \item $M + N$ massive Grassmann-even $P^{\alpha^+}_+$ and Grassmann-odd $P^{\alpha^-}_-$ superfields of gauge charge $-2$,
\end{enumerate*}
we can simply adapt the calculation leading up to \cite[eq.~(15.167)]{Hori:2003ic}  for a single $\mathrm{U}(1)$ gauge group, to arrive at a relation that $t$ must obey, from which one can determine where a non-compact branch of $\MSUSY$ appears in the quantum theory.

Specifically, in integrating out the massive fields, we would have an effective twisted superpotential for $\Sigma$ given by
\begin{equation}
  \label{eq:quadrics:twisted superpotential}
  \begin{aligned}
    \widetilde{W}
    = \Sigma \Biggl(
      & 2 \pi \mathbf{i} t(\mu)
      - \sum_{s = 1}^{2M} \left[
        \log \left( \frac{\Sigma}{\mu} \right) - 1
      \right]
      + \sum_{\alpha^+ = 1}^M 2 \left[
        \log \left( - \frac{2 \Sigma}{\mu} \right) - 1
      \right]
      \\
      &
      + \sum_{t = 1}^{2N} \left[
        \log \left( \frac{\Sigma}{\mu} \right) - 1
      \right]
      - \sum_{\alpha^- = 1}^N 2 \left[
        \log \left( - \frac{2 \Sigma}{\mu} \right) - 1
      \right]
    \Biggr)
    \, .
  \end{aligned}
\end{equation}
Then, via the same analysis which led us from \eqref{eq:U(1|1) phase transition:W-eff} to \eqref{eq:U(1|1) phase transition:coulomb branch:exponent:scale-invariant}, we have
\begin{equation}
  4^{M-N} = e^{- 2 \pi \mathbf{i} t}
  \, .
\end{equation}

From this relation that $t$ must obey, it must mean that in general, the solutions are such that $t \neq 0$.

In conclusion, in the quantum theory, $\MSUSY$ actually develops a new non-compact branch not at $r = 0$, but at some typically nonzero imaginary value of $t = \mathbf{i} r$!
Therefore, away from nonzero $r = - \mathbf{i} t$, the phase transition in \autoref{subsec: Phase Transition N Quadrics in SGr} when we go from $r > 0$ to $r < 0$ passing through $r = 0$, would actually be \emph{smooth} and therefore well-defined.

As such we would have a non-catastrophic topology change of the low energy NLSM's target space from (I) a CY complete intersection of $M + N$ quadrics in $Gr_{1|0}(\mathbb{C}^{2M|2N})$ to (II) a CY double cover $\widetilde{Gr_{1|0}(\mathbb{C}^{M|N})}$  branched over a hypersurface of degree $2M - 2N$ in $Gr_{1|0}(\mathbb{C}^{M|N})$, or from (II$'$) a CY double cover $\widetilde{Gr_{1|0}(\mathbb{C}^{N|M})}$ branched over a hypersurface of degree $2N - 2M$ in $Gr_{1|0}(\mathbb{C}^{N|M})$ to (I$'$) a CY complete intersection of $N + M$ quadrics in $Gr_{1|0}(\mathbb{C}^{2N|2M})$.

\subsection{A Super-Grassmannian Generalization of a Homological Projective Duality for CY Quadrics in Mathematics \label{subsec: Super HPD}}

\subtitle{A CY Quadrics to CY Branched Double Cover Transition}

In \autoref{subsec: Phase Transition N Quadrics in SGr}, we saw that in Branch 1, at $r > 0$ and $r < 0$, the low energy quantum NLSM on (I) a CY complete intersection of $M$ even and $N$ odd quadrics $G^+_{\alpha^+} = 0$ ($\alpha^+ \in \{1, \dots, M\}$) and $G^-_{\alpha^-} = 0$ ($\alpha^- \in \{1, \dots, N\}$) in $Gr_{1|0}(\mathbb{C}^{2M|2N})$, and (II) a CY double cover $\widetilde{Gr_{1|0}(\mathbb{C}^{M|N})}$ branched over the hypersurface $\flatfrac{\det[B^+(p)]}{\pf[C^-(p)]} = 0$ of degree $2M - 2N$ in $Gr_{1|0}(\mathbb{C}^{M|N})$, respectively, are just different phases of the \emph{same} underlying quantum GLSM with superpotential.

Thus, the low energy quantum NLSMs are actually related, whence (I) and (II) are also \emph{related} and even \emph{smoothly-connected} as we go from $r > 0$ to $r< 0$ (because the singularity at $r = 0$ is non-catastrophic, as explained in \autoref{subsec: Worldsheet Instantons for U(1|1) GLSM with W}).

\subtitle{A CY Branched Double Cover to CY Quadrics Transition}

In \autoref{subsec: Phase Transition N Quadrics in SGr}, we  saw that in Branch 2, at $r > 0$ and $r < 0$, the low energy quantum NLSM on (II$'$) a CY double cover $\widetilde{Gr_{1|0}(\mathbb{C}^{N|M})}$ branched over the hypersurface $\flatfrac{\det[B^+(p_{(\mathbf{\Pi})})]}{\det[C^-(p_{(\mathbf{\Pi})}))]} = 0$ in $Gr_{1|0}(\mathbb{C}^{N|M})$, and (I$'$) a CY complete intersection of $N$ even and $M$ odd quadrics $G^+_{\alpha^- (\mathbf{\Pi})} = 0$ ($\alpha^- \in \{1, \dots, N\}$) and $G^-_{\alpha^+ (\mathbf{\Pi})} = 0$ ($\alpha^+ \in \{1, \dots, M\}$) in $Gr_{1|0}(\mathbb{C}^{2N|2M})$, respectively, are just different phases of the \emph{same} underlying quantum GLSM with superpotential.

Thus, the low energy quantum NLSMs are actually related, whence (II$'$) and (I$'$) are also \emph{related} and even \emph{smoothly-connected} as we go from $r > 0$ to $r< 0$ (because the singularity at $r = 0$ is non-catastrophic, as explained in \autoref{subsec: Worldsheet Instantons for U(1|1) GLSM with W}).

\subtitle{A Super-Grassmannian Generalization of a Homological Projective Duality for CY Quadrics in Mathematics}

At any rate, in carrying out the reduction to the familiar $\mathrm{U}(1)$ GLSM with superpotential (via \autoref{subsec: transition reduced U(1) quadrics}), we find that with regard to the corresponding low energy quantum NLSM, in place of (I) and (II), we would have
(\textbf{i}) a CY complete intersection of $M$ quadrics $G^+_{\alpha^+} = 0$ in $\CP^{2M-1}$,
and
(\textbf{ii}) a CY double cover $\widetilde{\CP^{M-1}}$ branched over a hypersurface $\det[B^+(p)] = 0$ of degree $2M$ in $\CP^{M-1}$.

Clearly, (\textbf{i}) and (\textbf{ii}) remain \emph{related} and even \emph{smoothly-connected} as we go from $r> 0$ to $r < 0$ (because the singularity at $r = 0$ remains non-catastrophic).

This related and smooth conifold transition from (\textbf{i}) to (\textbf{ii}), is, as explained in~\cite[$\S$2.8]{Caldararu:2010ljp}, an example of a homological projective duality for CY quadrics by Kuznetsov-Perry~\cite{kuznetsov-2021-homol-projec}.

Thus, in deriving the above related and smooth CY quadrics to CY branched double cover transition from (I) to (II), we have a super-Grassmannian generalization of a well-known homological projective duality for CY quadrics in mathematics!

In fact, in deriving the related and smooth CY branched double cover to CY quadrics transition from (II$'$) to (I$'$), we also have a parity-reversed mirror version of this generalization!

\appendix
\section{Field Theory Calculations}
\label{app:u-1-1}

In this appendix, we shall furnish the explicit field theory calculations that support our physical claims in the main body of the paper.

\subsection{Deriving the Potential of the \texorpdfstring{$\mathrm{U(1|1)}$}{U(1|1)} GLSM}
\label{app:u-1-1:glsm}

The potential of the $\mathrm{U}(1)$ GLSM in \eqref{eq:u-1-1:potential} is obtained by integrating out the auxiliary fields.
In particular, the auxiliary $D$ field is integrated out via its equations of motion.
Let us show how this is done.

By explicitly writing out the terms involving the $D$ field in the Lagrangian \eqref{eq:u-1-1:lagrangian:potential}, we have
\begin{equation}
  \label{eq:u-1-1:lagrangian:potential:D terms}
  \begin{aligned}
    L^{\mathrm{U}(1|1)}_{\text{pot}}(D)
    &= \Str \left[
      \frac{1}{2e^2} D^2
      + D \left(
        \sum_{s = 1}^m \phi_s \bar{\phi}^s
        - \sum_{t = 1}^n \psi_t \bar{\psi}^t
        - r \mathbb{I}_{1|1}
      \right)
    \right]
    \\
    &= (-1)^{\varsigma(i)} \left[
      \frac{1}{2e^2} \tensor{D}{^i_j} \tensor{D}{^j_i}
      + \tensor{D}{^i_j}
      \left(
        \sum_{s = 1}^m \tensor{\phi}{^j_s} \tensor{\bar{\phi}}{_i^s}
        - \sum_{t = 1}^n \tensor{\psi}{^j_t} \tensor{\bar{\psi}}{_i^t}
        - r \tensor{\delta}{^j_i}
      \right)
    \right]
    \, ,
  \end{aligned}
\end{equation}
where Einstein summation over repeated $i, j$ indices (including the outermost Grassmann parity $\varsigma(i)$ term) is implied.

The EOM of $\tensor{D}{^i_j}$ is
\begin{equation}
  \label{eq:u-1-1:lagrangian:potential:D terms:eom}
  \frac{1}{e^2}(-1)^{\varsigma(i)} \tensor{D}{^j_i}
  + (-1)^{\varsigma(i)} \left[
    \sum_{s = 1}^m \tensor{\phi}{^j_s} \tensor{\bar{\phi}}{_i^s}
    - \sum_{t = 1}^n \tensor{\psi}{^j_t} \tensor{\bar{\psi}}{_i^t}
    - r \tensor{\delta}{^j_i}
  \right]
  = 0
  \, .
\end{equation}
That is to say, when $i = j$,
\begin{equation}
  \label{eq:u-1-1:lagrangian:potential:D terms:eom:diagonal}
  (-1)^{\varsigma(i)} \tensor{D}{^j_i}
  = - e^2 (-1)^{\varsigma(i)} \left[
    \sum_{s = 1}^m \tensor{\phi}{^j_s} \tensor{\bar{\phi}}{_i^s}
    - \sum_{t = 1}^n \tensor{\psi}{^j_t} \tensor{\bar{\psi}}{_i^t}
    - r \tensor{\delta}{^j_i}
  \right]
  \, ,
\end{equation}
and when $i \neq j$,
\begin{equation}
  \label{eq:u-1-1:lagrangian:potential:D terms:eom:off-diagonal}
  (-1)^{\varsigma(i)} \tensor{D}{^j_i}
  = - e^2 (-1)^{\varsigma(i)} \left[
    \sum_{s = 1}^m \tensor{\phi}{^j_s} \tensor{\bar{\phi}}{_i^s}
    - \sum_{t = 1}^n \tensor{\psi}{^j_t} \tensor{\bar{\psi}}{_i^t}
  \right]
  \, .
\end{equation}
By first substituting \eqref{eq:u-1-1:lagrangian:potential:D terms:eom:diagonal} and \eqref{eq:u-1-1:lagrangian:potential:D terms:eom:off-diagonal} in \eqref{eq:u-1-1:lagrangian:potential:D terms}, and then putting the result back in \eqref{eq:u-1-1:lagrangian:potential}, we find that the potential energy $U^{\mathrm{U}(1|1)}_{\text{pot}}$ will take the form
\begin{equation}
  \label{eq:u-1-1:potential:full}
  \begin{aligned}
    U^{\mathrm{U}(1|1)}_{\text{pot}}
    &= - L^{\mathrm{U}(1|1)}_{\text{pot}}
    \\
    &= \frac{e^2}{2} \sum_{i = j} (-1)^{\varsigma(i)} \left[
      \sum_{s = 1}^m \tensor{\phi}{^j_s} \tensor{\bar{\phi}}{_i^s}
      - \sum_{t = 1}^n \tensor{\psi}{^j_t} \tensor{\bar{\psi}}{_i^t}
      - r \tensor{\delta}{^j_i}
    \right]^2
    + \frac{e^2}{2} \sum_{i \neq j} (-1)^{\varsigma(i)} \left[
      \sum_{s = 1}^m \tensor{\phi}{^j_s} \tensor{\bar{\phi}}{_i^s}
      - \sum_{t = 1}^n \tensor{\psi}{^j_t} \tensor{\bar{\psi}}{_i^t}
    \right]^2
    \\
    &\quad
    + \frac{1}{2e^2} \Str \comm{\sigma}{\bar{\sigma}}^2
    + \sum_{s = 1}^m \bar{\phi}^s \acomm{\bar{\sigma}}{\sigma} \phi_s
    + \sum_{t = 1}^n \bar{\psi}^t \acomm{\bar{\sigma}}{\sigma} \psi_t
    \, .
  \end{aligned}
\end{equation}
Combining the summations in the first line of the last equality, we get \eqref{eq:u-1-1:potential}.

\subsection{Deriving the On-shell Constraints of the Gauge Field in the Low-energy Limit of the \texorpdfstring{$\mathrm{U}(1|1)$}{U(1|1)} GLSM}
\label{app:u-1-1:on-shell constraints}

The on-shell constraints of the gauge field $v_{\mu}$ are obtained by solving its EOM in \eqref{eq:u-1-1:nlsm:eom of v}, which is a system of four linear equations because $i, j \in \{1, \bar{1}\}$.

To this end, let us first introduce the following useful shorthands:
\begin{equation}
  \label{eq:u-1-1:nlsm:shorthands}
  \begin{aligned}
    \tensor{M}{^i_j}
    &\coloneq \sum_s \tensor{\phi}{^i_s} \tensor{\bar{\phi}}{_j^s} - \sum_t \tensor{\psi}{^i_t} \tensor{\bar{\psi}}{_j^t}
    \, ,
    \\
    \alpha
    &\coloneq - \tensor{J}{^{\mu}^1_{\bar{1}}} \tensor{M}{^{\bar{1}}_1} + \tensor{J}{^{\mu}^{\bar{1}}_1} \tensor{M}{^1_{\bar{1}}}
    \, ,
    \qquad
    \beta
    \coloneq \tensor{M}{^1_{\bar{1}}} \tensor{M}{^{\bar{1}}_1}
    \, ,
    \qquad
    \Gamma
    \coloneq \tensor{M}{^1_1} + \tensor{M}{^{\bar{1}}_{\bar{1}}}
    \, ,
    \\
    \Delta
    &\coloneq \frac{1}{2 \tensor{M}{^1_1} \tensor{M}{^{\bar{1}}_{\bar{1}}}} \left[
      1
      - \frac{
        \left(
          \tensor{M}{^1_1}
          - \tensor{M}{^{\bar{1}}_{\bar{1}}}
        \right) \beta
      }{\Gamma \tensor{M}{^1_1} \tensor{M}{^{\bar{1}}_{\bar{1}}}}
    \right]
    \, .
  \end{aligned}
\end{equation}

Using these shorthands, and the relation between $\tensor{J}{^{\mu}_i^j}$ and $\tensor{J}{^{\mu}^j_i}$ in \eqref{eq:u-1-1:nlsm:current matrix}, the four linear equations in \eqref{eq:u-1-1:nlsm:eom of v} can be written as
\begin{align}
  2 \tensor{v}{^{\mu}^1_1} \tensor{M}{^1_1}
  + \tensor{v}{^{\mu}^1_{\bar{1}}} \tensor{M}{^{\bar{1}}_1}
  - \tensor{v}{^{\mu}^{\bar{1}}_1} \tensor{M}{^1_{\bar{1}}}
  &= \tensor{J}{^{\mu}^1_1}
  \, ,
  \label{eq:u-1-1:nlsm:eom of v:1-1}
  \\
  \tensor{v}{^{\mu}^{\bar{1}}_1} \left(
    \tensor{M}{^1_1}
    + \tensor{M}{^{\bar{1}}_{\bar{1}}}
  \right)
  + \left(
    \tensor{v}{^{\mu}^{\bar{1}}_{\bar{1}}}
    + \tensor{v}{^{\mu}^1_1}
  \right) \tensor{M}{^{\bar{1}}_1}
  &= \tensor{J}{^{\mu}^{\bar{1}}_1}
  \, ,
  \label{eq:u-1-1:nlsm:eom of v:1bar-1}
  \\
  \tensor{v}{^{\mu}^1_{\bar{1}}} \left(
    \tensor{M}{^1_1}
    + \tensor{M}{^{\bar{1}}_{\bar{1}}}
  \right)
  + \left(
    \tensor{v}{^{\mu}^{\bar{1}}_{\bar{1}}}
    + \tensor{v}{^{\mu}^1_1}
  \right) \tensor{M}{^1_{\bar{1}}}
  &= \tensor{J}{^{\mu}^1_{\bar{1}}}
  \, ,
  \label{eq:u-1-1:nlsm:eom of v:1-1bar}
  \\
  2 \tensor{v}{^{\mu}^{\bar{1}}_{\bar{1}}} \tensor{M}{^{\bar{1}}_{\bar{1}}}
  - \tensor{v}{^{\mu}^1_{\bar{1}}} \tensor{M}{^{\bar{1}}_1}
  + \tensor{v}{^{\mu}^{\bar{1}}_1} \tensor{M}{^1_{\bar{1}}}
  &= \tensor{J}{^{\mu}^{\bar{1}}_{\bar{1}}}
  \, .
  \label{eq:u-1-1:nlsm:eom of v:1bar-1bar}
\end{align}

From \eqref{eq:u-1-1:nlsm:eom of v:1bar-1} and \eqref{eq:u-1-1:nlsm:eom of v:1-1bar}, we get
\begin{equation}
  \label{eq:u-1-1:nlsm:eom of v:off-diagonal:interim}
  \tensor{v}{^{\mu}^{\bar{1}}_1}
  = \flatfrac{
    \left[
      \tensor{J}{^{\mu}^{\bar{1}}_1}
      - \left(
        \tensor{v}{^{\mu}^{\bar{1}}_{\bar{1}}}
        + \tensor{v}{^{\mu}^1_1}
      \right) \tensor{M}{^{\bar{1}}_1}
    \right]
  }{\Gamma}
  \, ,
  \qquad
  \tensor{v}{^{\mu}^1_{\bar{1}}}
  = \flatfrac{
    \left[
      \tensor{J}{^{\mu}^1_{\bar{1}}}
      - \left(
        \tensor{v}{^{\mu}^{\bar{1}}_{\bar{1}}}
        + \tensor{v}{^{\mu}^1_1}
      \right) \tensor{M}{^1_{\bar{1}}}
    \right]
  }{\Gamma}
  \, ,
\end{equation}
assuming $\Gamma \neq 0$, i.e., that $\tensor{M}{^1_1} + \tensor{M}{^{\bar{1}}_{\bar{1}}} \neq 0$.

By substituting \eqref{eq:u-1-1:nlsm:eom of v:off-diagonal:interim} into \eqref{eq:u-1-1:nlsm:eom of v:1-1} and \eqref{eq:u-1-1:nlsm:eom of v:1bar-1bar}, we can solve for the on-shell contraints of $\tensor{v}{^{\mu}^1_1}$ and $\tensor{v}{^{\mu}^{\bar{1}}_{\bar{1}}}$ as a system of two linear equations that we can also express as
\begin{equation}
  \label{eq:u-1-1:nlsm:eom of v:1-1 and 1bar-1bar}
  \mqty(
  \tensor{M}{^1_1} - \flatfrac{\beta}{\Gamma}
  &
  - \flatfrac{\beta}{\Gamma}
  \\
  \flatfrac{\beta}{\Gamma}
  &
  \tensor{M}{^{\bar{1}}_{\bar{1}}} + \flatfrac{\beta}{\Gamma}
  ) \mqty(%
  \tensor{v}{^{\mu}^1_1}
  \\
  \tensor{v}{^{\mu}^{\bar{1}}_{\bar{1}}}
  )
  = \frac{1}{2} \mqty(%
  \tensor{J}{^{\mu}^1_1} + \flatfrac{\alpha}{\Gamma}
  \\
  \tensor{J}{^{\mu}^{\bar{1}}_{\bar{1}}} - \flatfrac{\alpha}{\Gamma}
  )
  \, ,
\end{equation}
where the matrix is a regular invertible matrix if and only if $\tensor{M}{^1_1} \neq 0 \neq \tensor{M}{^{\bar{1}}_{\bar{1}}}$, whence the inverse of its determinant is $2 \Delta$.

Lastly, by putting the solutions for $\tensor{v}{^{\mu}^1_1}$ and $\tensor{v}{^{\mu}^{\bar{1}}_{\bar{1}}}$ from \eqref{eq:u-1-1:nlsm:eom of v:1-1 and 1bar-1bar} back into \eqref{eq:u-1-1:nlsm:eom of v:off-diagonal:interim}, we obtain the on-shell constraints of the  remaining $\tensor{v}{^{\mu}^{\bar 1}_1}$ and $\tensor{v}{^{\mu}^{{1}}_{\bar{1}}}$ gauge fields.

Altogether, the on-shell constraints of the gauge field obtained from its EOM \eqref{eq:u-1-1:nlsm:eom of v} are
\begin{equation}
  \label{eq:u-1-1:nlsm:solutions of v}
  \begin{aligned}
    \tensor{v}{^{\mu}^1_1}
    &= \Delta \left(
      \tensor{M}{^{\bar{1}}_{\bar{1}}} \tensor{J}{^{\mu}^1_1}
      + \flatfrac{%
        \left[
          \tensor{M}{^{\bar{1}}_{\bar{1}}} \alpha
          + \beta (\tensor{J}{^{\mu}^1_1} + \tensor{J}{^{\mu}^{\bar{1}}_{\bar{1}}})
        \right]
      }{\Gamma}
    \right)
    \, ,
    \\
    \tensor{v}{^{\mu}^{\bar{1}}_1}
    &= \frac{\tensor{J}{^{\mu}^{\bar{1}}_1}}{\Gamma}
    - \Delta \left(
      \tensor{M}{^{\bar{1}}_{\bar{1}}} \tensor{J}{^{\mu}^1_1}
      + \tensor{M}{^1_1} \tensor{J}{^{\mu}^{\bar{1}}_{\bar{1}}}
      - \left[ \tensor{M}{^1_1} - \tensor{M}{^{\bar{1}}_{\bar{1}}} \right] \flatfrac{\alpha}{\Gamma}
    \right) \frac{\tensor{M}{^{\bar{1}}_1}}{\Gamma}
    \, ,
    \\
    \tensor{v}{^{\mu}^1_{\bar{1}}}
    &= \frac{\tensor{J}{^{\mu}^1_{\bar{1}}}}{\Gamma}
    - \Delta \left(
      \tensor{M}{^{\bar{1}}_{\bar{1}}} \tensor{J}{^{\mu}^1_1}
      + \tensor{M}{^1_1} \tensor{J}{^{\mu}^{\bar{1}}_{\bar{1}}}
      - \left[ \tensor{M}{^1_1} - \tensor{M}{^{\bar{1}}_{\bar{1}}} \right] \flatfrac{\alpha}{\Gamma}
    \right) \frac{\tensor{M}{^1_{\bar{1}}}}{\Gamma}
    \, ,
    \\
    \tensor{v}{^{\mu}^{\bar{1}}_{\bar{1}}}
    &= \Delta \left(
      \tensor{M}{^1_1} \tensor{J}{^{\mu}^{\bar{1}}_{\bar{1}}}
      - \flatfrac{%
        \left[
          \tensor{M}{^1_1} \alpha
          + \beta (\tensor{J}{^{\mu}^1_1} + \tensor{J}{^{\mu}^{\bar{1}}_{\bar{1}}})
        \right]
      }{\Gamma}
    \right)
    \, .
  \end{aligned}
\end{equation}

\section{Mathematical Verification of the Calabi-Yau Condition for Supermanifolds}
\label{app:verify cy}

In this appendix, we shall mathematically verify the Calabi-Yau (CY) condition for various supermanifolds that were physically derived in the main body of the paper.

\subsection{The Calabi-Yau Condition for Super-Grassmannians}
\label{app:verify cy:supergrassmannians}

In \autoref{subsec: applications to math U(1|1)}, we physically derived a \emph{novel} mathematical result that super-Grassmannians of the type ${Gr}_{1|1}(\mathbb{C}^{m|n})$ are CY supermanifolds if and only if $m = n$.

We will now mathematically prove and therefore verify that in fact,   all super-Grassmannians of the type ${Gr}_{p|q}(\mathbb{C}^{m|n})$, \emph{regardless of the values of $(p, q)$}, have a trivial Berezinian sheaf and are thus CY supermanifolds~\cite[def.~1.20]{noja-2016-topic-algeb}, if and only if $m = n$.

First, note that a super-Grassmannian $\mathcal{G} = {Gr}_{p|q}(\mathbb{C}^{m|n})$ has a universal exact sequence \cite[eq.~(3.93)]{noja-2016-topic-algeb}
\begin{equation}
  \label{eq:u-1-1:quantum:grassmannian universal sequence}
  \begin{tikzcd}[%
    ampersand replacement=\&
    ]
    0
    \arrow[r]
    \&
    \mathcal{S}_{\mathcal{G}}
    \arrow[r]
    \&
    \mathbb{C}^{m|n} \otimes \mathcal{O}_{\mathcal{G}}
    \arrow[r]
    \&
    \mathcal{Q}_{\mathcal{G}}
    \arrow[r]
    \&
    0
    \, ,
  \end{tikzcd}
\end{equation}
where
(i) $\mathcal{O}_{\mathcal{G}}$ is the structure sheaf;
(ii) $\mathcal{S}_{\mathcal{G}}$ is the tautological sheaf on $\mathcal{G}$ \cite[def.~3.3]{noja-2016-topic-algeb}, i.e., the sheaf of locally-free $\mathcal{O}_{\mathcal{G}}$-modules of rank $p|q$;
and (iii) the quotient sheaf $\mathcal{Q}_{\mathcal{G}}$ of rank $(m-p)|(n-q)$ is dual to the sheaf orthogonal to $\mathcal{S}_{\mathcal{G}}$.

Second, the Berezinian sheaf of a supermanifold is the Berezinian of its cotangent sheaf \cite[def.~1.18]{noja-2016-topic-algeb}.
For $\mathcal{G}$, its tangent sheaf can be determined from \eqref{eq:u-1-1:quantum:grassmannian universal sequence} to be $\mathcal{T}_{\mathcal{G}} \coloneq \Hom(\mathcal{S}_{\mathcal{G}}, \mathcal{Q}_{\mathcal{G}}) \cong S^{\vee}_{\mathcal{G}} \otimes \mathcal{Q}_{\mathcal{G}}$ \cite[thm.~4.3.11]{Manin:1988ds}, whence its cotangent sheaf is $\Omega^1_{\mathcal{G}} \coloneq \Hom(\mathcal{T}_{\mathcal{G}}, \mathcal{O}_{\mathcal{G}}) \cong \mathcal{Q}^{\vee}_{\mathcal{G}} \otimes \mathcal{S}_{\mathcal{G}}$, where ``${}^\vee$'' denotes a dual sheaf.
Therefore, the Berezinian sheaf of $\mathcal{G}$ can be computed as
\begin{equation}
  \label{eq:u-1-1:quantum:berezinian sheaf}
  \Ber (\mathcal{G})
  \cong \Ber (\mathcal{Q}^{\vee}_{\mathcal{G}} \otimes \mathcal{S}_{\mathcal{G}})
  \cong \Ber (\mathcal{Q}^{\vee}_{\mathcal{G}})^{(m-p)-(n-q)} \otimes_{\mathcal{G}} \Ber (\mathcal{S}_{\mathcal{G}})^{p-q}
  \, .
\end{equation}

Third, note that \eqref{eq:u-1-1:quantum:grassmannian universal sequence} implies a canonical isomorphism between $\Ber(\mathcal{S}_{\mathcal{G}})$ and $\Ber(\mathcal{Q}_{\mathcal{G}})^{-1}$ \cite[$\S$1.11]{deligne-1999-class-field-super}, and thus with $\Ber(\mathcal{Q}^{\vee}_{\mathcal{G}})$, i.e.,
\begin{equation}
  \label{eq:u-1-1:quantum:canonical isomorphism of universal sequence}
  \Ber(\mathcal{S}_{\mathcal{G}}) \otimes_{\mathcal{G}} \Ber(\mathcal{Q}_{\mathcal{G}}) \cong \Ber(\mathbb{C}^{m|n} \otimes \mathcal{O}_{\mathcal{G}}) \cong \mathcal{O}_{\mathcal{G}}
  \, ,
  \quad
  \implies
  \quad
  \Ber(\mathcal{S}_{\mathcal{G}}) \cong \Ber(\mathcal{Q}_{\mathcal{G}})^{-1} \cong \Ber(\mathcal{Q}^{\vee}_{\mathcal{G}})
  \, .
\end{equation}

Therefore, applying \eqref{eq:u-1-1:quantum:canonical isomorphism of universal sequence} to \eqref{eq:u-1-1:quantum:berezinian sheaf}, we get
\begin{equation}
  \label{eq:u-1-1:quantum:berezinian of supergrassmannian}
  \Ber(\mathcal{G})
  \cong \Ber(\mathcal{S}_{\mathcal{G}})^{m-n}
  \, ,
\end{equation}
which is indeed trivial when $m = n$. Hence, regardless of the values of $(p,q)$, a super-Grassmannian is a CY supermanifold when $m = n$,  as claimed.

Applying this result to projective superspaces $\mathbb{CP}^{m-1|n} \cong {Gr}_{1|0}(\mathbb{C}^{m|n})$, we find that they are CY supermanifolds if and only if $m = n$, which agrees with the literature \cite[eg.~1.3]{noja-2016-topic-algeb}.

\subsection{The Calabi-Yau Condition for Weighted Projective Superspaces}
\label{app:verify cy:wcp}

In \autoref{subsec: applications to math U(1|1)^N}, we physically derived a \emph{novel} mathematical result that weighted projective superspaces of the type $\WCP^{m-1|n}_{(\vec{Q},|\vec{\mathsf{Q}})}$ are CY supermanifolds if and only if $\Sigma_{s = 1}^m Q_s = \Sigma_{\alpha = 1}^n \mathsf{Q}_\alpha$,\footnote{%
  \label{ft:verify cy:wcp:explain use of greek characters}%
  In this appendix section, we shall use Greek letters $\alpha, \beta$ to label the indices on the $\mathsf{Q}$ weights, which correspond to the flavor indices $t$ on the Grassmann-odd chiral superfields in the underlying GLSM in \autoref{sec: U(1|1)^N GLSM}.
}
where some of the non-negative $Q_s$'s or $\mathsf{Q}_\alpha$'s may be zero.

The CY condition for weighted projective superspaces is not new in the physics literature; its first appearance was in \cite{sethi-1994-super-rigid}.
That said, its mathematical verification is mostly absent in the mathematics literature except for the case of $\WCP^{1|1}_{(1, 1|2)}$ in \cite[eq.~(3.15)]{noja-2017-one-dimen}.

We will now mathematically prove and therefore verify this condition for our above general weighted projective superspace.

\subtitle{$\WCP^{m^{\prime}-1|n^{\prime}}_{(\vec{Q^{\prime}}|\vec{\mathsf{Q}^{\prime}})}$ with Solely Nonzero Weights}

If $\WCP^{m-1|n}_{(\vec{Q},|\vec{\mathsf{Q}})}$ is such that there are $p$ and $q$ number of entries in $\vec{Q}$ and $\vec{\mathsf{Q}}$ which are zero, it would be isomorphic to $\mathbb{C}^{p|q} \times \WCP^{m^{\prime} - 1|n^{\prime}}_{(\vec{Q^{\prime}}|\vec{\mathsf{Q}^{\prime}})}$, where (i) $m^{\prime} \coloneq m - p$, (ii) $n^{\prime} \coloneq n - q$, and (iii) $(\vec{Q^{\prime}}|\vec{\mathsf{Q}^{\prime}})$ are solely nonzero weights. Then,  $\WCP^{m-1|n}_{(\vec{Q}|\vec{\mathsf{Q}})}$ would clearly be a CY supermanifold if and only if $\WCP^{m^{\prime} - 1|n^{\prime}}_{(\vec{Q^{\prime}}|\vec{\mathsf{Q}^{\prime}})}$ is a CY supermanifold. Therefore, in this case, we just have to prove the CY condition for $\WCP^{m^{\prime}-1|n^{\prime}}_{(\vec{Q^{\prime}}|\vec{\mathsf{Q}^{\prime}})}$.

If $\WCP^{m-1|n}_{(\vec{Q},|\vec{\mathsf{Q}})}$ is such that there are no entries in $\vec{Q}$ or $\vec{\mathsf{Q}}$ which are zero, it would simply coincide with $\WCP^{m^{\prime}-1|n^{\prime}}_{(\vec{Q^{\prime}}|\vec{\mathsf{Q}^{\prime}})}$ where $m' = m$ and $n' = n$. Therefore, in this case, we also just have to prove the CY condition for $\WCP^{m^{\prime}-1|n^{\prime}}_{(\vec{Q^{\prime}}|\vec{\mathsf{Q}^{\prime}})}$.

In short, to prove the CY condition for $\WCP^{m-1|n}_{(\vec{Q},|\vec{\mathsf{Q}})}$ with arbitrary non-negative weights that may be zero, it suffices for us to prove the CY condition for $\WCP^{m^{\prime}-1|n^{\prime}}_{(\vec{Q^{\prime}}|\vec{\mathsf{Q}^{\prime}})}$ with solely nonzero weights.

\subtitle{Relabeling $\WCP^{m^{\prime}-1|n^{\prime}}_{(\vec{Q^{\prime}}|\vec{\mathsf{Q}^{\prime}})}$ as $\WCP^{m-1|n}_{(\vec{Q},|\vec{\mathsf{Q}})}$ with Solely Nonzero Weights}

In order to ease notation and avoid clutter, we shall relabel $\WCP^{m^{\prime}-1|n^{\prime}}_{(\vec{Q^{\prime}}|\vec{\mathsf{Q}^{\prime}})}$ as $\WCP^{m-1|n}_{(\vec{Q},|\vec{\mathsf{Q}})}$ with solely nonzero weights $(\vec{Q},|\vec{\mathsf{Q}})$.

\subtitle{Coordinates on $\WCP^{m-1|n}_{(\vec{Q},|\vec{\mathsf{Q}})}$}

The homogeneous coordinates on $\WCP^{m-1|n}_{(\vec{Q},|\vec{\mathsf{Q}})}$ are identified under the equivalence relation
\begin{equation}
  \label{eq:verify cy:wcp:homogeneous coordinates}
  (z^1, \dots, z^m|\theta^1, \dots, \theta^n)
  \sim
  (\lambda^{Q_1} z^1, \dots, \lambda^{Q_m} z^m|\lambda^{\mathsf{Q}_1} \theta^1, \dots, \lambda^{\mathsf{Q}_n} \theta^n)
  \, ,
\end{equation}
for $\lambda \in \mathbb{C}^{\times}$, i.e., for $\lambda$ a nonzero complex number.
There are $m$ coordinate patches with $z^t \neq 0$ in the $t^{\text{th}}$ patch, from which we can define the inhomogeneous coordinates $(\tensor{w}{^1_{\{t\}}}, \dots, \tensor{w}{^m_{\{t\}}}|\tensor{\eta}{^1_{\{t\}}}, \dots, \tensor{\eta}{^n_{\{t\}}})$ as\footnote{%
  In this definition, the inhomogeneous coordinates will include a ``redundant'' coordinate whose value is $1$ in the $t^{\text{th}}$ component.
  For example, if $t = 1$, the inhomogeneous coordinates according to our definition would be $(1, \tensor{w}{^2_{\{1\}}}, \dots, \tensor{w}{^m_{\{1\}}}|\tensor{\eta}{^1_{\{1\}}}, \dots, \tensor{\eta}{^n_{\{1\}}})$, where if we exclude the first component, the remaining $m-1|n$ coordinates, i.e., $(\tensor{w}{^2_{\{1\}}}, \dots, \tensor{w}{^m_{\{1\}}}|\tensor{\eta}{^1_{\{1\}}}, \dots, \tensor{\eta}{^n_{\{1\}}})$, are the local coordinates on the affine patch and are often referred to as inhomogeneous coordinates in the literature.
  Nevertheless, we shall keep this ``redundant'' coordinate in our definition as it is more insightful for our purposes, as we shall see.
}
\begin{equation}
  \label{eq:verify cy:wcp:inhomogeneous coordinates}
  \tensor{w}{^s_{\{t\}}}
  \coloneq \frac{z^s}{(z^t)^{\flatfrac{Q_s}{Q_t}}}
  \, ,
  \qquad
  \tensor{\eta}{^{\alpha}_{\{t\}}}
  \coloneq \frac{\theta^{\alpha}}{(z^t)^{\flatfrac{\mathsf{Q}_{\alpha}}{Q_t}}}
  \, ,
\end{equation}
where Latin letters $\{s, t\} \in \{1, \dots, m\}$ and Greek letter(s) $\alpha \in \{1, \dots, n\}$.

The inhomogeneous coordinates on the $t^{\text{th}}$ patch are related to the inhomogeneous coordinates on the $q^{\text{th}}$ patch as
\begin{equation}
  \label{eq:verify cy:wcp:relation between patches}
  \tensor{w}{^s_{\{t\}}}
  = \frac{\tensor{w}{^s_{\{q\}}}}{(\tensor{w}{^t_{\{q\}}})^{Q_s/Q_t}}
  + \tensor{\delta}{^s_t} \left( \tensor{w}{^q_{\{q\}}} - 1 \right)
  \, ,
  \qquad
  \tensor{\eta}{^{\alpha}_{\{t\}}}
  = \frac{\tensor{\eta}{^{\alpha}_{\{q\}}}}{(\tensor{w}{^t_{\{q\}}})^{\mathsf{Q}_{\alpha}/Q_t}}
  \, .
\end{equation}

\subtitle{The Jacobian Matrix}

Consider a coordinate transformation $\Xi_{\{t, q\}}$ from the $t^{\text{th}}$ patch to the $q^{\text{th}}$ patch such that $(\tensor{w}{^s_{\{t\}}}, \tensor{\eta}{^{\alpha}_{\{t\}}}) \mapsto (\tensor{w}{^p_{\{q\}}}, \tensor{\eta}{^{\beta}_{\{q\}}})$.
The Jacobian matrix of $\Xi_{\{t, q\}}$ is an $m|n \times m|n$-supermatrix with components
\begin{equation}
  \label{eq:verify cy:wcp:jacobian matrix}
  \text{Jac}(\Xi_{\{t, q\}})
  = \mqty(A_{\{t, q\}} & B_{\{t, q\}} \\ C_{\{t, q\}} & D_{\{t, q\}})
  \, ,
\end{equation}
where
\begin{equation}
  \label{eq:verify cy:wcp:jacobian matrix:components}
  \begin{aligned}
    \tensor{(A_{\{t, q\}})}{^s_p}
    &\coloneq \pdv{\tensor{w}{^s_{\{t\}}}}{\tensor{w}{^p_{\{q\}}}}
    = \frac{\tensor{\delta}{^s_p}}{(\tensor{w}{^t_{\{q\}}})^{Q_s/Q_t}}
    - \frac{Q_s}{Q_t} \frac{\tensor{w}{^s_{\{q\}}} \tensor{\delta}{^t_p}}{(\tensor{w}{^t_{\{q\}}})^{Q_s/Q_t + 1}}
    + \tensor{\delta}{^s_t} \tensor{\delta}{^q_p}
    \, ,
    \\
    \tensor{(B_{\{t, q\}})}{^s_{\beta}}
    &\coloneq \pdv{\tensor{w}{^s_{\{t\}}}}{\tensor{\eta}{^{\beta}_{\{q\}}}}
    = 0
    \, ,
    \\
    \tensor{(C_{\{t, q\}})}{^{\alpha}_p}
    &\coloneq \pdv{\tensor{\eta}{^{\alpha}_{\{t\}}}}{\tensor{w}{^p_{\{q\}}}}
    = - \frac{\mathsf{Q}_{\alpha}}{Q_t} \frac{\tensor{\eta}{^{\alpha}_{\{p\}}} \tensor{\delta}{^t_p}}{(\tensor{w}{^t_{\{q\}}})^{\mathsf{Q}_{\alpha}/Q_t + 1}}
    \, ,
    \\
    \tensor{(D_{\{t, q\}})}{^{\alpha}_{\beta}}
    &\coloneq \pdv{\tensor{\eta}{^{\alpha}_{\{t\}}}}{\tensor{\eta}{^{\beta}_{\{q\}}}}
    = \frac{\tensor{\delta}{^{\alpha}_{\beta}}}{(\tensor{w}{^t_{\{q\}}})^{\mathsf{Q}_{\alpha}/Q_t}}
    \, .
  \end{aligned}
\end{equation}

\subtitle{The Berezinian of the Jacobian Matrix}

Let us compute the ``determinant'' of the Jacobian matrix.
Since it is a supermatrix, the analogue of a ``determinant'' is a superdeterminant, also known as a Berezinian.
The Berezinian of $\text{Jac}(\Xi_{\{t, q\}})$ is
\begin{equation}
  \label{eq:verify cy:wcp:berezinian:definition}
  \text{Ber} \left( \text{Jac}(\Xi_{\{t, q\}}) \right)
  = \frac{\det \left(A_{\{t, q\}} - B_{\{t, q\}} D_{\{t, q\}}^{-1} C_{\{t, q\}} \right)}{\det \left(D_{\{t, q\}}\right)}
  = \det \left(A_{\{t, q\}}\right) \det \left(D_{\{t, q\}}\right)^{-1}
  \, ,
\end{equation}
where $\det(X)$ is the determinant of a matrix $X$, and the final equality is because $B = 0$ from \eqref{eq:verify cy:wcp:jacobian matrix:components}.
This will be evaluated by computing $\det(A)$ and $\det(D)$ separately.

To compute $\det(D)$, note that $D$ is a diagonal matrix; its determinant is simply a product of its elements.
From \eqref{eq:verify cy:wcp:jacobian matrix:components}, we compute that
\begin{equation}
  \label{eq:verify cy:wcp:D:determinant}
  \det(D) = \prod_{\alpha = 1}^n \frac{1}{(\tensor{w}{^t_{\{q\}}})^{\mathsf{Q}_{\alpha}/Q_t}}
  = (\tensor{w}{^t_{\{q\}}})^{- \frac{1}{Q_t} \left( \sum_{\alpha = 1}^n \mathsf{Q}_{\alpha} \right)}
  \, .
\end{equation}

To compute $\det(A)$, let us first look at the components of $\tensor{(A_{\{t, q\}})}{^s_p}$ in \eqref{eq:verify cy:wcp:jacobian matrix:components}.
When $s = p$,
\begin{equation}
  \label{eq:verify cy:wcp:A:s eq p}
  \tensor{(A_{\{t, q\}})}{^s_s}
  =
  \begin{cases}
    0
    &\qif s = t
    \, ,
    \\
    \flatfrac{1}{(\tensor{w}{^t_{\{q\}}})^{Q_s/Q_t}}
    &\qif s \neq t
    \, .
  \end{cases}
\end{equation}
On the other hand, when $s \neq p$,
\begin{equation}
  \label{eq:verify cy:wcp:A:s neq p}
  \tensor{(A_{\{t, q\}})}{^s_p}
  =
  \begin{cases}
    1
    &\qif s = t \text{ and } p = q
    \, ,
    \\
    - \frac{Q_s}{Q_t} \flatfrac{\tensor{w}{^s_{\{q\}}}}{(\tensor{w}{^t_{\{q\}}})^{Q_s/Q_t + 1}}
    &\qif p = t
    \, ,
    \\
    0
    &\quad \text{otherwise}
    \, .
  \end{cases}
\end{equation}
The matrix of $A_{\{t, q\}}$ therefore looks like
\begin{equation}
  \label{eq:verify cy:wcp:A}
  \begin{pNiceMatrix}[first-row,first-col]
    & \Cdots & q-1 & q & \Cdots & t & t + 1 & \Cdots
    \\
    \Vdots
    & \ddots & 0 & 0 & 0 & \vdots & 0 & 0
    \\
    q-1
    & 0 & \tensor{(A_{\{t, q\}})}{^{q - 1}_{q - 1}} & 0 & 0 & \tensor{(A_{\{t, q\}})}{^{q - 1}_t} & 0 & 0
    \\
    \phantom{\quad} q \quad
    & 0 & 0 & \tensor{(A_{\{t, q\}})}{^q_q} & 0 & \tensor{(A_{\{t, q\}})}{^q_t} & 0 & 0
    \\
    \Vdots
    & 0 & 0 & 0 & \ddots & \vdots & 0 & 0
    \\
    \phantom{\quad} t \quad
    & 0 & 0 & \tensor{(A_{\{t, q\}})}{^t_q} & 0 & \tensor{(A_{\{t, q\}})}{^t_t} & 0 & 0
    \\
    t+1
    & 0 & 0 & 0 & 0 & \tensor{(A_{\{t, q\}})}{^{t + 1}_t} & \tensor{(A_{\{t, q\}})}{^{t + 1}_{t + 1}} & 0
    \\
    \Vdots
    & 0 & 0 & 0 & 0 & \vdots & 0 & \ddots
  \end{pNiceMatrix}
  \, ,
\end{equation}
where the off-diagonal zeros (determined from \eqref{eq:verify cy:wcp:A:s neq p}) are explicitly written.

Using standard matrix algebra tricks, we find that\footnote{%
  \label{ft:verify cy:wcp:order of q and t}%
  In \eqref{eq:verify cy:wcp:A}, we had implicitly assumed that $q < t$.
  However, if we had chosen $q > t$, using standard matrix algebra tricks, we will still arrive at the same result.
  Therefore, our result holds for arbitrary non-equal choices of $(q, t)$.
}
\begin{equation}
  \label{eq:verify cy:wcp:A:determinant}
  \det(A)
  \equiv \det \left[
    \mqty( \tensor{(A_{\{t, q\}})}{^q_q} & \tensor{(A_{\{t, q\}})}{^q_t} \\ \tensor{(A_{\{t, q\}})}{^t_q} & \tensor{(A_{\{t, q\}})}{^t_t} )
    \oplus A_{\text{diag}}
  \right]
  = \det \mqty( \tensor{(A_{\{t, q\}})}{^q_q} & \tensor{(A_{\{t, q\}})}{^q_t} \\ \tensor{(A_{\{t, q\}})}{^t_q} & \tensor{(A_{\{t, q\}})}{^t_t} )
  \det(A_{\text{diag}})
  \, ,
\end{equation}
where $A_{\text{diag}}$ is a diagonal matrix whose entries are the remaining diagonal elements of $A_{\{t, q\}}$.
Substituting the values from \eqref{eq:verify cy:wcp:A:s eq p} and \eqref{eq:verify cy:wcp:A:s neq p} in \eqref{eq:verify cy:wcp:A:determinant}, we get
\begin{equation}
  \label{eq:verify cy:wcp:A:determinant:evaluated}
  \begin{aligned}
    \det(A)
    &= \det \mqty(
    \frac{1}{(\tensor{w}{^q_{\{q\}}})^{Q_q/Q_t}} & - \frac{Q_q}{Q_t} \frac{\tensor{w}{^q_{\{q\}}}}{(\tensor{w}{^t_{\{q\}}})^{(Q_q/Q_t) + 1}}
    \\
    1 & 0
    )
    \prod_{s \in (\mathfrak{M} \setminus \{t, q\} )} \frac{1}{(\tensor{w}{^t_{\{q\}}})^{Q_s/Q_t}}
    \\
    &= \frac{Q_q}{Q_t} \frac{\tensor{w}{^q_{\{q\}}}}{(\tensor{w}{^t_{\{q\}}})^{(Q_q/Q_t) + 1}} \prod_{s \in (\mathfrak{M} \setminus \{t, q\} )} \frac{1}{(\tensor{w}{^t_{\{q\}}})^{Q_s/Q_t}}
    \\
    &= \frac{Q_q}{Q_t} (\tensor{w}{^t_{\{q\}}})^{- \frac{1}{Q_t} \left( \sum_{s \in \mathfrak{M}} Q_s \right)}
    \, ,
  \end{aligned}
\end{equation}
where
\begin{enumerate*}
\item $\mathfrak{M} \coloneq \{1, \dots, m\}$ is a set,

\item $\mathfrak{M} \setminus \{x\}$ means to remove element(s) $x$ from $\mathfrak{M}$, and

\item we have made use of the fact that $\tensor{w}{^q_{\{q\}}} = 1$ (from its definition in \eqref{eq:verify cy:wcp:inhomogeneous coordinates}) to arrive at the final equality.

\end{enumerate*}

Finally, substituting \eqref{eq:verify cy:wcp:D:determinant} and \eqref{eq:verify cy:wcp:A:determinant:evaluated} in \eqref{eq:verify cy:wcp:berezinian:definition}, we get
\begin{equation}
  \label{eq:verify cy:wcp:berezinian}
  \text{Ber} \left( \text{Jac}(\Xi_{\{t, q\}}) \right)
  = \frac{Q_q}{Q_t} (\tensor{w}{^t_{\{q\}}})^{- \frac{1}{Q_t} \left( \sum_{s = 1}^m Q_s - \sum_{\alpha = 1}^n \mathsf{Q}_{\alpha} \right)}
  \, .
\end{equation}

\subtitle{The CY Condition for $\WCP^{m-1|n}_{(\vec{Q},|\vec{\mathsf{Q}})}$}

For a supermanifold to be CY, its Berezinian sheaf needs to be a trivial sheaf \cite[def.~1.20]{noja-2016-topic-algeb}.
What this means for $\WCP^{m-1|n}_{(\vec{Q},|\vec{\mathsf{Q}})}$ is that the Berezinian of the Jacobian matrix from one patch to another needs to be trivial (up to a constant scaling factor).
We can see from \eqref{eq:verify cy:wcp:berezinian} that for this to happen, the exponent of $\tensor{w}{^t_{\{q\}}}$ needs to vanish, i.e.,
\begin{equation}
  \label{eq:verify cy:wcp:cy condition}
  \sum_{s = 1}^m Q_s = \sum_{\alpha = 1}^n \mathsf{Q}_{\alpha}
  \, .
\end{equation}

Hence, a weighted projective superspace of the type $\WCP^{m-1|n}_{(\vec{Q},|\vec{\mathsf{Q}})}$ is a CY supermanifold if and only if its weights satisfy \eqref{eq:verify cy:wcp:cy condition}, as claimed.

\subtitle{An Alternative Proof}

Let us provide a shorter, alternative proof of the CY condition for $\mathcal{W} = \WCP^{m-1|n}_{(\vec{Q},|\vec{\mathsf{Q}})}$.

First, consider a generalization of the Euler sequence on weighted projective spaces \cite[e.g.~3.1]{pena-2024-codim-one} to weighted \emph{super}projective spaces, i.e.,\footnote{%
  \label{ft:verify cy:wcp:generalization of euler sequence}%
  The generalization of the Euler sequence on unweighted projective spaces to unweighted \emph{super}projective spaces was done in a similar manner by Cacciatori-Noja in \cite[e.q.~(79)]{Cacciatori:2017qyd}.
  In particular, if we set the weights $(\vec{Q}, \vec{\mathsf{Q}}) = (\vec{1}, \vec{1})$, we will recover their generalized Euler sequence \emph{loc.~cit.}
}
\begin{equation}
  \label{eq:verify cy:wcp:exact sequence}
  \begin{tikzcd}
    0
    \arrow[r]
    &
    \mathcal{O}_{\mathcal{W}}
    \arrow[r]
    &
    \left[
      \bigoplus_{s = 1}^m \mathcal{O}_{\mathcal{W}}(Q_s)
    \right]
    \oplus
    \left[
      \bigoplus_{\alpha = 1}^n \mathbf{\Pi} \mathcal{O}_{\mathcal{W}}(\mathsf{Q}_{\alpha})
    \right]
    \arrow[r]
    &
    \mathcal{T}_{\mathcal{W}}
    \arrow[r]
    &
    0
    \, ,
  \end{tikzcd}
\end{equation}
where $\mathcal{O}_{\mathcal{W}}$ and $\mathcal{T}_{\mathcal{W}}$ are the (locally trivial) structure sheaf and tangent sheaf on $\mathcal{W}$, respectively.

Second, dualizing this exact sequence, we get
\begin{equation}
  \label{eq:verify cy:wcp:exact sequence:dual}
  \begin{tikzcd}
    0
    \arrow[r]
    &
    \Omega^1_{\mathcal{W}}
    \arrow[r]
    &
    \left[
      \bigoplus_{s = 1}^m \mathcal{O}_{\mathcal{W}}(-Q_s)
    \right]
    \oplus
    \left[
      \bigoplus_{\alpha = 1}^n \mathbf{\Pi} \mathcal{O}_{\mathcal{W}}(-\mathsf{Q}_{\alpha})
    \right]
    \arrow[r]
    &
    \mathcal{O}_{\mathcal{W}}
    \arrow[r]
    &
    0
    \, ,
  \end{tikzcd}
\end{equation}
where $\Omega^1_{\mathcal{W}}$ is the cotangent sheaf of $\mathcal{W}$.

Third, note that since the exact sequence \eqref{eq:verify cy:wcp:exact sequence:dual} splits locally and that $\Ber(\mathcal{O}_{\mathcal{W}})$ is trivial, we compute, from the canonical isomorphism between the Berezinians \cite[$\S$1.11]{deligne-1999-class-field-super} (\emph{\`{a}~la} \eqref{eq:u-1-1:quantum:canonical isomorphism of universal sequence}), that
\begin{equation}
  \label{eq:verify cy:wcp:berezinian:compute}
  \begin{aligned}
    \Ber \left( \Omega^1_{\mathcal{W}} \right)
    &\cong
    \Ber \left(
      \left[
        \bigoplus_{s = 1}^m \mathcal{O}_{\mathcal{W}}(-Q_s)
      \right]
      \oplus
      \left[
        \bigoplus_{\alpha = 1}^n \mathbf{\Pi} \mathcal{O}_{\mathcal{W}}(-\mathsf{Q}_{\alpha})
      \right]
    \right)
    \\
    &\cong
    \left[
      \bigotimes_{s = 1}^m
      \Ber \left(
        \mathcal{O}_{\mathcal{W}}(-Q_s)
      \right)
    \right]
    \otimes_{\mathcal{W}}
    \left[
      \bigotimes_{\alpha = 1}^n
      \Ber \left(
        \mathbf{\Pi} \mathcal{O}_{\mathcal{W}}(-\mathsf{Q}_{\alpha})
      \right)
    \right]
    \\
    &\cong \mathcal{O}_{\mathcal{W}} \left(
      - \sum_{s = 1}^m Q_s
    \right)
    \otimes_{\mathcal{W}}
    \mathcal{O}_{\mathcal{W}} \left(
      \sum_{\alpha = 1}^n \mathsf{Q}_{\alpha}
    \right)
    \\
    &\cong \mathcal{O}_{\mathcal{W}} \left(
      - \sum_{s = 1}^m Q_s + \sum_{\alpha = 1}^n \mathsf{Q}_{\alpha}
    \right)
    \, .
  \end{aligned}
\end{equation}

Lastly, recall that for a supermanifold to be CY, its Berezinian sheaf, i.e., the Berezinian of its cotangent sheaf \cite[def.~1.18]{noja-2016-topic-algeb}, has to be trivial.
From \eqref{eq:verify cy:wcp:berezinian:compute}, we can see that it is indeed trivial only when \eqref{eq:verify cy:wcp:cy condition} is satisfied.

Hence we are also able to verify, from this alternative proof, that a weighted projective superspace of the type $\WCP^{m-1|n}_{(\vec{Q},|\vec{\mathsf{Q}})}$ is a CY supermanifold if and only if its weights satisfy \eqref{eq:verify cy:wcp:cy condition}, as claimed.

\subsection{The Calabi-Yau Condition for Complete Intersections of Hypersurfaces in Super-Grassmannians}
\label{app:verify cy:hypersurfaces in Gr}

In \autoref{subsec: applications to math U(1|1) with W}, we physically derived a \emph{novel} mathematical result that a complete intersection
\begin{equation}
  \label{eq:verify cy:hypersurfaces in Gr:intersection}
  \mathscr{G}
  \coloneq \left( \bigcap_{\alpha^+ = 1}^{l_+} \left\{G^+_{\alpha^+} = 0\right\} \right)
  \cap \left( \bigcap_{\alpha^- = 1}^{l_-} \left\{G^-_{\alpha^-} = 0\right\} \right)
  \,
\end{equation}
of $l_+$ even and $l_-$ odd hypersurfaces $G^+_{\alpha^{+}} = 0$ and $G^-_{\alpha^-} = 0$ of even and odd degrees $q^+_{\alpha^+}$ and $q^-_{\alpha^-}$ in a super-Grassmannian of the type $Gr_{1|1}(\mathbb{C}^{m|n})$ is a CY supermanifold if and only if
\begin{equation}
  \label{eq:verify cy:hypersurfaces in Gr:cy condition}
  m - n
  = \sum_{\alpha^+ = 1}^{l_+} q^+_{\alpha^+}
  - \sum_{\alpha^- = 1}^{l_-} q^-_{\alpha^-}
  \, .
\end{equation}

We will now mathematically prove and therefore verify that in fact, all such $\mathscr{G}$'s in super-Grassmannians of the type $Gr_{p|q}(\mathbb{C}^{m|n})$, \emph{regardless of the values of $(p, q)$}, have a trivial Berezinian sheaf and are thus CY supermanifolds, if and only if \eqref{eq:verify cy:hypersurfaces in Gr:cy condition} is satisfied.

First, denote by $\mathcal{G}$ a super-Grassmannian of the type $Gr_{p|q}(\mathbb{C}^{m|n})$, and by $\mathcal{I}_{\mathscr{G}}$ the ideal sheaf of $\mathscr{G} \subset \mathcal{G}$.
Note that for (a complete intersection of) hypersurfaces $Y$ in a manifold $M$, there is an exact sequence which relates $\mathcal{I}_Y$ to the cotangent sheaves on $Y$ and $M$, i.e., the conormal exact sequence.
In particular, the conormal exact sequence that we have involving $\mathcal{I}_{\mathscr{G}}$ is
\begin{equation}
  \label{eq:verify cy:hypersurfaces in Gr:conormal sequence}
  \begin{tikzcd}
    0
    \arrow[r]
    &
    \mathcal{N}^{\vee}_{\mathscr{G}}
    \arrow[r]
    &
    \eval{\Omega^1_{\mathcal{G}}}_{\mathscr{G}}
    \arrow[r]
    &
    \Omega^1_{\mathscr{G}}
    \arrow[r]
    &
    0
    \, ,
  \end{tikzcd}
\end{equation}
where
\begin{enumerate*}
  \item $\mathcal{N}^{\vee}_{\mathscr{G}} \coloneq \flatfrac{\mathcal{I}_{\mathscr{G}}}{\mathcal{I}_{\mathscr{G}}^2}$ is the conormal sheaf on $\mathscr{G}$,
  \item $\eval{\Omega^1_{\mathcal{G}}}_{\mathscr{G}}$ is the cotangent sheaf on $\mathcal{G}$ restricted to $\mathscr{G}$, and
  \item $\Omega^1_{\mathscr{G}}$ is the cotangent sheaf on $\mathscr{G}$.
\end{enumerate*}

Second, recall that the Berezinian sheaf of a supermanifold is the Berezinian of its cotangent sheaf \cite[def.~1.18]{noja-2016-topic-algeb}.
Since the sequence \eqref{eq:verify cy:hypersurfaces in Gr:conormal sequence} splits locally as $\eval{\Omega^1_{\mathcal{G}}}_{\mathscr{G}} \cong \Omega^1_{\mathscr{G}} \oplus \mathcal{N}^{\vee}_{\mathscr{G}}$, we can compute the Berezinian sheaf of $\mathscr{G}$ as
\begin{equation}
  \label{eq:verify cy:hypersurfaces in Gr:berezinian via split sequence}
  \Ber(\mathscr{G})
  \coloneq \Ber(\Omega^1_{\mathscr{G}})
  \cong \Ber({\Omega^1_{\mathcal{G}}}\big\vert_{\mathscr{G}}) \otimes_{\mathscr{G}} \Ber(\mathcal{N}^{\vee}_{\mathscr{G}})^{-1}
  = \eval{\Ber(\Omega^1_{\mathcal{G}})}_{\mathscr{G}} \otimes_{\mathscr{G}} \Ber(\mathcal{N}^{\vee}_{\mathscr{G}})^{-1}
  \, .
\end{equation}

Third, since $\mathscr{G}$ is a complete intersection of $l_+$ even and $l_-$ odd hypersurfaces of even and odd degrees $q^+_{\alpha^+}$ and $q^-_{\alpha^-}$, the conormal sheaf can be expressed as
\begin{equation}
  \label{eq:verify cy:hypersurfaces in Gr:conormal sheaf}
  \mathcal{N}^{\vee}_{\mathscr{G}}
  \cong \left[
    \bigoplus_{\alpha^+ = 1}^{l_+} \eval{\mathcal{O}_{\mathcal{G}}(- q^+_{\alpha^+})}_{\mathscr{G}}
  \right] \oplus \left[
    \bigoplus_{\alpha^- = 1}^{l_-} \eval{\mathbf{\Pi} \mathcal{O}_{\mathcal{G}}(- q^-_{\alpha^-})}_{\mathscr{G}}
  \right]
  \, ,
\end{equation}
As Berezinians are multiplicative, we compute that
\begin{equation}
  \label{eq:verify cy:hypersurfaces in Gr:conormal sheaf:berezinian}
  \begin{aligned}
    \Ber(\mathcal{N}^{\vee}_{\mathscr{G}})
    &\cong \Ber \left(
      \left[
        \bigoplus_{\alpha^+ = 1}^{l_+} \eval{\mathcal{O}_{\mathcal{G}}(- q^+_{\alpha^+})}_{\mathscr{G}}
      \right] \oplus \left[
        \bigoplus_{\alpha^- = 1}^{l_-} \eval{\mathbf{\Pi} \mathcal{O}_{\mathcal{G}}(- q^-_{\alpha^-})}_{\mathscr{G}}
      \right]
    \right)
    \\
    &\cong \eval{\mathcal{O}_{\mathcal{G}} \left(
        - \sum_{\alpha^+ = 1}^{l_+} q^+_{\alpha^+}
        + \sum_{\alpha^- = 1}^{l_-} q^-_{\alpha^-}
      \right)}_{\mathscr{G}}
    \, .
  \end{aligned}
\end{equation}

Fourth, applying \eqref{eq:verify cy:hypersurfaces in Gr:conormal sheaf:berezinian} to \eqref{eq:verify cy:hypersurfaces in Gr:berezinian via split sequence}, and making use of $\Ber(\Omega^1_{\mathcal{G}}) \eqcolon \Ber(\mathcal{G})$ previously computed in \eqref{eq:u-1-1:quantum:berezinian of supergrassmannian}, we get
\begin{equation}
  \label{eq:verify cy:hypersurfaces in Gr:berezinian:substituted}
  \Ber(\mathscr{G})
  \cong \eval{ \left\{
      \Ber(\mathcal{S}_{\mathcal{G}})^{m-n}
      \otimes_{\mathcal{G}} \mathcal{O}_{\mathcal{G}} \left(
        \sum_{\alpha^+ = 1}^{l_+} q^+_{\alpha^+}
        - \sum_{\alpha^- = 1}^{l_-} q^-_{\alpha^-}
      \right)
    \right\}
  }_{\mathscr{G}}
  \, .
\end{equation}

Fifth, we have $\Ber(\mathcal{S}_{\mathcal{G}}) \cong \mathcal{O}_{\mathcal{G}}(-1)$, where $\mathcal{O}_{\mathcal{G}}(-1)$ is an invertible sheaf on $\mathcal{G}$. We can understand why as follows. First, recall that in a regular Grassmannian $\mathsf{G} = Gr_p(\mathbb{C}^m)$, the invertible sheaf $\mathcal{O}_{\mathsf{G}}(-1)$ is isomorphic to the determinant of its tautological sheaf, i.e., $\mathcal{O}_{\mathsf{G}}(-1) \cong \det(\mathcal{S}_{\mathsf{G}})$ (a good review explaining this isomorphism of the underlying vector bundles is given in \cite[$\S$5.4]{goetmark-2010-koppel-formul-grass}).
Then, for a \emph{super}-Grassmannian $\mathcal{G}$, as a \emph{super}-extension of a Grassmannian, the invertible sheaf $\mathcal{O}_{\mathcal{G}}(-1)$ ought to be isomorphic to the \emph{super}determinant, or equivalently, the Berezinian, of its tautological sheaf, i.e., $\mathcal{O}_{\mathcal{G}}(-1) \cong \Ber(\mathcal{S}_{\mathcal{G}})$.\footnote{%
  \label{ft:verify cy:hypersurfaces in Gr:Ber SG as invertible sheaf}%
  We can show that this \emph{super}-generalization of $\mathcal{O}_{\mathsf{G}}(-1) \cong \det(\mathcal{S}_{\mathsf{G}})$  reduces to the following known examples:
  \begin{enumerate}
    \item Consider $\mathcal{G}$ to be a projective superspace, i.e., $\mathcal{G} = Gr_{1|0}(\mathbb{C}^{m|n}) \cong \mathbb{CP}^{m-1|n}$.
    From \eqref{eq:u-1-1:quantum:berezinian of supergrassmannian} we know that the Berezinian sheaf of $\mathcal{G}$ is $\Ber(\mathcal{S}_{\mathcal{G}})^{m-n}$.
    Comparing with \cite[thm.~2.5]{noja-2016-topic-algeb}, we find that, indeed, $\mathcal{O}_{\mathcal{G}}(-1) \cong \Ber(\mathcal{S}_{\mathcal{G}})$ for projective superspaces.
    A more fundamental explanation as to why this is so can be derived via the universal exact sequence \eqref{eq:u-1-1:quantum:grassmannian universal sequence} and Euler sequence \cite[eq.~(79)]{Cacciatori:2017qyd} on projective superspaces.
    From these, one can show that $\mathcal{O}_{\mathcal{G}}(-1) \cong \mathcal{S}_{\mathcal{G}}$,\footnotemark{}
    but since $\mathcal{S}_{\mathcal{G}}$ is a rank-$1|0$ sheaf, $\Ber(\mathcal{S}_{\mathcal{G}})$ reduces to $\det(\mathcal{S}_{\mathcal{G}}) = \mathcal{S}_{\mathcal{G}}$, and thus $\mathcal{O}_{\mathcal{G}}(-1) \cong \Ber(\mathcal{S}_{\mathcal{G}})$.

    \item Consider $\mathcal{G}$ in a fully even ambient superspace $\mathbb{C}^{m|0}$ whence it can only be the space of Grassmann-even planes in $\mathbb{C}^{m|0}$, i.e., consider $\mathcal{G} = Gr_{p|0}(\mathbb{C}^{m|0}) \cong Gr_p(\mathbb{C}^m) = \mathsf{G}$ a regular Grassmannian.
    Its tautological sheaf is a rank-$p|0$ sheaf, whence $\Ber(\mathcal{S}_{\mathcal{G}})$ reduces to $\det(\mathcal{S}_{\mathcal{G}}) \cong \det(\mathcal{S}_{\mathsf{G}})$.
    Thus, the isomorphism $\mathcal{O}_{\mathcal{G}}(-1) \cong \Ber(\mathcal{S}_{\mathcal{G}})$ reduces to $\mathcal{O}_{\mathsf{G}}(-1) \cong \det(\mathcal{S}_{\mathsf{G}})$, as it should.
  \end{enumerate}
  }
This means that we can re-express \eqref{eq:verify cy:hypersurfaces in Gr:berezinian:substituted} as
\begin{equation}
  \label{eq:verify cy:hypersurfaces in Gr:berezinian}
  \Ber(\mathscr{G})
  \cong \eval{
    \mathcal{O}_{\mathcal{G}} \left(
      - m + n
      + \sum_{\alpha^+ = 1}^{l_+} q^+_{\alpha^+}
      - \sum_{\alpha^- = 1}^{l_-} q^-_{\alpha^-}
    \right)
  }_{\mathscr{G}}
  \, .
\end{equation}
\footnotetext{%
  \label{ft:verify cy:hypersurfaces in Gr:invertible sheaf of cp}%
  Twisting the universal exact sequence \eqref{eq:u-1-1:quantum:grassmannian universal sequence} on projective superspaces by the sheaf $\mathcal{S}^{\vee}_{\mathcal{G}}$ dual to $\mathcal{S}_{\mathcal{G}}$, we get
  \begin{equation*}
    \label{eq:verify cy:hypersurfaces in Gr:universal exact sequence:twisted}
    \begin{tikzcd}[ampersand replacement=\&]
      0
      \arrow[r]
      \&
      \mathcal{O}_{\mathcal{G}}
      \arrow[r]
      \&
      \mathbb{C}^{m|n} \otimes \mathcal{S}^{\vee}_{\mathcal{G}} \otimes \mathcal{O}_{\mathcal{G}}
      \arrow[r]
      \&
      \mathcal{T}_{\mathcal{G}}
      \arrow[r]
      \&
      0
    \end{tikzcd}
    \, ,
  \end{equation*}
  where $\mathcal{T}_{\mathcal{G}} \coloneq \mathcal{S}^{\vee}_{\mathcal{G}} \otimes \mathcal{Q}_{\mathcal{G}}$ is the tangent sheaf on $\mathcal{G}$.
  Comparing with the Euler sequence on projective superspaces \cite[eq.~(79)]{Cacciatori:2017qyd}, i.e., \eqref{eq:verify cy:wcp:exact sequence} with $Q_s = 1$ and $\mathsf{Q}_{\alpha} = 1$, we see that this implies $\mathcal{S}^{\vee}_{\mathcal{G}} \otimes \mathcal{O}_{\mathcal{G}} \cong \mathcal{O}_{\mathcal{G}}(1)$, or equivalently, $\mathcal{S}^{\vee}_{\mathcal{G}} \cong \mathcal{O}_{\mathcal{G}}(1)$.
  Dualizing this, we get $\mathcal{S}_{\mathcal{G}} \cong \mathcal{O}_{\mathcal{G}}(-1)$.
}

Lastly, recall that for a supermanifold to be CY, its Berezinian sheaf has to be trivial.
From \eqref{eq:verify cy:hypersurfaces in Gr:berezinian}, we see that it is trivial when \eqref{eq:verify cy:hypersurfaces in Gr:cy condition} is satisfied.

Hence, a complete intersection of $l_+$ even and $l_-$ odd hypersurfaces of even and odd degrees $q^+_{\alpha^+}$ and $q^-_{\alpha^-}$ in a super-Grassmannian of the type $Gr_{p|q}(\mathbb{C}^{m|n})$, regardless of the values of $(p, q)$, is a CY supermanifold if and only if \eqref{eq:verify cy:hypersurfaces in Gr:cy condition} is satisfied, as claimed.

\subsection{The Calabi-Yau Condition for Complete Intersections of Hypersurfaces in Products of Super-Grassmannians}
\label{app:verify cy:cicy in products of SG}

In \autoref{sec:U(1|1)-N with W:scale-invariance}, we physically derived a \emph{novel} mathematical result that a complete intersection
\begin{equation}
  \label{eq:verify cy:cicy in products of SG:hypersurfaces}
  \mathscr{G} \coloneq \Big\{G^+ = 0\Big\} \cap \Big\{G^- = 0\Big\}
  \,
\end{equation}
of an even and odd hypersurface $G^+ = 0$ and $G^- = 0$, with even and odd bidegrees $(\hat{m}_1, \hat{m}_2)$ and $(\hat{n}_1, \hat{n}_2)$, in a product of two super-Grassmannians of the type $Gr_{1|1}(\mathbb{C}^{m_1|n_1}) \times Gr_{1|1}(\mathbb{C}^{m_2|n_2})$ is a CY supermanifold if and only if
\begin{equation}
  \label{eq:verify cy:cicy in products of SG:cy condition}
  m_k - n_k = \hat{m}_k - \hat{n}_k
  \, \qcomma
  k \in \{1, \dots, \ell\}
  \, ,
\end{equation}
where $\ell = 2$ in \autoref{sec:U(1|1)-N with W:scale-invariance}.

The \emph{even} bidegrees $(\hat{m}_1, \hat{m}_2)$ of the $G^+$ polynomial refer to it being an \emph{even} bihomogeneous polynomial of
(i) degree $\hat{m}_1$ in a function of the coordinates on the first super-Grassmannian $Gr_{1|1}(\mathbb{C}^{m_1|n_1})$,
and (ii) degree $\hat{m}_2$ in a function of the coordinates on the second super-Grassmannian $Gr_{1|1}(\mathbb{C}^{m_2|n_2})$.

The \emph{odd} bidegrees $(\hat{n}_1, \hat{n}_2)$ of the $G^-$ polynomial refer to it being an \emph{odd} bihomogeneous polynomial of
(i) degree $\hat{n}_1$ in a function of the coordinates on the first super-Grassmannian $Gr_{1|1}(\mathbb{C}^{m_1|n_1})$,
and (ii) degree $\hat{n}_2$ in a function of the coordinates on the second super-Grassmannian $Gr_{1|1}(\mathbb{C}^{m_2|n_2})$.

Consider now a product of super-Grassmannians $\prod_{k = 1}^{\ell} Gr_{p_k|q_k}(\mathbb{C}^{m_k|n_k})$, containing an even hypersurface $G^+ = 0$ with even multidegrees $(\hat{m}_1, \dots, \hat{m}_{\ell})$, and an odd hypersurface $G^- = 0$ with odd multidegrees $(\hat{n}_1, \dots, \hat{n}_{\ell})$, where the multidegreeness of the $G^{\pm}$ polynomials are straightforward generalizations of the above description, which is the case for $\ell = 2$.

We will now mathematically prove and therefore verify that in fact, all such $\mathscr{G}$'s in products of super-Grassmannians of the type $\prod_{k = 1}^{\ell} Gr_{p_k|q_k}(\mathbb{C}^{m_k|n_k})$, \emph{regardless of the values of $(p_k, q_k)$ and $\ell$}, have a trivial Berezinian sheaf and are thus CY supermanifolds, if and only if \eqref{eq:verify cy:cicy in products of SG:cy condition} is satisfied.

As our analysis will follow that which was done in \autoref{app:verify cy:hypersurfaces in Gr}, we will omit some intermediate steps for brevity.

First, let us denote by $\mathcal{G}_k$, the $k^{\text{th}}$ super-Grassmannian of type $Gr_{p_k|q_k}(\mathbb{C}^{m_k|n_k})$, and by $\mathfrak{G}$, the product $\prod_{k = 1}^{\ell} \mathcal{G}_k$.

Second, since $\mathscr{G}$ is a complete intersection of an even and odd hypersurface of even and odd multidegrees $(\hat{m}_1, \dots, \hat{m}_{\ell})$ and $(\hat{n}_1, \dots, \hat{n}_{\ell})$, its conormal sheaf can be expressed as
\begin{equation}
  \label{eq:verify cy:cicy in product of SG:conormal sheaf}
  \mathcal{N}^{\vee}_{\mathscr{G}}
  \cong \mathcal{O}_{\mathfrak{G}} \bigl(- \hat{m}_1, \dots, - \hat{m}_{\ell}\bigr)
  \oplus \mathbf{\Pi} \mathcal{O}_{\mathfrak{G}} \bigl(- \hat{n}_1, \dots, - \hat{n}_{\ell}\bigr)
  \, ,
\end{equation}
where $\mathcal{O}_{\mathfrak{G}}(\underbrace{-1, \dots, -1}_{\ell}) \cong \mathcal{O}_{\mathcal{G}_1}(-1) \boxtimes \dots \boxtimes \mathcal{O}_{\mathcal{G}_{\ell}}(-1)$ is the external tensor product of invertible sheaves, with each $\mathcal{O}_{\mathcal{G}_k}(-1) \cong \Ber(\mathcal{S}_{\mathcal{G}_k})$ (see \autoref{app:verify cy:hypersurfaces in Gr}).
We can compute the Berezinian of $\mathcal{N}^{\vee}_{\mathscr{G}}$ as
\begin{equation}
  \label{eq:verify cy:cicy in product of SG:berezinian of conormal sheaf}
  \begin{aligned}
    \Ber(\mathcal{N}^{\vee}_{\mathscr{G}})
    &\cong \Ber \left(
      \eval{\mathcal{O}_{\mathfrak{G}} \bigl(- \hat{m}_1, \dots, - \hat{m}_{\ell}\bigr)}_{\mathscr{G}}
      \oplus \eval{\mathbf{\Pi} \mathcal{O}_{\mathfrak{G}} \bigl(- \hat{n}_1, \dots, - \hat{n}_{\ell}\bigr)}_{\mathscr{G}}
    \right)
    \\
    &\cong \eval{\mathcal{O}_{\mathfrak{G}} \bigl( - \hat{m}_1 + \hat{n}_1, \dots, - \hat{m}_{\ell} + \hat{n}_{\ell} \bigr)}_{\mathscr{G}}
    \, .
  \end{aligned}
\end{equation}

Third, using the result of \autoref{app:verify cy:supergrassmannians}, we can compute the Berezinian sheaf of $\mathfrak{G}$ as\footnote{%
  \label{ft:verify cy:cicy in product of SG:cy condition for product of SG}%
  An obvious consequence of this computation is that a product of super-Grassmannians is CY if and only if each super-Grassmannian multiplicand is CY.
}
\begin{equation}
  \label{eq:verify cy:cicy in product of SG:berezinian of product SG}
  \Ber(\mathfrak{G})
  = \Ber \left( \prod_{k = 1}^{\ell} \mathcal{G}_k \right)
  \cong \mathop{\boxtimes}_{k = 1}^{\ell} \Ber(\mathcal{G}_k)
  \cong \mathop{\boxtimes}_{k = 1}^{\ell} \mathcal{O}_{\mathcal{G}_k} \bigl(- m_k + n_k\bigr)
  \cong \mathcal{O}_{\mathfrak{G}}\bigl(- m_1 + n_1, \dots, - m_{\ell} + n_{\ell}\bigr)
  \, .
\end{equation}

Fourth, recalling that $\Ber(\mathfrak{G}) \coloneq \Ber(\Omega^1_{\mathfrak{G}})$, and substituting \eqref{eq:verify cy:cicy in product of SG:berezinian of conormal sheaf} and \eqref{eq:verify cy:cicy in product of SG:berezinian of product SG} in \eqref{eq:verify cy:hypersurfaces in Gr:berezinian via split sequence}, we get
\begin{equation}
  \label{eq:verify cy:cicy in product of SG:berezinian}
  \begin{aligned}
    \Ber(\mathscr{G})
    &\cong \eval{\biggl\{
        \mathcal{O}_{\mathfrak{G}}\bigl(- m_1 + n_1, \dots, - m_{\ell} + n_{\ell}\bigr)
        \otimes_{\mathfrak{G}}
        \mathcal{O}_{\mathfrak{G}} \bigr( \hat{m}_1 - \hat{n}_1, \dots, \hat{m}_{\ell} - \hat{n}_{\ell} \bigr)
      \biggr\}}_{\mathscr{G}}
    \\
    &\cong \eval{\mathcal{O}_{\mathfrak{G}}\bigl( - m_1 + n_1 + \hat{m}_1 - \hat{n}_1, \dots, - m_{\ell} + n_{\ell} + \hat{m}_{\ell} - \hat{n}_{\ell} \bigr)}_{\mathscr{G}}
    \\
    &\cong  \mathop{\boxtimes}_{k = 1}^{\ell} \eval{ \mathcal{O}_{\mathcal{G}_k} \bigl(- m_k + n_k + \hat{m}_k - \hat{n}_k \bigr) }_{\mathscr{G}}
    \, .
  \end{aligned}
\end{equation}

Lastly, recall that for a supermanifold to be CY, its Berezinian sheaf has to be trivial.
From \eqref{eq:verify cy:cicy in product of SG:berezinian}, we see that it is trivial only when the expression within the bracket of the last term therein vanishes for all $k$, i.e., when \eqref{eq:verify cy:cicy in products of SG:cy condition} is satisfied.

Hence, a complete intersection of an even and odd hypersurface of even and odd multidegrees $(\hat{m}_1, \dots, \hat{m}_{\ell})$ and $(\hat{n}_1, \dots, \hat{n}_{\ell})$ in a product of $\ell$ super-Grassmannians of the type $\prod_{k = 1}^{\ell} Gr_{p_k|q_k}(\mathbb{C}^{m_k|n_k})$, regardless of the values of $(p_k, q_k)$ and $\ell$, is a CY supermanifold if and only if \eqref{eq:verify cy:cicy in products of SG:cy condition} is satisfied, as claimed.

\subsection{The Calabi-Yau Condition for Supervector Bundles over ``Weighted Super-Grassmannians''}
\label{app:verify cy:supervector over WGr}

In \autoref{sec:U(1|1) phase transition:scale-invariance}, we physically derived a \emph{novel} mathematical result that a supervector bundle over a ``weighted super-Grassmannian'' of the type
\begin{equation}
  \label{eq:verify cy:supervector over WGr:bundle space}
  \mathcal{S}^{1|1}_{Q, \mathsf{Q}, r}
  = \bigoplus_{s'' = 1}^{m''} \mathscr{L}_{s''}^{Q_{s''}}
    \oplus \bigoplus_{t'' = 1}^{n''} \mathbf{\Pi} \mathscr{L}_{t''}^{\mathsf{Q}_{t''}}
    \rightarrow \text{W}Gr_{1|1} \left( \mathbb{C}^{m'|n'}_{(\vec{Q'}|\vec{\mathsf{Q}'})} \right)
  \, ,
\end{equation}
where
(i) $r$ is a positive size of the ``weighted super-Grassmannian'',
(ii) $(Q_{s'}, \mathsf{Q}_{t'})$ are non-negative integers,
and (iii) $(Q_{s''}, \mathsf{Q}_{t''})$ are non-positive integers,
is a CY supermanifold if and only if
\begin{equation}
  \label{eq:verify cy:supervector over WGr:cy condition}
  \sum_{s' = 1}^{m'} Q_{s'} + \sum_{s'' = 1}^{m''} Q_{s''}
  = \sum_{t' = 1}^{n'} \mathsf{Q}_{t'} + \sum_{t'' = 1}^{n''} \mathsf{Q}_{t''}
  \, .
\end{equation}

We will now mathematically prove and therefore verify that in fact, supervector bundles $\mathcal{S}^{p|q}_{Q, \mathsf{Q}, r}$ over all ``weighted super-Grassmannians'' of the type $\text{W} Gr_{p|q}\left( \mathbb{C}^{m'|n'}_{(\vec{Q'}|\vec{\mathsf{Q}'})} \right)$, \emph{regardless of the values of $(p ,q)$}, have a trivial Berezinian sheaf and are thus CY supermanifolds \cite[def.~1.20]{noja-2016-topic-algeb}, if and only if \eqref{eq:verify cy:supervector over WGr:cy condition} is satisfied.

\subtitle{The Berezinian Sheaf of ``Weighted Super-Grassmannians''}

Let us compute the Berezinian sheaf of a ``weighted super-Grassmannian'' $\mathcal{W}\mathcal{G} = \text{W} Gr_{p|q}\left( \mathbb{C}^{m'|n'}_{(\vec{Q'}|\vec{\mathsf{Q}'})} \right)$, as we will eventually need it.
The computation follows largely the computation of the Berezinian sheaf of weighted projective superspaces from the alternative proof in \autoref{app:verify cy:wcp}.

First, we consider a generalization of the Euler sequence on weighted projective superspaces \eqref{eq:verify cy:wcp:exact sequence} to ``weighted super-Grassmannians'', i.e.,
\begin{equation}
  \label{eq:verify cy:supervector over WGr:euler sequence}
  \begin{tikzcd}
    0
    \arrow[r]
    &
    \mathcal{O}_{\mathcal{W}\mathcal{G}}
    \arrow[r]
    &
    \left[ \bigoplus_{s' = 1}^{m'} \mathcal{O}_{\mathcal{W}\mathcal{G}}(Q_{s'}) \right]
    \oplus
    \left[ \bigoplus_{t' = 1}^{n'} \mathbf{\Pi} \mathcal{O}_{\mathcal{W}\mathcal{G}}(\mathsf{Q}_{t'}) \right]
    \arrow[r]
    &
    \mathcal{T}_{\mathcal{W}\mathcal{G}}
    \arrow[r]
    &
    0
    \, ,
  \end{tikzcd}
\end{equation}
where $\mathcal{O}_{\mathcal{W}\mathcal{G}}$ and $\mathcal{T}_{\mathcal{W}\mathcal{G}}$ are the (locally trivial) structure sheaf of rank $p|q$ and tangent sheaf on $\mathcal{W}\mathcal{G}$, respectively.

Second, dualizing this exact sequence, we get
\begin{equation}
  \label{eq:verify cy:supervector over WGr:dual euler sequence}
  \begin{tikzcd}
    0
    \arrow[r]
    &
    \Omega^1_{\mathcal{W}\mathcal{G}}
    \arrow[r]
    &
    \left[ \bigoplus_{s' = 1}^{m'} \mathcal{O}_{\mathcal{W}\mathcal{G}}(-Q_{s'}) \right]
    \oplus
    \left[ \bigoplus_{t' = 1}^{n'} \mathbf{\Pi} \mathcal{O}_{\mathcal{W}\mathcal{G}}(-\mathsf{Q}_{t'}) \right]
    \arrow[r]
    &
    \mathcal{O}_{\mathcal{W}\mathcal{G}}
    \arrow[r]
    &
    0
    \, ,
  \end{tikzcd}
\end{equation}
where $\Omega^1_{\mathcal{W}\mathcal{G}}$ is the cotangent sheaf on $\mathcal{W}\mathcal{G}$.

Third, noting that the exact sequence \eqref{eq:verify cy:supervector over WGr:dual euler sequence} splits locally and that $\Ber(\mathcal{O}_{\mathcal{W}\mathcal{G}})$ is trivial, we compute, \emph{\`{a}~la}~\eqref{eq:verify cy:wcp:berezinian:compute}, that
\begin{equation}
  \label{eq:verify cy:supervector over WGr:berezinian sheaf:WGr}
  \begin{aligned}
    \Ber(\Omega^1_{\mathcal{W}\mathcal{G}})
    &\cong \Ber \left(
      \left[ \bigoplus_{s' = 1}^{m'} \mathcal{O}_{\mathcal{W}\mathcal{G}}(-Q_{s'}) \right]
      \oplus
      \left[ \bigoplus_{t' = 1}^{n'} \mathbf{\Pi} \mathcal{O}_{\mathcal{W}\mathcal{G}}(-\mathsf{Q}_{t'}) \right]
    \right)
    \cong \mathcal{O}_{\mathcal{W}\mathcal{G}} \left(
      - \sum_{s' = 1}^{m'} Q_{s'}
      + \sum_{t' = 1}^{n'} \mathsf{Q}_{t'}
    \right)
    \, .
  \end{aligned}
\end{equation}

\subtitle{The CY Condition for $\mathcal{S}^{p|q}_{Q, \mathsf{Q}, r}$}

Since $\mathscr{S} = \mathcal{S}^{p|q}_{Q, \mathsf{Q}, r}$ is (the total space of) a supervector bundle over $\mathcal{W}\mathcal{G}$, its cotangent sheaf sits in the following exact sequence
\begin{equation}
  \label{eq:verify cy:supervector over WGr:cotangent sequence}
  \begin{tikzcd}
    0
    \arrow[r]
    &
    \pi^{*} \Omega^1_{\mathcal{W}\mathcal{G}}
    \arrow[r]
    &
    \Omega^1_{\mathscr{S}}
    \arrow[r]
    &
    \pi^{*} \mathcal{E}^{\vee}
    \arrow[r]
    &
    0
    \, ,
  \end{tikzcd}
\end{equation}
where $\pi: \mathscr{S} \rightarrow \mathcal{W}\mathcal{G}$ is the bundle projection, and $\mathcal{E}^{\vee}$ is the sheaf dual to the sheaf associated to the supervector bundle over $\mathcal{W}\mathcal{G}$, which, from \eqref{eq:verify cy:supervector over WGr:bundle space}, is
\begin{equation}
  \label{eq:verify cy:supervector over WGr:dual sheaf}
  \mathcal{E}^{\vee} = \left[
    \bigoplus_{s'' = 1}^{m''}
    \mathcal{O}_{\mathcal{W}\mathcal{G}}(- Q_{s''})
  \right]
  \oplus \left[
    \bigoplus_{t'' = 1}^{n''}
    \mathbf{\Pi} \mathcal{O}_{\mathcal{W}\mathcal{G}}(- Q_{t''})
  \right]
\end{equation}

Noting that this exact sequence \eqref{eq:verify cy:supervector over WGr:cotangent sequence} splits locally as $\Omega^1_{\mathscr{S}} \cong \pi^{*} \Omega^1_{\mathcal{W}\mathcal{G}} \oplus \pi^{*} \mathcal{E}^{\vee}$, we compute the Berezinian sheaf of $\mathscr{S}$ as
\begin{equation}
  \label{eq:verify cy:supervector over WGr:berezinian sheaf}
  \begin{aligned}
    \Ber (\mathscr{S})
    &\coloneq \Ber (\Omega^1_{\mathscr{S}})
    \\
    &\cong \Ber \left( \pi^{*} \Omega^1_{\mathcal{W}\mathcal{G}} \oplus \pi^{*} \mathcal{E}^{\vee} \right)
    \\
    &\cong \pi^{*} \left[ \Ber(\Omega^1_{\mathcal{W}\mathcal{G}}) \otimes_{\mathcal{W}\mathcal{G}} \Ber(\mathcal{E}^{\vee}) \right]
    \\
    &\cong \pi^{*} \left[
      \mathcal{O}_{\mathcal{W}\mathcal{G}} \left( - \sum_{s' = 1}^{m'} Q_{s'} + \sum_{t' = 1}^{n'} \mathsf{Q}_{t'} \right)
      \otimes_{\mathcal{W}\mathcal{G}} \mathcal{O}_{\mathcal{W}\mathcal{G}} \left( - \sum_{s'' = 1}^{m''} Q_{s''} + \sum_{t'' = 1}^{n''} \mathsf{Q}_{t''} \right)
    \right]
    \\
    &\cong \pi^{*} \mathcal{O}_{\mathcal{W}\mathcal{G}} \left(
      - \sum_{s' = 1}^{m'} Q_{s'} - \sum_{s'' = 1}^{m''} Q_{s''}
      + \sum_{t' = 1}^{n'} \mathsf{Q}_{t'} + \sum_{t'' = 1}^{n''} \mathsf{Q}_{t''}
    \right)
    \, .
  \end{aligned}
\end{equation}

Recall that for a supermanifold to be CY, its Berezinian sheaf has to be trivial.
From \eqref{eq:verify cy:supervector over WGr:berezinian sheaf}, we see that it is trivial when \eqref{eq:verify cy:supervector over WGr:cy condition} is satisfied.

Hence, a supervector bundle over a ``weighted super-Grassmannian'' of the type $\text{W} Gr_{p|q}\left( \mathbb{C}^{m'|n'}_{(\vec{Q'}|\vec{\mathsf{Q}'})} \right)$, regardless of the values of $(p ,q)$, is a CY supermanifold if and only if \eqref{eq:verify cy:supervector over WGr:cy condition} is satisfied, as claimed.

\subsection{The Calabi-Yau Condition for Double Covers of Super-Grassmannians Branched Over a Hypersurface}
\label{sec:verify cy:double cover}

In \autoref{sec:quadrics:cy}, we physically derived a \emph{novel} mathematical result that a double cover of a super-Grassmannian of the type $Gr_{1|0}(\mathbb{C}^{m|n})$ branched over an even hypersurface of degree $2Q$ in $Gr_{1|0}(\mathbb{C}^{m|n})$, is a CY supermanifold if and only if
\begin{equation}
  \label{eq:verify cy:double cover:cy condition}
  m - n = Q
  \, .
\end{equation}

We will now mathematically prove and therefore verify that in fact, double covers of all super-Grassmannians of the type $Gr_{p|q}(\mathbb{C}^{m|n})$ branched over an even hypersurface of degree $2Q$ in $Gr_{p|q}(\mathbb{C}^{m|n})$, \emph{regardless of the values of $(p, q)$}, have a trivial Berezinian sheaf and are thus CY supermanifolds \cite[def.~1.20]{noja-2016-topic-algeb}, if and only if \eqref{eq:verify cy:double cover:cy condition} is satisfied.

To this end, first, let us denote by
\begin{enumerate*}
  \item $\mathcal{G}$, the super-Grassmannian of type $Gr_{p|q}(\mathbb{C}^{m|n})$;

  \item $\mathcal{H}$, the hypersurface of degree $2Q$ in $\mathcal{G}$;

  and

  \item $\widetilde{\mathcal{G}}$, the double cover of $\mathcal{G}$ branched over $\mathcal{H} \subset \mathcal{G}$.
\end{enumerate*}

Second, note that there exists a projection map $\pi: \widetilde{\mathcal{G}} \rightarrow \mathcal{G}$, which is a 2-to-1 map for all points $\pi^{*}(x) \in \widetilde{\mathcal{G}}$ when $x \notin \mathcal{H}$, and a 1-to-1 map for all points $\pi^{*}(x) \in \widetilde{\mathcal{G}}$ when $x \in \mathcal{H}$.

Third, note that since the double cover $\widetilde{\mathcal{G}}$ is branched over $\mathcal{H} \subset \mathcal{G}$, the tangent sheaves on $\mathcal{G}$ and $\widetilde{\mathcal{G}}$ sit in the following exact sequence \cite[$\S$4]{pardini-1991-abelian-cover}
\begin{equation}
  \label{eq:verify cy:double cover:normal exact sequence}
  \begin{tikzcd}
    0
    \arrow[r]
    &
    \mathcal{T}_{\widetilde{\mathcal{G}}}
    \arrow[r]
    &
    \pi^{*} \mathcal{T}_{\mathcal{G}}
    \arrow[r]
    &
    \mathcal{N}_{\flatfrac{\mathcal{H}}{2}}
    \arrow[r]
    &
    0
    \, ,
  \end{tikzcd}
\end{equation}
where $\mathcal{N}_{\flatfrac{\mathcal{H}}{2}}$ is associated to the normal sheaf $\mathcal{N}_{\mathcal{H}}$ on $\mathcal{H}$, and whose dual $\mathcal{N}^{\vee}_{\flatfrac{\mathcal{H}}{2}}$ is such that
\begin{equation}
  \label{eq:verify cy:double cover:conormal sheaf}
  \Ber(\mathcal{N}^{\vee}_{\flatfrac{\mathcal{H}}{2}}) \otimes_{\mathcal{G}} \Ber(\mathcal{N}^{\vee}_{\flatfrac{\mathcal{H}}{2}})
  \cong \Ber(\mathcal{N}^{\vee}_{\mathcal{H}})
  \cong \mathcal{O}_{\mathcal{G}} (- 2Q)
  \, ,
\end{equation}
where the final relation is obtained via a derivation \emph{\`{a}~la} \eqref{eq:verify cy:hypersurfaces in Gr:conormal sheaf:berezinian}.

Dualizing the exact sequence \eqref{eq:verify cy:double cover:normal exact sequence}, we get
\begin{equation}
  \label{eq:verify cy:double cover:conormal exact sequence}
  \begin{tikzcd}
    0
    \arrow[r]
    &
    \mathcal{N}^{\vee}_{\flatfrac{\mathcal{H}}{2}}
    \arrow[r]
    &
    \pi^{*} \Omega^1_{\mathcal{G}}
    \arrow[r]
    &
    \Omega^1_{\widetilde{\mathcal{G}}}
    \arrow[r]
    &
    0
    \, .
  \end{tikzcd}
\end{equation}

Fourth, noting that the exact sequence \eqref{eq:verify cy:double cover:conormal exact sequence} splits locally as $\pi^{*} \Omega^1_{\mathcal{G}} \cong \Omega^1_{\widetilde{\mathcal{G}}} \oplus \mathcal{N}^{\vee}_{\flatfrac{\mathcal{H}}{2}}$, we compute the Berezinian sheaf of $\widetilde{\mathcal{G}}$ as
\begin{equation}
  \label{eq:verify cy:double cover:berezinian sheaf}
  \begin{aligned}
    \Ber \bigl( \widetilde{\mathcal{G}} \bigr)
    &\coloneq \Ber \bigl( \Omega^1_{\widetilde{\mathcal{G}}} \bigr)
    \\
    &\cong \pi^{*} \left[ \Ber(\mathcal{G}) \right]
    \otimes_{\widetilde{\mathcal{G}}} \pi^{*} \left[ \Ber(\mathcal{N}^{\vee}_{\flatfrac{\mathcal{H}}{2}})^{-1} \right]
    \\
    &\cong \pi^{*} \left[
      \Ber(\mathcal{G})
      \otimes_{\mathcal{G}} \Ber(\mathcal{N}^{\vee}_{\flatfrac{\mathcal{H}}{2}})^{-1} \right]
    \\
    &\cong \pi^{*} \left[
      \mathcal{O}_{\mathcal{G}}(- m + n)
      \otimes_{\mathcal{G}} \mathcal{O}_G(Q)
    \right]
    \\
    &\cong \pi^{*} \mathcal{O}_{\mathcal{G}}(- m + n + Q)
    \, .
  \end{aligned}
\end{equation}

Lastly, recall that for a supermanifold to be CY, its Berezinian sheaf has to be trivial.
From \eqref{eq:verify cy:double cover:berezinian sheaf}, we see that it is trivial when \eqref{eq:verify cy:double cover:cy condition} is satisfied.

Hence, a double cover of a super-Grassmannian of the type $Gr_{p|q}(\mathbb{C}^{m|n})$ branched over an even hypersurface of degree $2Q$ in $Gr_{p|q}(\mathbb{C}^{m|n})$, regardless of the values of $(p, q)$, is a CY supermanifold if and only if \eqref{eq:verify cy:double cover:cy condition} is satisfied, as claimed.

Notice that by setting $n$ to 0 whence $q$ necessarily becomes 0 as well, and $p$ to 1, we recover the well-known result that a double cover of a projective space $\CP^{m-1}$ branched over a hypersurface of degree $2Q$ in $\CP^{m-1}$ is CY if and only if \eqref{eq:verify cy:double cover:cy condition} (with $n = 0$) is satisfied \cite[$\S$2.8]{Caldararu:2010ljp}.

\section{Super-Pl\"{u}cker Coordinates of a Super-Grassmannian}
\label{app:plucker}

In this appendix, we will show how to write the super-Pl\"{u}cker coordinates of super-Grassmannians, with concrete examples, using the technology of super-Pl\"{u}cker embedding developed by Shemyakova-Voronov in \cite{Shemyakova_2022, shemyakova-2022-super-pluec}.

\subsection{Supermatrices and ``Wrong'' Supermatrices, and their Superdeterminants}
\label{app:plucker:prelim}

\subtitle{Supermatrices and Parity Reversals}

Let us first define the technical tools that we will need.
Consider an even supermatrix of the type
\begin{equation}
  \label{eq:plucker:prelim:supermatrix}
  \mathcal{X} = \left(
    \begin{array}{c|c}
      A & B
      \\ \hline
      C & D
    \end{array}
  \right)
  \, ,
\end{equation}
i.e., (i) the diagonal $A$ and $D$ are even matrices and (ii) the off-diagonal $B$ and $C$ are odd (nilpotent) matrices.

We define the \emph{left action} of the parity operator $\mathbf{\Pi}$ as
\begin{equation}
  \label{eq:plucker:prelim:supermatrix:left parity}
  \mathbf{\Pi} \mathcal{X}
  \coloneq \mqty(\admat[0]{\mathbb{I}}{\mathbb{I}}) \mathcal{X}
  = \left(
    \begin{array}{c|c}
      C & D
      \\ \hline
      A & B
    \end{array}
  \right)
  \, ,
\end{equation}
and the \emph{right action} as
\begin{equation}
  \label{eq:plucker:prelim:supermatrix:right parity}
  \mathcal{X} \mathbf{\Pi}
  \coloneq \mathcal{X} \mqty(\admat[0]{\mathbb{I}}{\mathbb{I}})
  = \left(
    \begin{array}{c|c}
      B & A
      \\ \hline
      D & C
    \end{array}
  \right)
  \, .
\end{equation}
Notice that both $\mathbf{\Pi} \mathcal{X}$ and $\mathcal{X} \mathbf{\Pi}$ are odd supermatrices in that their diagonals are odd matrices, while their off-diagonals are even matrices.

We define the \emph{parity reversal} of $M$ as
\begin{equation}
  \label{eq:plucker:prelim:supermatrix:parity conj}
  \mathcal{X}^{\mathbf{\Pi}}
  \coloneq \mathbf{\Pi} \mathcal{X} \mathbf{\Pi}
  = \left(
    \begin{array}{c|c}
      D & C
      \\ \hline
      B & A
    \end{array}
  \right)
  \, ,
\end{equation}
which, like $\mathcal{X}$, is an even supermatrix.

\subtitle{Superdeterminant and Inverse Superdeterminant of an Even Supermatrix}

If $D$ is invertible, then we can define the \emph{superdeterminant} of $\mathcal{X}$ as
\begin{equation}
  \label{eq:plucker:prelim:sdet}
  \sdet \mathcal{X} \coloneq \frac{\det(A - B D^{-1} C)}{\det(D)}
  \, .
\end{equation}
If $D$ is not invertible, the superdeterminant is ill-defined.

If $A$ is invertible, then we can define the \emph{inverse superdeterminant} of $\mathcal{X}$ as
\begin{equation}
  \label{eq:plucker:prelim:inverse sdet}
  \sdet^{*} \mathcal{X} \coloneq \sdet \mathcal{X}^{\mathbf{\Pi}} = \frac{\det(D - C A^{-1} B)}{\det(A)}
  \, .
\end{equation}
If $A$ is not invertible, the inverse superdeterminant is ill-defined.

If $A$ and $D$ are invertible, then both the superdeterminant and inverse superdeterminant are well defined; effectively, we have
\begin{equation}
  \label{eq:plucker:prelim:sdet and inverse sdet}
  \sdet \equiv (\sdet^{*})^{-1}
  \, ,
  \qqtext{whence}
  \sdet \mathcal{X} = (\sdet^{*} \mathcal{X})^{-1}
  \, .
\end{equation}

\subtitle{``Wrong'' Supermatrices}

``Wrong'' supermatrices, first introduced in \cite{li-2017-differ-operat}, are supermatrices with a \emph{single} column supervector replaced with another of the opposite parity.
In particular, a Type-I (resp. Type-II) ``wrong'' supermatrix replaces an odd (resp. even) column supervector with an even (resp. odd) column supervector.
As such, they are not standard supermatrices as their Grassmann parities are ill-defined.

A Type-I ``wrong'' supermatrix \cite[def.~2.1]{razzaq-2026-fundam-theor} would be of the type
\begin{equation}
  \label{eq:plucker:prelim:wrong supermatrix:type i}
  \mathcal{X}_{\text{(I)}} = \left(
    \begin{array}{c|c}
      A' & B
      \\ \hline
      C' & D
    \end{array}
  \right)
  \, ,
\end{equation}
and a Type-II ``wrong'' supermatrix \cite[def.~2.3]{razzaq-2026-fundam-theor} would be of the type
\begin{equation}
  \label{eq:plucker:prelim:wrong supermatrix:type ii}
  \mathcal{X}_{\text{(II)}} = \left(
    \begin{array}{c|c}
      A & B'
      \\ \hline
      C & D'
    \end{array}
  \right)
  \, ,
\end{equation}
where
\begin{enumerate*}
  \item $A'$ and $D'$ are even matrices with a \emph{single} even column vector replaced with an odd column vector, and

  \item $B'$ and $C'$ are odd matrices with a \emph{single} column vector replaced with a single odd column vector replaced with an even column vector.
\end{enumerate*}

\subtitle{Superdeterminants of ``Wrong'' Supermatrices}

As $A'$ (resp. $D'$) is non-invertible, the inverse superdeterminant (resp. superdeterminant) of $\mathcal{X}_{\text{(I)}}$ (resp. $\mathcal{X}_{\text{(II)}}$) is ill-defined.
However, the superdeterminant (resp. inverse superdeterminant) is well-defined, i.e.,
\begin{equation}
  \label{eq:plucker:prelim:wrong supermatrix:berezinians}
  \sdet \mathcal{X}_{\text{(I)}}
  = \frac{\det(A' - B D^{-1} C')}{\det(D)}
  \qand
  \sdet^{*} \mathcal{X}_{\text{(II)}}
  = \frac{\det(D' - C A^{-1} B')}{\det(A)}
  \, .
\end{equation}
Notice that unlike the (inverse) superdeterminant of a ``correct'' supermatrix (i.e., \eqref{eq:plucker:prelim:sdet} and \eqref{eq:plucker:prelim:inverse sdet}), the (inverse) superdeterminant of a ``wrong'' supermatrix is Grassmann-odd.

\subtitle{Transformations of the Superdeterminants}

Consider a transformation $\mathcal{X} \rightarrow M \mathcal{X}$ by an even supermatrix $M$.
Exploiting the multiplicativity of superdeterminants, the superdeterminant and inverse superdeterminant of $\mathcal{X}$ transform straightforwardly as
\begin{equation}
  \label{eq:plucker:prelim:transform:berezinian}
  \sdet \mathcal{X}
  \rightarrow \sdet(M \mathcal{X})
  = \sdet M \cdot \sdet \mathcal{X}
  \, ,
  \qquad
  \sdet^{*} \mathcal{X}
  \rightarrow \sdet^{*} (M \mathcal{X})
  = \sdet^{*} M \cdot \sdet^{*} \mathcal{X}
  \, .
\end{equation}

\subsection{Super-Pl\"{u}cker Coordinates of \texorpdfstring{$Gr_{1|1}(\mathbb{C}^{2|3})$}{Gr(1|1, 2|3)}}
\label{sec:plucker:Gr 1-1 2-3}

Let us first work with a concrete example of $Gr_{1|1}(\mathbb{C}^{2|3})$.
Such a space is spanned by 1 Grassmann-even and 1 Grassmann-odd supervector in $\mathbb{C}^{2|3}$, and its homogeneous coordinates can be expressed as a $(1|1 \times 2|3)$ supermatrix just like the $Z$ supermatrix in \autoref{ft:u-1-1:supergroup action} with $m = 2$ and $n = 3$, i.e.,
\begin{equation}
  \label{eq:plucker:Gr 1-1 2-3:grassmannian homogeneous coordinates}
  Z =
  \left(
    \begin{array}{cc|ccc}
      \tensor{\phi}{^1_1} & \tensor{\phi}{^1_2} & \tensor{\psi}{^1_1} & \tensor{\psi}{^1_2} & \tensor{\psi}{^1_3}
      \\ \hline
      \tensor{\phi}{^{\bar{1}}_1} & \tensor{\phi}{^{\bar{1}}_2} & \tensor{\psi}{^{\bar{1}}_1} & \tensor{\psi}{^{\bar{1}}_2} & \tensor{\psi}{^{\bar{1}}_3}
    \end{array}
  \right)
  \, ,
\end{equation}
where $(\tensor{\phi}{^1_{*}}, \tensor{\phi}{^{\bar{1}}_{*}})$ are Grassmann-even and Grassmann-odd components of an even $1|1$-supervector $\phi_{*}$, and $(\tensor{\psi}{^1_{*}}, \tensor{\psi}{^{\bar{1}}_{*}})$ are Grassmann-odd and Grassmann-even components of an odd $1|1$-supervector $\psi_{*}$.
As seen in \autoref{sec: U(1|1) and SGr}, both $\phi_{*}$ and $\psi_{*}$ can be regarded as physical fields in the fundamental representation of $\mathrm{U}(1|1)$.

Denote by $Y$ the super-Pl\"{u}cker coordinates corresponding to $Z$, which are actually coordinates on a weighted projective superspace with weights $(+1, -1)$ \cite{shemyakova-2024-weigh-projec}.
The coordinates are defined via $1|1$-minors, i.e., $(1|1 \times 1|1)$ submatrices, of \eqref{eq:plucker:Gr 1-1 2-3:grassmannian homogeneous coordinates}.\footnote{%
  \label{ft:plucker:full rank submatrices}%
  In general, for $Gr_{p|q}(\mathbb{C}^{m|n})$, the Pl\"{u}cker coordinates would be defined via full-rank, i.e., $(p|q \times p|q)$, submatrices of the $(p|q \times m|n)$ supermatrix of homogeneous coordinates.
}

\subtitle{The Submatrices of $Z$}

There are three types of submatrices that we can extract from \eqref{eq:plucker:Gr 1-1 2-3:grassmannian homogeneous coordinates}; an even supermatrix, a Type-I ``wrong'' supermatrix, and a Type-II ``wrong'' supermatrix.
The even supermatrix is
\begin{equation}
  \label{eq:plucker:Gr 1-1 2-3:even supermatrix}
  Z_{st}
  = \left(
    \begin{array}{c|c}
      \tensor{\phi}{^1_s} & \tensor{\psi}{^1_t}
      \\ \hline
      \tensor{\phi}{^{\bar{1}}_s} & \tensor{\psi}{^{\bar{1}}_t}
    \end{array}
  \right)
  \, ,
\end{equation}
where $s \in \{1, 2\}$ and $t \in \{1, 2, 3\}$ run through the even and odd indices on the ambient $\mathbb{C}^{2|3}$, respectively.
The Type-I ``wrong'' supermatrix is
\begin{equation}
  \label{eq:plucker:Gr 1-1 2-3:type-I supermatrix}
  Z_{s_1 s_2}
  = \left(
    \begin{array}{c|c}
      \tensor{\phi}{^1_{s_1}} & \tensor{\phi}{^1_{s_2}}
      \\ \hline
      \tensor{\phi}{^{\bar{1}}_{s_1}} & \tensor{\phi}{^{\bar{1}}_{s_2}}
    \end{array}
  \right)
  \, ,
\end{equation}
and the Type-II ``wrong'' supermatrix is
\begin{equation}
  \label{eq:plucker:Gr 1-1 2-3:type-II supermatrix}
  Z_{t_1 t_2}
  = \left(
    \begin{array}{c|c}
      \tensor{\psi}{^1_{t_1}} & \tensor{\psi}{^1_{t_2}}
      \\ \hline
      \tensor{\psi}{^{\bar{1}}_{t_1}} & \tensor{\psi}{^{\bar{1}}_{t_2}}
    \end{array}
  \right)
  \, .
\end{equation}

\subtitle{The Super-Pl\"{u}cker Coordinates of $Gr_{1|1}(\mathbb{C}^{2|3})$}

The super-Pl\"{u}cker coordinates of $Gr_{1|1}(\mathbb{C}^{2|3})$ are defined via the various superdeterminants of these matrices.
In particular, the superdeterminant and inverse superdeterminant of \eqref{eq:plucker:Gr 1-1 2-3:even supermatrix} will define two Grassmann-even super-Pl\"{u}cker coordinates;
and the inverse superdeterminant and superdeterminant of \eqref{eq:plucker:Gr 1-1 2-3:type-I supermatrix} and \eqref{eq:plucker:Gr 1-1 2-3:type-II supermatrix}, respectively, will define two Grassmann-odd super-Pl\"{u}cker coordinates.
Concretely, they are\footnote{%
  \label{ft:plucker:omit determinant notation}%
  The numerator and denominator of the fractions in the following expression are actually determinants.
  However, as they are determinants of $1 \times 1$ matrices, the notation is redundant, and thus is omitted.
}
\begin{equation}
  \label{eq:plucker:Gr 1-1 2-3:plucker coordinates}
  \begin{aligned}
    Y_{st}^+
    &\coloneq \sdet(Z_{st})
    = \frac{\tensor{\phi}{^1_s} - \tensor{\psi}{^1_t} \left[ \tensor{\psi}{^{\bar{1}}_t} \right]^{-1} \tensor{\phi}{^{\bar{1}}_s}}{\tensor{\psi}{^{\bar{1}}_t}}
    \, ,
    &\quad
    Y_{st}^-
    &\coloneq \sdet^{*}(Z_{st})
    = \frac{\tensor{\psi}{^{\bar{1}}_t} -  \tensor{\phi}{^{\bar{1}}_s} \left[ \tensor{\phi}{^1_s} \right]^{-1} \tensor{\psi}{^1_t}}{\tensor{\phi}{^1_s}}
    \, ,
    \\
    Y^-_{s_1 s_2}
    &\coloneq \sdet^{*}(Z_{s_1 s_2})
    = \frac{\tensor{\phi}{^{\bar{1}}_{s_2}} - \tensor{\phi}{^{\bar{1}}_{s_1}} \left[ \tensor{\phi}{^1_{s_1}} \right]^{-1} \tensor{\phi}{^1_{s_2}}}{\tensor{\phi}{^1_{s_1}}}
    \, ,
    &\quad
    Y^+_{t_1 t_2}
    &\coloneq \sdet(Z_{t_1 t_2})
    = \frac{\tensor{\psi}{^1_{t_1}} - \tensor{\psi}{^1_{t_2}} \left[ \tensor{\psi}{^{\bar{1}}_{t_2}} \right]^{-1} \tensor{\psi}{^{\bar{1}}_{t_1}}}{\tensor{\psi}{^{\bar{1}}_{t_2}}}
    \, .
  \end{aligned}
\end{equation}

As $s \in \{1, 2\}$ and $t \in \{1, 2 ,3\}$, there are $2 \times 3 = 6$ Grassmann-even super-Pl\"{u}cker coordinates $Y^+_{st}$ and $Y^-_{st}$ each.

Notice that $Y^-_{s_1 s_2} = 0$ and $Y^+_{t_1 t_2} = 0$ when $s_1 = s_2$ and $t_1 = t_2$, respectively.
Hence, there are $P^2_2 = 2$ and $P^3_2 = 6$ Grassmann-odd super-Pl\"{u}cker coordinates $Y^-_{s_1 s_2}$ and $Y^+_{t_1 t_2}$, respectively, where $P^a_b \coloneq \frac{a!}{(a-b)!}$.
The ``$+$'' and ``$-$'' in the superscript will be explained shortly.

\subtitle{Supergauge Transformations, and Weighted Coordinates on a Weighted Projective Superspace}

Being the homogeneous coordinates of a super-Grassmannian $Gr_{1|1}(\mathbb{C}^{2|3})$, the $Z$ supermatrix \eqref{eq:plucker:Gr 1-1 2-3:grassmannian homogeneous coordinates} has a $\mathrm{U}(1|1)$ supergauge group action (as described in \autoref{ft:u-1-1:supergroup action} with $m = 2$ and $n = 3$).
Likewise, so would the full-rank $(1|1 \times 1|1)$ submatrices \eqref{eq:plucker:Gr 1-1 2-3:even supermatrix}--\eqref{eq:plucker:Gr 1-1 2-3:type-II supermatrix}, i.e.,
\begin{equation}
  \label{eq:plucker:Gr 1-1 2-3:tfm of Z}
  \begin{aligned}
    Z_{st}
    &\rightarrow Z'_{st} = M Z_{st}
    \, ,
    &\qquad
    Z_{ts}
    &\rightarrow Z'_{ts} = M Z_{ts}
    \, ,
    \\
    Z_{s_1 s_2}
    &\rightarrow Z'_{s_1 s_2} = M Z_{s_1 s_2}
    \, ,
    &\qquad
    Z_{t_1 t_2}
    &\rightarrow Z'_{t_1 t_2} = M Z_{t_1 t_2}
    \, ,
  \end{aligned}
\end{equation}
for $M \in \mathrm{U}(1|1)$.
The supergauge group action then implies that the super-Pl\"{u}cker coordinates defined in \eqref{eq:plucker:Gr 1-1 2-3:plucker coordinates} will transform, via \eqref{eq:plucker:prelim:transform:berezinian},\footnote{%
  Recall from \autoref{P_alpha U(1) transform} that $\sdet M$ for an $M \in \mathrm{U}(1|1)$ is always a phase factor; thus, it is invertible.
  Hence, via \eqref{eq:plucker:prelim:sdet and inverse sdet}, we have $\sdet^{*} M = (\sdet M)^{-1}$ in the right equation in \eqref{eq:plucker:prelim:transform:berezinian}.
}
as
\begin{equation}
  \label{eq:plucker:Gr 1-1 2-3:tfm of Y}
  \begin{aligned}
    Y^+_{st}
    &\rightarrow \sdet M \cdot Y^+_{st}
    \, ,
    &\qquad
    Y^-_{st}
    &\rightarrow (\sdet M)^{-1} \cdot Y^-_{st}
    \, ,
    \\
    Y^-_{s_1 s_2}
    &\rightarrow (\sdet M)^{-1} \cdot Y^-_{s_1 s_2}
    \, ,
    &\qquad
    Y^+_{t_1 t_2}
    &\rightarrow \sdet M \cdot Y^+_{t_1 t_2}
    \, .
  \end{aligned}
\end{equation}

As the physical states, i.e., the $\mathrm{U}(1|1)$ fields $\phi_s$ and $\psi_t$, are physically equivalent up to a $\mathrm{U}(1|1)$ supergauge transformation, we thus have an identification of the super-Pl\"{u}cker coordinates under the following equivalence relation:
\begin{equation}
  \label{eq:plucker:Gr 1-1 2-3:wcp coordinates}
  \left( Y^+_{st}, Y^-_{st} \middle\vert Y^+_{t_1 t_2}, Y^-_{s_1 s_2} \right)
  \sim \left( \lambda^{+1} Y^+_{st}, \lambda^{-1} Y^-_{st} \middle\vert \lambda^{+1} Y^+_{t_1 t_2}, \lambda^{-1} Y^-_{s_1 s_2} \right)
  \, ,
\end{equation}
where $\lambda = \sdet M = (\sdet^{*} M)^{-1} \in \mathbb{C}^{\times}$.

Notice that \eqref{eq:plucker:Gr 1-1 2-3:wcp coordinates} is the equivalence relation for coordinates on a weighted projective superspace.
In particular, the super-Pl\"{u}cker coordinates with ``$+$'' in the superscript are coordinates with weight $+1$, while those with ``$-$'' in the superscript are coordinates with weight $-1$.

In total, the $6|6$ super-Pl\"{u}cker coordinates with weight $+1$ and $6|2$ super-Pl\"{u}cker coordinates with weight $-1$, are coordinates on a weighted projective superspace that $Gr_{1|1}(\mathbb{C}^{2|3})$ is embedded in.

\subtitle{The Super-Pl\"{u}cker Coordinates of the Parity-Reversed $Gr_{1|1}(\mathbb{C}^{3|2})$}

Let us consider the parity-reversal of $Gr_{1|1}(\mathbb{C}^{2|3})$, i.e., $Gr_{1|1}(\mathbb{C}^{3|2})$.
The homogeneous coordinates of such a space can be expressed as a $(1|1 \times 3|2)$ supermatrix $\widetilde{Z}$, which is equivalent to the parity reversal of the $Z$ supermatrix \eqref{eq:plucker:Gr 1-1 2-3:grassmannian homogeneous coordinates}, i.e.,
\begin{equation}
  \label{eq:plucker:Gr 1-1 2-3:grassmannian homogeneous coordinates:reversed}
  \widetilde{Z}
  \equiv Z^{\mathbf{\Pi}}
  = \left(
    \begin{array}{ccc|cc}
      \tensor{\psi}{^{\bar{1}}_1} & \tensor{\psi}{^{\bar{1}}_2} & \tensor{\psi}{^{\bar{1}}_3} & \tensor{\phi}{^{\bar{1}}_1} & \tensor{\phi}{^{\bar{1}}_2}
      \\ \hline
      \tensor{\psi}{^1_1} & \tensor{\psi}{^1_2} & \tensor{\psi}{^1_3} & \tensor{\phi}{^1_1} & \tensor{\phi}{^1_2}
    \end{array}
  \right)
  \, .
\end{equation}
The super-Pl\"{u}cker coordinates $\widetilde{Y}$ corresponding to $\widetilde{Z}$ are defined via its $1|1$-minors, which are just parity reversals of the $1|1$-minors of $Z$ in \eqref{eq:plucker:Gr 1-1 2-3:even supermatrix}--\eqref{eq:plucker:Gr 1-1 2-3:type-II supermatrix}.
Repeating the computation, we find that $\widetilde{Y}$ is related to the super-Pl\"{u}cker coordinates $Y$ in \eqref{eq:plucker:Gr 1-1 2-3:plucker coordinates} as
\begin{equation}
  \label{eq:plucker:Gr 1-1 2-3:plucker coordinates:reversed}
  \begin{aligned}
    \widetilde{Y}^+_{ts}
    &\coloneq \sdet(\widetilde{Z}_{ts})
    = \sdet(Z_{st}^{\mathbf{\Pi}})
    = Y^-_{st}
    \, ,
    &\qquad
    \widetilde{Y}^-_{ts}
    &\coloneq \sdet^{*}(\widetilde{Z}_{ts})
    = \sdet^{*}(Z^{\mathbf{\Pi}}_{st})
    = Y^+_{st}
    \, ,
    \\
    \widetilde{Y}^-_{t_2 t_1}
    &\coloneq \sdet^{*}(\widetilde{Z}_{t_2 t_1})
    = \sdet^{*}(Z^{\mathbf{\Pi}}_{t_1 t_2})
    = Y^+_{t_1 t_2}
    \, ,
    &\qquad
    \widetilde{Y}^+_{s_2 s_1}
    &\coloneq \sdet(\widetilde{Z}_{s_2 s_1})
    = \sdet(Z^{\mathbf{\Pi}}_{s_1 s_2})
    = Y^-_{s_1 s_2}
    \, .
  \end{aligned}
\end{equation}

Notice that there is a one-to-one correspondence between the $Y$'s and $\widetilde{Y}$'s, and that the correspondence changes the sign in the superscript.
Recall that the sign in the superscript denotes the weight of the coordinate on the weighted projective superspace that the super-Grassmannian is embedded in.
Hence, the $\widetilde{Y}$'s are negatively-weighted (in the sense that $+1 \times -1 = -1$ and $-1 \times -1 = +1$) $Y$'s.

In other words, the super-Pl\"{u}cker coordinates of the parity-reversed $Gr_{1|1}(\mathbb{C}^{3|2})$ are the same as the super-Pl\"{u}cker coordinates of $Gr_{1|1}(\mathbb{C}^{2|3})$, but with the signs of the weights flipped.

\subsection{Super-Pl\"{u}cker Coordinates of \texorpdfstring{$Gr_{p|q}(\mathbb{C}^{m|n})$}{Gr(p|q; m|n)}}
\label{sec:plucker:Gr}

Now that we know how super-Pl\"{u}cker Coordinates of $Gr_{1|1}(\mathbb{C}^{2|3})$ are defined, we can extend it to a more general super-Grassmannian of the type $Gr_{p|q}(\mathbb{C}^{m|n})$, where $p$ and $q \neq 0$.
Such a space is spanned by $p$ Grassmann-even and $q$ Grassmann-odd supervectors in $\mathbb{C}^{m|n}$, and its homogeneous coordinates can be expressed as a $(p|q \times m|n)$ matrix, i.e.,
\begin{equation}
  \label{eq:plucker:Gr:grassmannian homogeneous coordinates}
  Z = \left(
    \begin{array}{cccc|cccc}
      \tensor{\phi}{^1_1} & \tensor{\phi}{^1_2} & \cdots & \tensor{\phi}{^1_m} & \tensor{\psi}{^1_1} & \tensor{\psi}{^1_2} & \cdots & \tensor{\psi}{^1_n}
      \\
      \tensor{\phi}{^2_1} & \tensor{\phi}{^2_2} & \cdots & \tensor{\phi}{^2_m} & \tensor{\psi}{^2_1} & \tensor{\psi}{^2_2} & \cdots & \tensor{\psi}{^2_n}
      \\
      \vdots & \vdots & \ddots & \vdots & \vdots & \vdots & \ddots & \vdots
      \\
      \tensor{\phi}{^p_1} & \tensor{\phi}{^p_2} & \cdots & \tensor{\phi}{^p_m} & \tensor{\psi}{^p_1} & \tensor{\psi}{^p_2} & \cdots & \tensor{\psi}{^p_n}
      \\ \hline
      \tensor{\phi}{^{\bar{1}}_1} & \tensor{\phi}{^{\bar{1}}_2} & \cdots & \tensor{\phi}{^{\bar{1}}_m} & \tensor{\psi}{^{\bar{1}}_1} & \tensor{\psi}{^{\bar{1}}_2} & \cdots & \tensor{\psi}{^{\bar{1}}_n}
      \\
      \tensor{\phi}{^{\bar{2}}_1} & \tensor{\phi}{^{\bar{2}}_2} & \cdots & \tensor{\phi}{^{\bar{2}}_m} & \tensor{\psi}{^{\bar{2}}_1} & \tensor{\psi}{^{\bar{2}}_2} & \cdots & \tensor{\psi}{^{\bar{2}}_n}
      \\
      \vdots & \vdots & \ddots & \vdots & \vdots & \vdots & \ddots & \vdots
      \\
      \tensor{\phi}{^{\bar{q}}_1} & \tensor{\phi}{^{\bar{q}}_2} & \cdots & \tensor{\phi}{^{\bar{q}}_m} & \tensor{\psi}{^{\bar{q}}_1} & \tensor{\psi}{^{\bar{q}}_2} & \cdots & \tensor{\psi}{^{\bar{q}}_n}
    \end{array}
  \right)
  \, .
\end{equation}
The super-Pl\"{u}cker coordinates $Y_{\alpha\beta}$ will be defined via $p|q$-minors, i.e., $(p|q \times p|q)$ submatrices, of \eqref{eq:plucker:Gr:grassmannian homogeneous coordinates}.

\subtitle{The Submatrices of $Z$}

There are once again three types of submatrices: an even supermatrix, a Type-I ``wrong'' supermatrix, and a Type-II ``wrong'' supermatrix.
In particular, the even supermatrix is
\begin{equation}
  \label{eq:plucker:Gr p-q m-n:even supermatrix}
  Z_{s_1 \dots s_p \; t_1 \dots t_q}
  = \left(
    \begin{array}{cccc|cccc}
      \tensor{\phi}{^1_{s_1}} & \tensor{\phi}{^1_{s_2}} & \cdots & \tensor{\phi}{^1_{s_p}} & \tensor{\psi}{^1_{t_1}} & \tensor{\psi}{^1_{t_2}} & \cdots & \tensor{\psi}{^1_{t_q}}
      \\
      \tensor{\phi}{^2_{s_1}} & \tensor{\phi}{^2_{s_2}} & \cdots & \tensor{\phi}{^2_{s_p}} & \tensor{\psi}{^2_{t_1}} & \tensor{\psi}{^2_{t_2}} & \cdots & \tensor{\psi}{^2_{t_q}}
      \\
      \vdots & \vdots & \ddots & \vdots & \vdots & \vdots & \ddots & \vdots
      \\
      \tensor{\phi}{^p_{s_1}} & \tensor{\phi}{^p_{s_2}} & \cdots & \tensor{\phi}{^p_{s_p}} & \tensor{\psi}{^p_{t_1}} & \tensor{\psi}{^p_{t_2}} & \cdots & \tensor{\psi}{^p_{t_q}}
      \\ \hline
      \tensor{\phi}{^{\bar{1}}_{s_1}} & \tensor{\phi}{^{\bar{1}}_{s_2}} & \cdots & \tensor{\phi}{^{\bar{1}}_{s_p}} & \tensor{\psi}{^{\bar{1}}_{t_1}} & \tensor{\psi}{^{\bar{1}}_{t_2}} & \cdots & \tensor{\psi}{^{\bar{1}}_{t_q}}
      \\
      \tensor{\phi}{^{\bar{2}}_{s_1}} & \tensor{\phi}{^{\bar{2}}_{s_2}} & \cdots & \tensor{\phi}{^{\bar{2}}_{s_p}} & \tensor{\psi}{^{\bar{2}}_{t_1}} & \tensor{\psi}{^{\bar{2}}_{t_2}} & \cdots & \tensor{\psi}{^{\bar{2}}_{t_q}}
      \\
      \vdots & \vdots & \ddots & \vdots & \vdots & \vdots & \ddots & \vdots
      \\
      \tensor{\phi}{^{\bar{q}}_{s_1}} & \tensor{\phi}{^{\bar{q}}_{s_2}} & \cdots & \tensor{\phi}{^{\bar{q}}_{s_p}} & \tensor{\psi}{^{\bar{q}}_{t_1}} & \tensor{\psi}{^{\bar{q}}_{t_2}} & \cdots & \tensor{\psi}{^{\bar{q}}_{t_q}}
    \end{array}
  \right)
  \, ,
\end{equation}
where $s_{*} \in \{1, \dots, m\}$ and $t_{*} \in \{1, \dots, n\}$ run through the even and odd indices on the ambient $\mathbb{C}^{m|n}$, respectively.
The Type-I ``wrong'' supermatrix is
\begin{equation}
  \label{eq:plucker:Gr p-q m-n:type-I supermatrix}
  Z_{s_1 \dots s_p \; t_1 \dots s_i \dots t_q}
  = \left(
    \begin{array}{cccc|cccccccc}
      \tensor{\phi}{^1_{s_1}} & \tensor{\phi}{^1_{s_2}} & \cdots & \tensor{\phi}{^1_{s_p}} & \tensor{\psi}{^1_{t_1}} & \tensor{\psi}{^1_{t_2}} & \cdots & \tensor{\psi}{^1_{t_{i-1}}} & \tensor{\phi}{^1_{s_i}}  & \tensor{\psi}{^1_{t_{i+1}}} & \cdots & \tensor{\psi}{^1_{t_q}}
      \\
      \tensor{\phi}{^2_{s_1}} & \tensor{\phi}{^2_{s_2}} & \cdots & \tensor{\phi}{^2_{s_p}} & \tensor{\psi}{^2_{t_1}} & \tensor{\psi}{^2_{t_2}} & \cdots & \tensor{\psi}{^2_{t_{i-1}}} & \tensor{\phi}{^2_{s_i}}  & \tensor{\psi}{^2_{t_{i+1}}} & \cdots & \tensor{\psi}{^2_{t_q}}
      \\
      \vdots & \vdots & \ddots & \vdots & \vdots & \vdots & \ddots & \vdots & \vdots & \vdots & \ddots & \vdots
      \\
      \tensor{\phi}{^p_{s_1}} & \tensor{\phi}{^p_{s_2}} & \cdots & \tensor{\phi}{^p_{s_p}} & \tensor{\psi}{^p_{t_1}} & \tensor{\psi}{^p_{t_2}} & \cdots & \tensor{\psi}{^p_{t_{i-1}}} & \tensor{\phi}{^p_{s_i}}  & \tensor{\psi}{^p_{t_{i+1}}} & \cdots & \tensor{\psi}{^p_{t_q}}
      \\ \hline
      \tensor{\phi}{^{\bar{1}}_{s_1}} & \tensor{\phi}{^{\bar{1}}_{s_2}} & \cdots & \tensor{\phi}{^{\bar{1}}_{s_p}} & \tensor{\psi}{^{\bar{1}}_{t_1}} & \tensor{\psi}{^{\bar{1}}_{t_2}} & \cdots & \tensor{\psi}{^{\bar{1}}_{t_{i-1}}} & \tensor{\phi}{^{\bar{1}}_{s_i}}  & \tensor{\psi}{^{\bar{1}}_{t_{i+1}}} & \cdots & \tensor{\psi}{^{\bar{1}}_{t_q}}
      \\
      \tensor{\phi}{^{\bar{2}}_{s_1}} & \tensor{\phi}{^{\bar{2}}_{s_2}} & \cdots & \tensor{\phi}{^{\bar{2}}_{s_p}} & \tensor{\psi}{^{\bar{2}}_{t_1}} & \tensor{\psi}{^{\bar{2}}_{t_2}} & \cdots & \tensor{\psi}{^{\bar{2}}_{t_{i-1}}} & \tensor{\phi}{^{\bar{2}}_{s_i}}  & \tensor{\psi}{^{\bar{2}}_{t_{i+1}}} & \cdots & \tensor{\psi}{^{\bar{2}}_{t_q}}
      \\
      \vdots & \vdots & \ddots & \vdots & \vdots & \vdots & \ddots & \vdots & \vdots & \vdots & \ddots & \vdots
      \\
      \tensor{\phi}{^{\bar{q}}_{s_1}} & \tensor{\phi}{^{\bar{q}}_{s_2}} & \cdots & \tensor{\phi}{^{\bar{q}}_{s_p}} & \tensor{\psi}{^{\bar{q}}_{t_1}} & \tensor{\psi}{^{\bar{q}}_{t_2}} & \cdots & \tensor{\psi}{^{\bar{q}}_{t_{i-1}}} & \tensor{\phi}{^{\bar{q}}_{s_i}}  & \tensor{\psi}{^{\bar{q}}_{t_{i+1}}} & \cdots & \tensor{\psi}{^{\bar{q}}_{t_q}}
    \end{array}
  \right)
  \, ,
\end{equation}
and the Type-II ``wrong'' supermatrix is
\begin{equation}
  \label{eq:plucker:Gr p-q m-n:type-II supermatrix}
  Z_{s_1 \dots t_i \dots s_p \; t_1 \dots t_q}
  = \left(
    \begin{array}{cccccccc|cccc}
      \tensor{\phi}{^1_{s_1}} & \tensor{\phi}{^1_{s_2}} & \cdots & \tensor{\phi}{^1_{s_{i-1}}} & \tensor{\psi}{^1_{t_i}}  & \tensor{\phi}{^1_{s_{i+1}}} & \cdots & \tensor{\phi}{^1_{s_p}} & \tensor{\psi}{^1_{t_1}} & \tensor{\psi}{^1_{t_2}} & \cdots & \tensor{\psi}{^1_{t_q}}
      \\
      \tensor{\phi}{^2_{s_1}} & \tensor{\phi}{^2_{s_2}} & \cdots & \tensor{\phi}{^2_{s_{i-1}}} & \tensor{\psi}{^2_{t_i}}  & \tensor{\phi}{^2_{s_{i+1}}} & \cdots & \tensor{\phi}{^2_{s_p}} & \tensor{\psi}{^2_{t_1}} & \tensor{\psi}{^2_{t_2}} & \cdots & \tensor{\psi}{^2_{t_q}}
      \\
      \vdots & \vdots & \ddots & \vdots & \vdots & \vdots & \ddots & \vdots & \vdots & \vdots & \ddots & \vdots
      \\
      \tensor{\phi}{^p_{s_1}} & \tensor{\phi}{^p_{s_2}} & \cdots & \tensor{\phi}{^p_{s_{i-1}}} & \tensor{\psi}{^p_{t_i}}  & \tensor{\phi}{^p_{s_{i+1}}} & \cdots & \tensor{\phi}{^p_{s_p}} & \tensor{\psi}{^p_{t_1}} & \tensor{\psi}{^p_{t_2}} & \cdots & \tensor{\psi}{^p_{t_q}}
      \\ \hline
      \tensor{\phi}{^{\bar{1}}_{s_1}} & \tensor{\phi}{^{\bar{1}}_{s_2}} & \cdots & \tensor{\phi}{^{\bar{1}}_{s_{i-1}}} & \tensor{\psi}{^{\bar{1}}_{t_i}}  & \tensor{\phi}{^{\bar{1}}_{s_{i+1}}} & \cdots & \tensor{\phi}{^{\bar{1}}_{s_p}} & \tensor{\psi}{^{\bar{1}}_{t_1}} & \tensor{\psi}{^{\bar{1}}_{t_2}} & \cdots & \tensor{\psi}{^{\bar{1}}_{t_q}}
      \\
      \tensor{\phi}{^{\bar{2}}_{s_1}} & \tensor{\phi}{^{\bar{2}}_{s_2}} & \cdots & \tensor{\phi}{^{\bar{2}}_{s_{i-1}}} & \tensor{\psi}{^{\bar{2}}_{t_i}}  & \tensor{\phi}{^{\bar{2}}_{s_{i+1}}} & \cdots & \tensor{\phi}{^{\bar{2}}_{s_p}} & \tensor{\psi}{^{\bar{2}}_{t_1}} & \tensor{\psi}{^{\bar{2}}_{t_2}} & \cdots & \tensor{\psi}{^{\bar{2}}_{t_q}}
      \\
      \vdots & \vdots & \ddots & \vdots & \vdots & \vdots & \ddots & \vdots & \vdots & \vdots & \ddots & \vdots
      \\
      \tensor{\phi}{^{\bar{q}}_{s_1}} & \tensor{\phi}{^{\bar{q}}_{s_2}} & \cdots & \tensor{\phi}{^{\bar{q}}_{s_{i-1}}} & \tensor{\psi}{^{\bar{q}}_{t_i}}  & \tensor{\phi}{^{\bar{q}}_{s_{i+1}}} & \cdots & \tensor{\phi}{^{\bar{q}}_{s_p}} & \tensor{\psi}{^{\bar{q}}_{t_1}} & \tensor{\psi}{^{\bar{q}}_{t_2}} & \cdots & \tensor{\psi}{^{\bar{q}}_{t_q}}
    \end{array}
  \right)
  \, .
\end{equation}

\subtitle{The Super-Pl\"{u}cker Coordinates of $Gr_{p|q}(\mathbb{C}^{m|n})$}

From \eqref{eq:plucker:Gr p-q m-n:even supermatrix}--\eqref{eq:plucker:Gr p-q m-n:type-II supermatrix}, the super-Pl\"{u}cker coordinates of $Gr_{p|q}(\mathbb{C}^{m|n})$ are defined as
\begin{equation}
  \label{eq:plucker:Gr p-q m-n:plucker coordinates}
  \begin{aligned}
    Y^+_{s_1 \dots s_p \; t_1 \dots t_q}
    &\coloneq \sdet (Z_{s_1 \dots s_p \; t_1 \dots t_q})
    \, ,
    &\qquad
    Y^-_{s_1 \dots s_p \; t_1 \dots t_q}
    &\coloneq \sdet^{*} (Z_{s_1 \dots s_p \; t_1 \dots t_q})
    \, ,
    \\
    Y^-_{s_1 \dots s_p \; t_1 \dots s_i \dots t_q}
    &\coloneq \sdet^{*} (Z_{s_1 \dots s_p \; t_1 \dots s_i \dots t_q})
    \, ,
    &\qquad
    Y^+_{s_1 \dots t_i \dots s_p \; t_1 \dots t_q}
    &\coloneq \sdet (Z_{s_1 \dots t_i \dots s_p \; t_1 \dots t_q})
    \, ,
  \end{aligned}
\end{equation}

\subtitle{Supergauge Transformations, and Weighted Coordinates on a Weighted Projective Superspace}

Note that the $Z$ supermatrix \eqref{eq:plucker:Gr:grassmannian homogeneous coordinates} has a $\mathrm{U}(p|q)$ supergauge group action, similar to the $\mathrm{U}(1|1)$ supergauge group action in \autoref{sec:plucker:Gr 1-1 2-3}.
Likewise, so would the full-rank $(p|q \times p|q)$ submatrices \eqref{eq:plucker:Gr p-q m-n:even supermatrix}--\eqref{eq:plucker:Gr p-q m-n:type-II supermatrix}.
Thus, the super-Pl\"{u}cker coordinates defined in \eqref{eq:plucker:Gr p-q m-n:plucker coordinates} will transform, like in \eqref{eq:plucker:Gr 1-1 2-3:tfm of Y}, as
\begin{equation}
  \label{eq:plucker:Gr p-q m-n:tfm of Y}
  \begin{aligned}
    Y^+_{s_1 \dots s_p \; t_1 \dots t_q}
    &\rightarrow \sdet M \cdot Y^+_{s_1 \dots s_p \; t_1 \dots t_q}
    \, ,
    &\qquad
    Y^-_{s_1 \dots s_p \; t_1 \dots t_q}
    &\rightarrow (\sdet M)^{-1} \cdot Y^-_{s_1 \dots s_p \; t_1 \dots t_q}
    \, ,
    \\
    Y^-_{s_1 \dots s_p \; t_1 \dots s_i \dots t_q}
    &\rightarrow (\sdet M)^{-1} \cdot Y^-_{s_1 \dots s_p \; t_1 \dots s_i \dots t_q}
    \, ,
    &\qquad
    Y^+_{s_1 \dots t_i \dots s_p \; t_1 \dots t_q}
    &\rightarrow \sdet M \cdot Y^+_{s_1 \dots t_i \dots s_p \; t_1 \dots t_q}
    \, ,
  \end{aligned}
\end{equation}
where $M \in \mathrm{U}(p|q)$.

As the physical states, i.e., the $\mathrm{U}(p|q)$ fields $\phi_s$ and $\psi_t$, are physically equivalent up to a $\mathrm{U}(p|q)$ supergauge transformation, we thus have an identification of the super-Pl\"{u}cker coordinates under the following equivalence relation:
\begin{equation}
  \label{eq:plucker:Gr p-q m-n:wcp coordinates}
  \begin{gathered}
    \left(
      Y^+_{s_1 \dots s_p \; t_1 \dots t_q}, Y^-_{s_1 \dots s_p \; t_1 \dots t_q}
      \middle\vert
      Y^+_{s_1 \dots t_i \dots s_p \; t_1 \dots t_q}, Y^-_{s_1 \dots s_p \; t_1 \dots s_i \dots t_q}
    \right)
    \\
    \sim
    \left(
      \lambda^{+1} Y^+_{s_1 \dots s_p \; t_1 \dots t_q}, \lambda^{-1} Y^-_{s_1 \dots s_p \; t_1 \dots t_q}
      \middle\vert
      \lambda^{+1} Y^+_{s_1 \dots t_i \dots s_p \; t_1 \dots t_q}, \lambda^{-1} Y^-_{s_1 \dots s_p \; t_1 \dots s_i \dots t_q}
    \right)
    \, ,
  \end{gathered}
\end{equation}
where $\lambda = \sdet M = (\sdet^{*} M)$.
In other words, the $Y^+$'s and $Y^-$'s are, just like in \autoref{sec:plucker:Gr 1-1 2-3}, coordinates on a weighted projective superspace, where coordinates with superscript ``$+$'' are of weight $+1$, while the those with superscript ``$-$'' are of weight $-1$.

In total, there are $(P^m_p \times P^n_q)|(P^m_{p-1} \times P^n_{q+1})$ coordinates with weight $+1$ and $(P^m_p \times P^n_q)|(P^m_{p+1} \times P^n_{q-1})$ coordinates with weight $-1$, on a weighted projective space that $Gr_{p|q}(\mathbb{C}^{m|n})$ is embedded in.

\subtitle{The Super-Pl\"{u}cker Coordinates of the Parity-Reversed $Gr_{q|p}(\mathbb{C}^{n|m})$}

Let us consider the parity-reversal of $Gr_{p|q}(\mathbb{C}^{m|n})$, i.e., $Gr_{q|p}(\mathbb{C}^{n|m})$.
The homogeneous coordinates of such a space can be expressed as a $(q|p \times n|m)$ supermatrix $\widetilde{Z}$, which is equivalent to the parity reversal of the $Z$ supermatrix \eqref{eq:plucker:Gr:grassmannian homogeneous coordinates}.
The super-Pl\"{u}cker coordinates $\widetilde{Y}$ corresponding to $\widetilde{Z}$ are defined via its $q|p$-minors, which are equivalent to the parity reversals of the $p|q$-minors of $Z$ in \eqref{eq:plucker:Gr p-q m-n:even supermatrix}--\eqref{eq:plucker:Gr p-q m-n:type-II supermatrix}.
Repeating the computation, we find that $\widetilde{Y}$ is related to the super-Pl\"{u}cker coordinates $Y$ in \eqref{eq:plucker:Gr p-q m-n:plucker coordinates} as
\begin{equation}
  \label{eq:plucker:Gr p-q m-n:plucker coordinates:reversed}
  \begin{aligned}
    \widetilde{Y}^+_{t_1 \dots t_q \; s_1 \dots s_p}
    &= Y^-_{s_1 \dots s_p \; t_1 \dots t_q}
    \, ,
    &\qquad
    \widetilde{Y}^-_{t_1 \dots t_q \; s_1 \dots s_p}
    &= Y^+_{s_1 \dots s_p \; t_1 \dots t_q}
    \, ,
    \\
    \widetilde{Y}^-_{t_1 \dots t_q \; s_1 \dots t_i \dots s_p}
    &= Y^+_{s_1 \dots t_i \dots s_p \; t_1 \dots t_q}
    \, ,
    &\qquad
    \widetilde{Y}^+_{t_1 \dots s_i \dots t_q \; s_1 \dots s_p}
    &= Y^-_{s_1 \dots s_p \; t_1 \dots s_i \dots t_q}
    \, .
  \end{aligned}
\end{equation}

Just like in \eqref{eq:plucker:Gr 1-1 2-3:plucker coordinates:reversed}, notice that there is a one-to-one correspondence between the $Y$'s and $\widetilde{Y}$'s, and that the correspondence changes the sign in the superscript.
Hence, the $\widetilde{Y}$'s are negatively-weighted $Y$'s.

In other words, the super-Pl\"{u}cker coordinates of the parity-reversed $Gr_{q|p}(\mathbb{C}^{n|m})$ are the same as the super-Pl\"{u}cker coordinates of $Gr_{p|q}(\mathbb{C}^{m|n})$, but with the signs of the weights flipped.

\subsection{Super-Pl\"{u}cker Coordinates of \texorpdfstring{$Gr_{p|0}(\mathbb{C}^{m|n})$}{Gr(p|0; m|n)}}
\label{sec:plucker:Gr p-0 m-n}

Let us now describe the super-Pl\"{u}cker coordinates of a super-Grassmannian with purely even subspaces in $\mathbb{C}^{m|n}$, i.e., a super-Grassmannian of the type $Gr_{p|0}(\mathbb{C}^{m|n})$ where $p \neq 0$.\footnote{%
  The description in this subsection also applies similarly to super-Grassmannians of the type $Gr_{0|q}(\mathbb{C}^{m|n})$.
}
Such a space is spanned by $p$ Grassmann-even supervectors in $\mathbb{C}^{m|n}$, and its homogeneous coordinates can be expressed as a $(p|0 \times m|n)$ ``supermatrix'', i.e.,
\begin{equation}
  \label{eq:plucker:gr p-0 m-n:homogeneous coordinates}
  Z = \left(
    \begin{array}{cccc|cccc}
      \tensor{\phi}{^1_1} & \tensor{\phi}{^1_2} & \cdots & \tensor{\phi}{^1_m} & \tensor{\psi}{^1_1} & \tensor{\psi}{^1_2} & \cdots & \tensor{\psi}{^1_n}
      \\
      \tensor{\phi}{^2_1} & \tensor{\phi}{^2_2} & \cdots & \tensor{\phi}{^2_m} & \tensor{\psi}{^2_1} & \tensor{\psi}{^2_2} & \cdots & \tensor{\psi}{^2_n}
      \\
      \vdots & \vdots & \ddots & \vdots & \vdots & \vdots & \ddots & \vdots
      \\
      \tensor{\phi}{^p_1} & \tensor{\phi}{^p_2} & \cdots & \tensor{\phi}{^p_m} & \tensor{\psi}{^p_1} & \tensor{\psi}{^p_2} & \cdots & \tensor{\psi}{^p_n}
    \end{array}
  \right)
  \, .
\end{equation}
Notice that superdeterminants are ill-defined on such a ``supermatrix''; as such, the super-Pl\"{u}cker coordinates $Y_{\alpha\beta}$ will be formulated differently, i.e., without superdeterminants.
However, they can still be defined using the same principle of some invariant of full-rank, i.e., $(p|0 \times p|0)$, submatrices of \eqref{eq:plucker:gr p-0 m-n:homogeneous coordinates} (see \autoref{ft:plucker:full rank submatrices}).

\subtitle{The Submatrices of $Z$}

There are four types of submatrices:
a submatrix of fully Grassmann-even components,
a submatrix of fully Grassmann-odd components,
a Type-I ``wrong'' submatrix with mostly Grassmann-odd components,
and a Type-II ``wrong'' submatrix with mostly Grassmann-even components.

In particular, the fully Grassmann-even submatrix is
\begin{equation}
  \label{eq:plucker:Gr p-0 m-n:even supermatrix}
  Z_{s_1 \dots s_p}
  = \mqty(
  \tensor{\phi}{^1_{s_1}} & \tensor{\phi}{^1_{s_2}} & \cdots & \tensor{\phi}{^1_{s_p}}
  \\
  \tensor{\phi}{^2_{s_1}} & \tensor{\phi}{^2_{s_2}} & \cdots & \tensor{\phi}{^2_{s_p}}
  \\
  \vdots & \vdots & \ddots & \vdots
  \\
  \tensor{\phi}{^p_{s_1}} & \tensor{\phi}{^p_{s_2}} & \cdots & \tensor{\phi}{^p_{s_p}}
  )
  \, ,
\end{equation}
the fully Grassmann-odd submatrix is
\begin{equation}
  \label{eq:plucker:Gr p-0 m-n:odd supermatrix}
  Z_{t_1 \dots t_p}
  = \mqty(
  \tensor{\psi}{^1_{t_1}} & \tensor{\psi}{^1_{t_2}} & \cdots & \tensor{\psi}{^1_{t_p}}
  \\
  \tensor{\psi}{^2_{t_1}} & \tensor{\psi}{^2_{t_2}} & \cdots & \tensor{\psi}{^2_{t_p}}
  \\
  \vdots & \vdots & \ddots & \vdots
  \\
  \tensor{\psi}{^p_{t_1}} & \tensor{\psi}{^p_{t_2}} & \cdots & \tensor{\psi}{^p_{t_p}}
  )
  \, ,
\end{equation}
the Type-I ``wrong'' submatrix is
\begin{equation}
  \label{eq:plucker:Gr p-0 m-n:type-I supermatrix}
  Z_{t_1 \dots s_i \dots t_p}
  = \mqty(
  \tensor{\psi}{^1_{t_1}} & \tensor{\psi}{^1_{t_2}} & \cdots & \tensor{\psi}{^1_{t_{i-1}}} & \tensor{\phi}{^1_{s_i}}  & \tensor{\psi}{^1_{t_{i+1}}} & \cdots & \tensor{\psi}{^1_{t_p}}
  \\
  \tensor{\psi}{^2_{t_1}} & \tensor{\psi}{^2_{t_2}} & \cdots & \tensor{\psi}{^2_{t_{i-1}}} & \tensor{\phi}{^2_{s_i}}  & \tensor{\psi}{^2_{t_{i+1}}} & \cdots & \tensor{\psi}{^2_{t_q}}
  \\
  \vdots & \vdots & \ddots & \vdots & \vdots & \vdots & \ddots & \vdots
  \\
  \tensor{\psi}{^p_{t_1}} & \tensor{\psi}{^p_{t_2}} & \cdots & \tensor{\psi}{^p_{t_{i-1}}} & \tensor{\phi}{^p_{s_i}}  & \tensor{\psi}{^p_{t_{i+1}}} & \cdots & \tensor{\psi}{^p_{t_q}}
  )
  \, ,
\end{equation}
and the Type-II ``wrong'' submatrix is
\begin{equation}
  \label{eq:plucker:Gr p-0 m-n:type-II supermatrix}
  Z_{s_1 \dots t_i \dots s_p}
  = \mqty(
  \tensor{\phi}{^1_{s_1}} & \tensor{\phi}{^1_{s_2}} & \cdots & \tensor{\phi}{^1_{s_{i-1}}} & \tensor{\psi}{^1_{t_i}}  & \tensor{\phi}{^1_{s_{i+1}}} & \cdots & \tensor{\phi}{^1_{s_p}}
  \\
  \tensor{\phi}{^2_{s_1}} & \tensor{\phi}{^2_{s_2}} & \cdots & \tensor{\phi}{^2_{s_{i-1}}} & \tensor{\psi}{^2_{t_i}}  & \tensor{\phi}{^2_{s_{i+1}}} & \cdots & \tensor{\phi}{^2_{s_p}}
  \\
  \vdots & \vdots & \ddots & \vdots & \vdots & \vdots & \ddots & \vdots
  \\
  \tensor{\phi}{^p_{s_1}} & \tensor{\phi}{^p_{s_2}} & \cdots & \tensor{\phi}{^p_{s_{i-1}}} & \tensor{\psi}{^p_{t_i}}  & \tensor{\phi}{^p_{s_{i+1}}} & \cdots & \tensor{\phi}{^p_{s_p}}
  )
  \, .
\end{equation}
Here, $s_{*} \in \{1, \dots, m\}$ and $t_{*} \in \{1, \dots, n\}$ run through the even and odd indices on the ambient $\mathbb{C}^{m|n}$, respectively.\footnote{%
  \label{ft:plucker:condition for regular plucker}%
  The submatrices \eqref{eq:plucker:Gr p-0 m-n:odd supermatrix} and \eqref{eq:plucker:Gr p-0 m-n:type-I supermatrix} can exist only if $n \geq p$ and $n - 1 \geq p$, respectively.
  Otherwise, only the remaining two submatrices \eqref{eq:plucker:Gr p-0 m-n:even supermatrix} and \eqref{eq:plucker:Gr p-0 m-n:type-II supermatrix} are used to define the super-Pl\"{u}cker coordinates.
  And, if $n = 0$, we have only submatrix \eqref{eq:plucker:Gr p-0 m-n:even supermatrix}, which will be used to define the (regular) Pl\"{u}cker coordinates of the (regular) Grassmannian $G_{p|0}(\mathbb{C}^{m|0}) \cong G_p(\mathbb{C}^m)$ \cite[ch.~1,~$\S$5]{Griffiths:1994prl}.
}

\subtitle{The Super-Pl\"{u}cker Coordinates of $Gr_{p|0}(\mathbb{C}^{m|n})$}

Just as how the invariant of the full-rank submatrices \eqref{eq:plucker:Gr p-q m-n:even supermatrix}--\eqref{eq:plucker:Gr p-q m-n:type-II supermatrix} of a generic super-Grassmannian of the type $Gr_{p|q}(\mathbb{C}^{m|n})$ for $p$ and $q \neq 0$ --- which are \emph{super}matrices --- was the \emph{super}determinant, the invariant of the full-rank submatrices \eqref{eq:plucker:Gr p-0 m-n:even supermatrix}--\eqref{eq:plucker:Gr p-0 m-n:type-II supermatrix} of a super-Grassmannian of the type $Gr_{p|0}(\mathbb{C}^{m|n})$ for $p \neq 0$ --- which are regular matrices --- is the determinant.
Moreover, as the regular determinant of these submatrices \eqref{eq:plucker:Gr p-0 m-n:even supermatrix}--\eqref{eq:plucker:Gr p-0 m-n:type-II supermatrix} are well-defined even if their components are fully Grassmann-odd, or if they are of the ``wrong'' types, we do not need the inverse determinant to define the super-Pl\"{u}cker coordinates associated to $Z$.

In other words, the super-Pl\"{u}cker coordinates of a $Gr_{p|0}(\mathbb{C}^{m|n})$ with homogeneous coordinates $Z$ \eqref{eq:plucker:gr p-0 m-n:homogeneous coordinates} are defined as
\begin{equation}
  \label{eq:plucker:Gr p-0 m-n:plucker coordinates}
  \begin{aligned}
    Y^+_{\alpha_1 \dots \alpha_p}
    &\coloneq \det (Z_{\alpha_1 \dots \alpha_p})
    \, ,
    &\qquad
    Y^+_{\hat{\beta}_1 \dots \hat{\beta}_p}
    &\coloneq \det (Z_{\hat{\beta}_1 \dots \hat{\beta}_p})
    \, ,
    \\
    Y^+_{\hat{\beta}_1 \dots \beta_i \dots \hat{\beta}_p}
    &\coloneq \det (Z_{\hat{\beta}_1 \dots \beta_i \dots \hat{\beta}_p})
    \, ,
    &\qquad
    Y^+_{\alpha_1 \dots \hat{\alpha}_i \dots \alpha_p}
    &\coloneq \det (Z_{\alpha_1 \dots \hat{\alpha}_i \dots \alpha_p})
    \, .
  \end{aligned}
\end{equation}
Here, the ``$+$'' in the superscript denote the $+1$ weight of the coordinates on the weighted projective superspace that $Gr_{p|0}(\mathbb{C}^{m|n})$ is embedded in.
As the coordinates all have the same weight, $Gr_{p|0}(\mathbb{C}^{m|n})$ is actually embedded in an unweighted projective superspace \cite[$\S$3]{shemyakova-2022-super-pluec}.

\subtitle{The Super-Pl\"{u}cker Coordinates of $\mathbb{CP}^{m-1|n}$}

Let us specialize to $p = 1$, i.e., specialize to a projective superspace $Gr_{1|0}(\mathbb{C}^{m|n}) \cong \mathbb{CP}^{m-1|n}$.
The matrix $Z$ \eqref{eq:plucker:gr p-0 m-n:homogeneous coordinates} is now a $(1|0 \times m|n)$ matrix, and the super-Pl\"{u}cker coordinates are defined via determinants of full-rank, i.e., $(1|0 \times 1|0)$, submatrices of $Z$.
These are simply the components of the $Z$ matrix!

In other words, the super-Pl\"{u}cker coordinates of $\mathbb{CP}^{m-1|n}$ are simply the coordinates on $\mathbb{CP}^{m-1|n}$, as they should be.

\printbibliography

\end{document}